\documentclass{aa}
\usepackage{lscape}
\usepackage{hyperref}

\usepackage{graphicx}
\usepackage{subcaption}
\usepackage{txfonts}
\usepackage{natbib}
\usepackage{longtable}

\newcommand{\oii}{[\ion{O}{ii}]}
\newcommand{\oiiab}{[\ion{O}{ii}]$\lambda\lambda$3727,3729}
\newcommand{\oiia}{[\ion{O}{ii}]$\lambda$3727}
\newcommand{\oiib}{[\ion{O}{ii}]$\lambda$3729}
\newcommand{\oiii}{[\ion{O}{iii}]}

\newcommand{\oiiia}{[\ion{O}{iii}]$\lambda$5007}

\newcommand{\hb}{H$\beta$}

\begin{document} 

   \title{The Tully-Fisher relation in dense groups at $z \sim 0.7$ in the MAGIC survey
   \thanks{Tables \ref{tab:morphokin} and \ref{tab:dynamics} are also available at the CDS via anonymous ftp to \url{cdsarc.u-strasbg.fr} (\url{130.79.128.5}) or via \url{http://cdsarc.u-strasbg.fr/viz-bin/cat/J/A+A/647/A152}}$^{\text{,}}$\thanks{Based on observations made with ESO telescopes at the Paranal Observatory under programs 094.A-0247, 095.A-0118, 096.A-0596, 097.A-0254, 099.A-0246, 100.A-0607, 101.A-0282.}
   }
   \titlerunning{TFR in dense groups $z \sim 0.7$ in the MAGIC survey}

\author{
    Valentina Abril-Melgarejo\inst{1}
    \and
    Benoît Epinat\inst{1, 2}
    \and
    Wilfried Mercier \inst{2}
    \and
    Thierry Contini\inst{2}
    \and
    Leindert A. Boogaard\inst{3}
    \and
    Jarle Brinchmann\inst{3, 4}
    \and
    Hayley Finley\inst{2}
    \and
    Léo Michel-Dansac\inst{5}
    \and
    Emmy Ventou \inst{2}
    \and
    Philippe Amram\inst{1}
    \and
    Davor Krajnovi\'{c}\inst{6}
    \and
    Guillaume Mahler\inst{7}
    \and
    Juan C. B. Pineda\inst{8}
    \and
    Johan Richard\inst{5}
  }

\institute{
    Aix Marseille Univ, CNRS, CNES, LAM, Laboratoire d'Astrophysique de Marseille, Marseille, France\\
    \email{valentina.abril@lam.fr}
    \and
    Institut de Recherche en Astrophysique et Planétologie (IRAP), Université de Toulouse, CNRS, UPS, CNES, Toulouse, France
    \and
    Leiden Observatory, Leiden University, PO Box 9513, NL-2300 RA Leiden, The Netherlands
    \and
    Instituto de Astrof{\'\i}sica e Ci{\^e}ncias do Espaço, Universidade do Porto, CAUP, Rua das Estrelas, PT4150-762 Porto, Portugal
    \and
    Univ Lyon, Univ Lyon1, Ens de Lyon, CNRS, Centre de Recherche Astrophysique de Lyon UMR5574, F-69230, Saint-Genis-Laval, France
    \and
    Leibniz-Institut für Astrophysik Potsdam (AIP), An der Sternwarte 16, D-14482 Potsdam, Germany
    \and
    Department of Astronomy, University of Michigan, 1085 South University Ave, Ann Arbor, MI 48109, USA
    \and
    Escuela de F{\'\i}sica, Universidad Industrial de Santander, Bucaramanga 680002, Colombia
}

   \date{Received 2 July 2 2020 / Accepted 19 January 2021}

  \abstract
   {Galaxies in dense environments are subject to interactions and mechanisms that directly affect their evolution by lowering their gas fractions and consequently reducing their star-forming capacity earlier than their isolated counterparts.}
   {The aim of our project is to get new insights into the role of environment in the stellar and baryonic content of galaxies using a kinematic approach, through the study of the Tully-Fisher relation (TFR).}
   {We study a sample of galaxies in eight groups, over-dense by a factor larger than 25 with respect to the average projected density, spanning a redshift range of $0.5<z<0.8$ and located in ten pointings of the MAGIC MUSE Guaranteed Time Observations program. We perform a morpho-kinematics analysis of this sample and set up a selection based on galaxy size, \oiiab\ emission line doublet signal-to-noise ratio, bulge-to-disk ratio, and nuclear activity to construct a robust kinematic sample of 67 star-forming galaxies.}
   {We show that this selection considerably reduces the number of outliers in the TFR, which are predominantly dispersion-dominated galaxies.
   Similar to other studies, we find that including the velocity dispersion in the velocity budget mainly affects galaxies with low rotation velocities, reduces the scatter in the relation, increases its slope, and decreases its zero-point. Including gas masses is more significant for low-mass galaxies due to a larger gas fraction, and thus decreases the slope and increases the zero-point of the relation.
   Our results suggest a significant offset of the TFR zero-point between galaxies in low- and high-density environments, regardless of the kinematics estimator used. This can be interpreted as a decrease in either stellar mass by $\sim 0.05 - 0.3$ dex or an increase in rotation velocity by $\sim 0.02 - 0.06$ dex for galaxies in groups, depending on the samples used for comparison. We also studied the stellar and baryon mass fractions within stellar disks and found they both increase with stellar mass, the trend being more pronounced for the stellar component alone. These fractions do not exceed 50\%. We show that this evolution of the TFR is consistent either with a decrease in star formation or with a contraction of the mass distribution due to the environment. These two effects probably act together, with their relative contribution depending on the mass regime.}
   {}


   \maketitle
%

\section{Introduction}
\label{introduction}

$$\Delta(\log{M_*})\sim 0.2\text{~dex}$$

$$\Delta(\log{V})\sim 0.04\text{~dex}$$

Galaxies mainly assemble their mass and evolve inside dark matter halos (DMHs) via continuous accretion of cold gas \citep[e.g.,][]{Dekel+09} and by the merging of galaxies.
However, the way baryons are accreted on galaxies inside these DMHs along cosmic time is still a matter of debate. Additionally, it is not yet clear if DMHs evolve simultaneously with the baryonic content of galaxies or if they are first settled before baryons are accreted. Estimating dark matter mass inside galaxies is necessary to solve this, which is only possible through the study of galaxy dynamics.

The Tully-Fisher relation \citep[TFR,][]{Tully+77} can be used to answer these questions since it links the total mass content of a population of star-forming galaxies to their luminosity, or to their stellar content. If DMHs are already settled we expect a strong evolution of the TFR zero-point, whereas if the baryonic and dark matter contents evolve simultaneously, the evolution is expected to depend mainly on the gas fraction.

Significant efforts to solve this question have been made by studying how the TFR evolves with cosmic time using various samples of star-forming galaxies, but the evolution of this relation with redshift is still a matter of debate.
In the local Universe, many studies of the TFR have been performed on spiral galaxies \citep[e.g.,][]{Bell+01, Masters+06, Pizagno+07, Sorce+14, Gomez-Lopez+19}. However, one difficulty in comparing these studies comes from the heterogeneous datasets (\ion{H}{i} or ionized gas spectroscopy, magnitudes in diverse spectral bands) and methods (integral field and long-slit spectroscopy data, \ion{H}{i} widths, or velocity maps). Nevertheless, it seems clear that the TFR slope and zero-point depend on the mass range \citep{McGaugh+00, McGaugh05}.
\citet{Simons+15} show that there is a transition in the TFR at a stellar mass of around 10$^{9.5}$ M$_{\sun}$ at $0.1<z<0.4$.

The population of star-forming galaxies at intermediate to high redshift, during and after the peak of cosmic star formation \citep[$0.5<z<3$,][]{Madau+14}, has been the topic of several kinematics studies using integral field spectrographs on 10m class telescopes.
The first was obtained at $z\sim 0.6$ from the IMAGES (Intermediate-mass Galaxy Evolution Sequence) sample. \citet{Puech+08} found an evolution in both near infrared luminosity and stellar mass TFR zero-point, which implies that disks double their stellar content between $z\sim 0.6$ and $z=0$. On the other hand, taking into account the gas content and using the same sample, \citet{Puech+10} found no evolution in the baryonic TFR. They also found a high scatter in the TFR that they attributed to galaxies with perturbed or peculiar kinematics, mainly based on visual inspection of their velocity fields. At higher redshift, an evolution in the stellar mass TFR was also found by \citet{Cresci+09} using the SINS (Spectroscopic Imaging survey in the near-infrared with SINFONI) sample at $z\sim 2$, with a similar amplitude as IMAGES, and more marginally by \citet{Gnerucci+11} from the LSD (Lyman-break galaxies Stellar population and Dynamics) and AMAZE (Assessing the Mass-Abundance redshift Evolution) samples at $z\sim 3$. However, \citet{Vergani+12} did not find such an evolution using the MASSIV (Mass Assembly Survey with SINFONI in VVDS) sample at $z\sim 1.2$.
Other studies using long-slit spectroscopy on samples at $z\sim 1$ did not find any evolution of the stellar mass TFR either \citep[e.g.,][]{Miller+12, Pelliccia+17}.
All these samples contained fewer than $\sim 100$ galaxies.

Recently, the multiplexing power of the multi-object integral field unit spectrograph KMOS (K-band Multi Object Spectrograph) has led to new, much larger samples of around a thousand galaxies \citep{Wisnioski+15, Stott+16}. Still, despite samples being larger, the debate has not been closed. Indeed, using the KROSS (KMOS Redshift One Spectroscopic Survey) sample, \citet{Tiley+19} found no evolution of the stellar mass TFR between $z=0$ and $z=1$, whereas \citet{Ubler+17}, using the KMOS3D sample, found some evolution.
Indeed, in their first analysis of the KROSS sample \citet{Tiley+16a}, Tiley et al. did find an evolution as well, but in \citet{Tiley+19}, they refined their analysis by making a careful comparison at $z \sim 0$ with the SAMI (Sydney-AAO Multi-object Integral-field spectrograph) Galaxy Survey \citep[e.g.,][]{Bryant+15}.
This allowed them to minimize the potential methodological biases arising from sample selection, analysis methods, and data quality. Indeed, the differences in instrumental set-ups and the methodologies to extract kinematics and, more generally, to select the samples make comparisons problematic. Most of these intermediate redshift samples of star-forming galaxies have been preselected from large spectroscopic samples based on color or magnitude. Therefore, they are most often limited to the most massive galaxies (M$_*>10^{10}$ M$_{\sun}$), whereas most of the galaxies in the Universe are below this limit. A few TFR studies on smaller samples at $z>0.5$ have, however, considered galaxies down to low masses using either the MUSE (Multi Unit Spectrograph Explorer) integral field instrument \citep{Contini:2016} or the DEIMOS (DEep Imaging Multi-Object Spectrograph) multi-slit spectrograph \citep[][]{Simons+17}.

On the other hand, the environment also plays an important role in the mass assembly of galaxies and in the transformation of star-forming galaxies into passive ones. Indeed, it has been observed in the Local Universe that the quenching of star formation and the buildup of the red sequence happen earlier in dense environments than in the field \citep[e.g.,][]{Peng:2010, Muzzin:2012}. This could be induced by direct interactions among galaxies, by interactions between galaxies and the gravitational potential of the group or cluster, or by interactions with intra-group or cluster media, which either prevent gas accretion onto galaxies or remove their gas content \citep[e.g.,][]{Boselli+06}.
Environment might also have an impact on the baryonic content of underlying DMHs.
However, studying the kinematics of disks is more challenging in dense environments than in the field due to the reduced gas fraction. In the local Universe, the population of disks is larger in the outskirts of such structures, which probably means that they are just entering these structures or that they are less affected by the environment. This is supported by the fact that studies using local cluster galaxies do not show evidence for variation of the TFR with environment \citep[e.g.,][]{Masters+06, Masters+08}. In addition, a comparison of the TFR between galaxies in the field \citep{Torres-Flores+11} and in compact groups \citep{Torres-Flores+13}, where galaxy interactions are supposed to be more important, provides a similar conclusion.

At higher redshift, and most specifically at $z\sim 1$ when the cosmic star formation starts its decrease,
the densest structures have already started their relaxation but still contain a large fraction of star-forming galaxies \citep[e.g.,][]{Muzzin+13}, probably because galaxies had more gas at that time and
because the environmental processes that turn off star formation operate on fairly long timescales \citep[e.g.,][]{Cibinel+13, Wetzel+13}.
This means that disks might survive longer before being devoid of gas and could therefore be more severely impacted by the environment than in the local Universe.

However, despite the new large samples presented above, studying the impact of environment on the TFR is not yet possible at intermediate redshift since very little is known about the density field in which those galaxies reside, due to incomplete spectroscopic coverage. Some studies have, however, targeted a few dozen galaxies in groups and (proto-)clusters at intermediate redshift with KMOS and FORS2 \citep[e.g.,][]{Sobral+13, Perez-Martinez+17, Bohm+20}. So far, the most complete study of the TFR as a function of environment has been performed from long-slit spectroscopy observations by \citet{Pelliccia+19}, using a sample of 94 galaxies at $z\sim 1$, a fraction of which were members of clusters. They do not report significant modification of the TFR with environment.

Focusing on star-forming galaxies along the main sequence is essential to analyze and compare the TFR for similar populations of galaxies in various environments.
In this paper, we present the study of the spatially resolved ionized gas kinematics and of the TFR for a sample of star-forming galaxies in dense environments at intermediate redshifts ($z\sim 0.7$) from the MUSE gAlaxy Groups In Cosmos (MAGIC) dataset (Epinat et al., in prep), using data from MUSE \citep[][]{Bacon:2015}. At these redshifts, the MUSE field of view corresponds to a linear physical size of more than 400 kpc, which is the typical size for groups. This allows us to study the properties of star-forming galaxies in groups without any other preselection besides targeting known over-densities, leading to samples not limited in magnitude nor in color \citep{Contini:2016}.

The structure of the paper is as follows. In Sect. \ref{obs_dataset} we present the MUSE observations of the MAGIC program and the data reduction. Identification of groups and their properties, detailed group galaxies physical properties, including their morphological and kinematics modeling, and kinematic sample selection criteria are described in Sect. \ref{Phy_properties}. Section \ref{analysis} is focused on the detailed analysis of the TFRs and their comparison with reference samples, whereas the interpretation of the results is conducted in Sect. \ref{results} before concluding the analysis in Sect. \ref{Conclusions}. Throughout the paper, we assume a $\Lambda$CDM cosmology with H$_0=70$~km~s$^{-1}$~Mpc$^{-1}$, $\Omega_M=0.3,$ and $\Omega_\Lambda=0.7$. 

\section{MUSE observations in dense environments and data reduction}
\label{obs_dataset}

The sample of galaxies located in dense environment that is analyzed in this paper is a subsample of the MAGIC Survey (MUSE gAlaxy Groups In Cosmos).
We focus on eight groups of the COSMOS field \citep{Scoville:2007} selected in the COSMOS group catalog of \citet{Knobel+12} with redshifts $0.5<z<0.8$ observed during MUSE Guaranteed Time Observations (GTO) as part of an observing program focusing on the effect of the environment on galaxy evolution over the past 8 Gyrs (PI: T. Contini). All the details on group selection, observations and data reduction will be presented in the MAGIC survey paper (Epinat et al. in prep), so here we give a summary of the most important steps.

The observations presented in this paper were spread over seven periods. For each targeted field, observing blocks of four 900 seconds exposures including a small dithering pattern and a rotation of the field of 90\degr\ between each, were combined to obtain final datacubes with depths from one to ten hours. In order to be able to perform kinematics analysis, good seeing conditions providing a point spread function full width at half maximum (PSF FWHM) lower than 0.8\arcsec\ were required unless the adaptive optics system was used in the last observing runs.

The basic data reduction was applied for each OB separately using the MUSE standard pipeline \citep{Weilbacher+20}. Version v1.6 was used for all seeing limited observations except for CGr30 (v1.2), whereas v2.4 was used for AO observations.
A default sky subtraction was applied to each science exposure before aligning and combining them using stars in the field.
The Zurich Atmosphere Purge software \citep[ZAP;][]{Soto+16} was then applied to the final combined cube to further improve sky subtraction. Version 0.6 was used for CGr30, CGr34, and the one hour exposure cube on CGr84, version 1.1 was used for CGr28, the deepest cube on CGr84, and CGr114, whereas version 2.0 was used for the two other groups observed with AO.

In the end, each cube has a spatial sampling of 0.2\arcsec\ and a spectral sampling of 1.25~\AA\ over a 4750~\AA\ to 9350~\AA\ spectral range and an associated variance datacube is also produced.

The knowledge of both point spread function (PSF) and line spread function (LSF) are essential to derive accurate kinematic measurements.
In order to compute the MUSE PSF, we modeled the surface brightness profiles of all the stars in each field using the \textsc{Galfit} software \citep{Peng:2002}. Gaussian luminosity profiles were used to model all the stars since their observed profiles are quite symmetric. We have from three to five stars per field and since all the galaxies in a given group are within a narrow redshift range, we extracted narrow-band images of each star around the wavelength of the \oii\ doublet (used to extract kinematics, see Sect. \ref{kinematics}) redshifted at the median redshift of each group. We then computed the median FWHM of the stars for each group in each field to evaluate the corresponding PSF.
The median PSF FWHM value for all the fields is $\sim 0.66$\arcsec, the smallest PSF value corresponds to 0.58\arcsec\ and the largest to 0.74\arcsec.

We determined the LSF FWHM using the prescriptions from \cite{Bacon:2017} and \cite{Guerou:2017}, in which they analyzed the variation of the MUSE LSF with wavelength in the Hubble Ultra Deep Field and in the Hubble Deep Field South. The MUSE LSF is described as:
\begin{equation}
 {\rm FWHM} = \lambda^2 \times 5.866 \times 10^{-8} - \lambda \times 9.187 \times 10^{-4} + 6.040 \text{ ,}
 \label{lsf}
\end{equation}
where FWHM and $\lambda$ are both in Angstroms. The corresponding dispersion is then estimated assuming the LSF is Gaussian.

Table \ref{table_info_group} summarizes the main observational properties of the eight studied galaxy groups: group ID, coordinates of the center, exposure time per field, median redshift, MUSE PSF FWHM, total number of galaxy members and number of galaxies belonging to the kinematic sample (see Sect. \ref{Sample_Selection_Criteria}).
Medium-deep data ($>$ 4h) are used for most of the groups.

\begin{table*}[t]
\caption{General properties of the galaxy groups in the parent sample.}
\label{table_info_group}
\centering
\begin{tabular}{lcccccccc}
\hline
\hline
ID COSMOS & R.A. & Dec. & Exp. Time   & Redshift & PSF FWHM & Number of galaxies & M$_{\textrm{vir}}$  \\ 
Group   & (J2000) & (J2000)  & (h) &  & \arcsec  & (all/SED/MS/KS) & 10$^{13}$ M$_\sun$\\
(1) & (2) & (3)  & (4) & (5) & (6) & (7) & (8) \\
\hline
CGr28          &  150\degr13\arcmin32\arcsec\   &  1\degr48\arcmin42\arcsec\    &  $1$        & 0.530    & 0.654    & 10/10/9/3    &   7.1        \\ 
CGr30          &  150\degr08\arcmin30\arcsec\   &  2\degr04\arcmin01\arcsec\    &  $9.75$     & 0.725    & 0.700   & 44/39/33/15    &  6.5      \\
CGr32          &  149\degr55\arcmin19\arcsec\   &  2\degr31\arcmin16\arcsec\ &  $3\times(1+\textbf{3.35})$\tablefootmark{~(a)}  & 0.730    & 0.596-0.624-0.722 \tablefootmark{~(a)}   & 106/92/50/13  &   81.4         \\ 
CGr34          &  149\degr51\arcmin32\arcsec\   &  2\degr29\arcmin28\arcsec\    &  $5.25$     & 0.732    & 0.664   & 20/20/17/9   &  10.4     \\
CGr79          &  149\degr49\arcmin15\arcsec\   &  1\degr49\arcmin18\arcsec\    &  $\textbf{4.35}$     & 0.531    & 0.658   & 19/19/15/8   &  11.2     \\
CGr84          &  150\degr03\arcmin24\arcsec\   &  2\degr36\arcmin09\arcsec\    &  $5.25$ + $1$ \tablefootmark{~(b)}  & 0.697    &   0.620-0.578 \tablefootmark{~(b)}   & 31/26/21/8  &  8.2 \\
CGr84b         &  150\degr03\arcmin28\arcsec\   &  2\degr36\arcmin32\arcsec\    &  $5.25$ + $1$ \tablefootmark{~(b)}  & 0.681    &   0.620-0.578 \tablefootmark{~(b)}   & 35/32/25/9  & 8.8  \\
CGr114         &  149\degr59\arcmin50\arcsec\   &  2\degr15\arcmin33\arcsec\    &  $2.2$      & 0.659  &  0.740  & 12/12/8/2  &  3.5       \\
\hline
\textbf{Median}  &     &   &   & \textbf{0.689}   &   \textbf{0.656}   &   & \textbf{8.5} \\
\textbf{Total}  &     &   &  \textbf{41.9 } &    &      & \textbf{277/250/178/67}  &  \\
\hline
\end{tabular}
\tablefoot{
 (1) COSMOS Group ID. (2) and (3) J2000 coordinates of the centers of the galaxy groups. (4) Exposure time per field. Exposure using adaptive optics are marked in bold. (5) Median redshift of the group. (6) MUSE PSF FWHM of the narrow band image around the observed wavelength of the \oii\ doublet at the group redshift (7) Number of galaxies (i) all: total in each group; (ii) SED: with SED fitting information (SFR and stellar mass); (iii) MS: on the main sequence of star-forming galaxies; (iv) KS: in the final kinematic sample. (8) Virial mass of the groups.
\tablefoottext{a}{CGr32 has been observed within a mosaic of three adjacent MUSE fields having the same exposure time (1h without and 3.35h with adaptive optics). The three PSF FWHM values correspond to each field.}
\tablefoottext{b}{CGr84 and CGr84b are both observed in the same two adjacent MUSE fields. The two values of exposure times and of seeing correspond to each field.
}
}
\end{table*}

\section{Physical properties of galaxies in dense environments}
\label{Phy_properties}

\subsection{Redshift determination and group membership}

The groups have been targeted in the COSMOS field \citep{Scoville:2007}, therefore, all galaxies in the field have already been identified in broad-band photometry up to a limiting magnitude of $\sim 26$ at $3\sigma$ in the $z$++ band \citep[COSMOS2015;][]{Laigle:2016}.
The spectroscopic redshifts for all objects in the photometric COSMOS2015 catalog located inside the MUSE fields were estimated using the redshift finding algorithm MARZ \citep{Hinton:2016, Inami:2017} based on their absorption and emission spectral features.
At the redshift range of our groups (average $z\sim 0.7$), the most prominent emission lines are the \oiiab\ doublet, \oiii $\lambda$5007, and the Balmer lines starting with \hb . The main absorption lines are \ion{Ca}{ii} H$\lambda$3968.47, \ion{Ca}{ii} K$\lambda$3933.68, G band at 4100 \AA, and the Balmer absorption lines. For each source in each field a PSF-weighted spectrum was extracted \cite[as described in][]{Inami:2017} and then the strongest absorption and emission lines were identified giving a robust redshift determination. We attributed confidence flags for all these objects, following the procedure described in \citet{Inami:2017}.

We used secure spectroscopic redshifts to identify dense structures and galaxies within them using a friends of friends (FoF) algorithm, which will be described in the MAGIC survey paper (Epinat et al. in prep.).
This method is intended to assign the membership of galaxies to a certain group or cluster if they are below given thresholds of angular separation on the sky plane and of velocity separation in the redshift domain from the nearest neighbors. We used a projected separation of 450 kpc and a velocity separation of 500 km~s$^{-1}$ between neighbors, as suggested by \citet{Knobel+09}. Such conservative separations ensure that we do not miss any group galaxy in the process. Using this technique allowed us to find other structures than the targeted ones in the redshift range $0.5<z<0.8$. However, except in one case (CGr84 fields), the number of galaxies in secondary structures is small and we therefore concentrated our efforts on the densest structures containing at least ten members. This led us with a sample of 277 galaxies inside eight galaxy groups. Due to blending or due to the absence of photometry in the COSMOS2015 catalog, we discarded 27 galaxies from the analysis, leading to a parent sample of 250 galaxies in groups.

\subsection{Groups properties}
\label{group_properties}

The eight groups studied in this paper are quite dense and massive. Their virial mass was estimated from the velocity dispersion of members computed using the gapper method discussed in \citet{Beers+90} and used in \citet{Cucciati+10}:
\begin{equation}
 \sigma_v = \frac{\sqrt{\pi}}{N(N-1)} \sum^{N-1}_{i=1} i(N-i) (v_{i+1} - v_{i}) \text{ ,}
 \label{eq:gapper_disp}
\end{equation}
where $v_i$ are the velocities of the $N$ members sorted in ascending order, computed with respect to the median redshift of the group. The mass was then computed as:
\begin{equation}
M_{vir} = \frac{3\sqrt{3} \sigma_v^3}{11.4 G H(z)} \text{ ,}
\label{eq:mvir}
\end{equation}
where $G$ is the gravitational constant and $H(z)$ is the Hubble parameter at redshift $z$ \citep[see][]{Lemaux+12}.
These masses are provided in Table \ref{table_info_group}. They span a range between 3.5 and 81.4 $\times 10^{13}$~M$_\sun$, with a median mass of $8.5 \times 10^{13}$~M$_\sun$.
For some of the groups presented here, masses were also estimated using X-ray data from XMM and \textit{Chandra} \citep{Gozaliasl+19}. We checked the consistency between the two estimates and found an agreement within $\sim 0.5$ dex.
The most massive group contains more than 100 members and is more likely a cluster \citep[CGr32; cf.][]{Boselli+19}. We will, however, refer to all the structures as groups. More detail will be provided in the survey paper (Epinat et al. in prep). The typical projected galaxy density in the MUSE fields ranges between 65 and 220 galaxies per Mpc$^2$, with a median of around 100 galaxies per Mpc$^2$ (20 galaxies per squared arcminute).
The typical galaxy density ranges between 130 and 500 galaxies per Mpc$^3$, with a median of around 200 galaxies per Mpc$^3$ assuming that the third dimension equal the projected one. This is 200 times denser than the typical galaxy density that is around 0.5 galaxy per Mpc$^3$ \citep[e.g.,][]{Conselice+16} at $z\sim0.5-0.7$, converting comoving volume to proper one.
We also made similar typical galaxy density estimates from the MUSE data in order to have a consistent selection. Indeed, the galaxy density provided by \citet{Conselice+16} is based on photometric redshifts, whereas we only used galaxies with secure MUSE spectroscopic redshifts\footnote{Around half of the galaxies detected in our MUSE cubes have such secure redshifts.}. We considered all the galaxies between $z=0.5$ and $z=0.75$ with secure redshifts and we estimated the average density including and excluding the studied groups. We obtained a similar density as in \citet{Conselice+16} when we included the groups, and a value four times lower when groups were excluded. This result is not surprising, because of the different sample selections and since the study of \citet{Conselice+16} does include some groups. We can therefore consider that our groups are at least 200 times denser than the average density within the considered redshift range.
Since the sizes of the groups in the third dimension are not well constrained, we also compared the surface densities and found that our groups are on average 25 times denser than the field, assuming a typical redshift bin of 0.025 ($\sim 4500 - 5000$~km~s$^{-1}$) for the groups, which is an upper limit.
Given the size of the group sample studied here, we do not refine further the density estimate. We can nevertheless claim that our groups are much denser than the average environments encountered in the Universe.

\subsection{Global galaxy properties}
\label{Global_pro}

Using the extensive photometry available in the COSMOS field \citep{Laigle:2016}, stellar mass, star formation rate (SFR), and extinction were estimated for all the galaxies in the selected groups within MUSE fields. Apertures of 3\arcsec\ were used over 32 bands from the COSMOS2015 catalog to obtain photometric constrains on the stellar population synthesis (SPS) models.
Instead of using the properties derived from the purely photometric redshift catalog of \citet{Laigle:2016}, we remodeled the photometry taking advantage of the robust spectroscopic redshift measurements from MUSE spectra to introduce additional constraints into the SPS models. To model the spectra we used the spectral energy distribution (SED) fitting code FAST \citep{Kriek:2009} with a synthetic library generated by the SPS model of \citet{Conroy:2010}, assuming a \citet{Chabrier:2003} initial mass function (IMF), an exponentially declining SFR ($\text{SFR} \propto \exp{(-t/\tau)}$, with $8.5<\log{(\tau [\text{yr}^{-1}])}<10$), and a \citet{Calzetti+00} extinction law. We used the same method as in the previous studies by \citet{Epinat:2018} and \citet{Boselli+19} on the COSMOS Groups 30 and 32, respectively, using MUSE data, to derive the uncertainties on the SED by adding in quadrature a 0.05 dex uncertainty to each band that account for residual calibration uncertainties.

\begin{figure}
 \includegraphics[width=\columnwidth]{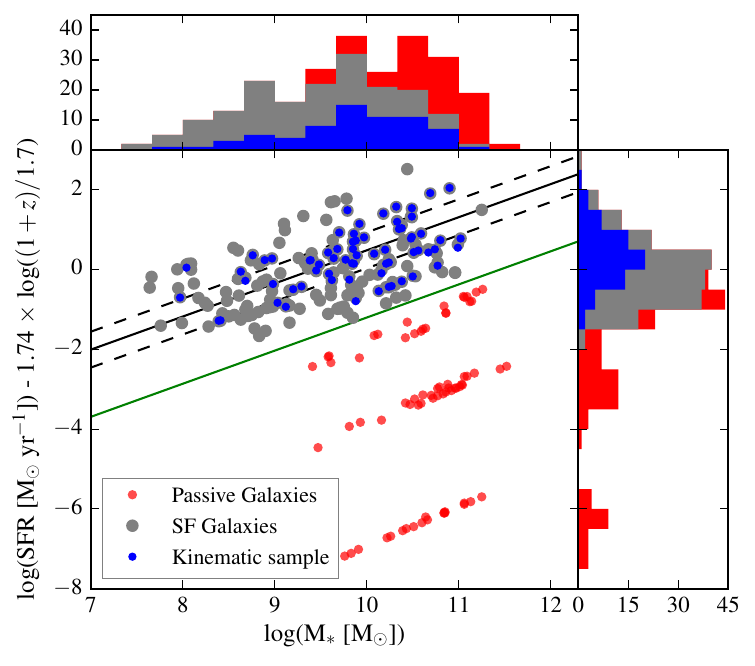}
 \caption{Distribution of the whole sample on the M$_*$ - SFR plane.
 The green line represents the separation defined in Eq. \ref{eq:sf_vs_rs} used to distinguish between galaxies on the red sequence (red dots) and galaxies along the main sequence of star-forming galaxies (gray dots).
 The solid black line represents the best fitting of SFR - M$_*$ empirical relation for star-forming galaxies derived by \citet{Boogaard:2018}, with SFR normalized to redshift $z=0.7$ to account for the evolution with redshift, whereas the dashed lines represent the 1-$\sigma$ intrinsic scatter of this relation.
 Blue dots correspond to the 67 galaxies in the final kinematic sample (S/N $\ge$ 40, R$_{eff}$/FWHM $\ge$ 0.5).
 The histograms show the stellar mass and SFR distributions for the passive galaxies (red), for the parent sample of star-forming galaxies (gray) and for the kinematic sample (blue).}
\label{Main_sequence_SFR}
\end{figure}

The distributions of stellar mass (M$_*$) and SFR extracted from the SED fitting are presented in Fig. \ref{Main_sequence_SFR}. The whole sample of galaxies in groups covers a wide range of masses from $\sim 10^{7.5} ~\rm M_{\sun}$ to $\sim 10^{11.5} ~\rm M_{\sun}$.
The parent sample includes a large fraction of galaxies lying on the main-sequence of star-forming galaxies and some passive galaxies.
In this study, we are interested in the kinematic properties of star-forming galaxies in groups. We therefore need to distinguish between passive and star-forming galaxies. To do that, we used the prescription of \citet{Boogaard:2018} to account for the SFR evolution with redshift at a given stellar mass and we identified galaxies along the main sequence of star-forming galaxies as those for which
\begin{equation}
\log{\rm (SFR)} -1.74 \times \log{((1+z)/1.7)} \ge 0.83 \times \log{\rm (M_*)} - 9.5 \text{ ,}
 \label{eq:sf_vs_rs}
\end{equation}
where M$_*$ is the stellar mass in M$_{\sun}$ and SFR is the star formation rate in M$_{\sun}$~yr$^{-1}$. This subsample contains 178 galaxies and is referred to as the parent sample of star-forming galaxies. It spreads over a wide range of masses ($\sim 10^{7.5}~\rm M_{\sun}$ to $\sim 10^{11.5}~\rm M_{\sun}$) and SFRs ($10^{-2}$~M$_{\sun}$~yr$^{-1}$ to $10^{2}$~M$_{\sun}$~yr$^{-1}$). Masses and SFRs for the kinematic sample (see Sect. \ref{Sample_Selection_Criteria}) are presented in Table \ref{tab:dynamics} of Appendix \ref{app:tables}.

\subsection{Morphological analysis}
\label{Morphology}

In order to constrain efficiently geometrical parameters required for the kinematics analysis, a homogeneous extraction of the surface brightness distribution using high resolution images is highly desirable.
In addition, studying galaxy morphology is instructive of their type and size.
Ideally one should probe the old stellar populations better traced by rest-frame red images. For the COSMOS sample, the publicly available images with the best spatial resolution are the \textit{Hubble} Space Telescope (HST) Advanced Camera for Surveys (ACS) images observed with the F814W filter, which corresponds to rest-frame wavelengths around 5000~\AA\ at $z\sim 0.7$, and that have a spatial resolution better than 0.1\arcsec~(corresponding to $\sim 700-600$ pc in the galaxies frame). These images were produced using the MultiDrizzle software \citep{Koekemoer:2007} on the COSMOS field \citep{Scoville:2007}, with a spatial sampling of 0.03\arcsec/pixel and a median exposure time of 2028 seconds.

Models for the surface brightness distribution of the stellar continuum and the determination of the geometrical parameters were performed for all the galaxies in groups on these HST-ACS images, using the data analysis algorithm \textsc{Galfit} \citep{Peng:2002}.
First, with the purpose of determining the PSF in these images, we identified 27 non saturated stars present over 14 fields corresponding to groups observed with MUSE in the COSMOS field before October 2018. These stars were modeled using a circular Moffat profile. We then built with \textsc{Galfit} the theoretical PSF taking the median value of each parameter derived from the 27 stars: $\rm FWHM=0.0852$\arcsec and $\beta=$1.9 (Moffat index).

We followed the same method as in \citet{Contini:2016} for galaxies in the Hubble Deep Field South, and in \citet{Epinat:2018} for the galaxies in the group COSMOS-Gr30: The galaxies in our sample were modeled with a composite of bulge and disk model profiles. The bulge is spheroidal and is described by a classical de Vaucouleurs profile (Sersic index $n=4$),
\begin{equation}
I(r)=I_{b}(r_e)e^{-7.67\left[(r/r_e)^{1/4}-1\right]} \text{ ,}
\end{equation}
where $r_e$ is the bulge effective radius and $I_{b}(r_e)$ is the bulge intensity at the effective radius. The disk is described by an exponential disk,
\begin{equation}
I(r)=I_{d}(0)e^{-r/R_{d}} \text{ ,}
\end{equation}
where $I_{d}(0)$ is the disk central intensity and $R_{d}$ the disk scale length. Both profiles share a common center. For the bulge the free parameters are the total magnitude and the effective radius. For the disk we let free the position angle (PA$_m$), axis ratio, total magnitude and scale length. 

When necessary, additional components were introduced to account for extra features (nearby faint galaxies in the field, strong bars, star-forming clumps, etc.) and avoid the bias they could cause in the resulting parameters. This was very obvious in some cases where strong star-forming clumps biased the center of the galaxy closer to the clump than to the actual center. Therefore, modeling them as individual components allowed us to have better geometrical determinations of the bulge and of the disk.
We performed morphological modeling the full parent sample, that includes passive galaxies, with the aim at statistically characterizing the galaxy groups and the ratio of passive to star-forming galaxies.
Morphological parameters for the kinematic sample (see Sect. \ref{Sample_Selection_Criteria}) are presented in Appendix \ref{app:tables}.

We compared our morphological decomposition with morphological catalogs in the COSMOS field to assess its robustness. Several morphological catalogs are publicly available and provide, among others, galaxy sizes, axis ratios and morphological types
(Tasca int catalog: \citealp{Tasca+09}, Tasca linee catalog: \citealp{Abraham+96} and Tasca SVMM catalog: \citealp{Huertas-Company+08}\footnote{\url{https://irsa.ipac.caltech.edu/data/COSMOS/gator_docs/cosmos_morph_tasca_colDescriptions.html}};
Cassata catalog \citealp{Cassata+07}\footnote{\url{https://irsa.ipac.caltech.edu/data/COSMOS/gator_docs/cosmos_morph_cassata_colDescriptions.html}};
Zurich catalog \citealp{Scarlata+07, Sargent+07}\footnote{\url{https://irsa.ipac.caltech.edu/data/COSMOS/gator_docs/cosmos_morph_zurich_colDescriptions.html}}).
In all these catalogs, morphological measurements based on \textsc{SExtractor} \citep{Bertin+96} segmentation maps are provided. Whereas our forward modeling approach corrects the size and the axis ratio for the PSF of the HST-ACS images, this is not the case for methods based on segmentation maps.
For this analysis we used data on four additional MUSE fields targeting groups at redshifts $0.3<z<0.5$, and we included the morphological analysis performed on field galaxies, leading to a total sample of 659 galaxies, in order to increase the intersection of our sample with that of the COSMOS catalogs. We present the comparison with the \citet{Cassata+07} catalog, for which the intersection with our sample is large (471 galaxies in common) and for which the dispersion between their effective radius measurements and ours is the smallest. It is however worth noticing that the trends presented hereafter are the same whatever the catalog used.

\begin{figure}
 \includegraphics[width=\columnwidth]{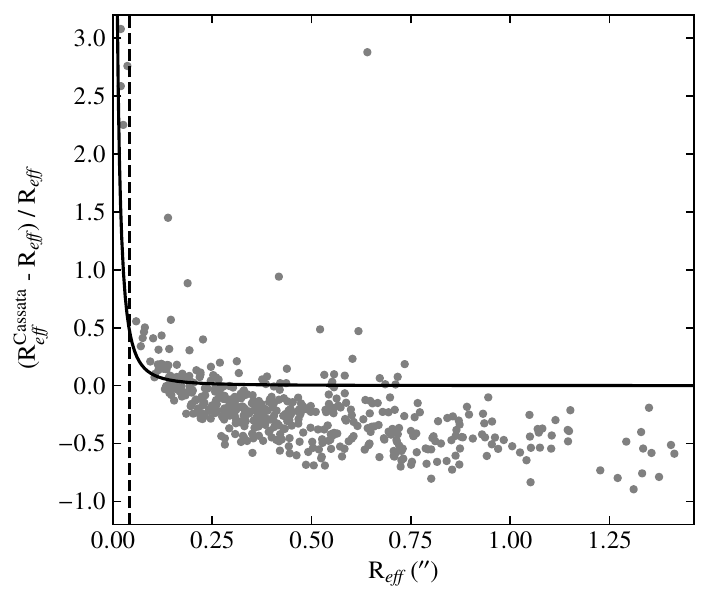}
 \caption{Relative difference between the global effective radius inferred from our \textsc{Galfit} models and the \citet{Cassata+07} morphological catalog as a function of the \textsc{Galfit} effective radius. The solid black line represents the limiting effective radius measurable when not taking into account the PSF, and the dashed vertical line marks the value of half the HST-ACS PSF FWHM.
 }
\label{effective_radius_cosmos}
\end{figure}

\begin{figure}
 \includegraphics[width=\columnwidth]{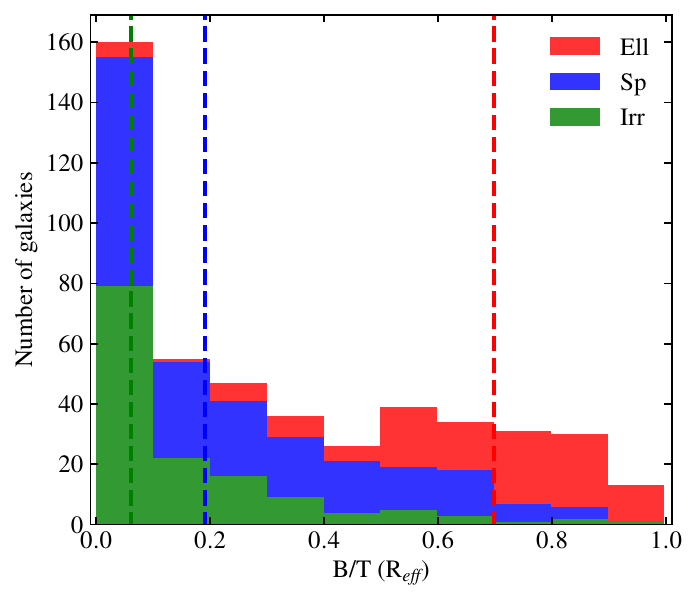}
 \caption{Distribution of the three morphological classes defined in the \citet{Cassata+07} catalog as a function of the bulge to total flux ratio estimated at the effective radius. Green, blue, and red stacked histograms correspond to galaxies classified, respectively, as irregular, spiral, and elliptical. The median values for those three classes are 0.06, 0.19, and 0.70, respectively, and are indicated by dashed colored vertical lines.
 }
\label{Histo_morpho_cosmos}
\end{figure}

\begin{figure}
 \includegraphics[width=\columnwidth]{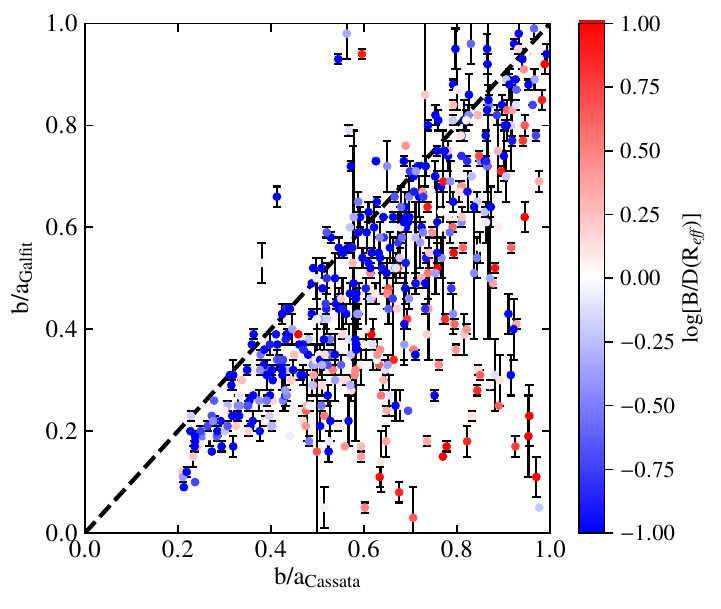}
 \caption{Axis ratio for the disk determined using \textsc{Galfit} as a function of the global axis ratio from the \citet{Cassata+07} catalog. The color indicates the luminosity ratio of the bulge with respect to the disk inside the effective radius, red and blue colors corresponding, respectively, to bulge- and disk-dominated galaxies.
 }
\label{axis_ratio}
\end{figure}

We first compared the effective radii (see Fig. \ref{effective_radius_cosmos}). Our method provides effective radii for both bulge and disk components. In order to do a fair comparison, we used our models to infer a global effective radius (R$_{eff}$).
\textsc{Galfit} effective radii are smaller than those in the COSMOS catalogs for galaxies with radii smaller than $\sim 0.2$\arcsec. The relative difference increases when the effective radius decreases, which clearly demonstrates that the difference is due to the fact that we compare a forward model that takes into account the PSF to \textsc{SExtractor} based measurements that are limited by the image spatial resolution. For larger radii, \textsc{Galfit} effective radii are larger but the relative difference is lower than 50\% and depends on the catalog used for comparison. In addition, this does not impact the total magnitude of our model that is in good agreement with that of those catalogs.
We also computed the bulge to disk ratio inside R$_{eff}$ and checked that this correlates with the COSMOS morphological classification (see Fig. \ref{Histo_morpho_cosmos}), in order to assess the validity of our bulge/disk decomposition. Clearly, elliptical galaxies have a larger fraction of bulge-dominated galaxies and spiral galaxies have a larger fraction of disk-dominated galaxies. We also see such an agreement in the distribution of galaxies inside the M$_*$ - SFR plot for the present parent sample of galaxies in groups.
Last, we compared the axis ratio from \textsc{Galfit} to the one determined by \citet{Cassata+07} (see Fig. \ref{axis_ratio}) and found that, on average, it is lower using \textsc{Galfit}. This is expected since the COSMOS catalog provides a global axis ratio while we determined it for the disk with \textsc{Galfit}.
We can see that most of the scatter comes from bulge-dominated galaxies. The agreement for disk-dominated galaxies is fairly good, despite \textsc{Galfit} still leads to lower values. For very low disk axis ratios, this can be explained by the fact that \textsc{Galfit} corrects the ratio for the HST-ACS PSF, whereas \textsc{SExtractor} does not.
These results underline the reliability of our morphological analysis that is homogeneous for the whole sample of galaxies.

\begin{figure}
 \includegraphics[width=\columnwidth]{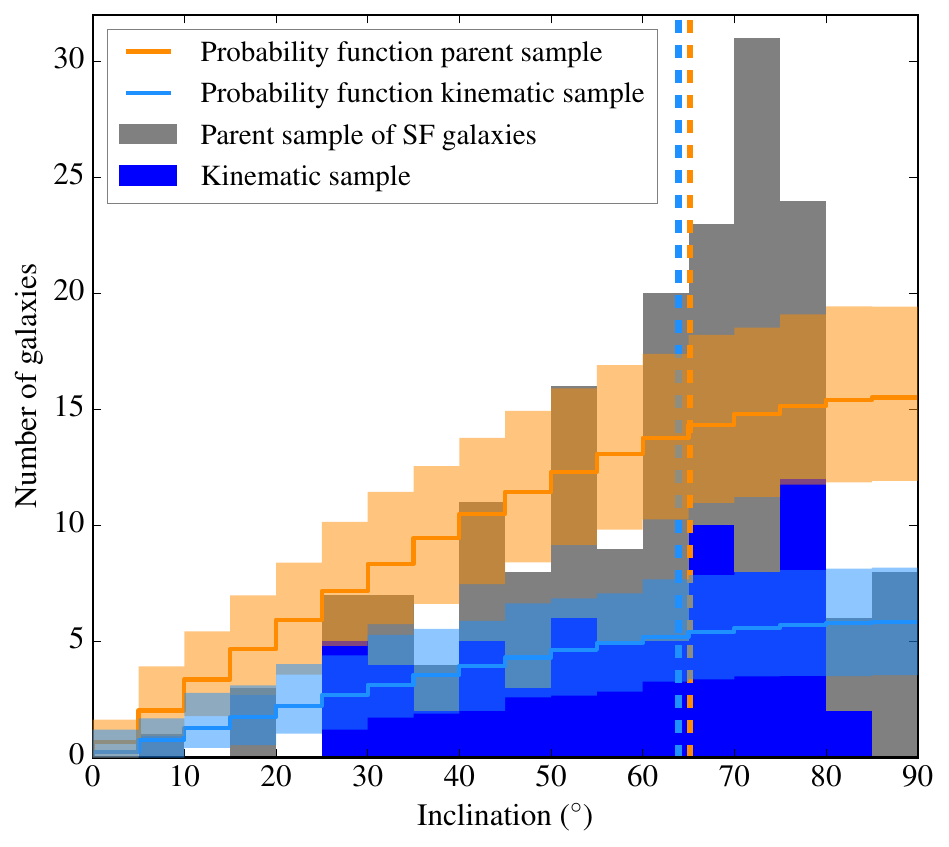}
 \caption{Distribution of disk inclinations for the parent sample on the main sequence of star-forming galaxies with a bin of 5\degr\ (gray). The orange curve and shadow area represent the expected theoretical distribution for random orientations and the $1\sigma$ uncertainties computed from 5000 realizations of such a random distribution with the same number of 178 galaxies as the parent sample.
 The distribution of the final kinematic sample (S/N $>$ 40, R$_{eff}$/FWHM $>$ 0.5) of 67 galaxies is also displayed (filled blue histogram) along with the associated theoretical distribution and the corresponding $1\sigma$ uncertainties (light blue curve and shadow area). The dashed orange and blue vertical lines indicate the median value for the parent ($i=65.2$\degr) and final kinematic ($i=63.9$\degr) samples, respectively.
 }
\label{Histo_ALL_INC5}
\end{figure}

By construction, our group sample should display a distribution compatible with random disk orientation since there is no galaxy preselection. In order to evaluate if there is any bias on the inclination parameter for our sample, we extracted the distributions of inclinations of our sample from our morphological modeling with \textsc{Galfit}, assuming razor-thin disks.
In this analysis, we only use the parent sample of star-forming galaxies (see Sect. \ref{Global_pro}) since red sequence galaxies are supposed to be ellipticals.
We compared our observed distribution with the theoretical distribution for randomly oriented disks. This theoretical distribution is derived from the fact that the probability to observe a thin disk with an orientation between $\theta_{1}$ and $\theta_{2}$ is equal to $| \cos\theta_{1}-\cos\theta_{2} |$.
The distribution of inclinations shown in Fig. \ref{Histo_ALL_INC5} illustrates the fact that the construction of our sample is done without any prior. The median value of inclination for our sample is 65.2\degr, which is close to the theoretical median inclination of 60\degr.
Our sample misses low inclination systems ($i<40$\degr) and highly inclined ones ($i>80$\degr) with respect to the expected distribution of inclinations for randomly oriented disks (orange curve). On the one hand, at low inclination, this may be due to the small number of galaxies expected from the theoretical distribution function. 
Moreover, despite we modeled separately some strong features in the light distributions, residual features, such as bars or arms, may have persisted, affecting the models of the disks and biasing the estimated inclination toward higher values.
On the other hand, the lack of edge-on galaxies, could be due to high levels of dust extinction that could reduce the detection of such galaxies. However, this could also be due to the fact that we assume a null thickness for the modeling, so for galaxies with thick disks we may underestimate their real inclination. Using an intrinsic axial ratio of the scale height to the scale length different than zero would mainly fill the histogram at high inclination without changing it much at low inclination.
For bulge-dominated galaxies, the morphological decomposition may lead to unconstrained disk parameters. We therefore also studied the inclination distribution after removing bulge-dominated galaxies from the distribution. This resulted in removing almost all galaxies with disk inclinations larger than 80\degr. Nevertheless, this did not impact much the rest of the distribution.
This shows that our bulge-disk decomposition provides meaningful measurements of disk inclinations for disk-dominated galaxies.

\subsection{Kinematics of the ionized gas}
\label{kinematics}

We extracted the spatial distribution and kinematics of the ionized gas for all galaxies in the parent sample\footnote{We were able to extract \oii\ kinematics for only six galaxies considered as passive, two of which show signs of AGN activity.}.
In this study, in order to have an homogeneous analysis, we focus on kinematics derived from the \oiiab\ doublet.
From the MUSE datacubes, we extracted the line flux, signal-to-noise ratio (S/N), velocity field and velocity dispersion map by modeling the \oiiab\ line doublet using the python code \textsc{Camel}\footnote{\texttt{https://bitbucket.org/bepinat/camel.git}} described in \cite{Epinat:2012}, which fits any emission line with a Gaussian profile and uses a polynomial continuum. For each galaxy we extracted a sub-datacube around its center with a total size of 30 $\times$ 30 pixels and, in order to increase the S/N per pixel without lessening the resolution, a 2D spatial Gaussian smoothing with a FWHM of two pixels was applied.
To perform the fitting around the \oii\ doublet spaxel by spaxel, we use a constant continuum. Each line was modeled separately by a Gaussian profile at different rest-frame wavelengths, but assuming the same velocity and velocity dispersion. The ratio between the flux of the two \oii\ doublet lines was constrained between $0.35 \le F_{\textrm{\oiia}}/F_{\textrm{\oiib}} \le 1.5$, according to the expected photo-ionization mechanism \citep{Osterbrock+06}.
We then run a cleaning routine to remove all the spaxels with S/N below a threshold of 5 or having a velocity dispersion lower than 0.8 times the dispersion corresponding to the spectral resolution (see Sect. \ref{obs_dataset}). The latter ensures to avoid fitting noise while keeping detection even of narrow lines that might be present.
After the automatic cleaning we visually inspected all the resulting velocity fields with the goal to remove isolated spaxels or spaxels with both low S/N and atypical velocity values with respect to their neighbors.

The kinematics of each galaxy was then modeled as a rotating disk in two dimensions. This model is taking into account properly the effect of the limited spatial resolution of observations. It produces a high resolution model and uses the observed line flux distribution to weight the velocity contribution within each spaxel of the low resolution velocity field after smoothing by the MUSE PSF, as described in \citet{Epinat+10}. We fixed the inclination ($i$) and center ($x, y$) to those derived from the morphological analysis of the high resolution HST-ACS images (see Sect. \ref{Morphology}) in order to suppress the degeneracy with the rotation velocity amplitude and with the systemic velocity, respectively. This degeneracy in the kinematic model is strong when the data is severely affected by beam smearing \citep{Epinat+10}. The rotation curve used in the model is linearly rising up to a constant plateau and has two parameters: $V_t$, the velocity of the plateau and $r_t$, the radius at which the plateau is reached:
\begin{equation}
V_r = V_t \times \frac{r}{r_t}\textrm{~~~when~~~} r\le r_t \textrm{~~~or~~~} V_r = V_t \textrm{~~~when~~~} r > r_t \text{ .}
\end{equation}
The other fitted parameters are the systemic redshift ($z_s$) and the kinematic position angle of the major axis (PA$_k$). This procedure fits the observed velocity field, taking the velocity uncertainty map into account to weight the contribution of each spaxel. It uses a $\chi^2$ minimization based on the Levenberg-Marquardt algorithm. It also produces a map of the beam smearing correction to be subtracted in quadrature to the observed velocity dispersion map. For the pixels where the measured velocity dispersion is lower than the LSF, the LSF corrected dispersion is set to zero. Similarly, when the LSF corrected dispersion is lower than the beam smearing correction, the beam smearing corrected dispersion map is null. This method has been used in several studies \citep{Epinat+09, Epinat+10, Epinat:2012, Vergani+12, Contini:2016}.

\begin{figure}
 \includegraphics[width=\linewidth]{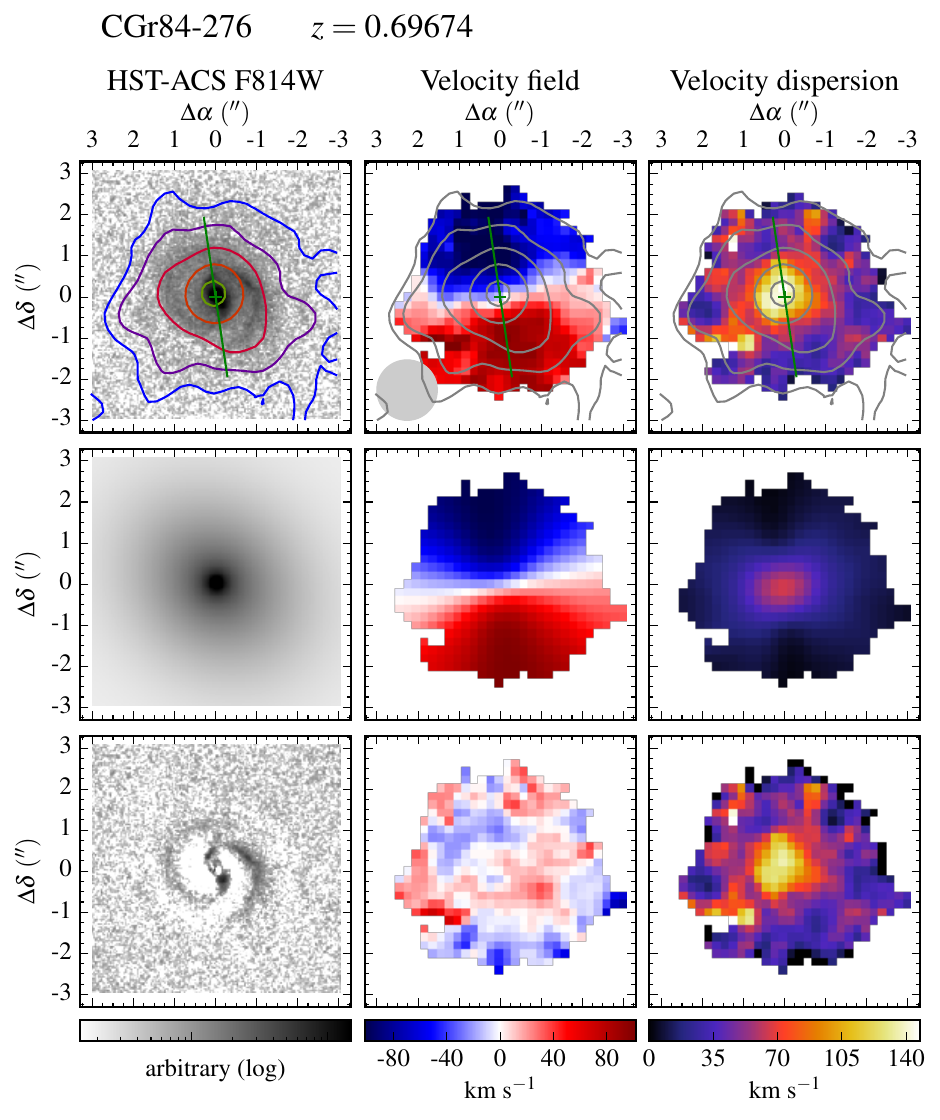}
 \caption{Example of maps and models for galaxy ID276 in CGr84. Top row, from left to right: HST-ACS F814W images, MUSE velocity fields and velocity dispersion maps corrected for spectral resolution. 
 Middle row: Associated models. The velocity dispersion (third column) corresponds to a beam smearing correction map. Bottom row: Residuals, except for the third column that shows the beam smearing corrected velocity dispersion map. On each observed map, the green cross indicates the center derived from the morphology, whereas the green segment indicates the kinematic major axis and has a length corresponding to $R_{22}$. The \oii\ flux distribution is shown with contours at levels of surface brightness $\Sigma$([\ion{O}{ii}]) = 2.5, 5.0, 10.0, 20.0, 40.0, 80.0 $\times 10^{-18}$~erg~s$^{-1}$~cm$^{-2}$~arcsec$^{-2}$. The MUSE spatial resolution is indicated with a gray disk of diameter FWHM in the bottom-left corner of the velocity field.}
 \label{maps_ex}
\end{figure}

The goal of this modeling is the derivation of $V_{r22}$, the rotation velocity at $R_{22} = 2.2 \times R_d$, and of $\sigma$, the velocity dispersion (see Sect. \ref{sec:fitting_methods}).
For the final kinematic sample (see Sect. \ref{Sample_Selection_Criteria}), the cleaned kinematic maps are more extended than $R_{22}$, except for two galaxies for which the data extends up to 96\%\ and 87\%\ of $R_{22}$. For the less extended one, the plateau is reached within the data. We therefore kept both in the analysis.
When the plateau is reached within the data but not reached within $R_{22}$, the uncertainty on the velocity is estimated from the uncertainty on $V_t$, $r_t$, and $R_d$.
When the plateau is not reached within the data, uncertainties on both $r_t$ and $V_t$ are large due to a degeneracy in the model. However, the slope might be well constrained. In those cases, the uncertainty on $V_{r22}$ is deduced from the one on $R_{22}$, estimated as the mean of the MUSE PSF standard deviation (FWHM $/2\sqrt{2\log{2}}$) and of the uncertainty on $2.2\times R_d$. In the other cases, we use the formal uncertainty on $V_t$.
The velocity dispersion is estimated on the beam smearing corrected velocity dispersion map as the median of the spaxels where the S/N is above 5. It therefore equals to zero when more than half of those spaxels have a measured line width lower than the quadratic combination of LSF plus beam smearing correction widths.
We present model parameters and their uncertainties for the kinematic sample in Table \ref{tab:morphokin}. Quantities derived from these models are stored in Table \ref{tab:dynamics}. Appendix \ref{app:maps} shows the maps obtained for the full kinematic sample, whereas Fig. \ref{maps_ex} shows one example of the maps obtained.

\subsection{Kinematic sample selection criteria}
\label{Sample_Selection_Criteria}

Our parent sample of 178 star-forming galaxies in groups covers a broad mass range from $\sim 10^{7.5}$~M$_{\sun}$ to $\sim 10^{11.5}$~M$_{\sun}$ (see Fig. \ref{Main_sequence_SFR}). Six galaxies of this parent sample are embedded in a large structure of ionized gas in the group CGr30 \citep{Epinat:2018}, which prevents their kinematics to be retrieved unambiguously from the ionized nebula. They are therefore removed from this analysis.
Galaxies with low SFR are not expected to provide detailed and accurate kinematics information from ionized gas emission lines. In addition, low-mass galaxies are expected to be small and may be potentially unresolved in our MUSE observations.
Therefore, a first visual inspection allowed us to identify the sample of galaxies with spatially resolved kinematics. We attributed flags corresponding to the quality of the velocity fields (VF) based on its extent and on the presence of a smooth velocity gradient. We ended up with eight flags: 0 when no signal is detected, 1 when the size of the cleaned VF does not exceed $4\times4$ pixels ($\sim 6\times 6$~kpc$^{2}$ at $z=0.7$), 2 and 3 when the VF is smaller than $8\times8$ pixels ($\sim 11\times 11$~kpc$^{2}$), 4 and 5 when it is smaller than $10\times10$ pixels ($\sim 14\times 14$~kpc$^{2}$), 6 and 7 when it is smaller than $13\times13$ pixels ($\sim 19\times 19$~kpc$^{2}$), and 8 when it is larger than 13 pixels . Flags 3, 5, 7, and 8 correspond to galaxies with clear velocity gradients whereas galaxies with flags 2, 4, and 6 do not show such gradients.
There are 28 flag 0 galaxies in the parent sample of star-forming galaxies.
These galaxies are probably galaxies with no further star formation and would appear as red sequence galaxies if we had used \oii\ flux to estimate the SFR. They are mainly located at the bottom of the main sequence, despite some of them are above.
One of them is a quasar, whereas neither active galactic nucleus (AGN) nor quasar templates were used, so its mass and SFR may be incorrect. For the others, since no clear sign of AGN was observed (see Sect. \ref{sec:impact_selection_error}), this could mean that they have been quenched recently.

\begin{figure}
\includegraphics[width=\linewidth]{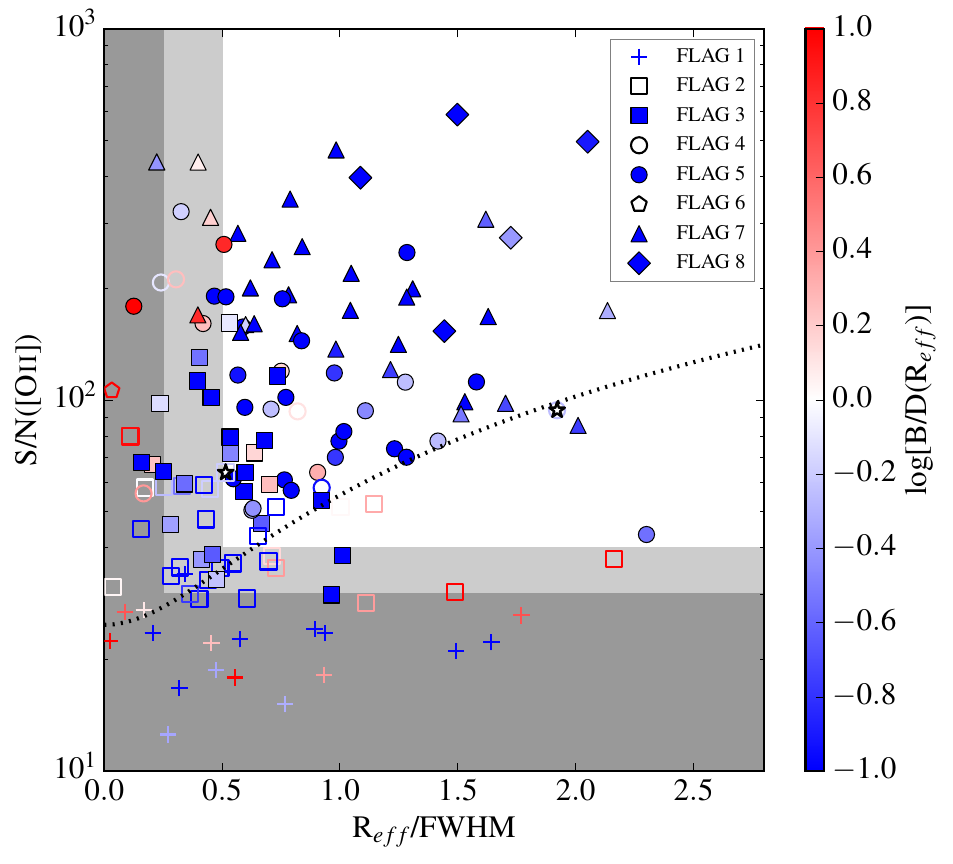}
\caption{
S/N of the total \oii\ flux as a function of the global effective radius divided by the MUSE PSF FWHM for the parent sample of galaxies located in the main sequence of star-forming galaxies.
The white area corresponds to the parameter space used for the strict selection (S/N $>$ 40 and R$_{eff}$/FWHM $>$ 0.5), whereas the light gray area also contains galaxies within the relaxed selection (S/N $>$ 30 and R$_{eff}$/FWHM $>$ 0.25) used in Sect. \ref{analysis} to study the impact of the selection on the analysis.
The dotted black line corresponds to the limit for a constant S/N of 8 over the effective radius.
The symbols correspond to the visual velocity field flags from 1 to 8. Crosses and empty symbols correspond to galaxies with no obvious velocity shear, whereas filled symbols correspond to galaxies with evidence for rotation.
The color indicates the ratio of the bulge luminosity with respect to the disk one inside the global effective radius, red and blue colors corresponding, respectively, to bulge- and disk-dominated galaxies.
The two problematic AGN hosts rejected in the final sample are identified with stars.
}
\label{Flux_SNR_Selection_Criteria}
\end{figure}

In order to have a more objective and physical selection, we used the combined information extracted from morphology, kinematics and SED fitting.
Since we need the kinematics of galaxies to be spatially resolved, a first criterion is used to quantify the size of the ionized gas distribution with respect to that of the spatial resolution. Since the limited spatial resolution of MUSE data might prevent to measure robustly the size of the \oii\ flux distribution, we define the first criterion as the ratio of the effective radius of the stellar distribution with respect to the size of the MUSE PSF.
It seems reasonable to assume that the ionized gas disk is closely related to the underlying stellar distribution\footnote{In the case of nonresonant lines}, which contains the newly formed stars that photo-ionize the gas \citep[e.g.,][]{Epinat+08, Vergani+12}. The higher HST-ACS spatial resolution makes measurements of stellar distributions much more accurate, especially for galaxies with optical sizes close to, or less extended than the seeing in MUSE data. Moreover, the extent of kinematics maps where the S/N is above 5 might overestimates the true extent of ionized gas disks and might gets larger than the MUSE PSF FWHM, especially for bright compact \oii\ emitters.
We define a second criterion as the S/N of the \oii\ doublet in the MUSE data in order to derive meaningful kinematic maps.
We estimated the flux in the \oii\ doublet using the cleaned line flux maps, where the individual S/N per pixel is at least 5 (see Sect. \ref{kinematics}). This ensures to optimize the aperture over which the flux is computed and to increase the S/N with respect to using integrated spectra that might show strong deviation to Gaussian lines due to the underlying galaxy kinematics. The noise was estimated as the quadratic sum of the uncertainty on the flux of individual pixels.
We plot the S/N as a function of the ratio between the global effective radius (R$_{eff}$) divided by the MUSE PSF FWHM in Fig. \ref{Flux_SNR_Selection_Criteria} and used various symbols depending on the visual flags. As expected, all the flag 1 galaxies have a S/N lower than 40 (17/18 have S/N $<$ 30) and are very small (10/18 have R$_{eff}$/FWHM $<$ 0.5). Most of the flag 2 galaxies are also either small or with rather low S/N. Only five of them (three excluding bulge-dominated galaxies) have disk effective radii larger than half the MUSE PSF FWHM and a S/N larger than 40. Around half flag 3 galaxies are above these thresholds, whereas higher flags are almost all within these constraints. There is only one flag 6 galaxy that is the quasar. It has a very small effective radius but has some diffuse extended emission around it. This galaxy is therefore outside the thresholds.
Based on this quantitative agreement with the visual classification, we conclude that these objective criteria are a good way of performing a robust selection.
The PSF FWHM is an estimate of the half light radius of a point source, it therefore makes sense that galaxies with disk effective radii lower than half the MUSE PSF FWHM are not resolved within the MUSE data.
On the other hand, assuming a constant surface brightness and a S/N above a threshold of 8 per pixel over a circular surface up to the effective radius, the global S/N might be above the dotted line shown in Fig. \ref{Flux_SNR_Selection_Criteria}. We see that using the S/N threshold of 40 leads to a very small difference. We checked each galaxy between these two limits. In most of the cases, the ionized gas is less extended than the stellar disk, or it is patchy, therefore not covering uniformly the stellar disk. In other cases, the galaxies are quite edge-on, leading to a lower spatial coverage than when assuming a circular galaxy. For only one galaxy, it is clear that its velocity field only covers the central region of a very extended disk. We therefore use the global S/N threshold as selection criterion.
Hereafter, we opt as a baseline for strict selection criteria with a S/N limit of 40 and a R$_{eff}$/FWHM limit of 0.5, which leads to a sample of 77 galaxies.
In Sect. \ref{analysis}, we also use relaxed selection criteria with a S/N limit of 30 and a R$_{eff}$/FWHM limit of 0.25, leading to a sample of 112 galaxies, in order to investigate the impact of selection on the analysis.

We further removed the eight galaxies for which the bulge to disk ratio within R$_{eff}$ is larger than unity (red symbols in Fig. \ref{Flux_SNR_Selection_Criteria}). Indeed, we found that those galaxies have less accurate morphological parameters, with usually very small disk scale lengths, and therefore inaccurate estimates of their stellar mass within $R_{22}$ (see Sect. \ref{sec:fitting_methods}) or bad estimates of the rotation velocity because it is inferred at too small radii. This interpretation is more relevant for low-mass systems for which we do not expect strong bulges. Nonetheless, this indicates that the morphology is not accurate due, most of the time, to galaxies being faint.

Last, we identified galaxies hosting AGN, for which the strong signal from the central region can dominate the ionized gas emission and prevent secure disk kinematics measurements by underestimating the rotation and overestimating the velocity dispersion. Their high ionization levels may also cause large uncertainties in the determination of both stellar mass and SFR.
In order to identify AGN inside our sample, we used the diagnostic diagram that combines the \oiiia /\hb\ and \oiiab /\hb\ line ratios, proposed by \citet{Lamareille:2010}, and the mass-excitation diagram that compares the \oiiia /\hb\ line ratio with the stellar mass \citep{Juneau:2011}. These two diagnostic diagrams are based on the emission lines available in the optical spectra of intermediate redshift galaxies.
We obtained seven AGN candidates in the kinematic sample identified as such in at least one of these two diagnostic diagrams.
We then visually inspected their integrated spectra to look for broaden features and/or high-level emission lines typical of AGN activity like [\ion{Ne}{iii}]$\lambda$3868 and [\ion{Mg}{ii}]$\lambda$2800.
We finally identified three secure AGN (CGr30-71, CGr32-268 and CGr32-454) and two ambiguous cases (CGr32-132 and CGr32-345) in the kinematic sample. The ambiguous objects can be due to low-level AGN activity or star-forming galaxies with multiple kinematic components.
Among the secure AGN, we excluded CGr32-268 and CGr32-454 (identified with stars in Fig. \ref{Flux_SNR_Selection_Criteria}) because their AGN affected the velocity fields of the host galaxy.
Using these two additional selection criteria leads to a final kinematic sample containing 67 galaxies.
Half of the galaxies removed with these criteria were classified as potentially nonrotating, leaving only three such galaxies within the final sample.
In the next section, we will discuss the impact of the thresholds on the TFR.

The distributions corresponding to the final kinematic sample are overplotted on the parent sample distributions in Fig. \ref{Main_sequence_SFR}. The kinematic sample is covering the galaxies with the largest SFR.
Nevertheless, the median SFR (1.95~M$_{\sun}$~yr$^{-1}$) is only twice as large as the median SFR of the parent sample of star-forming galaxies (0.85~M$_{\sun}$~yr$^{-1}$).
It covers the same mass range as the parent sample of star-forming galaxies but has a lower fraction of low-mass galaxies ($<10^{8.5}~\rm M_{\sun}$) than the parent sample since they are less extended than more massive ones.
The median stellar mass is $7.6\times 10^{9}$~M$_{\sun}$, which is slightly higher than the median stellar mass of the parent sample of star-forming galaxies of $4.4\times 10^{9}$~M$_{\sun}$.
In the end, the kinematic sample represents around 38\% of the parent sample in the main sequence. However, removing galaxies with very low ionized gas fluxes that may be passive (flags 0 and 1), and excluding the quasar and the galaxies embedded in the extended nebula in CGr30, the kinematic sample represents more than 50\% of the population of star-forming galaxies. Most of the galaxies that are not in the kinematic sample are low-mass ones because they are too small to be properly resolved.

Finally, we also checked that the inclination distribution of the final kinematic sample is still compatible with that of a randomly selected sample (see Fig. \ref{Histo_ALL_INC5}). The kinematic sample better follows a random orientation distribution. This is probably due to the removal of the smallest galaxies and of bulge-dominated galaxies for which morphology may be more difficult to retrieve with accuracy. This means that the inclination is not affecting much our ability to retrieve kinematics information.
We thus decide not to put any constraint on inclination since very few galaxies have an inclination below 30\degr.
For some of these galaxies, inclination may be overestimated, leading to an underestimation rather than to an overestimation of their rotation velocity. Only one galaxy (CGr84b-23) seems to be potentially in this case.

\section{Stellar and baryonic mass Tully-Fisher relations}
\label{analysis}

Our aim is to study the impact of environment on the TFR by comparing this relation obtained for the MAGIC group sample to the one derived for other samples of either field (KMOS3D, \citealp{Ubler+17}; KROSS, \citealp{Tiley+19}) or cluster (ORELSE, \citealp{Pelliccia+19}) galaxies over a similar redshift range.
The whole KMOS3D sample covers a wider redshift range. In the following analysis, we focus on the lowest redshift subsample with $z<1.1$. We also consider separately the KROSS rotation dominated ($V_{r22}/\sigma > 1$) and disky ($V_{r22}/\sigma > 3$) subsamples. We decided to compare our dataset to the KMOS3D and the KROSS samples since they are the two largest ones observed with integral field spectroscopy, and to the ORESLE sample because it is currently the only one exploring the impact of environment, despite it has been observed using long-slit spectroscopy. The average redshift is around 0.9 for all comparison samples used, slightly higher than the average redshift around 0.7 of the MAGIC sample.
We first ensure that methodology biases, due to uncertainties, fitting methods (Sect. \ref{sec:fitting_methods}), and sample selection (Sect. \ref{sec:impact_selection_error}), are minimized before analyzing and comparing the TFR for the MAGIC and the comparison samples with the same procedure.

\subsection{Fitting methods and uncertainties}
\label{sec:fitting_methods}

The TFR is a relation that connects either the magnitude or the mass of a galaxy to its rotation velocity. This relation is commonly fitted with the following expression:
\begin{equation}
 \log{M} = \alpha \left[\log{V} - \log{V_{ref}}\right] + \beta \text{ ,}
 \label{eq:tfr}
\end{equation}
where $\log{V_{ref}}$ is set to a non-null value to reduce the correlation between $\alpha$ and $\beta$ and therefore reduce uncertainties on the latter. To reach this goal, $\log{V_{ref}}$ has to be set accordingly to the sample, as the mean or median of the velocity distribution. In order to be able to make comparisons between various samples, we use $\log{(V_{ref}~[\text{km~s}^{-1}])}=2.2$, which is the median value of the rotation velocity for the final kinematic sample of 67 galaxies defined in Sect. \ref{Sample_Selection_Criteria}. We will study the cases where $M$ is either the stellar or the baryonic mass and where $V$ is either the rotation velocity or a velocity that includes both the rotation and the velocity dispersion components.

The velocity can either be a terminal, an asymptotic velocity or a velocity within a given radius. At intermediate redshift, it is common to measure the velocity at $R_{22} = 2.2\times R_d$ because this radius is usually within kinematic measurements coverage and because the maximum rotation is expected to be reached in most of the cases. For a purely exponential disk mass distribution, it is the radius where the maximum circular velocity is reached, but for real distributions, this is not necessarily the case and we could expect that the maximum rotation might quite often not be reached in low-mass galaxies, based on rotation curves observed in the local Universe \citep[e.g.,][]{Persic+96, Epinat+08}.

It is also common to add an asymmetric drift correction to the rotation velocity measurement in TFR studies in order to account for the gas pressure support \citep[e.g.,][]{Meurer+96, Burkert+10} and estimate the circular velocity that the gas would have in absence of pressure. This correction is usually performed assuming that the gas is in dynamical equilibrium, has axisymmetric kinematics, an exponential surface density and an isotropic velocity ellipsoid, that the velocity dispersion is constant with radius, and that it predominantly traces the pressure of the gas\footnote{This hypothesis may not be verified in practice due to the coarse spatial resolution that might prevent the correction of deviations from rotation by our model (see Sect. \ref{kinematics}) or of unresolved large-scale motions.}.
When no dispersion term is added, we will refer to rotation velocity $V_r$, whereas we will talk about corrected velocity $V_c$ when it is included.
For comparison purposes with the work of \citet{Ubler+17}, we use the corrected velocity defined as:
\begin{equation}
 V_c(r) = \sqrt{V_r(r)^2 + 2\sigma^2 \times \frac{r}{R_d}} \text{ .}
 \label{eq:vcirc}
\end{equation}
In this equation, we assume that the gas and stellar disks have the same scale length.
We decide to estimate the velocity at $R_{22}$, therefore:
\begin{equation}
 V_{c22} = V_c(R_{22}) = \sqrt{V_{r22}^2 + 4.4\sigma^2} \text{ .}
 \label{eq:vcirc22}
\end{equation}
The uncertainty on this velocity takes into account the uncertainty on both $V_{r22}=V_r(R_{22})$ and $\sigma$. The velocity dispersion and its associated uncertainty are estimated, respectively, as the median and as the standard deviation of the beam smearing corrected dispersion map (see Sect. \ref{kinematics}).
Uncertainties on the corrected velocity are therefore larger than those on the rotation velocity alone.
Equation \ref{eq:vcirc} was derived by \citet{Burkert+10}, assuming that the velocity dispersion does not depend on the height with respect to the disk plane.
Whereas this assumption may be unrealistic, because it theoretically predicts gas disks with an exponentially growing thickness with radius, other authors \citep[e.g.,][]{Meurer+96, Pelliccia+19} assumed the disk scale height to be constant with radius, leading to a similar relation, but with a weight twice lower for the velocity dispersion (i.e., a factor of 2.2 in Eq. \ref{eq:vcirc22}). Such equations are also comparable to the combined velocity scale introduced by \citet{Weiner+06} in order to refine the agreement between kinematics inferred from integrated spectra and from spatially resolved spectroscopy. They found the best agreement using the parameter:
\begin{equation}
 S_{0.5}^2 = 0.5 V_r^2 + \sigma^2 \text{ .}
 \label{eq:s05}
\end{equation}
This parameter is very similar to half the squared velocity corrected for the asymmetric drift of \citet{Meurer+96}, estimated at $R_{22}$. Equation \ref{eq:vcirc} might therefore over-estimate the contribution of the dispersion.

Since the velocity is measured within a given size, it is also necessary to estimate masses at the same radius. Stellar masses at intermediate redshifts are derived from photometric measurements inside circular apertures. In principle this has to be taken into account. In \citet{Pelliccia+17, Pelliccia+19}, a global lowering of the stellar mass by a factor of 1.54 was inferred assuming that the apertures are large with respect to galaxy sizes and that their mass distribution follows that of an exponential disk. For the MAGIC group sample, we have refined this correction since galaxies are not always well described by a disk alone and because large galaxies can be as large as the photometric apertures.
In order to have estimates of the stellar mass inside $R_{22}$, for each galaxy, we used the morphological decomposition performed with \textsc{Galfit} in order to estimate a correction defined as the ratio of the flux within $R_{22}$ in the galaxy plane over the flux within a 3\arcsec\ diameter circular aperture centered on the galaxy, using a Gaussian smoothing of 0.8\arcsec\ to mimic the methodology used to extract the photometry. On average, for the kinematic sample, the stellar masses at $R_{22}$ are 1.4 times smaller than inside the 3\arcsec\ diameter apertures used to extract photometry.
In some studies, the TFR is determined without considering any mass correction. We have therefore made an analysis of the difference we obtain depending on the correction (see Sect. \ref{sec:impact_selection_error}).

Several linear regressions or fitting methods are commonly used to adjust the TFR on galaxy samples.
The simplest methods are the Ordinary Least Square linear fits (direct, inverse or bissector) that consider one of the variable to depend on a fixed one. This method only takes into account the uncertainty on the dependent parameter. Usually, because the selection is performed on the mass or luminosity of galaxies, inverse fits, considering the velocity as the dependent variable, are used. However, uncertainties on both mass and velocity should be accounted for.
Therefore, more sophisticated methods have been developed, such as (i) the Orthogonal Distance Regression \citep{Boggs90}, that minimizes the distance of the data points orthogonal to the linear function, (ii) MPFITEXY\footnote{\url{https://github.com/williamsmj/mpfitexy}} \citep{Williams+10}, based on the Levenberg-Marquardt algorithm and that additionally takes into account, and eventually adjusts, the intrinsic scatter in the weighting scheme, (iii) HYPERFIT\footnote{\url{http://hyperfit.icrar.org}} \citep{Robotham+15}, which has the same capabilities as MPFITEXY, but which can use various optimization algorithms, including bayesian methods, or (iv) other bayesian methods.
The HYPERFIT method was used for the KROSS sample \citep{Tiley+19}, whereas MPFITEXY was used for the ORELSE \citep{Pelliccia+19} and KMOS3D \citep{Ubler+17} samples. \citet{Ubler+17} used both MPFITEXY and bayesian methods and showed that the agreement is good. We have further taken the public values available online of masses and velocities provided by \citet{Pelliccia+19} and \citet{Tiley+19} to study the agreement of the results obtained by HYPERFIT and MPFITEXY. We found in both ORELSE and KROSS a perfect agreement. We therefore use MPFITEXY in the following analysis, using the inverse linear fit approach with the adjustment of the intrinsic scatter to the relation, as done by other authors mentioned above. We use the formal uncertainties provided by the fit since they are comparable to those obtained using bootstrapping methods by other authors.

In order to make appropriate comparisons with other surveys and to avoid peculiar galaxies with very small uncertainties to drive the fit of the TFR, we introduce systematic uncertainties on both stellar masses and velocities.
For the uncertainties on stellar masses, we adopt a similar strategy as \citet{Pelliccia+19} for the MAGIC group sample by adding in quadrature a systematic uncertainty of 0.2 dex to the uncertainties delivered by the SED fits described in Sect. \ref{Global_pro}. This is rather similar to what is done in \citet{Tiley+19} on the KROSS sample where the uncertainty is constant and equal to 0.2 dex, or in \citet{Ubler+17}, where the uncertainty is around 0.15 dex. For the velocities, we similarly add in quadrature several systematic uncertainties (see Sect. \ref{sec:impact_selection_error}).
We use a systematic uncertainty of 20~km~s$^{-1}$ as a reference.
In order to ensure a proper comparison with other samples, we use the same unique methodology to fit TFR for all the samples.

Lastly, the disk thickness used to infer inclination impacts the deprojected rotation velocity. We use a null thickness whereas \citet{Pelliccia+19}, \citet{Tiley+19} and \citet{Ubler+17} use an intrinsic axial ratio of the scale height to the scale length $q_0$ of 0.19, 0.20, and 0.25 for all galaxies in the ORELSE, KROSS, and KMOS3D samples, respectively. Using such values reduces $\log{(V_r)}$ by $\log{\left(\sqrt{1-q_0^2}\right)}=$ 0.008, 0.009, and 0.014 dex, respectively, for all galaxies, regardless of inclination. This therefore does not add any dispersion in the relations and only modifies the zero-point, by about +0.03-0.05 dex on the mass. The impact on the corrected velocity $V_c$ is lower since the correction only applies on the rotation velocity and not on the velocity dispersion term. These offsets are small compared to the offset we observe between MAGIC and other samples (see Table \ref{tab:diff_TFR}). We further stress that in these studies, no bulge-disk decomposition is performed, which means that the considered thickness also accounts for the bulge. It therefore makes sense to use a lower thickness in our study. Using $q_0=0.1$ instead of $q_0=0$  would lead to a decrease in velocity by $\sim 0.002$ dex only (increase in mass zero-point by $\sim 0.01$ dex), which is negligible. We therefore use a null disk thickness for MAGIC to compare with other studies.

\subsection{Impact of selection, uncertainties, and aperture correction on the stellar mass Tully-Fisher relation}
\label{sec:impact_selection_error}

In this subsection, we study the impact of sample selection, of uncertainties, and of the stellar mass aperture correction on the stellar mass TFR (smTFR) using the rotation velocity. In many studies published so far, the selection is based on the $V_r/\sigma$ ratio \citep[e.g.,][]{Ubler+17, Tiley+19, Pelliccia+19}. Without such a selection, a significant fraction of galaxies appears as outliers in the TFR, with many galaxies having lower rotation velocities than expected at a given stellar mass (or magnitude or baryonic mass).

\begin{table*}
  \caption{Fits of the stellar mass TFR using the rotation velocity for various MAGIC kinematic sample selections, various uncertainties on the velocity and with or without stellar mass correction.}
  \label{tab:smTFR}
  \begin{center}
 \begin{tabular}{cccccccccc}
\hline
\hline
Sample & Mass correction & $\log{(M_*\rm ~[M_\odot])}$ & Slope & $\Delta V$ [km~s$^{-1}$]& $\alpha$ & $\beta$ & $\sigma_{\rm int}$ & $\sigma_{\rm tot}$ & d.o.f./N$_{\rm gal}$ \\
(1) & (2) & (3) & (4) & (5) & (6) & (7) & (8) & (9) & (10) \\
\hline
Final       & Yes   & $>9.0$     & Free  & 20    & $ 4.03 \pm 0.63$ & $ 9.79 \pm 0.09$ &  0.43 &  0.55 &   53 / 55 \\
Final       & Yes   & $>9.0$     & Fixed & 20    &             3.61 & $ 9.82 \pm 0.07$ &  0.38 &  0.50 &   54 / 55 \\
\hline
Relaxed     & Yes   &            & Free  & 20    & $ 3.66 \pm 0.35$ & $ 9.93 \pm 0.07$ &  0.53 &  0.77 &  110 / 112 \\
Strict      & Yes   &            & Free  & 20    & $ 3.75 \pm 0.40$ & $ 9.86 \pm 0.07$ &  0.42 &  0.60 &   75 / 77 \\
Final       & Yes   &            & Free  & 20    & $ 3.81 \pm 0.43$ & $ 9.82 \pm 0.07$ &  0.41 &  0.56 &   65 / 67 \\
Final       & Yes   & $>8.5$     & Free  & 20    & $ 3.67 \pm 0.50$ & $ 9.83 \pm 0.07$ &  0.40 &  0.54 &   59 / 61 \\
Final       & Yes   & $>10.0$    & Free  & 20    & $ 3.49 \pm 1.17$ & $ 9.93 \pm 0.19$ &  0.39 &  0.47 &   26 / 28 \\
Relaxed     & Yes   &            & Fixed & 20    &             3.61 & $ 9.92 \pm 0.07$ &  0.52 &  0.76 &  111 / 112 \\
Strict      & Yes   &            & Fixed & 20    &             3.61 & $ 9.86 \pm 0.07$ &  0.40 &  0.58 &   76 / 77 \\
Final       & Yes   &            & Fixed & 20    &             3.61 & $ 9.83 \pm 0.07$ &  0.38 &  0.54 &   66 / 67 \\
Final       & Yes   & $>8.5$     & Fixed & 20    &             3.61 & $ 9.83 \pm 0.07$ &  0.39 &  0.53 &   60 / 61 \\
Final       & Yes   & $>10.0$    & Fixed & 20    &             3.61 & $ 9.91 \pm 0.09$ &  0.40 &  0.49 &   27 / 28 \\
\hline
Final       & Yes   & $>9.0$     & Free  &       & $ 3.69 \pm 0.54$ & $ 9.86 \pm 0.08$ &  0.47 &  0.53 &   53 / 55 \\
Final       & Yes   & $>9.0$     & Free  & 10    & $ 3.81 \pm 0.57$ & $ 9.84 \pm 0.08$ &  0.46 &  0.54 &   53 / 55 \\
Final       & Yes   & $>9.0$     & Fixed &       &             3.61 & $ 9.87 \pm 0.07$ &  0.46 &  0.52 &   54 / 55 \\
Final       & Yes   & $>9.0$     & Fixed & 10    &             3.61 & $ 9.85 \pm 0.07$ &  0.44 &  0.52 &   54 / 55 \\
\hline
Final       & No    & $>9.0$     & Free  & 20    & $ 3.56 \pm 0.55$ & $ 9.94 \pm 0.08$ &  0.38 &  0.49 &   53 / 55 \\
Final       & No    & $>9.0$     & Fixed & 20    &             3.61 & $ 9.94 \pm 0.07$ &  0.37 &  0.50 &   54 / 55 \\
\hline
 \end{tabular}
\tablefoot{
 The relaxed (strict) selection corresponds to all galaxies fulfilling S/N $>$ 30 (40) and R$_{eff}$/FWHM $>$ 0.25 (0.5), whereas problematic AGN and bulge-dominated galaxies are further removed from the strict selection in the final sample (defined in Sect. \ref{Sample_Selection_Criteria}). The two first rows correspond to the TFR fits used as reference. The fixed slope corresponds to the slope we derived for the KROSS rotation dominated sample. (1) Identification of the MAGIC group sample used. (2) Indication about mass correction with $R_{22}$. (3) Threshold used on the stellar mass. (4) Constraint on the slope. (5) Systematic uncertainty on velocity added in quadrature. (6) Value of the TFR slope. (7) Value of the TFR zero-point. (8) Intrinsic dispersion around the TFR obtained by enforcing the $\chi^2$ to be equal to unity. (9) Total dispersion around the TFR. (10) Number of degrees of freedom/number of galaxies.
 }
  \end{center}
\end{table*}

\begin{figure}
 \includegraphics[width=\columnwidth]{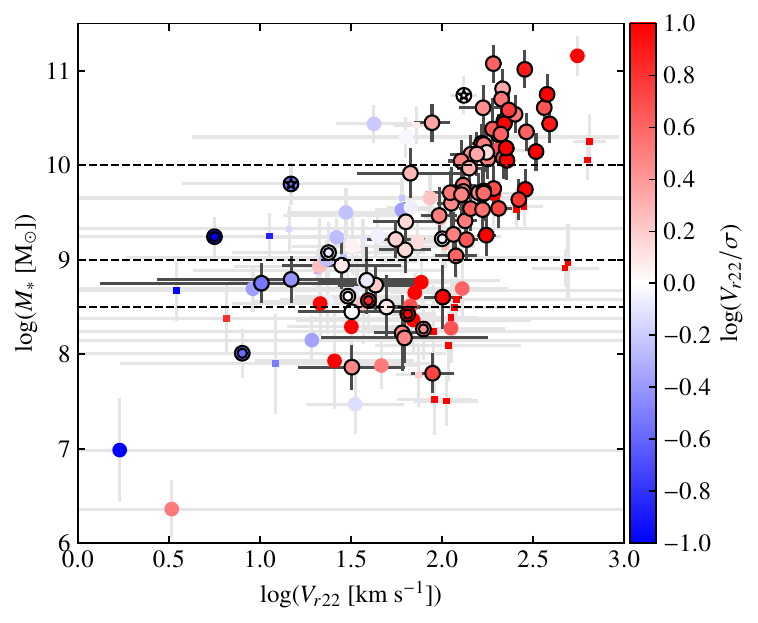}
 \caption{Stellar mass TFR for the MAGIC group sample, using the rotation velocity. The black circles correspond to galaxies fulfilling the strict conditions (S/N $>$ 40 and R$_{eff}$/FWHM $>$ 0.5), whereas circles without contours indicate additional galaxies with more relaxed conditions (S/N $>$ 30 and R$_{eff}$/FWHM $>$ 0.25). Small squares show all other galaxies on the main sequence with velocity flags larger than or equal to 2. The color indicates the values of $V_{r22}/\sigma$. The three horizontal dashed lines indicate the threshold in mass used.
 For galaxies in the kinematic sample, we also identify bulge-dominated galaxies (double circles) and problematic AGN (circled stars) that are discarded from the analysis of the final sample.
 }
 \label{fig:TFR_vs}
\end{figure}

In a first attempt to draw the smTFR, we included all the star-forming galaxies with velocity field flags larger than or equal to 2. In Fig. \ref{fig:TFR_vs}, where colors indicate the ratio of $V_{r22}/\sigma$, we see, as in other TFR studies, that most of the outliers have a low ratio. This does not necessarily mean that the velocity dispersion is large but rather that the rotation velocity is low. For low-mass systems, this can be due to the shape of the rotation curves or to a lack of spatial resolution. We further identified in this figure galaxies in the kinematic sample defined in Sect. \ref{Sample_Selection_Criteria} with both relaxed (S/N $>$ 30 and R$_{eff}$/FWHM $>$ 0.25) and strict (S/N $>$ 40 and R$_{eff}$/FWHM $>$ 0.5) selection conditions, as well as galaxies removed due to large bulge-to-disk ratio or AGN.
Our selection criteria mainly discard low-mass galaxies and most of the strong outliers to the relation, except at low stellar masses, without any prior on the kinematic measurements.
We therefore also studied the smTFR using various thresholds in mass.
A first threshold of $10^{8.5}$~M$_\sun$ seems necessary in terms of completeness and provides a sample of 61 galaxies. A second threshold of $10^{9}$~M$_\sun$ enables us to keep galaxies with more robust mass and velocity estimates and leads to 55 galaxies. Finally, a threshold of $10^{10}$~M$_\sun$ produces a sample of 28 galaxies that is more similar to the KROSS disky and KMOS3D samples.
For the $10^{9}$ ($10^{8.5}$)~M$_\sun$ limit, the agreement between morphological and kinematic position angles is better than 40\degr\ for all but one (two) galaxies.
Under the reasonable assumption that the ionized gas is rotating within the same plane as the stellar disk, this indicates that both morphology and kinematic measurements are robust.
These results are clearly in favor of observational biases to explain outliers to the smTFR rather than to a change of dynamical support.

We perform fits of the smTFR for all those various samples both without and with constraints on the slope (see Table \ref{tab:smTFR}, second panel). When no constraint is set on the slope, we observe that the slope and the zero-point are correlated but that both intrinsic and total scatter reduce when increasingly stringent selection criteria are used.
The absence of a monotonic trend on the slope seems to indicate that a few galaxies can have a significant impact on the slope. 
Because the choice of a proper $V_{ref}$ impacts the observed variations of the zero-point when the slope is free, we use fixed slopes to compare zero-points.
We decided to fix the slope to the one we find for the KROSS rotation dominated sample since it is the closest to the slope we find on average without constraint (see Sect. \ref{sec:smTFR} and Table \ref{tab:TFR}). Using this fixed slope still provides the same result for the scatter. The zero-point seems to slightly decrease by about 0.1 dex when outliers are removed, which is expected since outliers have on average large masses with respect to the expectations from their rotation velocity. These variations of the zero-point are within the uncertainties. Nevertheless, the zero-point increases by about 0.1 dex when we only keep galaxies with masses larger than $10^{10}$~M$_\sun$. The free slope for this subsample is lower than for the other ones but this could be due to the stellar mass range of this sample being of the same order of magnitude as the intrinsic scatter.
We decide to use, as a reference, the sample with a mass limit of $10^{9}$~M$_\sun$ that represents a good compromise between the accuracy of measurements and the statistics since it contains 55 galaxies. We reached similar conclusions using slopes fixed to the ones found using the KROSS disky and ORELSE samples.

We further studied the impact of the uncertainties on the velocity to the fits of the smTFR. To do that, we added in quadrature a systematic value of 10 or 20 km~s$^{-1}$ to the uncertainty derived from the kinematics modeling (see Table \ref{tab:smTFR}, third panel). When this systematic uncertainty increases, the slope increases whereas the zero-point decreases. Since the zero-point variation can be due to the slope variation, we also fixed the slope to that found for the KROSS rotation dominated sample. In that case, the decrease in the zero-point is less and is around 0.05 dex for the case of a systematic uncertainty of 20 km~s$^{-1}$, which is within the typical uncertainty of 0.07 dex. We checked that the impact is similar whatever the MAGIC subsample used. We also tried to add a relative uncertainty rather than an absolute one, but this provides the same behavior.
It is also interesting to notice that this does not affect significantly the intrinsic and total scatters. We decided to use the 20 km~s$^{-1}$ value because it is in better agreement with the typical uncertainties observed or used for other published samples.

Last, we also investigated the impact of the stellar mass aperture correction within the $R_{22}$ radius (see Table \ref{tab:smTFR}, bottom panel). Using a fixed slope shows that the mass correction reduces the zero-point by about 0.12 dex, which is expected since on average the mass is reduced by a factor of 1.4, corresponding to a correction of -0.15 dex. Correcting the masses also clearly increases the slope. This reflects the fact that the mass correction is larger for low-mass galaxies than for high-mass ones since the size of a galaxy correlates with its mass.
It is worth noticing that for this same reason, the zero-point decrease observed when we use more stringent sample selection constraints is stronger when the mass is not corrected, which means that correcting the mass is necessary to avoid over-estimating the zero-point variation with mass.
The scatter is not much affected when the slope is fixed whereas it is increased when it is free. The same conclusions are reached whatever the sample used.
In conclusion, we can expect a rise of the zero-point of around 0.25 dex at maximum depending on the uncertainties (0.05 dex), on the stellar mass correction (0.15 dex) and sample selection (0.05 dex) with respect to the reference we adopt in the following analysis.

\subsection{Stellar mass Tully-Fisher relation}
\label{sec:smTFR}

\begin{table*}
  \caption{Results of the fits of the various TFR for reference samples and for MAGIC using fixed and free slopes.
  }
  \label{tab:TFR}
  \begin{center}

 \begin{tabular}{cccccccccc}
\hline
\hline
Sample & $\log{(M_*\rm ~[M_\odot])}$ & TFR & Tracer & Slope & $\alpha$ & $\beta$ & $\sigma_{\rm int}$ & $\sigma_{\rm tot}$ & d.o.f/N$_{\rm gal}$ \\
(1) & (2) & (3) & (4) & (5) & (6) & (7) & (8) & (9) & (10) \\
\hline
\textbf{MAGIC} & $>9.0$     & smTFR    & $V_{r22}$  & Free  & $ 4.03 \pm 0.63$ & $ 9.79 \pm 0.09$ &  0.43 &  0.55 &   53 / 55 \\
ORELSE               &            & smTFR    & $V_{r22}$  & Free  & $ 3.18 \pm 0.41$ & $10.23 \pm 0.09$ &  0.48 &  0.63 &   75 / 77 \\
MAGIC       & $>9.0$     & smTFR    & $V_{r22}$  & Fixed &             3.18 & $ 9.85 \pm 0.06$ &  0.34 &  0.46 &   54 / 55 \\
KROSS rotdom         &            & smTFR    & $V_{r22}$  & Free  & $ 3.61 \pm 0.31$ & $10.61 \pm 0.07$ &  0.66 &  0.71 &  257 / 259 \\
MAGIC       & $>9.0$     & smTFR    & $V_{r22}$  & Fixed &             3.61 & $ 9.82 \pm 0.07$ &  0.38 &  0.50 &   54 / 55 \\
KROSS disky          &            & smTFR    & $V_{r22}$  & Free  & $ 4.42 \pm 0.43$ & $10.21 \pm 0.04$ &  0.34 &  0.45 &  110 / 112 \\
MAGIC       & $>9.0$     & smTFR    & $V_{r22}$  & Fixed &             4.42 & $ 9.76 \pm 0.08$ &  0.46 &  0.60 &   54 / 55 \\
\hline
\textbf{MAGIC} & $>9.0$     & smTFR    & $V_{c22}$  & Free  & $ 4.35 \pm 0.59$ & $ 9.51 \pm 0.10$ &  0.33 &  0.48 &   53 / 55 \\
MAGIC       & $>10.0$    & smTFR    & $V_{c22}$  & Free  & $ 3.97 \pm 1.32$ & $ 9.65 \pm 0.27$ &  0.37 &  0.47 &   26 / 28 \\
ORELSE               &            & smTFR    & $V_{c22}$  & Free  & $ 3.24 \pm 0.33$ & $ 9.96 \pm 0.06$ &  0.37 &  0.51 &   75 / 77 \\
MAGIC       & $>9.0$     & smTFR    & $V_{c22}$  & Fixed &             3.24 & $ 9.65 \pm 0.05$ &  0.27 &  0.41 &   54 / 55 \\
KMOS3D               &            & smTFR    & $V_{c22}$  & Free  & $ 2.91 \pm 0.28$ & $ 9.96 \pm 0.06$ &  0.12 &  0.27 &   63 / 65 \\
MAGIC       & $>9.0$     & smTFR    & $V_{c22}$  & Fixed &             2.91 & $ 9.69 \pm 0.05$ &  0.26 &  0.39 &   54 / 55 \\
MAGIC       & $>10.0$    & smTFR    & $V_{c22}$  & Fixed &             2.91 & $ 9.85 \pm 0.07$ &  0.26 &  0.38 &   27 / 28 \\
KMOS3D               &            & smTFR    & $V_{c22}$  & Fixed &             3.60\tablefootmark{~(a)} & $ 9.84 \pm 0.04$ &  0.17 &  0.32 &   64 / 65 \\
MAGIC       & $>9.0$     & smTFR    & $V_{c22}$  & Fixed &             3.60 & $ 9.60 \pm 0.06$ &  0.28 &  0.43 &   54 / 55 \\
MAGIC       & $>10.0$    & smTFR    & $V_{c22}$  & Fixed &             3.60 & $ 9.72 \pm 0.08$ &  0.32 &  0.44 &   27 / 28 \\
\hline
\textbf{MAGIC} & $>9.0$     & bmTFR   & $V_{r22}$  & Free  & $ 3.55 \pm 0.49$ & $ 9.99 \pm 0.07$ &  0.35 &  0.45 &   53 / 55 \\
\hline
\textbf{MAGIC} & $>9.0$     & bmTFR   & $V_{c22}$  & Free  & $ 3.74 \pm 0.45$ & $ 9.76 \pm 0.08$ &  0.25 &  0.38 &   53 / 55 \\
MAGIC       & $>10.0$    & bmTFR   & $V_{c22}$  & Free  & $ 3.64 \pm 1.06$ & $ 9.83 \pm 0.22$ &  0.31 &  0.41 &   26 / 28 \\
KMOS3D               &            & bmTFR   & $V_{c22}$  & Free  & $ 2.21 \pm 0.22$ & $10.26 \pm 0.05$ &  0.01 &  0.23 &   63 / 65 \\
MAGIC       & $>9.0$     & bmTFR   & $V_{c22}$  & Fixed &             2.21 & $ 9.95 \pm 0.04$ &  0.21 &  0.31 &   54 / 55 \\
MAGIC       & $>10.0$    & bmTFR   & $V_{c22}$  & Fixed &             2.21 & $10.11 \pm 0.06$ &  0.20 &  0.30 &   27 / 28 \\
KMOS3D               &            & bmTFR   & $V_{c22}$  & Fixed &             3.75\tablefootmark{~(b)} & $10.00 \pm 0.04$ &  0.19 &  0.33 &   64 / 65 \\
MAGIC       & $>9.0$     & bmTFR   & $V_{c22}$  & Fixed &             3.75 & $ 9.75 \pm 0.05$ &  0.24 &  0.38 &   54 / 55 \\
MAGIC       & $>10.0$    & bmTFR   & $V_{c22}$  & Fixed &             3.75 & $ 9.81 \pm 0.08$ &  0.31 &  0.42 &   27 / 28 \\
 \hline
 \end{tabular}
\tablefoot{
For KMOS3D, we use uncertainties on stellar mass of 0.2 dex rather than the 0.15 dex used in \citet{Ubler+17} in order to provide a better comparison to the other samples. The results for reference samples are provided without correction of the velocity due to the intrinsic axial ratio and without correction of the stellar mass. The latter correction would generate an offset of their zero-point of -0.15 dex.
  The average redshift is around 0.9 for all comparison samples used, slightly higher than the average redshift around 0.7 of the MAGIC final kinematic sample.
  (1) Identification of the sample used. Bold font indicates the MAGIC group TFR we consider as references. (2) Threshold used on the stellar mass. (3) Type of TFR studied. (4) Tracer used to infer the TFR. (5) Constraint on the slope. When fixed, the value of the slope used is the one indicated in column (6) and corresponds to the free slope of one of the comparison samples. (6) Value of the TFR slope. (7) Value of the TFR zero-point. (8) Intrinsic dispersion around the TFR obtained by enforcing the $\chi^2$ to be equal to unity. (9) Total dispersion around the TFR. (10) Number of degrees of freedom/number of galaxies.
\tablefoottext{a}{Value from \citet{Reyes+11} used in \citet{Ubler+17} to compare with local Universe.}
\tablefoottext{b}{Value from \citet{Lelli+16} used in \citet{Ubler+17} to compare with local Universe.}
}
  \end{center}
\end{table*}

\begin{table*}[h!]
  \caption{TFR zero-point difference between various reference samples and the MAGIC final kinematic sample, with a mass threshold of $10^9$~M$_\sun$, using similar slopes.
  }
  \label{tab:diff_TFR}
  \begin{center}
 \begin{tabular}{cccccccc}
\hline
\hline
Sample & TFR & Tracer & $\alpha$ & \multicolumn{2}{c}{$\Delta \log{M}$} & \multicolumn{2}{c}{$\Delta \log{V}$} \\
(1) & (2) & (3) & (4) & (5) & (6) & (7) & (8) \\
\hline
ORELSE                  & smTFR    & $V_{r22}$    &  3.18 &  0.38 & 0.23  & -0.119 & -0.072  \\
KROSS rotdom            & smTFR    & $V_{r22}$    &  3.61 &  0.79 & 0.64  & -0.219 & -0.177  \\
KROSS disky             & smTFR    & $V_{r22}$    &  4.42 &  0.45 & 0.30  & -0.102 & -0.068  \\
\hline
ORELSE                  & smTFR    & $V_{c22}$    &  3.24 &  0.31 & 0.16  & -0.096 & -0.049  \\
KMOS3D                  & smTFR    & $V_{c22}$    &  2.91 &  0.27 & 0.12  & -0.093 & -0.041  \\
KMOS3D                  & smTFR    & $V_{c22}$    &  3.60\tablefootmark{~(b)} &  0.24 & 0.09  & -0.067 & -0.025  \\
KMOS3D\tablefootmark{~(a)} & smTFR    & $V_{c22}$    &  2.91 &  0.11 & -0.04  & -0.038 & 0.014   \\
KMOS3D\tablefootmark{~(a)} & smTFR    & $V_{c22}$    &  3.60\tablefootmark{~(b)} &  0.12 & -0.03  & -0.033 & 0.008   \\
\hline
KMOS3D                  & bmTFR   & $V_{c22}$     &  2.21 &  0.31 & 0.16  & -0.140 & -0.072  \\
KMOS3D                  & bmTFR   & $V_{c22}$     &  3.75\tablefootmark{~(c)} &  0.26 & 0.11  & -0.069 & -0.029  \\
KMOS3D\tablefootmark{~(a)} & bmTFR   & $V_{c22}$     &  2.21 &  0.15 & 0.00   & -0.068 &  0.000  \\
KMOS3D\tablefootmark{~(a)} & bmTFR   & $V_{c22}$     &  3.75\tablefootmark{~(c)} &  0.19 & 0.04   & -0.051 & -0.011  \\
 \hline
 \end{tabular}
\tablefoot{
(1) Identification of the comparison sample. (2) Type of TFR studied. (3) Tracer used to infer the TFR. (4) Value of the slope used. Zero-point difference in mass / velocity: (5) / (7) with no correction; (6) / (8) taking into account the -0.15 dex offset for the correction of stellar masses within $R_{22}$.
\tablefoottext{a}{These lines correspond to the difference between the KMOS3D sample and the MAGIC sample with a mass threshold of $10^{10}$~M$_\sun$.}
\tablefoottext{b}{Value from \citet{Reyes+11} used in \citet{Ubler+17} to compare with local Universe.}
\tablefoottext{c}{Value from \citet{Lelli+16} used in \citet{Ubler+17} to compare with local Universe.}
}
  \end{center}
\end{table*}

\begin{figure}
 \includegraphics[width=\columnwidth]{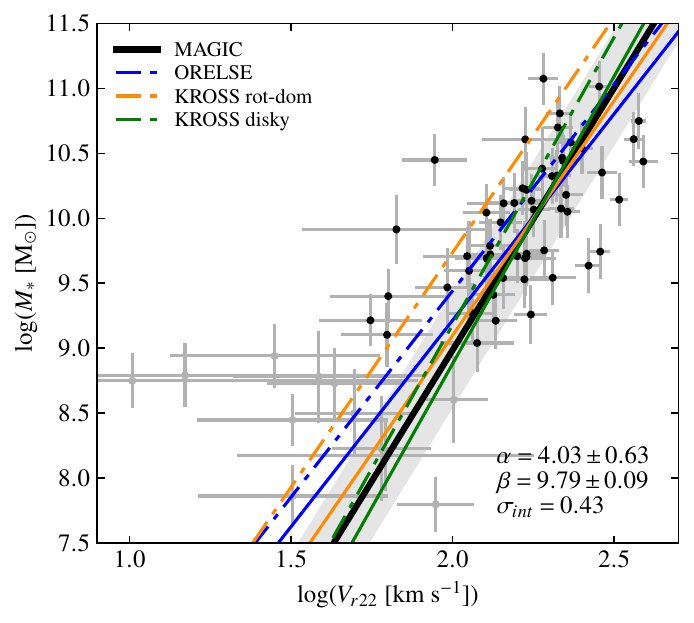}
 \caption{Stellar mass TFR for the MAGIC final kinematic sample, using the rotation velocity. The black dots indicate galaxies used for the reference subsample with stellar masses above $10^9$~M$_\sun$, whereas gray dots indicate the other galaxies for information. The black line corresponds to the fit to the MAGIC group sample with a free slope and the intrinsic scatter is represented with the gray shaded area. The parameters of this fit are indicated in the figure. Colored dotted-dashed lines correspond to fits on other samples, with an offset of -0.15 dex for the correction of the stellar mass. The slopes of these fits are used to perform additional fits of the smTFR on the MAGIC group sample. The corresponding fits are shown with continuous lines having the same colors as the samples used as reference for the slope.}
 \label{fig:smTFR_Vrot}
\end{figure}

We showed in the previous section that not correcting the mass for the limited photometric apertures possibly biases the slope of the relation in addition to obviously shifting the zero-point. The studies we are comparing to use similar initial mass functions to compute their stellar mass but do not make any correction for this aperture effect. We will assume that, at fixed slope, the main impact is a shift of the zero-point by 0.15 dex, although it depends on the size of the apertures used.
The results we are discussing here and in Sect. \ref{sec:bmTFR} are summarized in Tables \ref{tab:TFR} and \ref{tab:diff_TFR}.

We fitted the smTFR for the KROSS rotation dominated and disky samples as well as for the ORELSE sample with the same fitting routine as for the MAGIC group sample and letting the slope free.
We then fixed the slope for the MAGIC group sample to those reference slopes to infer a possible zero-point shift (see Fig. \ref{fig:smTFR_Vrot}).
We find a significant evolution of the zero-point toward lower values for the MAGIC group sample with respect to each reference sample.
The decrease in the zero-point is more significant compared to the KROSS samples \citep{Tiley+19} that are supposed to have less galaxies in dense environments than the ORELSE sample \citep{Pelliccia+19} dominated by cluster galaxies. The offset with the KROSS rotation dominated sample is about 0.79 dex, whereas it reduces to 0.45 dex with the KROSS disky sample. This is expected from the selection function of those two samples, which is based on two different thresholds of $V_{r22}/\sigma$ (1 and 3, respectively). Our sample selection seems to better match that of the rotation dominated sample since its average stellar masses is identical to that of the MAGIC final sample ($10^{10.0}$~M$_\sun$) and because they have a very similar standard deviation. It is however instructive to observe that the offset is still significant with respect to the KROSS disky sample that has an average stellar mass of $10^{10.2}$~M$_\sun$. The offset with respect to the ORELSE sample is 0.38 dex. The ORELSE and MAGIC group samples are also rather similar despite the ORELSE sample has a slightly lower average stellar mass, in better agreement with the MAGIC group sample with a cut in stellar mass of $10^{8.5}$~M$_\sun$, however, using this subsample still gives an offset of 0.37 dex. If we consider the correction of the mass for apertures, these shifts lower to 0.64 dex, 0.30 dex and 0.23 dex. These offsets are not within the uncertainties on the parameter.

The slope we obtain for the MAGIC group sample is in between the two KROSS samples.
If we use a stellar mass threshold of $10^{8.5}$~M$_\sun$, the slope $\alpha=3.67$ decreases and agrees well with that of the KROSS rotation dominated sample, and is in better agreement with that of the ORELSE sample, despite it remains larger. This difference could be due to the stellar mass correction that increases the slope (see Sect. \ref{sec:impact_selection_error} and Table \ref{tab:smTFR}). However, in any case, the intrinsic and total scatters for the MAGIC group sample are lower than for the KROSS rotation dominated and ORELSE samples, regardless of whether the slope is free or fixed.
These scatters are nevertheless higher than for the KROSS disky sample and are the highest when the slope is fixed to that of this sample.
For the latter sample, the slope is steep and the scatter is reduced due to the selection function that naturally removes slow rotators.

In order to ensure that the different selection does not bias the comparisons, we checked that the median value of $V_r(R_{22})/\sigma$ for our sample is not larger than that of the other samples that used a limit of either 1 or 3. The MAGIC final kinematic sample is clearly not biased toward very high values of this parameter since the lowest value is 0.3 and the median is 3.6. As a comparison, the median value for the ORELSE sample is 3.0.

\begin{figure}
 \includegraphics[width=\columnwidth]{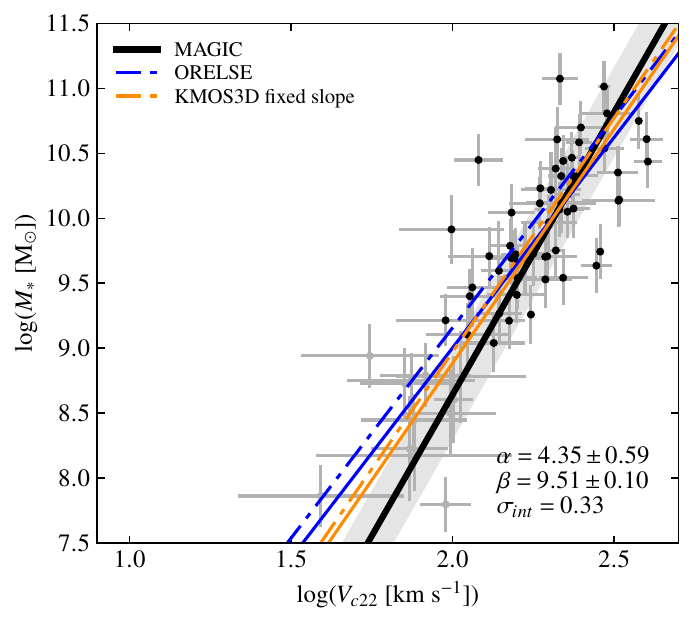}
 \caption{Stellar mass TFR for the MAGIC final kinematic sample, using the corrected velocity. See Fig. \ref{fig:smTFR_Vrot} for the description of lines and symbols. An offset of -0.15 dex has been applied for the comparison samples to account for the stellar mass correction within $R_{22}$.
 }
 \label{fig:smTFR_Vcirc}
\end{figure}

We did a similar analysis including the contribution of the dispersion term, that is, using the corrected rather than the rotation velocity. We used the KMOS3D \citep{Ubler+17} and the ORELSE samples, for which the dispersion contribution is available, as references for the slope. For the KMOS3D sample, we used both the fixed slope they employed in their analysis, taken from \citet{Reyes+11} in the local Universe, and the one resulting from the fit with a free slope we performed on their dataset (see Fig. \ref{fig:smTFR_Vcirc}).
The primary effect of the inclusion of the dispersion is to increase the rotation velocity, to increase the slope of the relation and to decrease the scatter around the smTFR even for galaxies below the mass threshold.
The median velocity dispersion in our sample is $\sim 40$~km~s$^{-1}$. Using it to infer the corrected velocity therefore mainly impacts galaxies with a low rotation velocity, which is the reason for the slope to be steeper and for the zero-point at $\log{(V_{ref}~[\textrm{km~s}^{-1}])}=2.2$ to be reduced.
Several studies assume that the ratio of the rotation velocity to the velocity dispersion inferred from the ionized gas indicates the dynamical support of galaxies \citep[see, e.g.,][]{Burkert+10, Wuyts+16, Turner+17a, Ubler+19}, although it is theoretically valid only for collisionless systems.
Whereas this is not the scope of this paper, we stress that a high velocity dispersion associated with a small disk size might mean that the galaxy ionized gas kinematics are not correctly resolved. Indeed, if a galaxy has an intrinsic rotation but its ionized gas distribution is not resolved, the observed line will be enlarged. In other words, the rotation will be encoded in the velocity dispersion. This is the typical beam smearing effect that affects mainly the central parts of galaxies at intermediate to high redshift.
Our analysis of the impact of selection on the smTFR seems to support this. Indeed, our final sample, which is constructed without any prior on the dynamical support, contains very few outlying and dispersion-dominated galaxies, whereas such galaxies are included when the constraints on galaxy sizes are relaxed, similarly to what is observed in other samples.
We also note that the uncertainties on velocity dispersion are usually large due (i) to the limitation of the beam smearing correction, (ii) to the limited spectral resolution of the various spectrographs used since $R=3000$ corresponds to an instrumental dispersion of 40~km~s$^{-1}$, and (iii) to the lesser accuracy to retrieve the second order moment with respect to the first order moment.

As for the previous smTFR determined using the rotation velocity, we find a significant decrease in the zero-point with respect to the other samples when a fixed slope is used. The offset with respect to the KMOS3D sample is 0.27 dex or 0.24 dex when we use either the value of the fit with a free slope we performed on the KMOS3D dataset, or the fixed slope from \citet{Lelli+16} that \citet{Ubler+17} used to compare with local Universe.
We stress that the KMOS3D sample is mainly composed of massive galaxies since its median stellar mass is $10^{10.5}$~M$_\sun$. Using a stellar mass threshold of $10^{10}$~M$_\sun$ rather than $10^9$~M$_\sun$ for the MAGIC sample reduces the offset by more than 0.1 dex.
Further assuming a correction for the stellar mass would make the zero-point for MAGIC compatible with the KMOS3D sample.
However, we emphasize that this MAGIC subsample contains only 27 galaxies. The offset for the ORELSE sample is 0.31 dex, rather similar to the offset found using the rotation velocity, reduced to 0.16 dex accounting for the mass correction. The better agreement with KMOS3D, which is supposed to be dominated by galaxies in low-density environment seems to point toward methodological differences, either in the mass estimation or in the kinematics extraction.

Similarly to the smTFR determined with the rotation velocity, the intrinsic and total scatters are lower in the MAGIC group sample than in the ORELSE sample. On the contrary, the scatter in the KMOS3D sample is lower than for the MAGIC group sample. This probably indicates strong differences in sample selections among the various samples discussed here.

\subsection{Baryonic mass Tully-Fisher relation}
\label{sec:bmTFR}

\begin{figure}
 \includegraphics[width=\columnwidth]{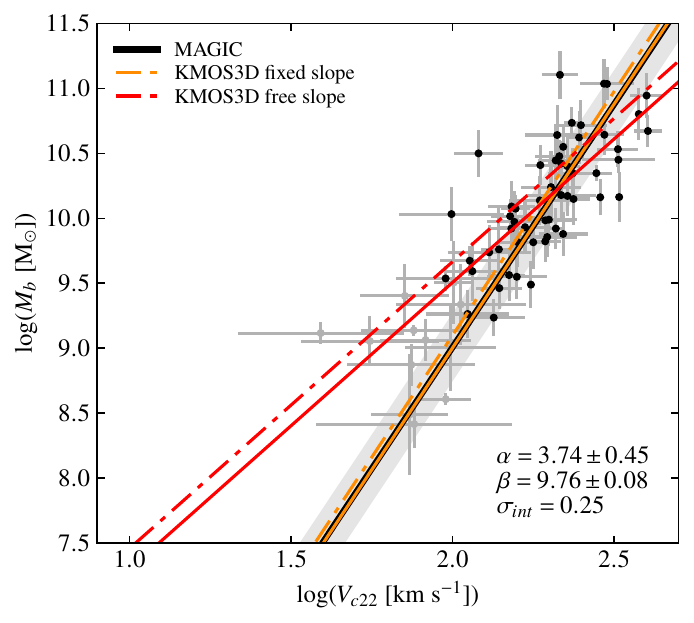}
 \caption{Baryonic mass TFR for the MAGIC final kinematic sample, using the corrected velocity. See Fig. \ref{fig:smTFR_Vrot} for the description of lines and symbols.
 An offset of -0.15 dex has been applied for the comparison samples to account for the stellar mass correction within $R_{22}$.
 }
 \label{fig:bmTFR_Vcirc}
\end{figure}

In order to complement the previous smTFR analysis, we also tried to include the gas mass in the mass budget in order to infer the baryonic mass TFR (bmTFR) because the gas content of galaxies is supposed to differ depending on their environment.
We have estimated gas masses for our sample from the Kennicutt-Schmidt relation between the SFR surface density and the gas surface density \citep{Kennicutt98b}, assuming the gas density to be constant over $R_{22}$, leading to:
\begin{equation}
M_{g} = (4\times 10^9 \times \textrm{SFR})^{5/7} \times (\pi R_{22}^2)^{2/7} \text{ ,}
 \label{gas_mass_sed}
\end{equation}
where $M_{g}$ is the mass of gas in M$_\sun$, $R_{22}$ is in pc, and SFR is in M$_\sun$~yr$^{-1}$. SFR can be deduced either from SED fitting or from \oii\ flux. For the latter, we derived the average SFR per square parsec using the \citep{Kennicutt98a} relation, assuming that the flux is emitted within $R_{22}$. Combining these laws, the relation between the \oii\ flux, $F_{[\ion{O}{ii}]}$, in erg s$^{-1}$ cm$^{-2}$, and the gas mass, $M_{g}$, in M$_\sun$ is:
\begin{equation}
M_{g} = (4.756\pm 1.944)\times 10^{-23} \times (\pi R_{22}^2)^{2/7} \times (4\pi D_{L}^2 F_{[\ion{O}{ii}]})^{5/7} \text{ ,}
 \label{gas_mass}
\end{equation}
where $D_{L}$ is the luminosity distance at the redshift of the source in cm, and $R_{22}$ is in pc. In this relation, we used the \oii\ flux corrected for galactic extinction, assuming the Milky Way dust attenuation curve from \citet{Cardelli+89}, and for internal dust-extinction, using the absorption deduced from the SED fitting and the \citet{Calzetti+00} dust attenuation law.
We find a good correlation between those two estimates. For our final kinematic sample, masses of gas inferred from the \oii\ doublet are $\sim 0.15$ dex larger than those inferred from SED fitting on average, the difference being larger at low mass. In Table \ref{tab:dynamics}, we provide the mass of gas inferred from SED fitting that is used as a reference in the following analysis, despite it corresponds to SFR integrated over longer timescales ($\sim 1$ Gyr) compared to SFR derived from the \oii\ doublet ($\sim 10$ Myr). Indeed, nebular lines might not be the best proxy to infer gas masses, especially in dense environments where ionization could arise from other mechanisms than photo ionization \citep[e.g.,][]{Epinat:2018, Boselli+19} but also because SFR could be enhanced temporarily by environment/interactions, therefore biasing gas mass measurements. Similarly, gas masses inferred from scaling relations \citep[e.g.,][]{Scoville+17, Tacconi+18},  might also be inappropriate for our sample since such scaling relations have been obtained independently from environment.

The impact of including gas masses is more pronounced for low-mass galaxies that have a larger gas fraction on average. This is reflected in the bmTFR fit, that has a less steep slope ($\alpha=3.55$) than for the smTFR fit ($\alpha=4.03$), in agreement with previous studies. Including gas mass also increases the overall mass, and therefore increases the zero-point. The samples we used for the comparison of the smTFR using the rotation velocity in Sect. \ref{sec:smTFR} do not provide both rotation velocities and baryonic masses. Nevertheless, the KMOS3D sample enables us to make a comparison of the bmTFR using the corrected velocity (see Fig. \ref{fig:bmTFR_Vcirc}). The inclusion of the velocity dispersion contribution induces a steeper slope, as for the smTFR. When we fix the slope, we find a zero-point offset with respect to the KMOS3D sample of 0.31 dex or 0.25 dex depending on the slope used. This difference is larger than for the smTFR because the difference between fixed and free slopes on the KMOS3D data is larger. We also checked that using a threshold of $10^{10}$~M$_\sun$ for the selection and no correction of the stellar mass for the MAGIC group sample leads to lower offsets of 0.15 dex ($\alpha=2.21$) and 0.19 dex ($\alpha=3.75$), which reverses the difference. This offset is slightly more important than the one for the smTFR, meaning that the MAGIC zero-point is lower than for other samples, which probably reflects a difference in the gas mass estimates. Using gas masses obtained from Eq. \ref{gas_mass} leads to lower free slopes and to lower offsets of the zero-point at fixed slope for the $10^{9}$~M$_\sun$ mass threshold, whereas it provides substantially similar results with the $10^{10}$~M$_\sun$ mass threshold.
As for the smTFR, the intrinsic and total scatters are larger for the MAGIC group sample than for the KMOS3D sample.

\section{Interpretation of the TFR evolution}
\label{results}

It seems clear from the previous section that for any TFR studied, the stellar mass for a given rotation velocity is lower for the MAGIC group sample than for other samples used as reference, by at least 0.1 dex, or reversely that the rotation velocity at a given stellar mass is larger for MAGIC. Among these samples, MAGIC is the only one that only contains galaxies in dense groups or clusters, the ORELSE sample being composed of both cluster and field galaxies. Our findings might therefore point to an effect of environment on the baryonic content of galaxies. In this section we are first investigating the baryon fraction before exploring two main hypotheses, either the quenching of star formation or the contraction of baryons, to interpret these results.

\subsection{Stellar and baryonic matter fraction}
\label{sec:bar_fraction}

\begin{figure}
 \includegraphics[width=\columnwidth]{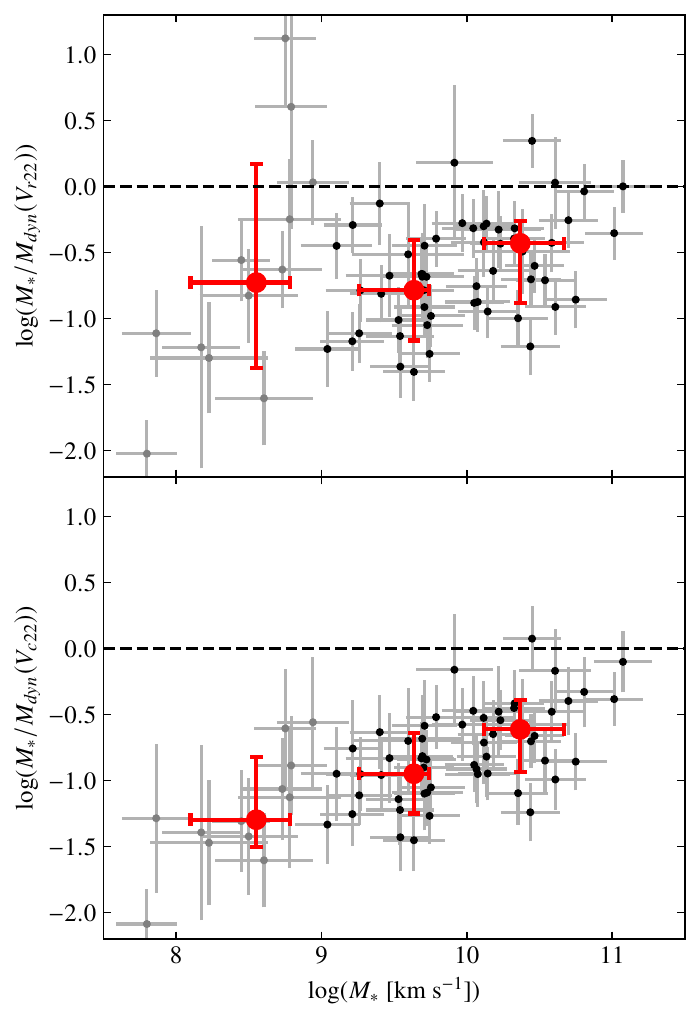}
 \caption{Stellar mass fraction as a function of the stellar mass, within $R_{22}$. Black dots correspond to the final MAGIC kinematic sample with stellar masses above $10^9$~M$_\sun$, whereas the gray dots are for galaxies with masses below this limit. The black horizontal dashed line marks the theoretical upper limit of a fraction unity. The three red dots correspond to median values for the three following stellar mass bins: $M_* \le 10^9$~M$_\sun$, $10^9$~M$_\sun$ $< M_* \le 10^{10}$~M$_\sun$ and $M_* > 10^{10}$~M$_\sun$. They contain, respectively, 12, 27, and 27 galaxies. The errors bars indicate the 16th and 84th percentiles in each bin. The dynamical mass is computed either using the rotation (top) or corrected (bottom) velocities.}
 \label{fig:smfrac}
\end{figure}

\begin{figure}
 \includegraphics[width=\columnwidth]{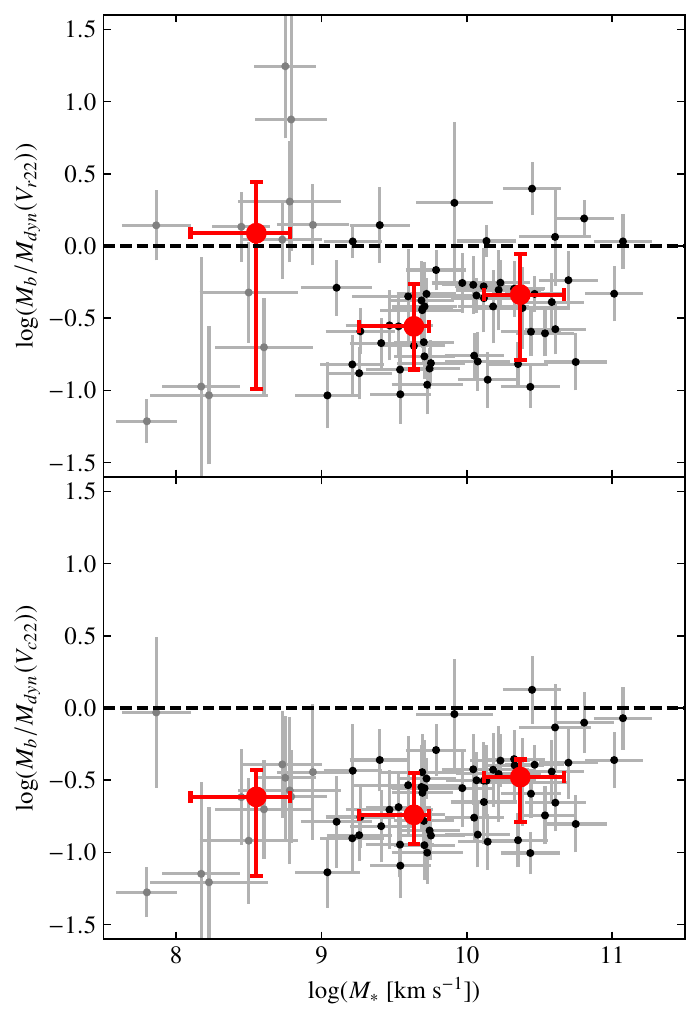}
 \caption{Baryonic mass fraction as a function of the stellar mass, within $R_{22}$. See Fig. \ref{fig:smfrac} for the description of symbols.}
 \label{fig:bmfrac2}
\end{figure}

In order to do a fair interpretation of the TFR variation with environment, understanding the stellar mass and baryonic mass fraction is necessary. Indeed, the dynamical mass might measure the total mass within a given radius and is related to both this considered radius and the velocity measured at this radius. If the mass within $R_{22}$ is dominated by baryons, the interpretation might differ from the hypothesis that the mass is dominated by dark matter.
We therefore estimated the dynamical mass for the galaxies in the MAGIC kinematic sample in order to search for a dependence of the stellar and baryonic content of galaxies in groups as a function of the stellar mass. 
In this work, we assume that at large distance, one can consider that the mass distribution is spherical:
\begin{equation}
 M_{dyn}(r) = \frac{r V(r)^2}{G} \text{ .}
 \label{eq:mdyn}
\end{equation}
We computed this mass at $R_{22}$ using both the rotation velocity and the corrected velocity that includes the velocity dispersion (Eq. \ref{eq:vcirc}). These dynamical masses are provided in Table \ref{tab:dynamics}. We then derived the stellar mass fraction within $R_{22}$ and the baryonic mass fraction within the same radius, using the stellar masses corrected for the photometric apertures (see Sect. \ref{sec:fitting_methods}) and assuming the gas mass is within this radius.
Figures \ref{fig:smfrac} and \ref{fig:bmfrac2} show these fractions as a function of stellar mass. Because we believe that galaxies with masses lower than $10^9$~M$_{\sun}$ have kinematic measurements less reliable than more massive ones,  we have split the sample in three stellar mass bins, one for masses below $10^9$~M$_{\sun}$, another for masses above $10^{10}$~M$_{\sun}$ and the last one for stellar masses in between those two values. We have measured median masses and mass fractions for those three bins and show them as red dots in these figures.

Whatever the velocity estimator used, the stellar mass fraction is more important at high mass, similarly to the results of \citet{Pelliccia+19} for both low and high-density environments. This trend is more pronounced when the velocity dispersion is accounted for since using the corrected velocity clearly increases the dynamical mass for low-mass systems.

When the gas content is included, the baryon fraction seems to increase with stellar mass. The baryon fraction seems however slightly higher for low-mass systems than for high-mass ones, nevertheless, the lowest stellar mass bin clearly has a large dispersion linked to dynamical measurements.
This possibly indicates either that the gas mass is, as expected, contributing more to the total baryonic mass in galaxies with low stellar mass or that the gas mass is overestimated for low stellar mass galaxies, at least within $R_{22}$.

Taking or not into account the velocity dispersion to compute the dynamical mass, the median baryon fractions change, although both estimates remain consistent within the error bars.
As already discussed in Sect. \ref{sec:smTFR}, it is not yet clear whether the velocity dispersion really traces the dynamical mass rather than local star formation or feedback mechanisms.
Figures \ref{fig:smfrac} and \ref{fig:bmfrac2} tend to indicate that dispersion should be included since more galaxies are close or even above the theoretical limit of a fraction of unity when using only rotation (upper panels). Nevertheless, only a few galaxies with stellar masses above $10^9$~M$_\sun$ are above this theoretical limit within the large associated uncertainties, one of which (CGr84b-23) having a low inclination, and therefore probably an underestimated rotation.
Moreover, for barely resolved objects that are more common at low mass where the impact of including velocity dispersion is the largest, velocity dispersion might contain some information on the rotation velocity not measurable from velocity fields.

We find a typical stellar mass fraction below 50\%, even for massive galaxies. The fraction in the low-mass regime can be as low as $\sim 10$\%. Similarly, we find a baryonic mass fraction lower than 50\%, even for massive galaxies. This indicates that, at low mass, kinematic measurements probe DMHs and that at high mass, they trace equally the potential of baryons and dark matter. This means that the interpretation of the TFR offset could differ between high-mass and low-mass systems. For low-mass systems, it seems clear that the baryonic content of galaxies in dense groups is less important for a given DMH, whereas for high-mass systems, velocities could be higher either due to a higher dark matter fraction than expected in this mass regime or because the baryon distribution in galaxies has been contracted, as expected from Eq. \ref{eq:mdyn} at fixed dynamical mass.

We cannot compare robustly the results found with our sample to others. Indeed, the work done for KROSS in \citet{Stott+16} is limited in terms of stellar mass range (above $10^{9.5}$~M$_\sun$) and does not discuss the evolution of the fraction of baryons as a function of the stellar mass but as a function of the dynamical mass.
The only comparison we could perform is with the stellar mass fraction computed in \citet{Pelliccia+19} on the ORELSE sample. However, many discrepancies between their study and the present one makes such a comparison difficult. Indeed, our size measurements rely on a bulge-disk decomposition, our stellar masses are corrected with individual corrections, the combination of rotation and dispersion are different and they used both dispersion and rotation-dominated galaxies.

Studying the impact of environment on the stellar and baryonic fractions would require a sample on which a similar methodology would be adopted for (i) the measurements of galaxy sizes, (ii) the kinematics modeling, (iii) the stellar mass estimates and corrections for photometric apertures and (iv) for sample selection. This will be the scope of a specific study using foreground and background galaxies observed in the MAGIC dataset, ensuring a robust comparison.

\subsection{Quenching timescale in groups}
\label{sec:quenching_time}

One impact of dense environment is a decrease in star formation in galaxies.
\citet{Grutzbauch+11} found that galaxies in overdensities, where density is larger by a factor of 5 with respect to the mean density, have their SFR lowered by a factor of $\sim 2-3$ up to $z=2$.
Similarly, \citet{Tomczak+19}, focusing on star-forming galaxies, found that there is a decrease of around $0.2-0.3$ dex of the SFR in the highest-density environments with respect to the lowest-density ones, more pronounced for galaxies with stellar masses between $\sim 10^{10}$ and $\sim 10^{11}$~M$_\sun$, this decrease being larger in the lowest redshift bin ($0.6<z<0.9$).  We find a similar trend with stellar mass for our final kinematic sample when comparing the SFR to the relation provided by \citet{Boogaard:2018} displayed in Fig. \ref{Main_sequence_SFR} in various mass bins.
Such a lowering of the SFR ($\Delta SFR$) might reduce the stellar mass content of DMHs by an amount $\Delta M_*$ that should depend on the time $\Delta T$ when galaxies entered into dense structures:
\begin{equation}
 \Delta T = \frac{\Delta M_*}{\Delta SFR} \text{ .}
 \label{eq:lifetime}
\end{equation}
Given the baryonic mass fraction derived for our sample in Sect. \ref{sec:bar_fraction}, we can assume that the velocity traces the large-scale mass of the underlying DMH. This is further reinforced by the fact that the plateau velocity is reached within $R_{22}$ for 48/77 galaxies of the kinematic sample with strict criteria, for 43/67 galaxies using the final kinematic sample (no bulge-dominated galaxies and no problematic AGN), and for 40/55 galaxies when the mass threshold of $10^9$~M$_\sun$ is applied, that is, from two thirds to three quarters of the sample for an increasingly stringent selection.
We can thus expect that the offset of the zero-point we observed in the smTFR in dense groups with respect to low-density environments results from an impact of the environment on SFR, and it is therefore possible to estimate the time at which the groups formed.
\begin{equation}
 \Delta T = \frac{\Delta(\log{M_*})}{\alpha \times \Delta(\log{SFR})} \text{ ,}
 \label{eq:time_mass}
\end{equation}
where we assume that the main sequence relation is linear between stellar mass and SFR with $SFR = \alpha \times M_*$, with $\alpha = 10^{-9.5}$~s$^{-1}$. This approximation provides a description of the relation provided in \citet{Boogaard:2018} better than 0.2 dex in the mass range between $10^9$ and $10^{11}$~M$_\sun$.

Assuming a lowering of the SFR by $\Delta(\log{SFR}) = 0.3$~dex, as suggested by the studies mentioned above, the offsets in zero-point of 0.64, 0.30, 0.23 and $\sim 0.1$ we found in Sects. \ref{sec:smTFR} and \ref{sec:bmTFR} (see Table \ref{tab:diff_TFR}) lead to times since galaxies entered the structures of 6.7 Gyr, 3.2 Gyr, 2.4 Gyr, and 1 Gyr for the KROSS rotation dominated, KROSS disky, ORELSE, and KMOS3D samples, respectively.
Depending on the sample, the difference is quite large. The result is unlikely for the KROSS rotation dominated sample since it would mean that groups formed right after the big bang. This discrepancy, even between the two KROSS subsamples, probably points to a methodology bias either in the sample selection or in the way kinematics are derived.
The comparison with KMOS3D would lead to a formation of groups at redshift around $z\sim 0.9$. It is worth emphasizing that the KMOS3D subsample we use as reference has a median redshift of 0.9. Nevertheless, we assume there is no significant evolution of the TFR between $z\sim 0.7$ and $z\sim 0.9$.
The difference with ORELSE could be interpreted in line with the environment selection. Indeed, in case quenching is most efficient in clusters, galaxies that enters the structures might stop their star formation in a much shorter amount of time. Therefore, star-forming galaxies observed in clusters might be galaxies that just entered the clusters. In addition, the ORELSE sample is not exclusively composed of cluster galaxies, which means that the results from this sample may be more similar to that for field galaxies. Making this assumption, the comparison with ORELSE would lead to a formation of groups from the MAGIC sample at redshift around $z\sim 1.2$.

In order to check whether these timescales are realistic, we have also estimated the typical depletion time for the galaxies in our sample, based on our gas mass and SFR estimates. The depletion time seems rather constant with stellar mass. This timescale is on the order of $\sim 1$ Gyr for masses above $10^9$~M$_\sun$. Therefore, if our interpretation is correct, this means that either gas accretion remains possible after galaxies enter groups or that they entered in groups with a larger gas fraction, by a factor of around 2.

It is interesting to notice that the zero-point offset is slightly less important for high-mass galaxies than for low-mass ones, by about 0.1 dex. This is compatible with these galaxies being less dark matter dominated, hence their rotation velocity is not only tracing the underlying DMH potential.

In any cases, these results are subject to large uncertainties on the lowering of the SFR. Ideally, one would need to compute this lowering from a reference sample of galaxies in low-density environments.

\subsection{Mass distribution contraction in groups}
\label{sec:contraction}

Another consequence of dense environments is that the distribution of baryons is more concentrated than in low-density environments. In the local Universe, several studies show evidence for this contraction, which is more pronounced for low-mass late-type galaxies \citep[e.g.,][]{Maltby+10, Fernandez-Lorenzo+13, Cebrian+14}. At higher redshift a similar trend is also found between field and cluster galaxies \citep[e.g.,][]{Kuchner+17, Matharu+19, Pelliccia+19}.
\citet{Matharu+19} found a contraction of 0.07 dex for late-type cluster galaxies at $z\sim 1$, consistent with size growth being inhibited for about 1 Gyr by cluster environment, whereas \citet{Pelliccia+19} found a more pronounced decrease of 0.13 dex for the ORELSE sample.
\citet{Kuchner+17} interpret the decrease in the contraction with mass as quenching timescale being longer for high-mass galaxies.

If the mass is dominated by baryons, or dark matter distribution contracts similarly to baryons, then one expects from Eq. \ref{eq:mdyn} that the rotation velocity is larger when a galaxy is contracted. If we assume the mass within $R_{22}$ remains identical, then we have
\begin{equation}
\Delta (\log{V}) = - 0.5 \Delta(\log{R_{22}}) \text{ .}
\label{eq:contraction}
\end{equation}
Depending on the expected contraction, the variation in velocity might be around 0.035 and 0.065 dex. This can be compared to the offset in zero-point of the TFR. This offset in mass can be converted into an offset in velocity using the corresponding slope for each comparison sample.
We found velocity offset of 0.177, 0.068, 0.025, and 0.072 for the KROSS rotation dominated, KROSS disky, KMSO3D, and ORELSE samples, respectively (see Table. \ref{tab:diff_TFR}).
Except for the KROSS rotation dominated sample, these offsets are compatible with contraction. The discrepancy with KROSS might be explained by a difference of methodology in either sample selection or kinematics extraction, as already proposed in Sect. \ref{sec:quenching_time}.
The difference found with respect to the ORELSE sample may be due to a differential contraction since ORELSE galaxies already have smaller radii than field \citep{Pelliccia+19} and because cluster galaxies should even be smaller than group ones. It can either be explained by this sample not containing only cluster galaxies, by the methodology to extract velocities or by DMHs in groups and clusters not being impacted similarly by environment.

Contraction of baryons is supposed to be more pronounced at low mass, which is also the regime where dark matter dominates. In this regime we therefore expect that kinematics trace the DMH distribution. The observed increase in velocity in groups may indicate that for a given baryonic mass, the ratio of the DMH mass over the radius is larger in groups than at low density, which could be induced either by dynamical processes contracting DMHs or by galaxies living in more massive halos in dense environments, or by a combination of both.
At high mass, contraction seems insufficient to explain the zero-point offset, except if we compare with KMOS3D. In addition, we assumed a similar galaxy contraction in groups and clusters whereas it is expected to be less important in groups. It is therefore difficult to explain the offset at high mass with contraction only. Proper mass models would be required to understand what component drives the large rotations depending on the mass of galaxies.

\section{Conclusions}
\label{Conclusions}

We investigated the impact of environment on the stellar mass and baryonic mass TFR using a sample of 67 star-forming galaxies located in eight groups at redshift $0.5 < z < 0.8$ with projected densities at least 25 times larger than the average galaxy densities in the same redshift range. These groups were observed as part of the 70h on-source MAGIC MUSE-GTO project (PI: T. Contini) inside the COSMOS field. We performed a bulge-disk decomposition on 250 galaxies in those groups to infer accurate bulge-to-disk ratios, galaxy sizes and projection parameters. We built a robust kinematic sample of galaxies on the main sequence of star-forming galaxies based on their size with respect to MUSE observations spatial resolution as well as on the signal-to-noise ratio of the \oii\ doublet used to derive galaxy kinematic maps. Some AGN and bulge-dominated galaxies were also discarded from our analysis. We extracted rotation velocities at $2.2$ times the disk scale lengths ($R_{22}$) as well as the velocity dispersion from these maps using kinematic models that take into account the limited spatial resolution of our observations.
Our selection lead to a sample spanning a wide range of stellar masses between $10^{8}$ and $10^{11}$~M$_\sun$, and mainly composed of rotation-dominated galaxies without requiring a prior condition on the ratio of rotation over dispersion velocities, in contrast with selection functions used in most of such kinematics studies. We derived both the stellar and baryonic mass TFR for our sample, studied the impact of selection and methodology, and compared it to KROSS, KMOS3D and ORELSE, major surveys at similar redshifts. We summarize here the main results.
\begin{itemize}
 \item Methodology biases can induce a change of the zero-point of 0.05 dex depending on how velocities uncertainties are estimated and of 0.05 dex depending on the selection criteria for the MAGIC sample, that is, by removing or including peculiar galaxies and by modifying the stellar mass range. Not correcting stellar masses to have estimates at the same radius where the velocity is measured introduces an additional increase of 0.15 dex. Methodology biases with respect to other studies have been reduced as much as possible and, whereas we have estimated stellar masses within $R_{22}$, we have taken into account this in the subsequent comparisons with other samples.
 \item Using the rotation velocity alone, the MAGIC groups best fit smTFR (bmTFR) computed assuming a reference velocity $\log{(V_{ref}~[\text{km~s}^{-1}])}=2.2$ has a slope $\alpha=4.03\pm0.63$ $(3.55\pm 0.49)$, a zero-point $\beta=9.79\pm0.09$ $(9.99\pm0.07)$, and an intrinsic scatter $\sigma_{\rm int}=0.43$ $(0.35)$. Taking also into account an asymmetric drift correction, we found that the best fit smTFR (bmTFR) has a slope $\alpha=4.35\pm0.59$ $(3.74\pm 0.45)$, a zero-point $\beta=9.51\pm0.10$ $(9.76\pm0.08)$, and an intrinsic scatter $\sigma_{\rm int}=0.33$ $(0.25)$.
 \item The inclusion of the velocity dispersion in the dynamics budget has a more significant impact on galaxies with low rotation velocities and reduces the intrinsic scatter of the various TFR. It increases the slope and decreases the zero-point, however the effect is the same for any sample.
 \item 
 Using the rotation velocity either alone or corrected for an asymmetric drift, smTFR slopes for the various samples are compatible within the fitting uncertainties.
 However, using fixed slopes, we systematically find a zero-point in terms of mass that is lower for the MAGIC sample than for the other samples used for comparison. This decrease in the zero-points is about $0.05-0.3$ dex depending on the sample used, except for the KROSS rotation dominated sample where it reaches 0.6 dex, which we interpret as a combination of sample selection and methodology biases. The offset is the lowest when comparing to KMOS3D, which targeted essentially galaxies more massive than 10$^{10}$~M$_\sun$. Finally, this evolution of the zero-point with environment contrasts with the result of \citet{Pelliccia+19} for the ORELSE sample. The non-evolution of the zero-point for this sample containing cluster galaxies may be due to the fact that this sample includes also field galaxies and to quenching being more efficient in clusters, which might bias the sample toward star-forming galaxies that just entered dense structures.
 \item Similarly, we studied the bmTFR relation, by including gas masses, estimated by reversing the Kennicutt-Schmidt law. The impact of gas mass is more pronounced for the low-mass regime than for the high-mass one because low-mass galaxies have a larger gas fraction.
 \item We find a reduction of the bmTFR zero-point similar to that of the smTFR for KMOS3D sample for which this comparison was possible.
\end{itemize}

We also derived stellar and baryonic mass fraction for the MAGIC group sample and found that the stellar mass fraction increases from around 10\% in the low-mass regime up to 50\% at maximum in the high-mass regime, in line with previous studies. A similar but weaker trend is observed for the baryon fraction that remains below 50\%.

We then interpreted the differences we observe in the TFR with two main hypotheses. On the one hand, we made the hypothesis that kinematics trace the DMH mass and therefore interpreted the difference in zero-point as resulting from quenching. Assuming a decrease in the SFR by 0.3 dex due to environment, we could infer that the bulk of galaxies we observe in groups would have entered the over-density between 1 to 3 Gyr ago.
On the other hand, we studied the impact of a contraction of the mass distribution within $R_{22}$ to infer the expected increase in velocity. The hypothesis that environment contracts the stellar content of galaxies by 0.07 to 0.13 dex, seems able to justify from half to the whole offset we observe in the TFR with respect to other samples. Since dark matter fraction is large, it would therefore indicate that dark matter is also contracted or that DMHs are more massive.
It is therefore likely that both contraction of the mass distribution and star-formation quenching participate in the observed differences in the TFR between dense groups and low-density environments and that the contribution of each mechanism depends on the mass regime.

Nevertheless, despite our efforts to minimize systematics, the results we obtained might still depend on the comparison samples and on the methodology used (i) to perform sample selection, (ii) to measure galaxy sizes, (iii) to infer stellar and gas masses within a given radius and (iv) to derive kinematics properties. Combining all these source of uncertainties make difficult to extract robust comparisons that could serve as the basis for an estimation of each process and link this to star-formation quenching mechanisms, such as ram-pressure stripping, gravitational interactions between galaxies or between galaxy and group potential wells, merging, or starvation.

We therefore need a comparison sample of field galaxies with a dataset very similar to the MAGIC group sample so that we can estimate within stellar mass bins the variation with environment of (i) the SFR, (ii) the disk scale length, (iii) the velocity, and therefore study the impact of environment on both the TFR and the baryon mass fraction as a function of mass. Such a sample will be built from the MAGIC dataset itself.
Ideally, mass models constrained with observed stellar distribution should be used to determine if halos have been contracted depending on the stellar mass or on the asymptotic halo mass.

\begin{acknowledgements}
We dedicate this article in memory of Hayley Finley.
This work has been carried out thanks to the support of the Ministry of Science, Technology and Innovation of Colombia (MINCIENCIAS) PhD fellowship program No. 756-2016.
This work has been carried out through the support of the ANR FOGHAR (ANR-13-BS05-0010-02), the OCEVU Labex  (ANR-11-LABX-0060), and the A*MIDEX project (ANR-11-IDEX-0001-02), which are funded by the ``Investissements d'avenir'' French government program managed by the ANR.
This work was supported by the Programme National Cosmology et Galaxies (PNCG) of CNRS/INSU with INP and IN2P3, co-funded by CEA and CNES.
JB acknowledges support by FCT/MCTES through national funds by the grant UID/FIS/04434/2019, UIDB/04434/2020 and UIDP/04434/2020 and through the Investigador FCT Contract No. IF/01654/2014/CP1215/CT0003.
JCBP acknowledges financial support of Vicerrector\'{\i}a de Investigaci\'{o}n y Extensi\'{o}n de la Universidad Industrial de Santander under project 2494.
This research has made use of \textsc{matplotlib} \citep{Hunter:2007}.
We acknowledge David Carton for his investment in the build-up of the project.
The authors thank the referee for the useful and constructive comments that enabled to significantly improve the paper.
\end{acknowledgements}

\bibliographystyle{aa}
\bibliography{TFR_MUSE_z07}

\appendix
\clearpage

\section{Morpho-kinematics maps for individual galaxies with S/N $\ge$ 40 and R$_{eff}$/FWHM $\ge$ 0.5}
\label{app:maps}

\subsection{Galaxies of the final kinematic sample}
\label{app:final_sample}

\begin{figure}[h]
\includegraphics[width=0.5\textwidth]{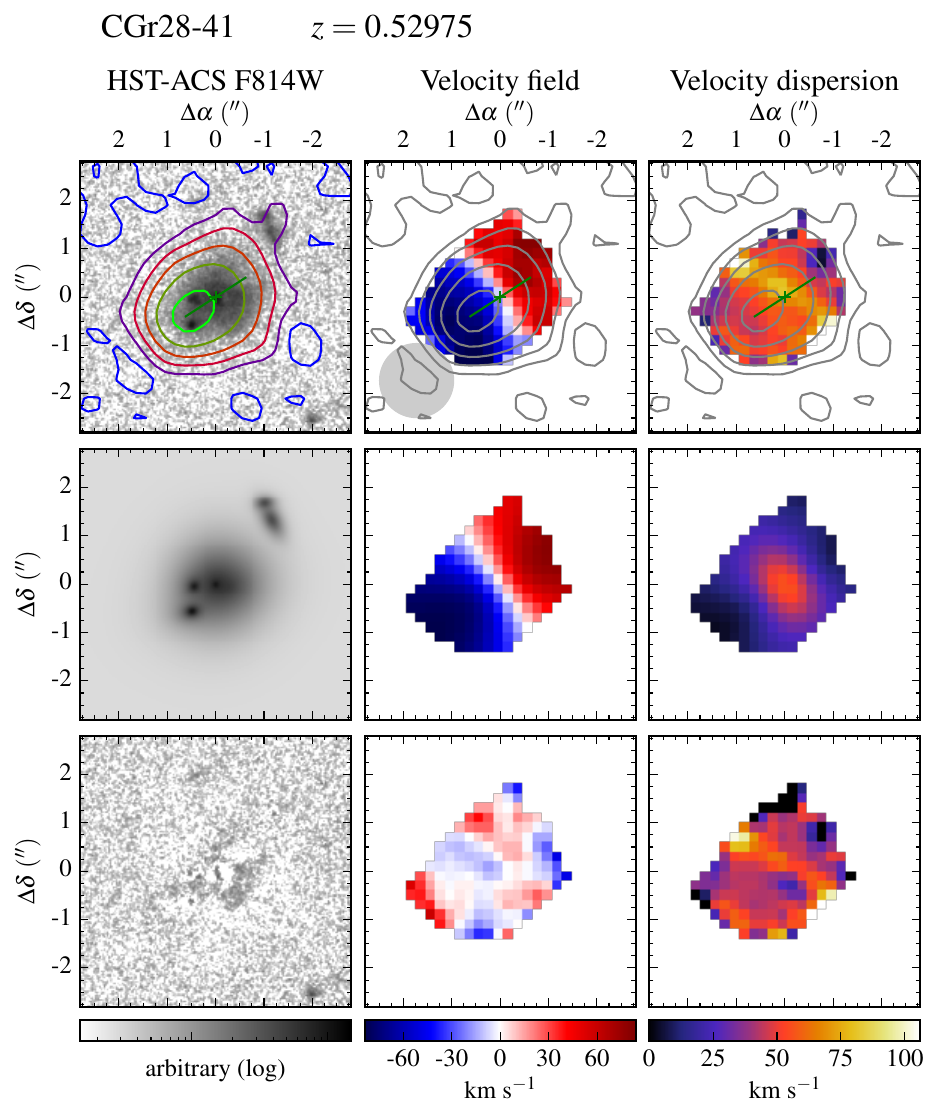}
\caption{Morpho-kinematics maps for galaxy CGr28-41. Top row, from left to right: HST-ACS F814W images, MUSE velocity fields and velocity dispersion maps corrected for spectral resolution.
 Middle row: Associated models. The velocity dispersion (third column) corresponds to a beam smearing correction map. Bottom row: Residuals, except for the third column that shows the beam smearing corrected velocity dispersion map. On each observed map, the green cross indicates the center derived from the morphology, whereas the green segment indicates the kinematic major axis and has a length corresponding to $R_{22}$. The [\ion{O}{ii}] flux distribution is shown with contours at levels of surface brightness $\Sigma$([\ion{O}{ii}]) = 2.5, 5.0, 10.0, 20.0, 40.0, 80.0 $\times 10^{-18}$~erg~s$^{-1}$~cm$^{-2}$~arcsec$^{-2}$. The MUSE spatial resolution is indicated with a gray disk of diameter FWHM in the bottom-left corner of the velocity field.
}
\label{Mopho_KINMap_CGr28-41} 
\end{figure} 
 
\begin{figure}
\includegraphics[width=0.5\textwidth]{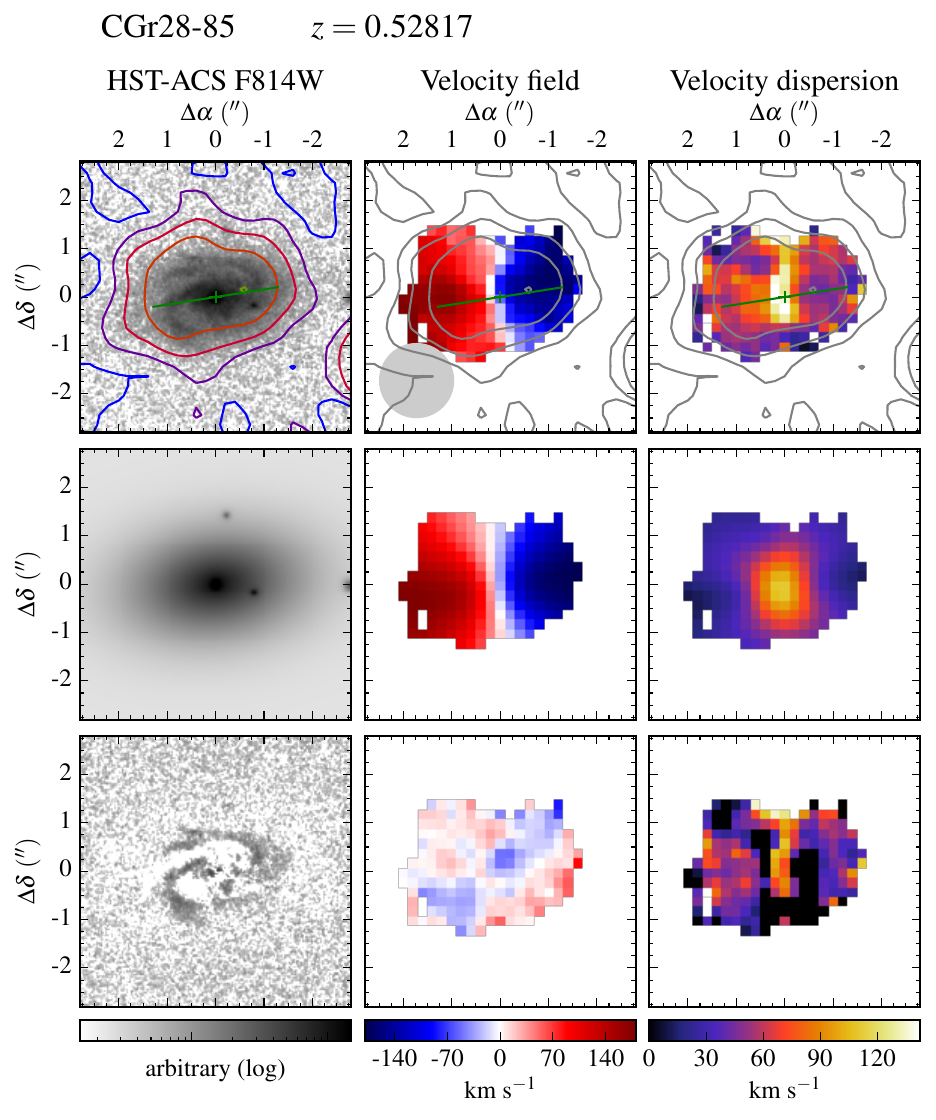}
\caption{Morpho-kinematics maps for galaxy CGr28-85. See caption of Fig. \ref{Mopho_KINMap_CGr28-41} for the description of figure.} 
\label{Mopho_KINMap_CGr28-85} 
\end{figure} 
 
\begin{figure}
\includegraphics[width=0.5\textwidth]{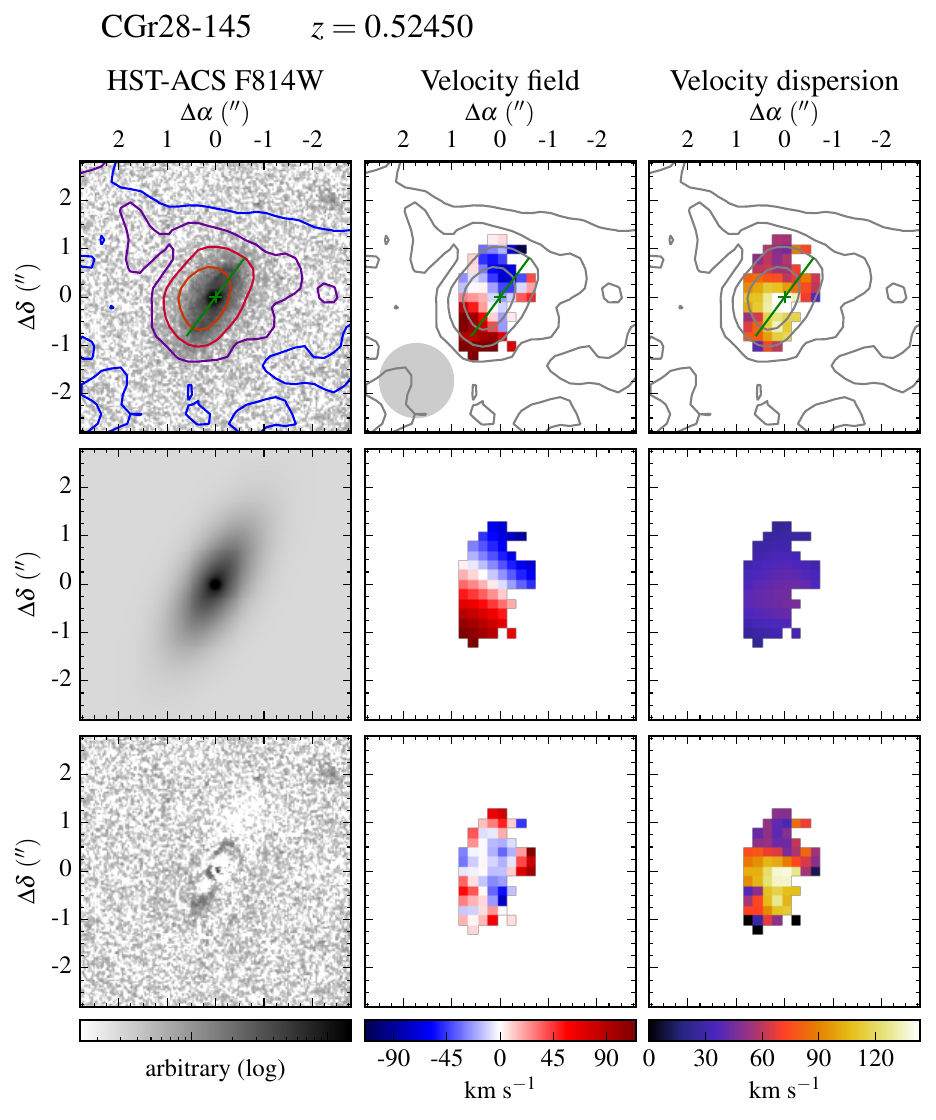}
\caption{Morpho-kinematics maps for galaxy CGr28-145. See caption of Fig. \ref{Mopho_KINMap_CGr28-41} for the description of figure.} 
\label{Mopho_KINMap_CGr28-145} 
\end{figure}

\clearpage

\begin{figure}
\includegraphics[width=0.5\textwidth]{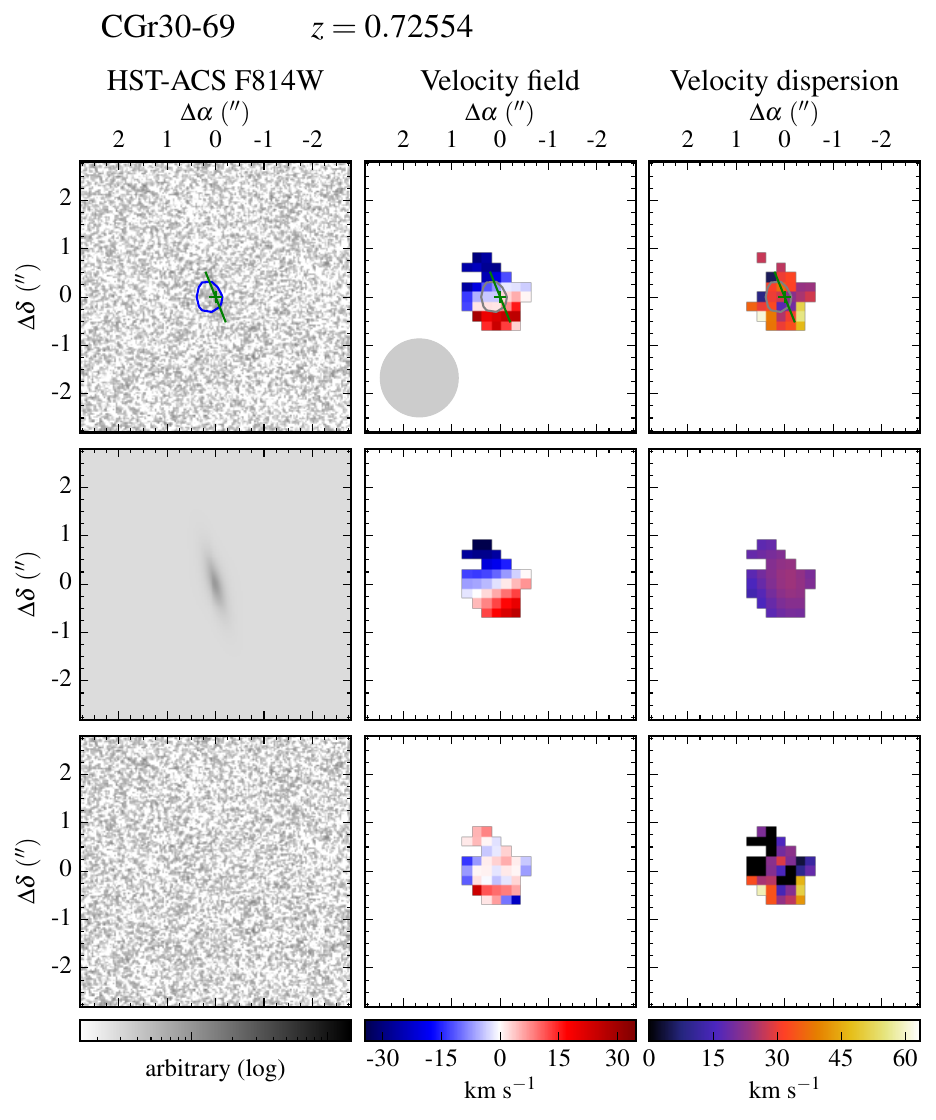}
\caption{Morpho-kinematics maps for galaxy CGr30-69. See caption of Fig. \ref{Mopho_KINMap_CGr28-41} for the description of figure.} 
\label{Mopho_KINMap_CGr30-69} 
\end{figure} 
 
\begin{figure}
\includegraphics[width=0.5\textwidth]{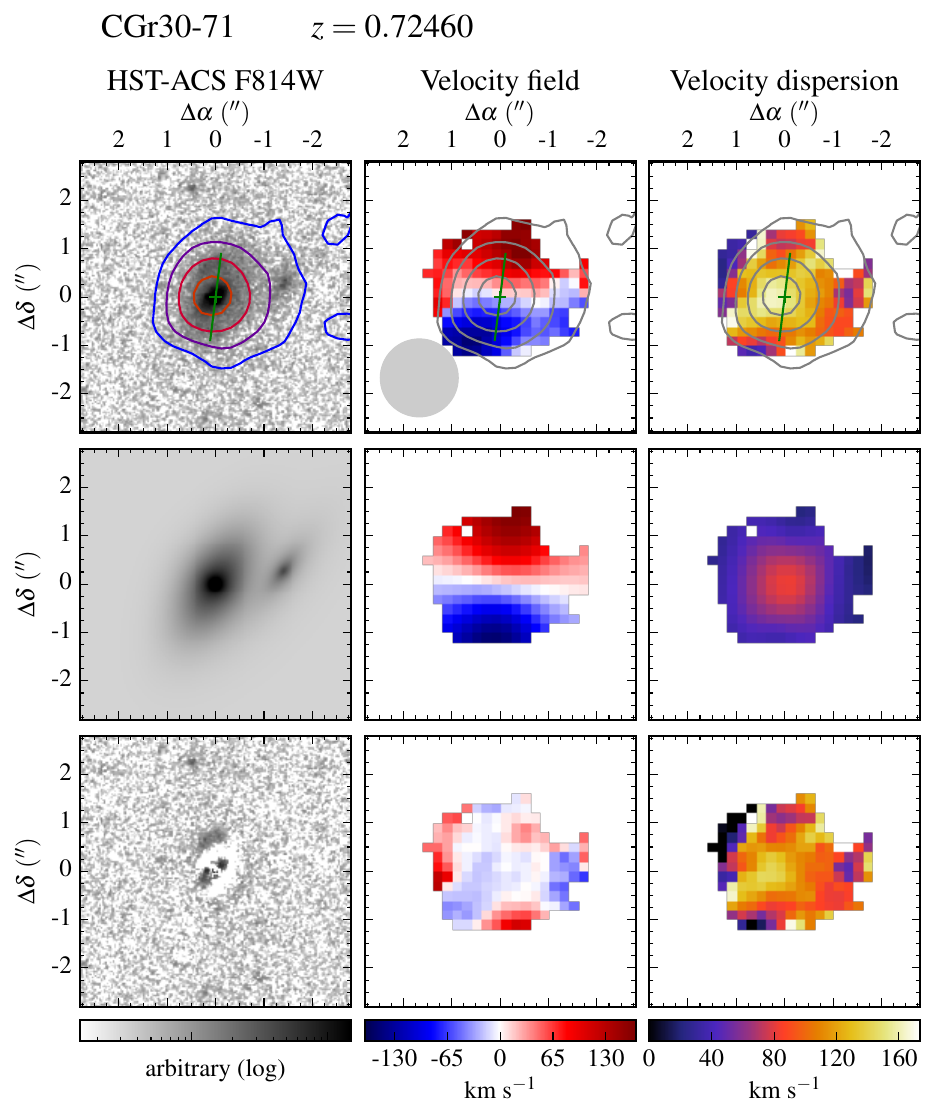}
\caption{Morpho-kinematics maps for galaxy CGr30-71. See caption of Fig. \ref{Mopho_KINMap_CGr28-41} for the description of figure.} 
\label{Mopho_KINMap_CGr30-71} 
\end{figure} 
 
\begin{figure}
\includegraphics[width=0.5\textwidth]{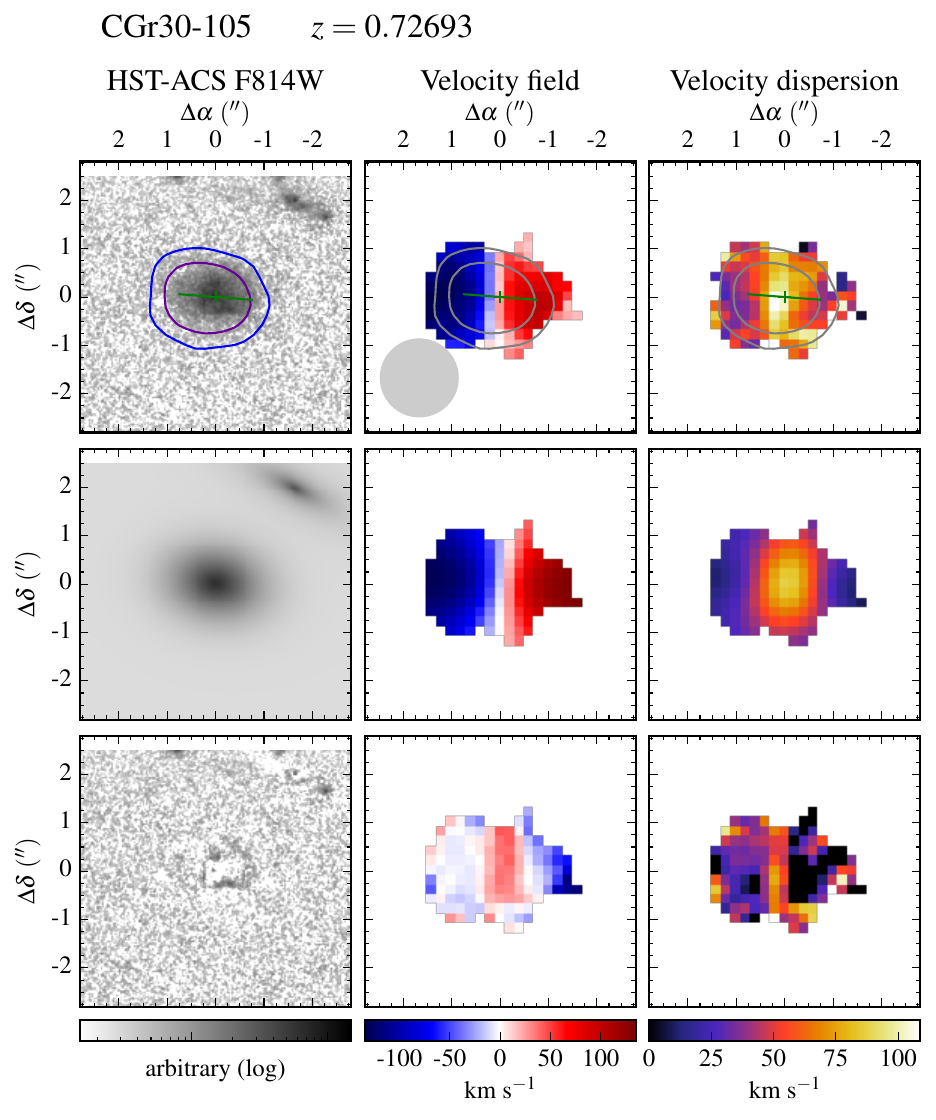}
\caption{Morpho-kinematics maps for galaxy CGr30-105. See caption of Fig. \ref{Mopho_KINMap_CGr28-41} for the description of figure.} 
\label{Mopho_KINMap_CGr30-105} 
\end{figure} 
 
\begin{figure}
\includegraphics[width=0.5\textwidth]{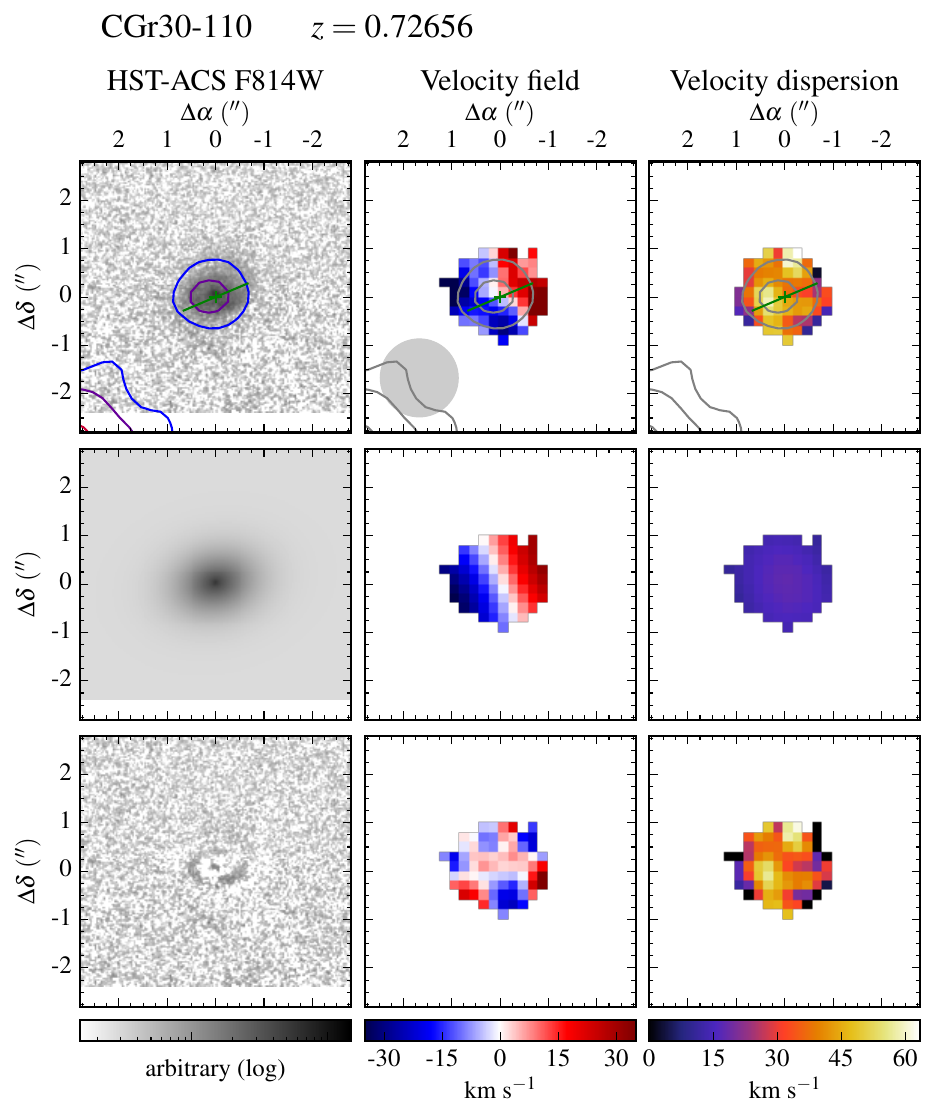}
\caption{Morpho-kinematics maps for galaxy CGr30-110. See caption of Fig. \ref{Mopho_KINMap_CGr28-41} for the description of figure.} 
\label{Mopho_KINMap_CGr30-110} 
\end{figure} 
 
\begin{figure}
\includegraphics[width=0.5\textwidth]{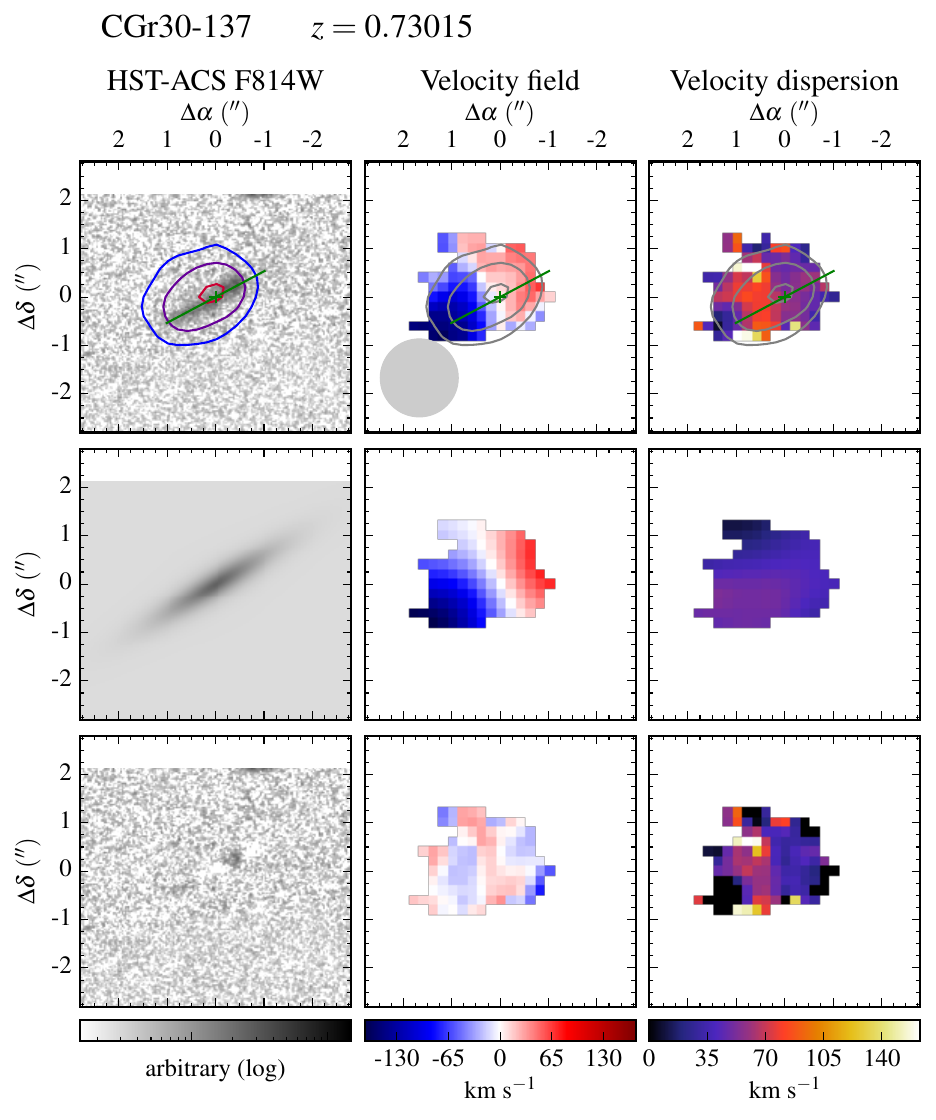}
\caption{Morpho-kinematics maps for galaxy CGr30-137. See caption of Fig. \ref{Mopho_KINMap_CGr28-41} for the description of figure.} 
\label{Mopho_KINMap_CGr30-137} 
\end{figure} 
 
\begin{figure}
\includegraphics[width=0.5\textwidth]{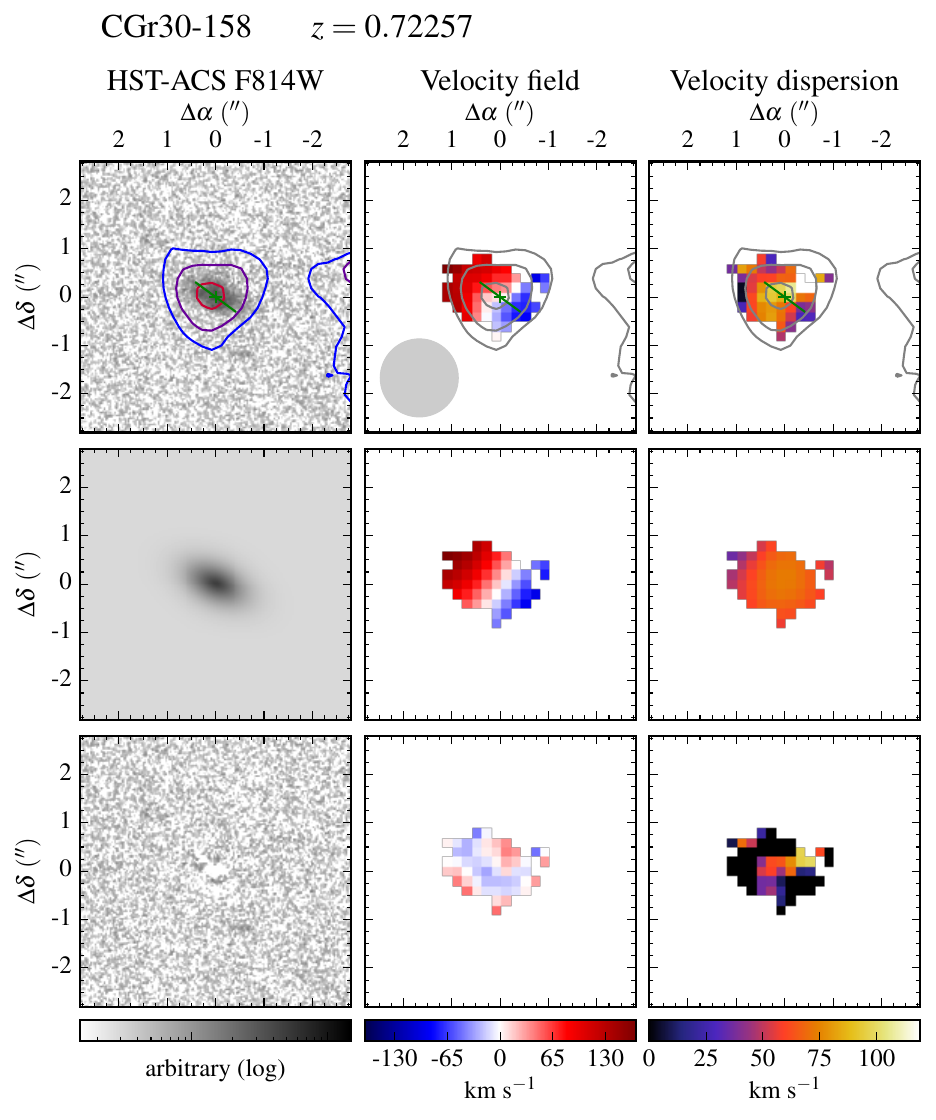}
\caption{Morpho-kinematics maps for galaxy CGr30-158. See caption of Fig. \ref{Mopho_KINMap_CGr28-41} for the description of figure.} 
\label{Mopho_KINMap_CGr30-158} 
\end{figure} 
 
\begin{figure}
\includegraphics[width=0.5\textwidth]{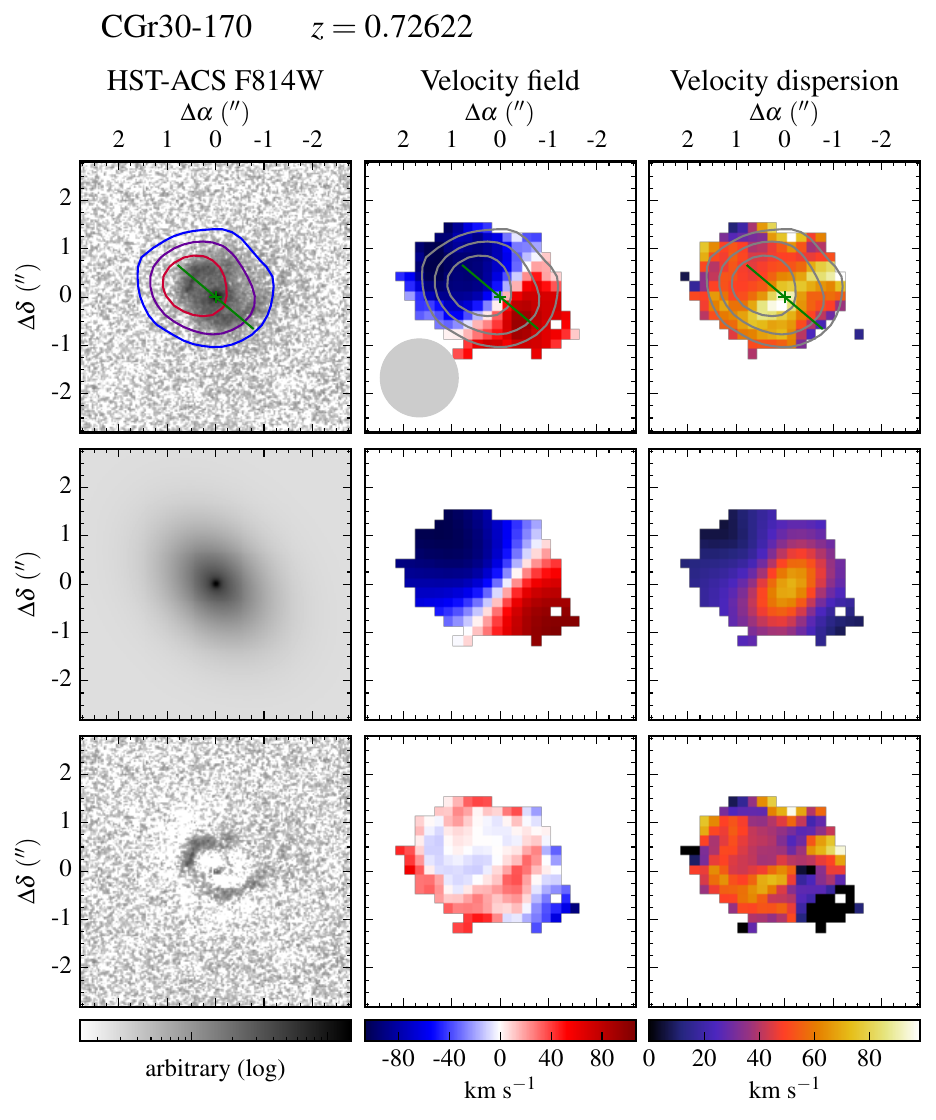}
\caption{Morpho-kinematics maps for galaxy CGr30-170. See caption of Fig. \ref{Mopho_KINMap_CGr28-41} for the description of figure.} 
\label{Mopho_KINMap_CGr30-170} 
\end{figure} 
 
\begin{figure}
\includegraphics[width=0.5\textwidth]{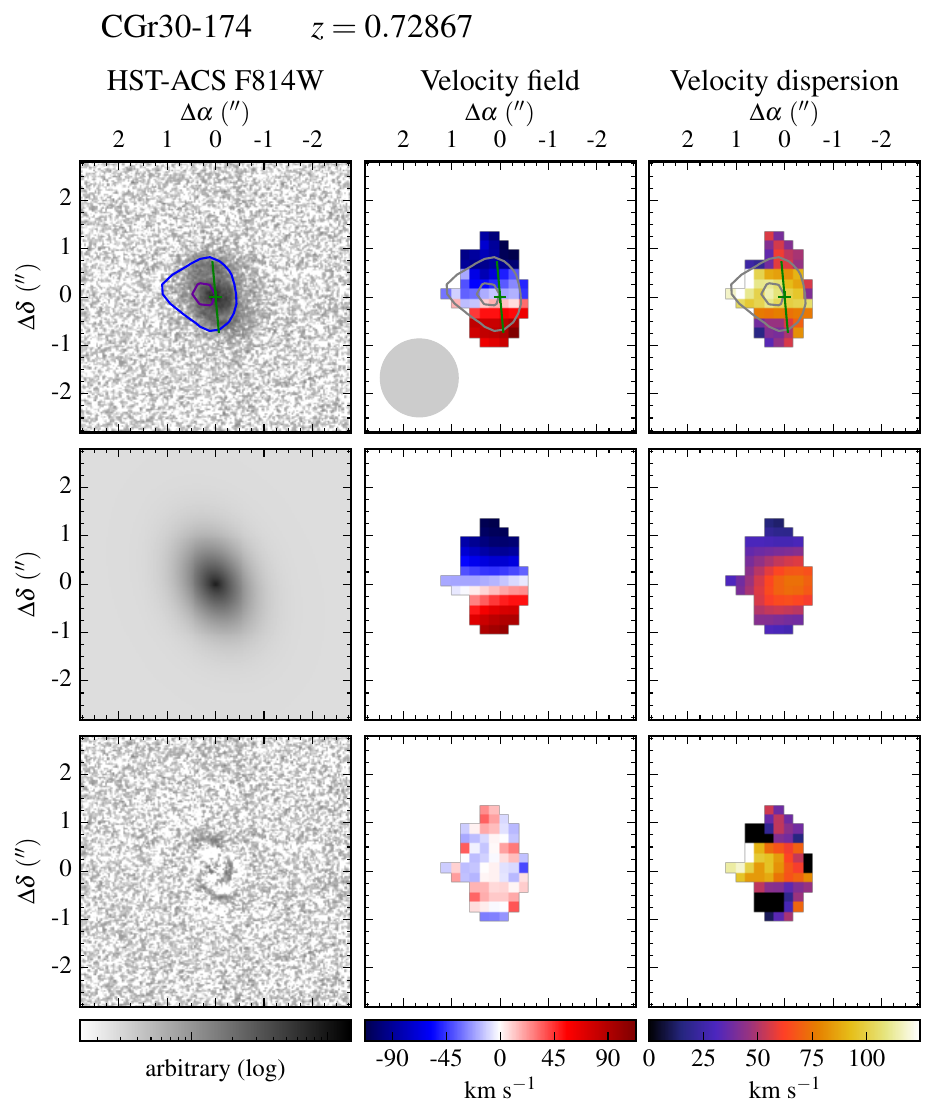}
\caption{Morpho-kinematics maps for galaxy CGr30-174. See caption of Fig. \ref{Mopho_KINMap_CGr28-41} for the description of figure.} 
\label{Mopho_KINMap_CGr30-174} 
\end{figure} 
 
\begin{figure}
\includegraphics[width=0.5\textwidth]{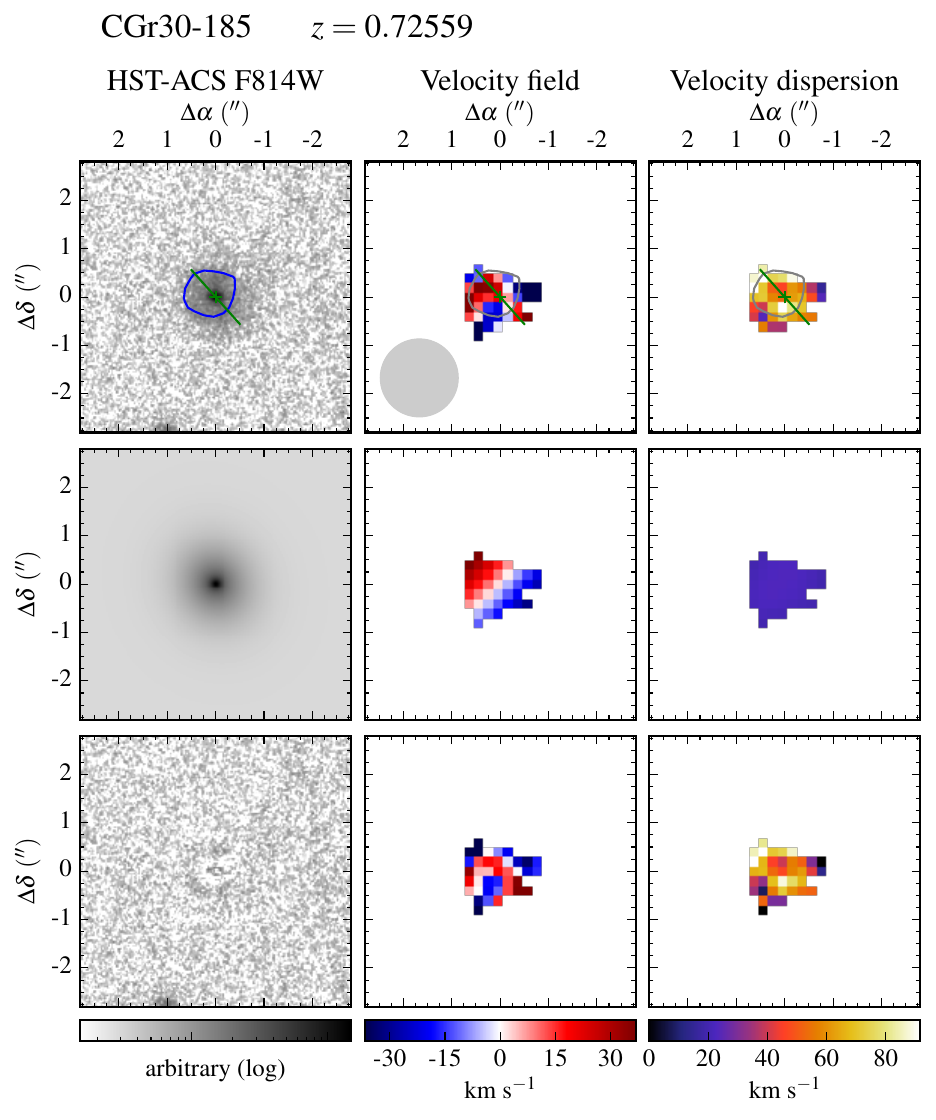}
\caption{Morpho-kinematics maps for galaxy CGr30-185. See caption of Fig. \ref{Mopho_KINMap_CGr28-41} for the description of figure.} 
\label{Mopho_KINMap_CGr30-185} 
\end{figure} 
 
\begin{figure}
\includegraphics[width=0.5\textwidth]{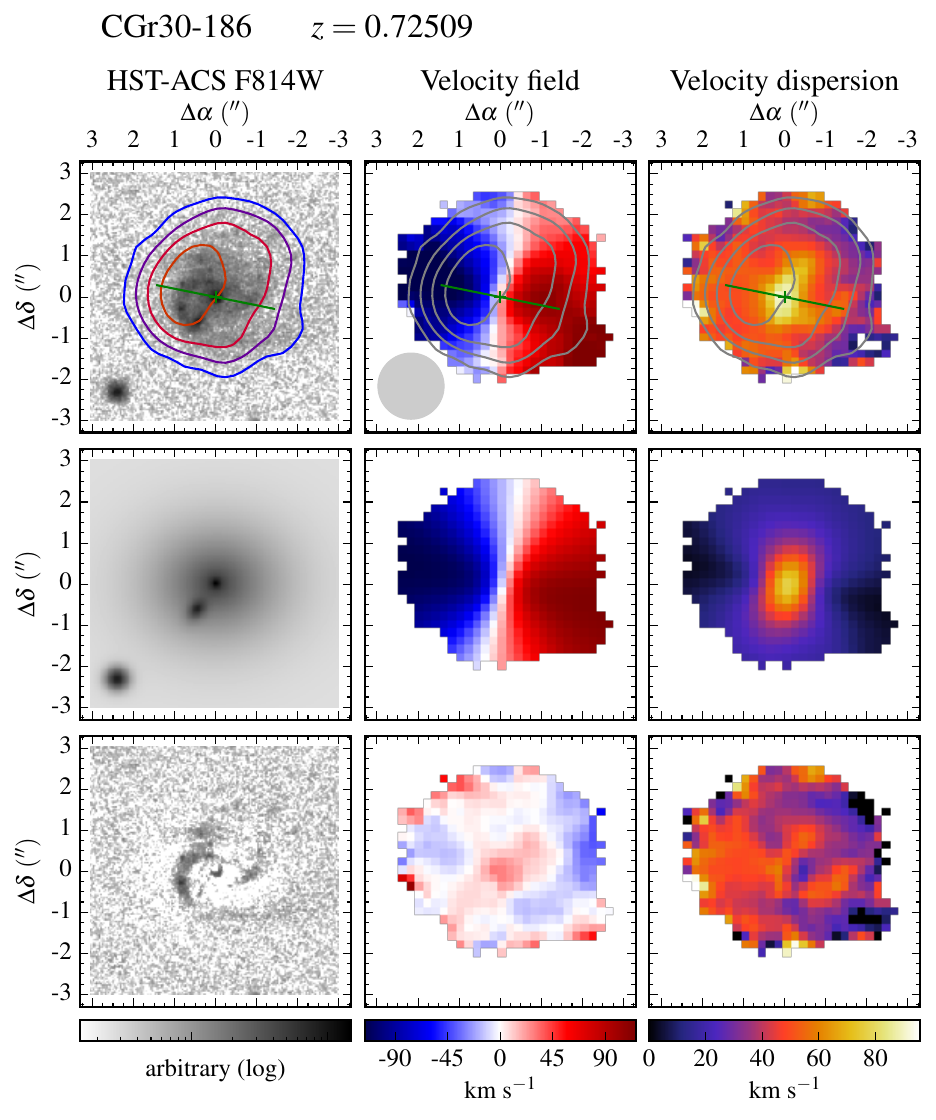}
\caption{Morpho-kinematics maps for galaxy CGr30-186. See caption of Fig. \ref{Mopho_KINMap_CGr28-41} for the description of figure.} 
\label{Mopho_KINMap_CGr30-186} 
\end{figure} 
 
\begin{figure}
\includegraphics[width=0.5\textwidth]{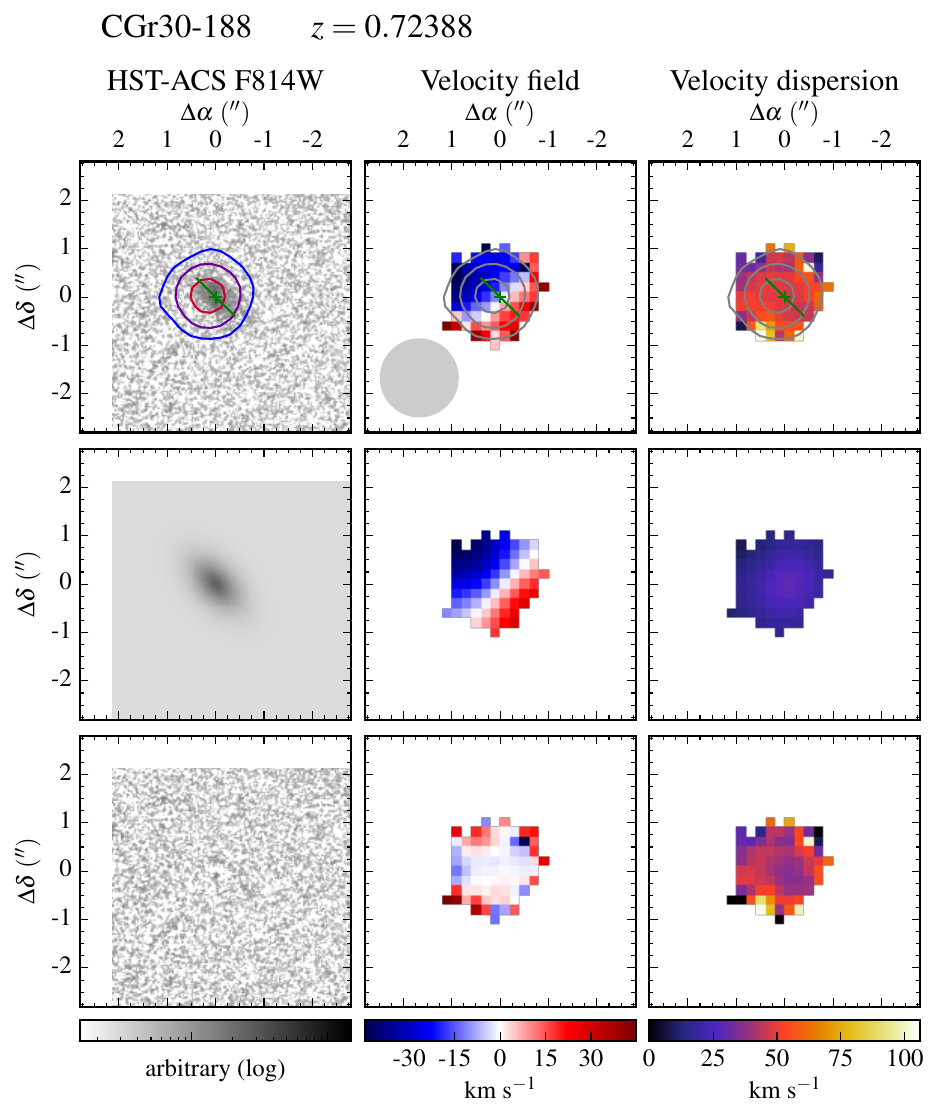}
\caption{Morpho-kinematics maps for galaxy CGr30-188. See caption of Fig. \ref{Mopho_KINMap_CGr28-41} for the description of figure.} 
\label{Mopho_KINMap_CGr30-188} 
\end{figure} 
 
\begin{figure}
\includegraphics[width=0.5\textwidth]{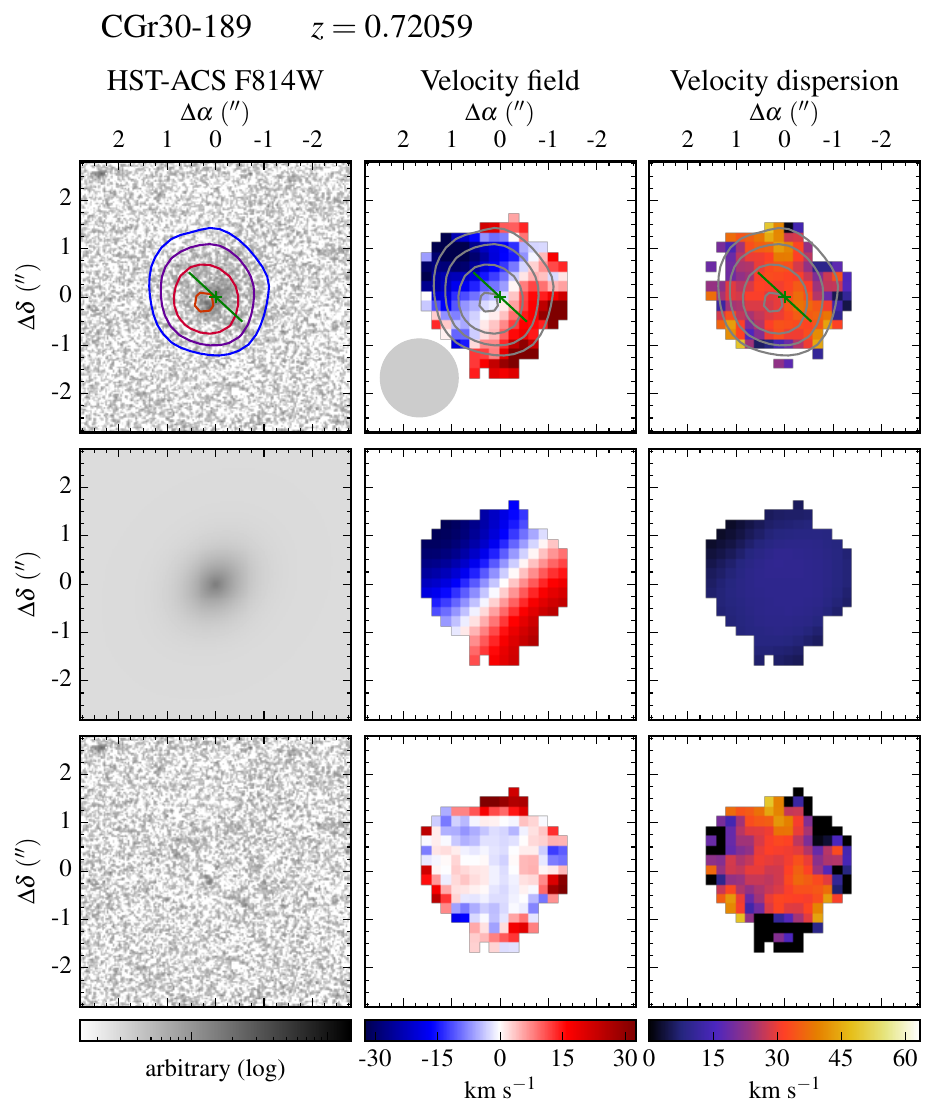}
\caption{Morpho-kinematics maps for galaxy CGr30-189. See caption of Fig. \ref{Mopho_KINMap_CGr28-41} for the description of figure.} 
\label{Mopho_KINMap_CGr30-189} 
\end{figure} 
 
\begin{figure}
\includegraphics[width=0.5\textwidth]{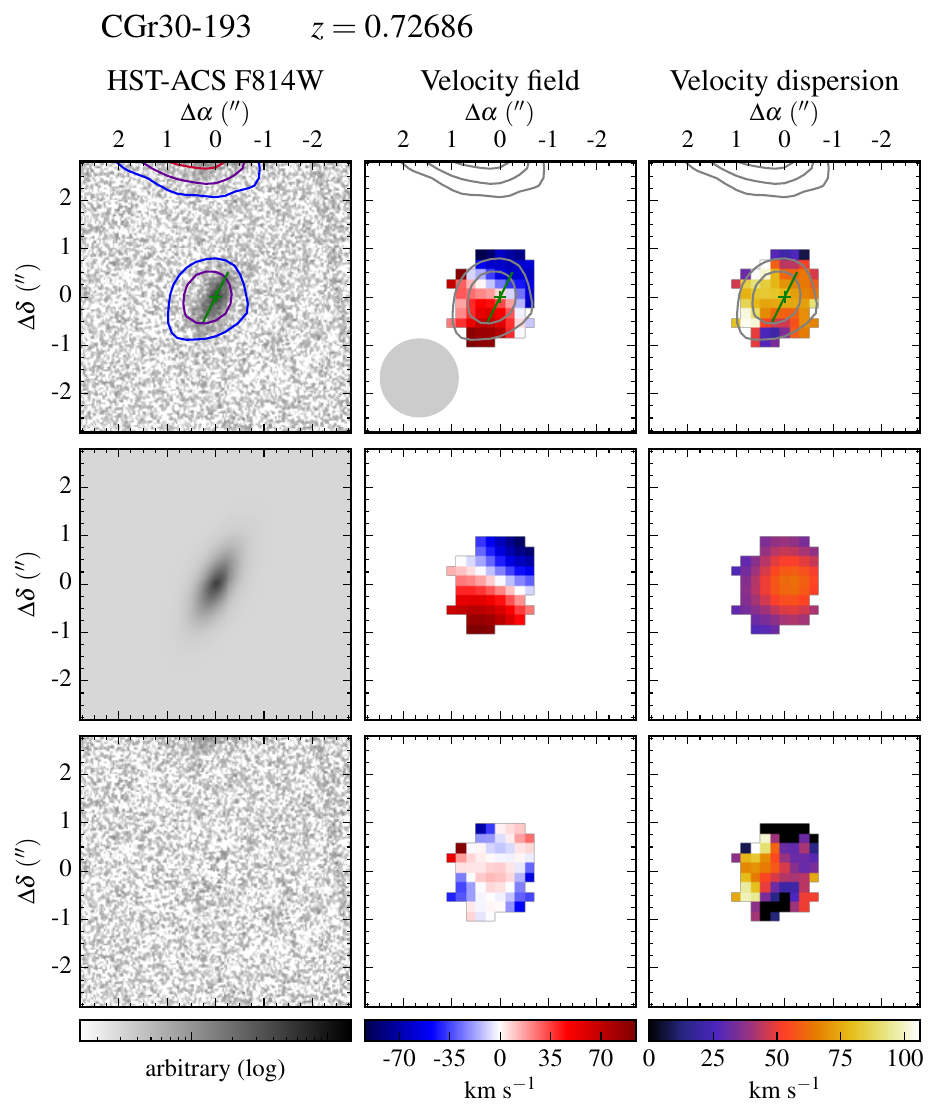}
\caption{Morpho-kinematics maps for galaxy CGr30-193. See caption of Fig. \ref{Mopho_KINMap_CGr28-41} for the description of figure.} 
\label{Mopho_KINMap_CGr30-193} 
\end{figure} 
 
\begin{figure}
\includegraphics[width=0.5\textwidth]{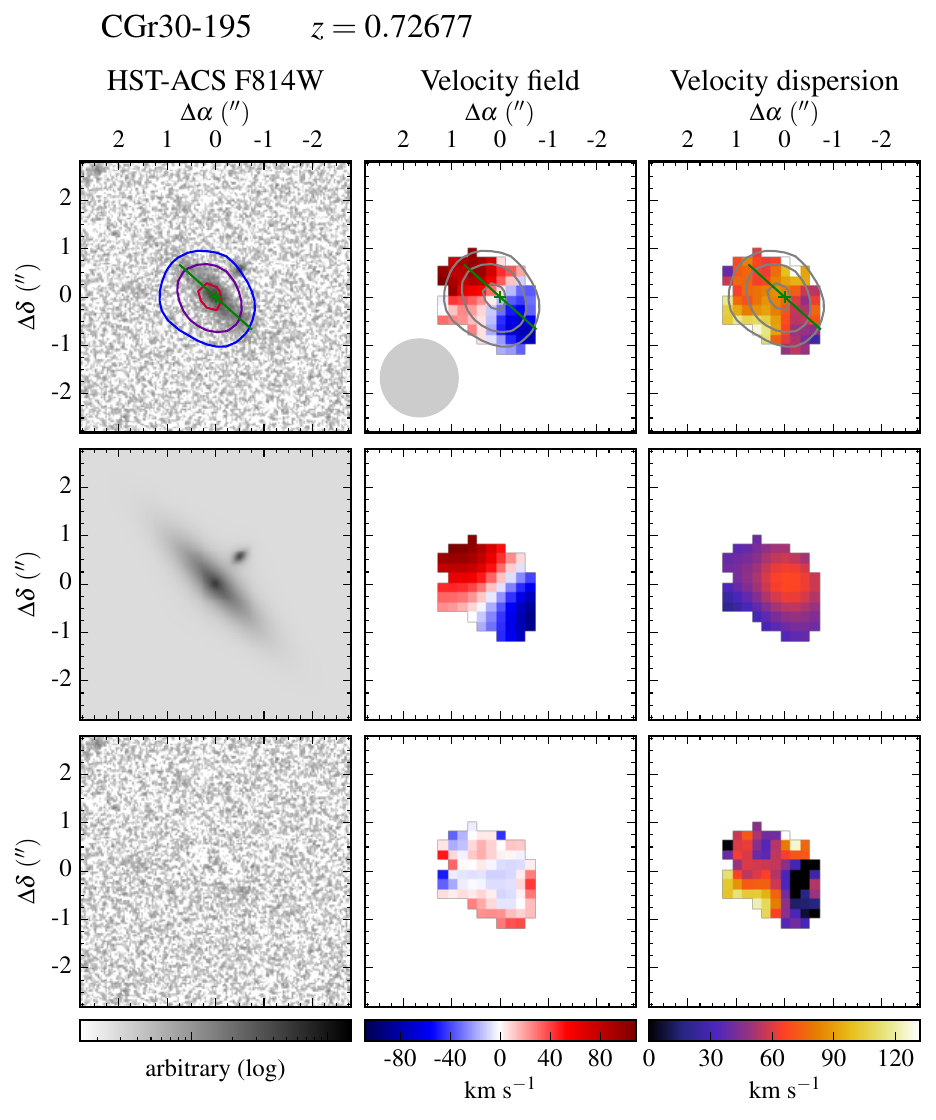}
\caption{Morpho-kinematics maps for galaxy CGr30-195. See caption of Fig. \ref{Mopho_KINMap_CGr28-41} for the description of figure.} 
\label{Mopho_KINMap_CGr30-195} 
\end{figure} 
 
\begin{figure}
\includegraphics[width=0.5\textwidth]{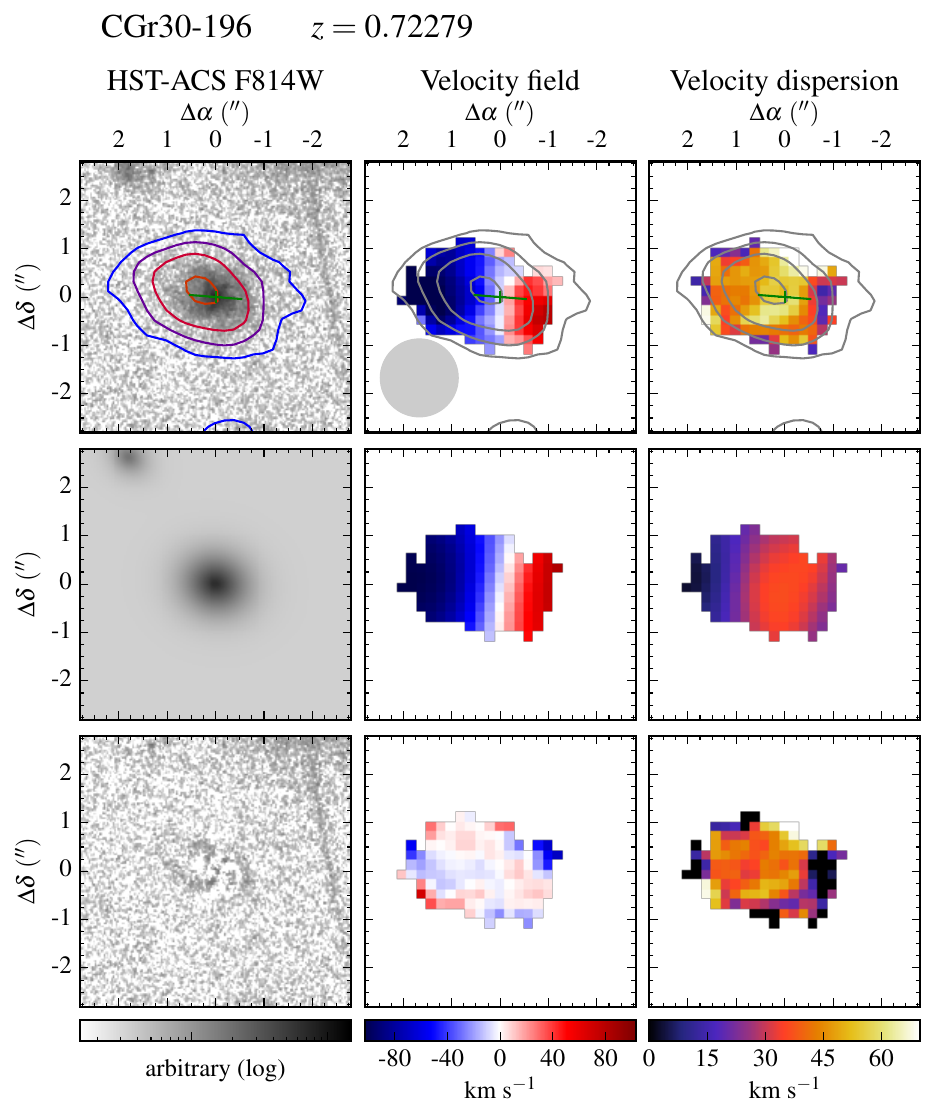}
\caption{Morpho-kinematics maps for galaxy CGr30-196. See caption of Fig. \ref{Mopho_KINMap_CGr28-41} for the description of figure.} 
\label{Mopho_KINMap_CGr30-196} 
\end{figure}

\clearpage

\begin{figure}
\includegraphics[width=0.5\textwidth]{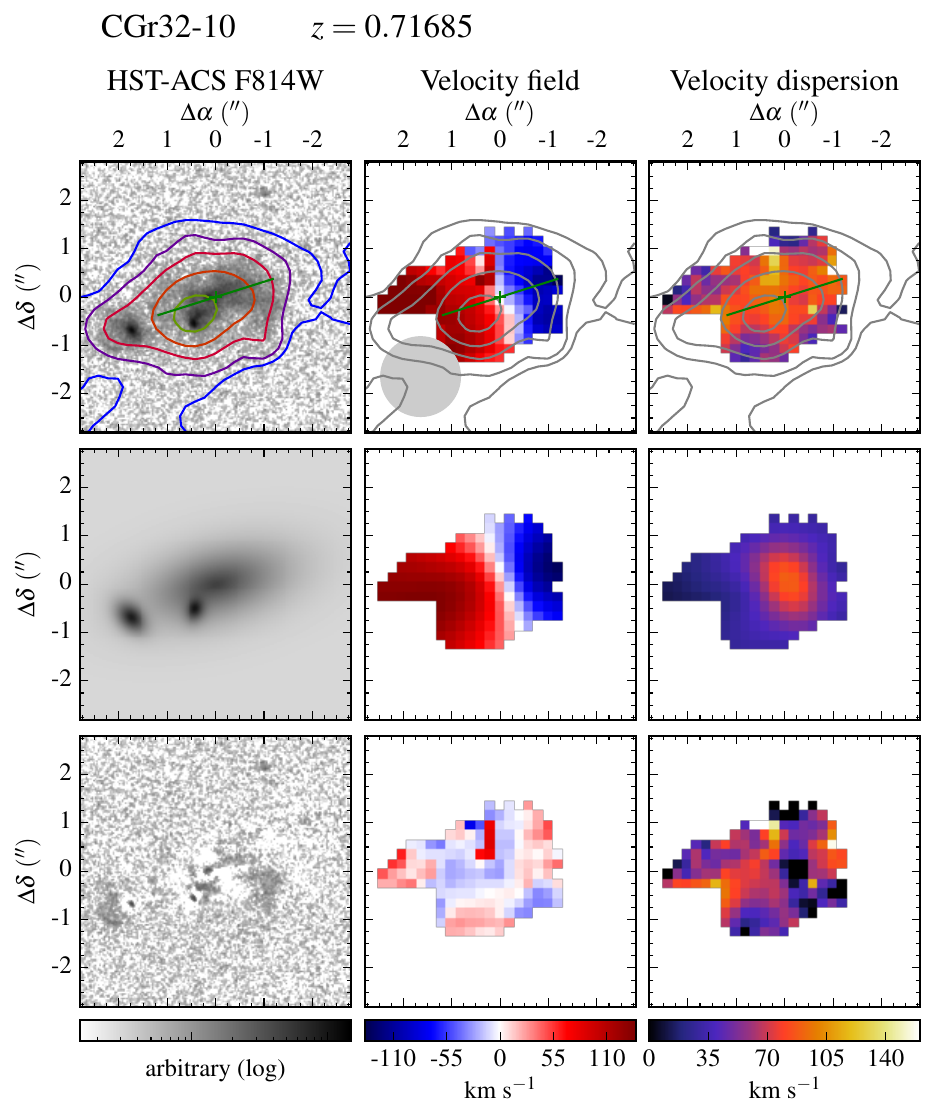}
\caption{Morpho-kinematics maps for galaxy CGr32-10. See caption of Fig. \ref{Mopho_KINMap_CGr28-41} for the description of figure.} 
\label{Mopho_KINMap_CGr32-10} 
\end{figure} 
 
\begin{figure}
\includegraphics[width=0.5\textwidth]{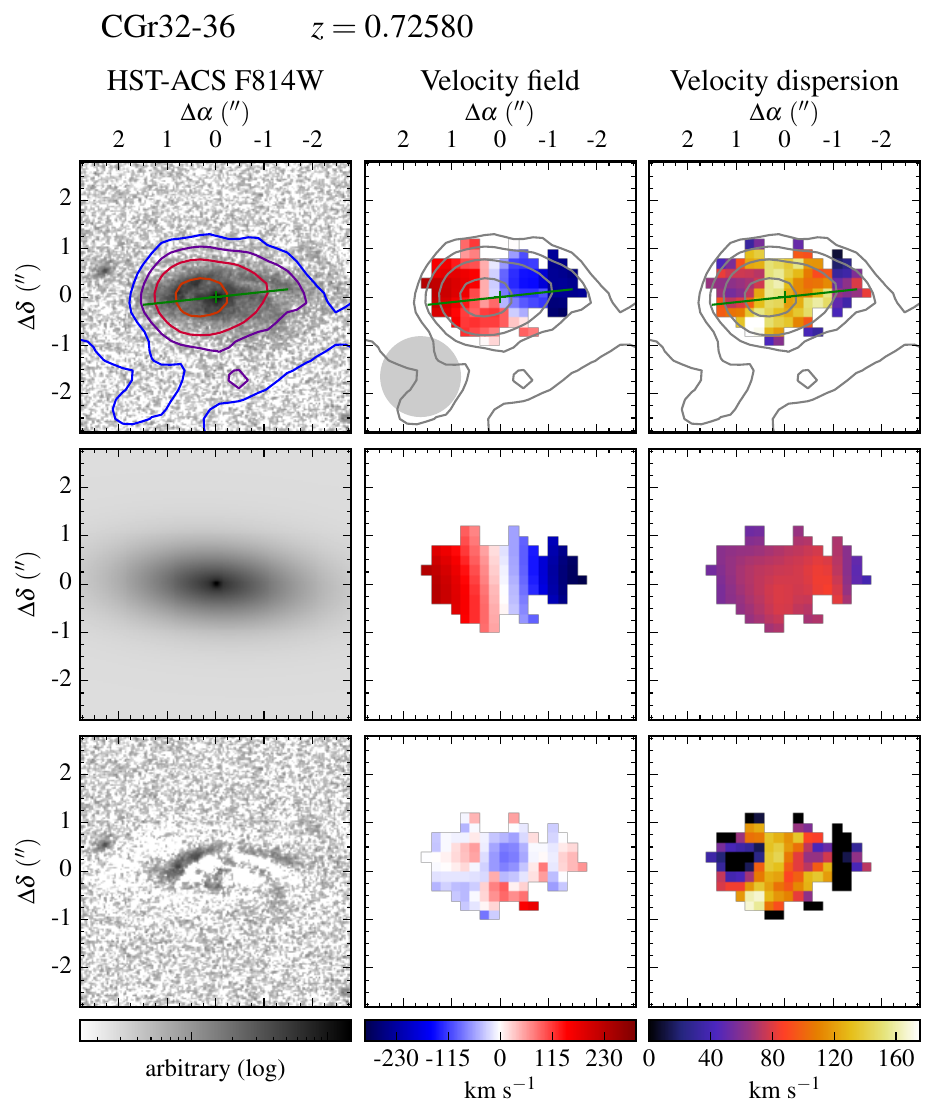}
\caption{Morpho-kinematics maps for galaxy CGr32-36. See caption of Fig. \ref{Mopho_KINMap_CGr28-41} for the description of figure.} 
\label{Mopho_KINMap_CGr32-36} 
\end{figure} 
 
\begin{figure}
\includegraphics[width=0.5\textwidth]{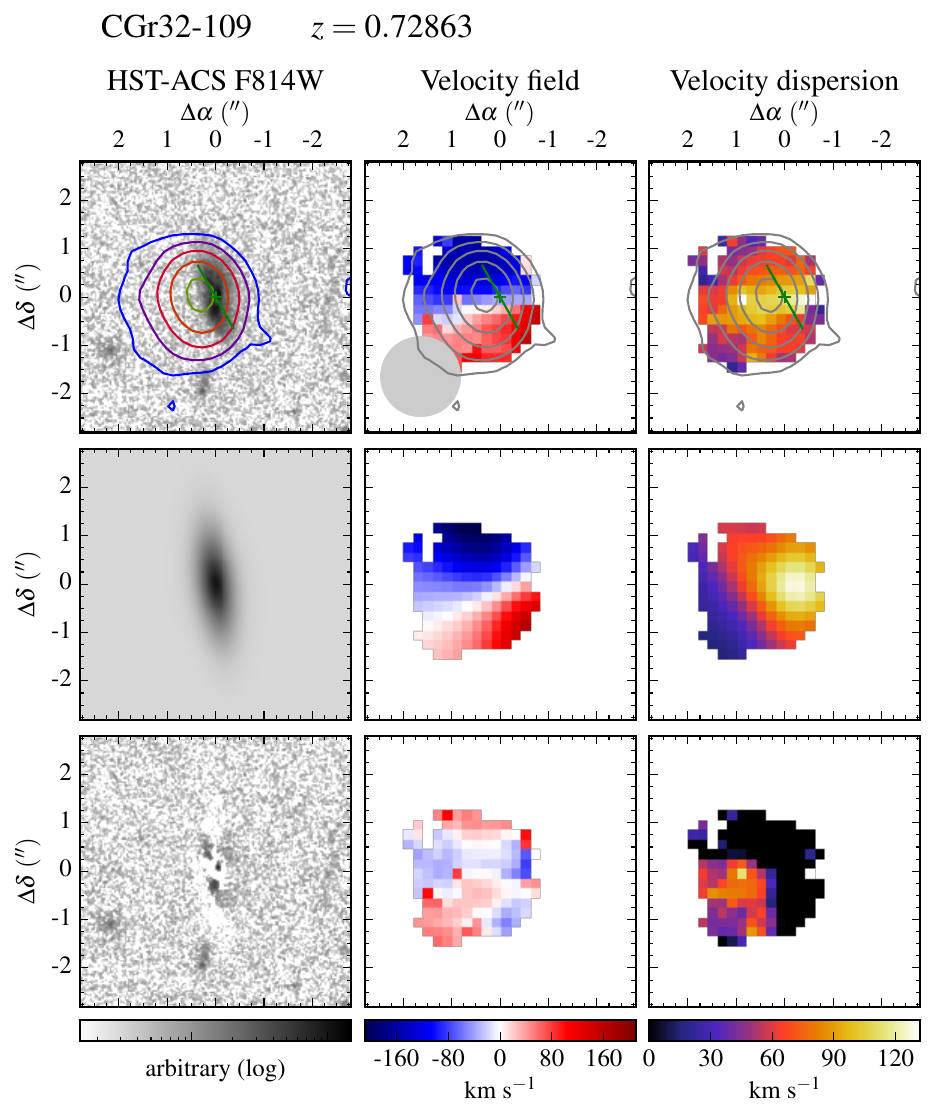}
\caption{Morpho-kinematics maps for galaxy CGr32-109. See caption of Fig. \ref{Mopho_KINMap_CGr28-41} for the description of figure.} 
\label{Mopho_KINMap_CGr32-109} 
\end{figure} 
 
\begin{figure}
\includegraphics[width=0.5\textwidth]{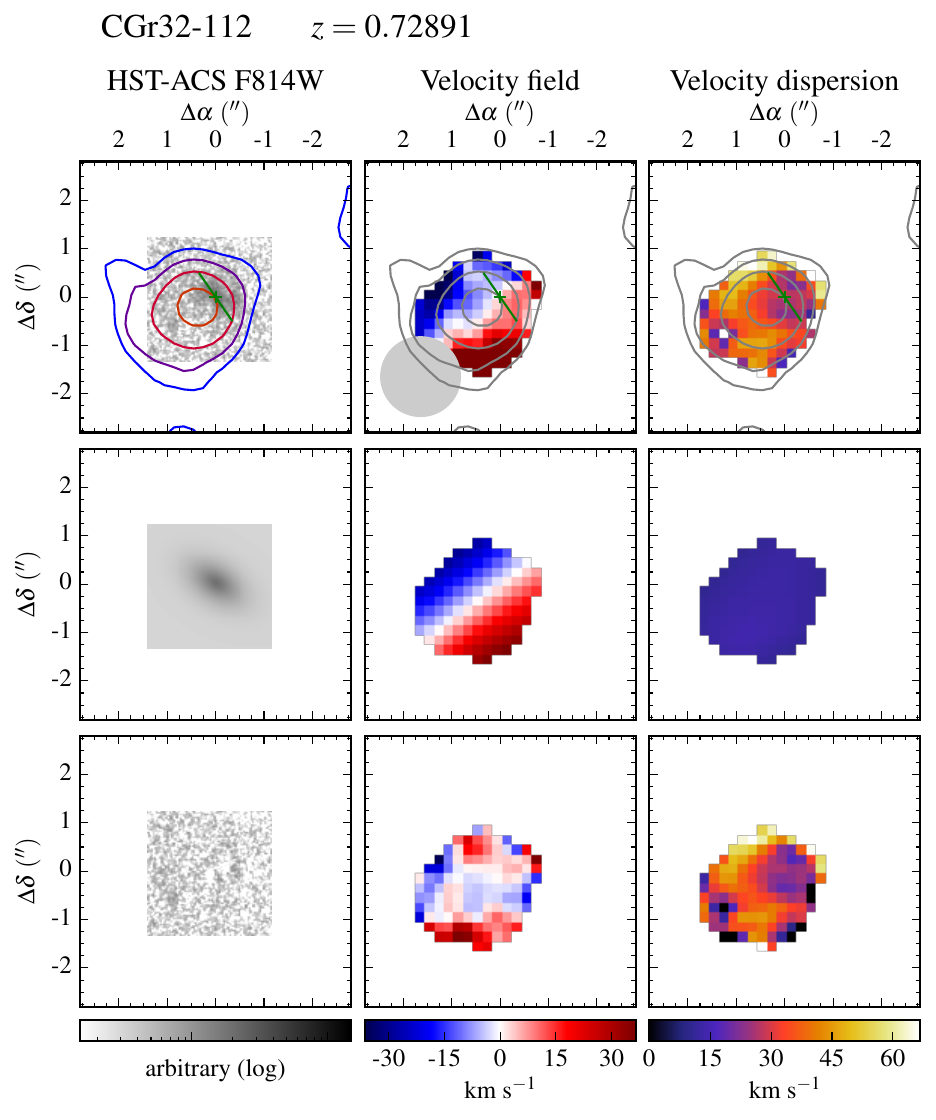}
\caption{Morpho-kinematics maps for galaxy CGr32-112. See caption of Fig. \ref{Mopho_KINMap_CGr28-41} for the description of figure.} 
\label{Mopho_KINMap_CGr32-112} 
\end{figure}

\begin{figure}
\includegraphics[width=0.5\textwidth]{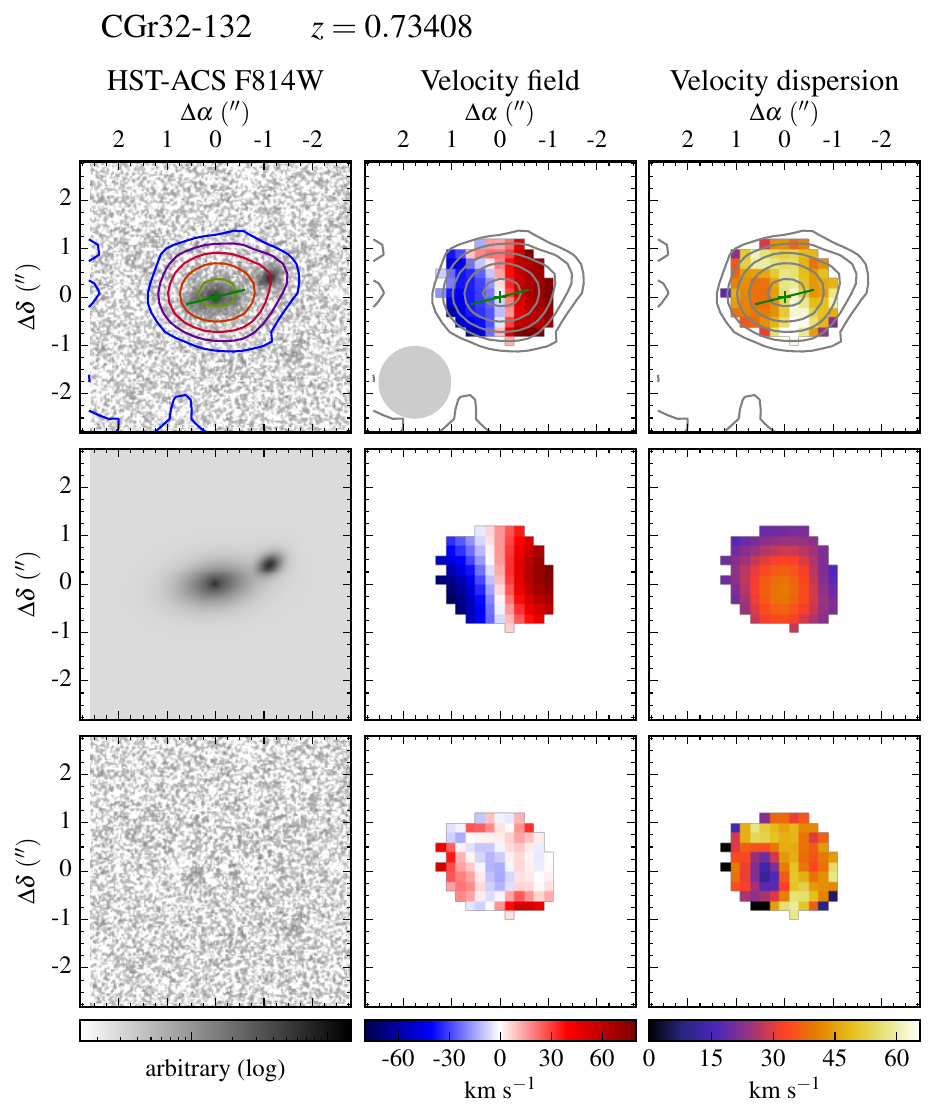}
\caption{Morpho-kinematics maps for galaxy CGr32-132. See caption of Fig. \ref{Mopho_KINMap_CGr28-41} for the description of figure.} 
\label{Mopho_KINMap_CGr32-132} 
\end{figure} 

\begin{figure}
\includegraphics[width=0.5\textwidth]{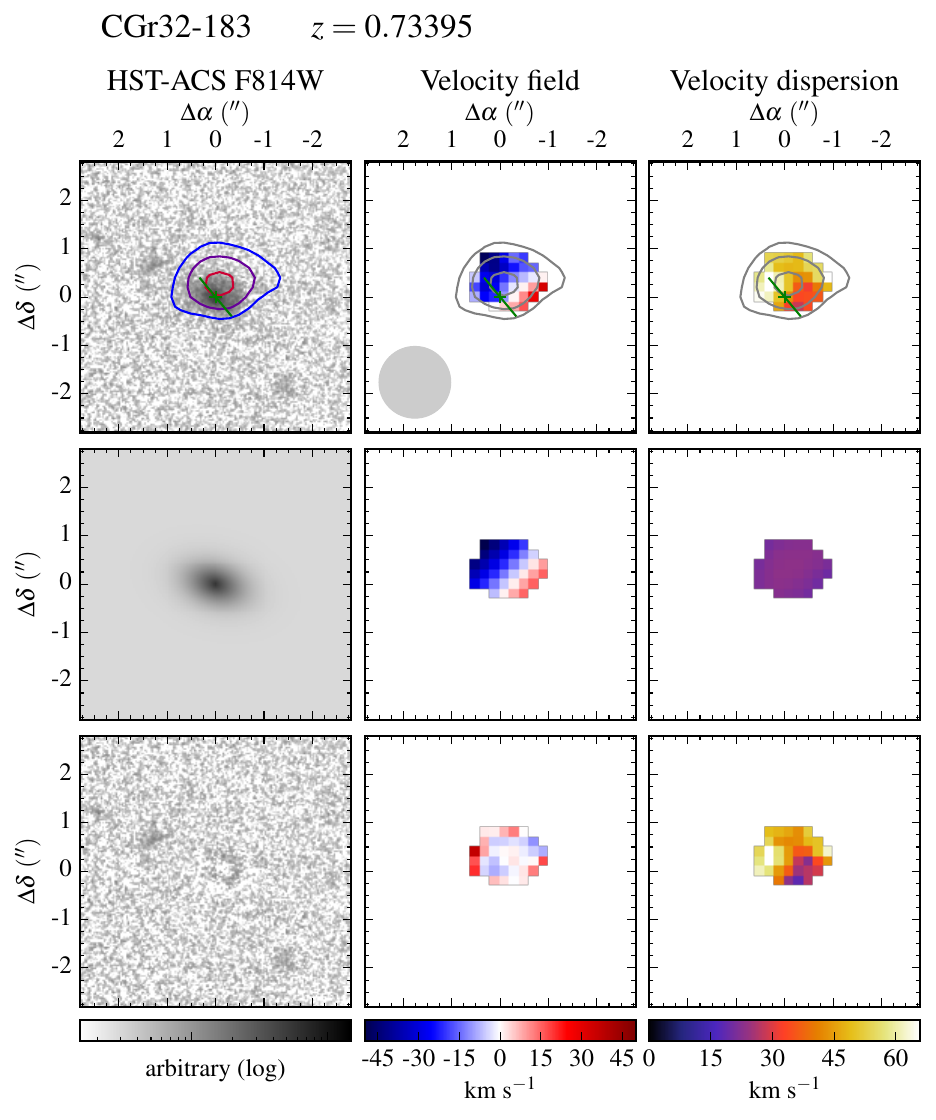}
\caption{Morpho-kinematics maps for galaxy CGr32-183. See caption of Fig. \ref{Mopho_KINMap_CGr28-41} for the description of figure.} 
\label{Mopho_KINMap_CGr32-183} 
\end{figure} 
 
\begin{figure}
\includegraphics[width=0.5\textwidth]{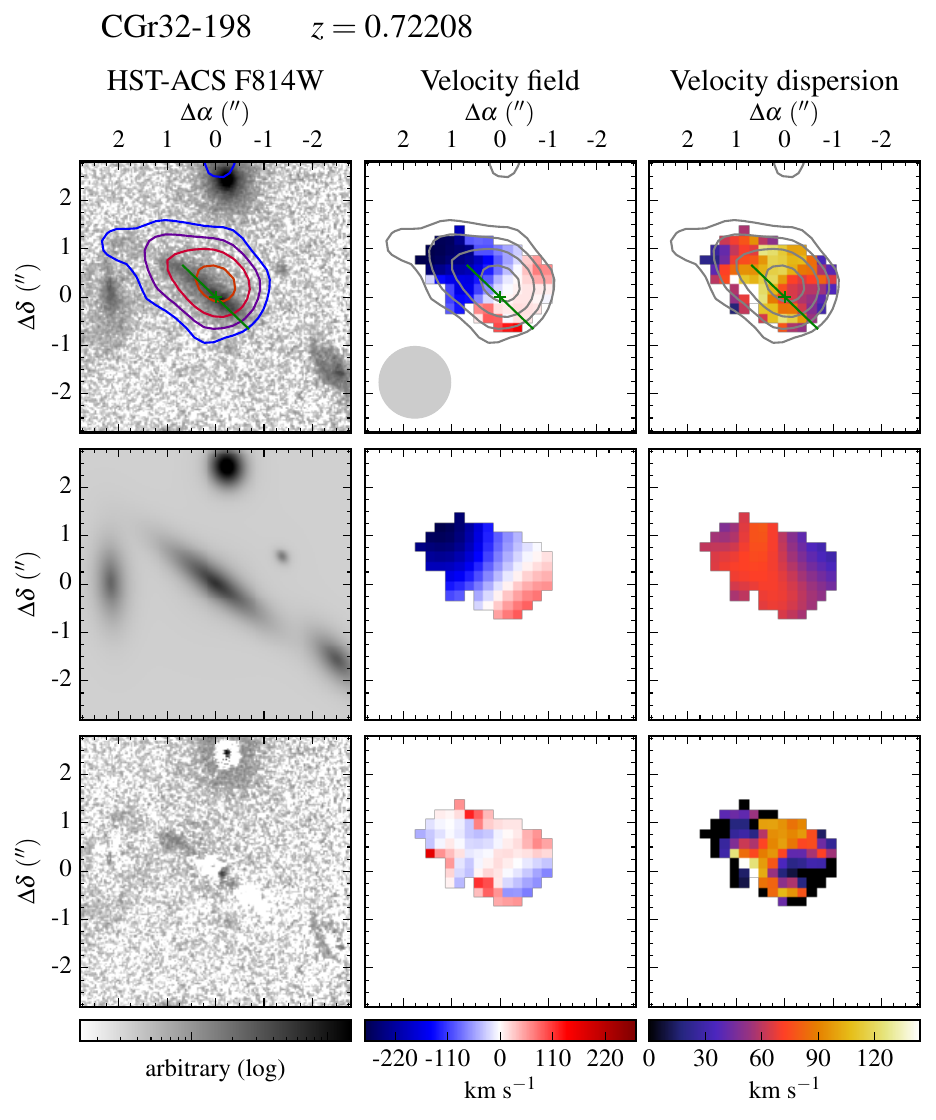}
\caption{Morpho-kinematics maps for galaxy CGr32-198. See caption of Fig. \ref{Mopho_KINMap_CGr28-41} for the description of figure.} 
\label{Mopho_KINMap_CGr32-198} 
\end{figure} 

\begin{figure}
\includegraphics[width=0.5\textwidth]{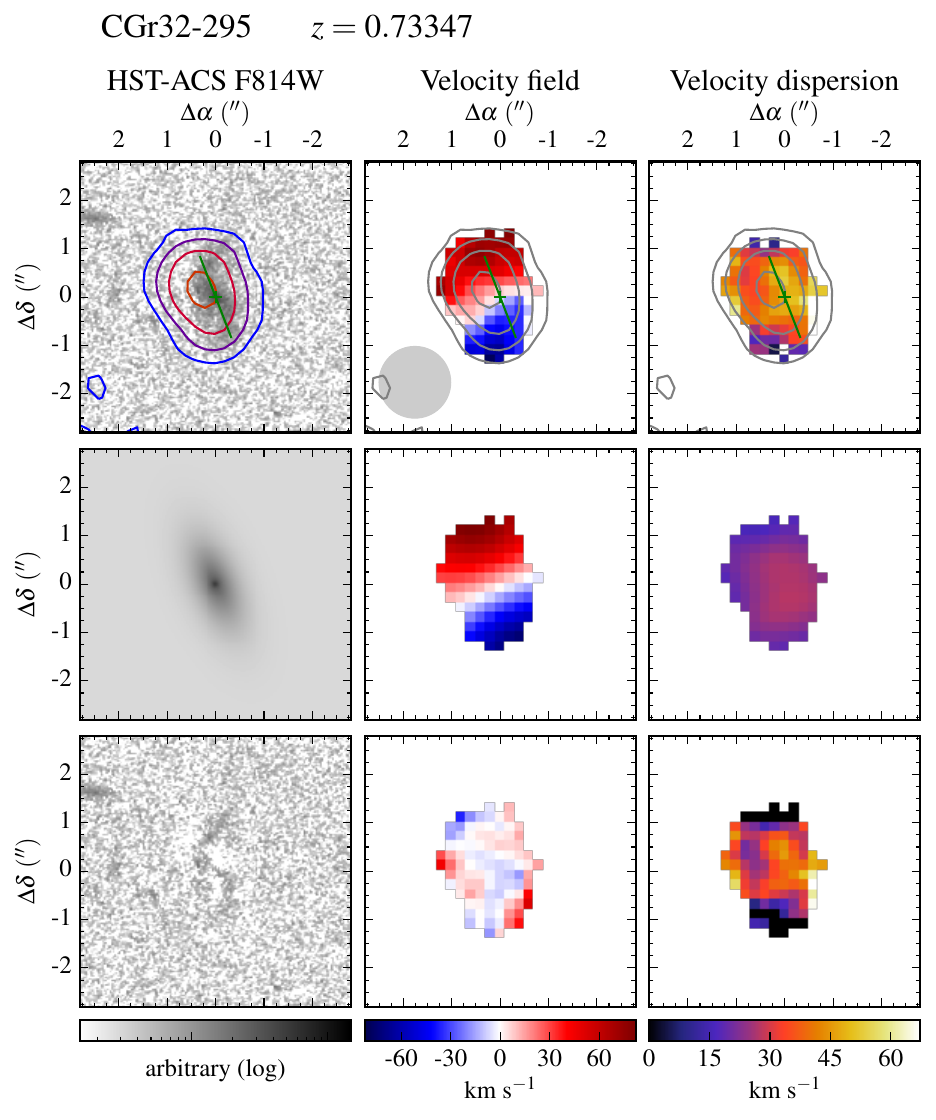}
\caption{Morpho-kinematics maps for galaxy CGr32-295. See caption of Fig. \ref{Mopho_KINMap_CGr28-41} for the description of figure.} 
\label{Mopho_KINMap_CGr32-295} 
\end{figure}

\begin{figure}
\includegraphics[width=0.5\textwidth]{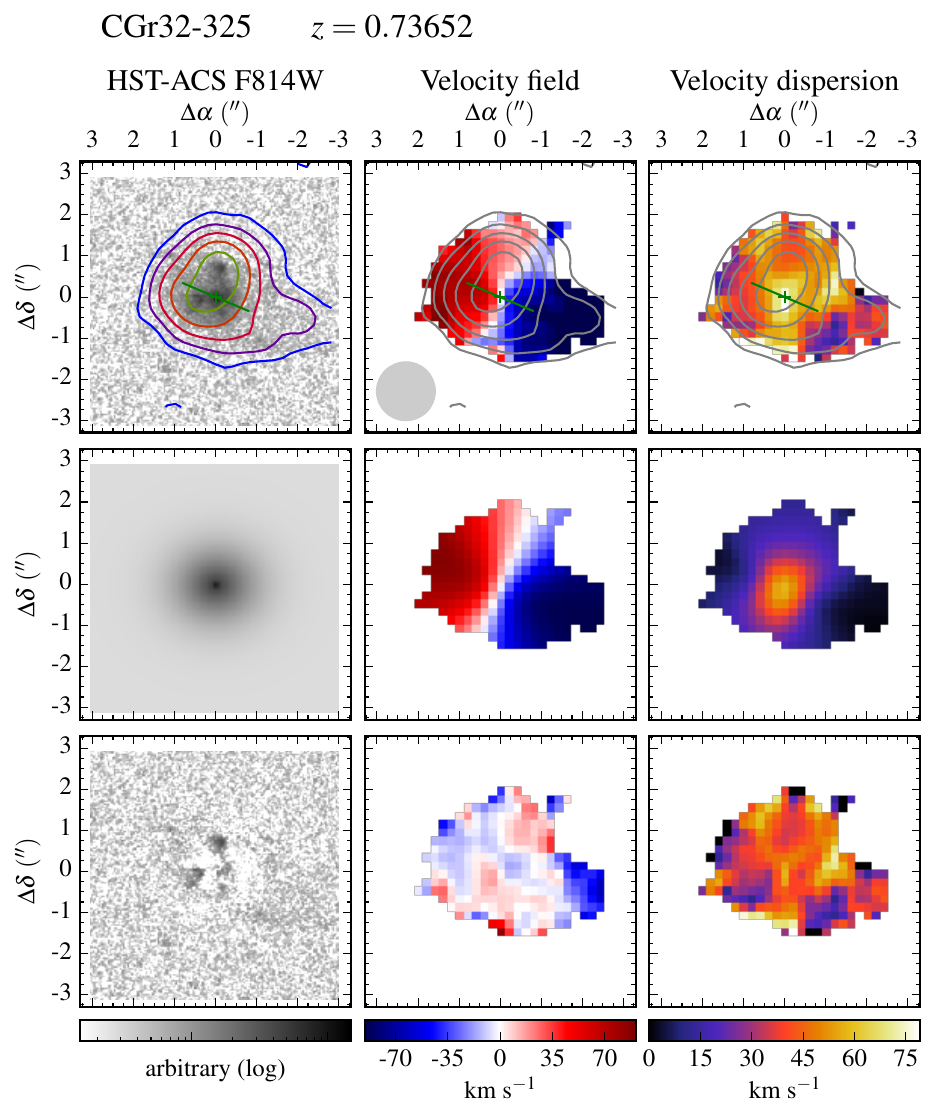}
\caption{Morpho-kinematics maps for galaxy CGr32-325. See caption of Fig. \ref{Mopho_KINMap_CGr28-41} for the description of figure.} 
\label{Mopho_KINMap_CGr32-325} 
\end{figure} 
 
\begin{figure}
\includegraphics[width=0.5\textwidth]{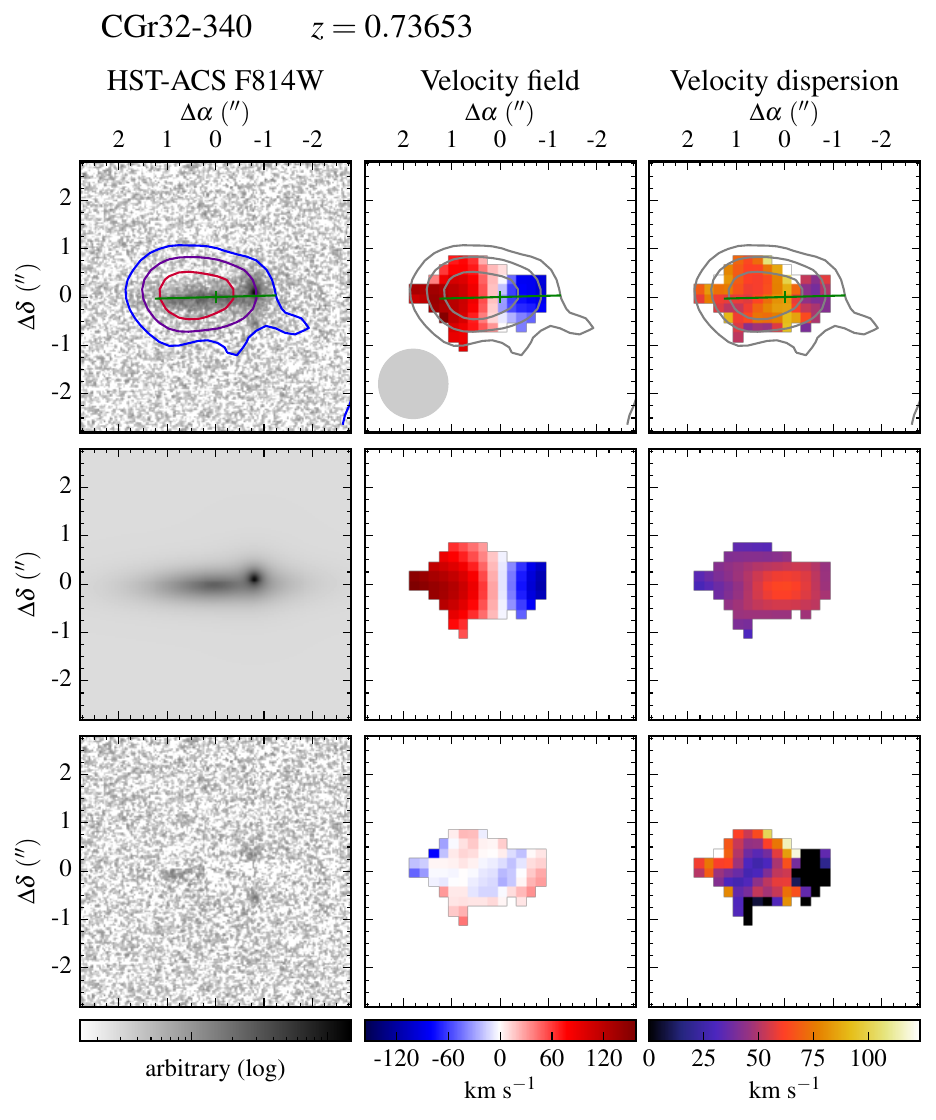}
\caption{Morpho-kinematics maps for galaxy CGr32-340. See caption of Fig. \ref{Mopho_KINMap_CGr28-41} for the description of figure.} 
\label{Mopho_KINMap_CGr32-340} 
\end{figure} 
 
\begin{figure}
\includegraphics[width=0.5\textwidth]{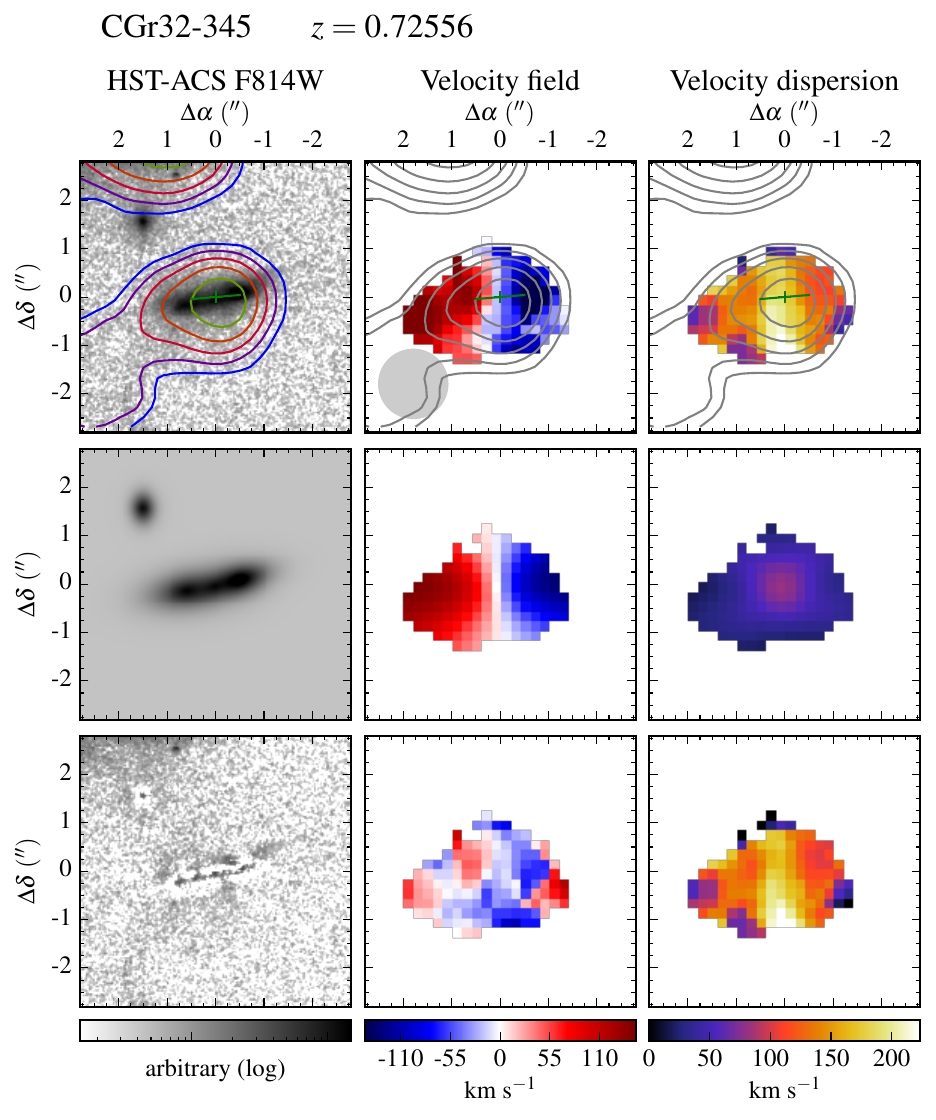}
\caption{Morpho-kinematics maps for galaxy CGr32-345. See caption of Fig. \ref{Mopho_KINMap_CGr28-41} for the description of figure.} 
\label{Mopho_KINMap_CGr32-345} 
\end{figure} 
 
\begin{figure}
\includegraphics[width=0.5\textwidth]{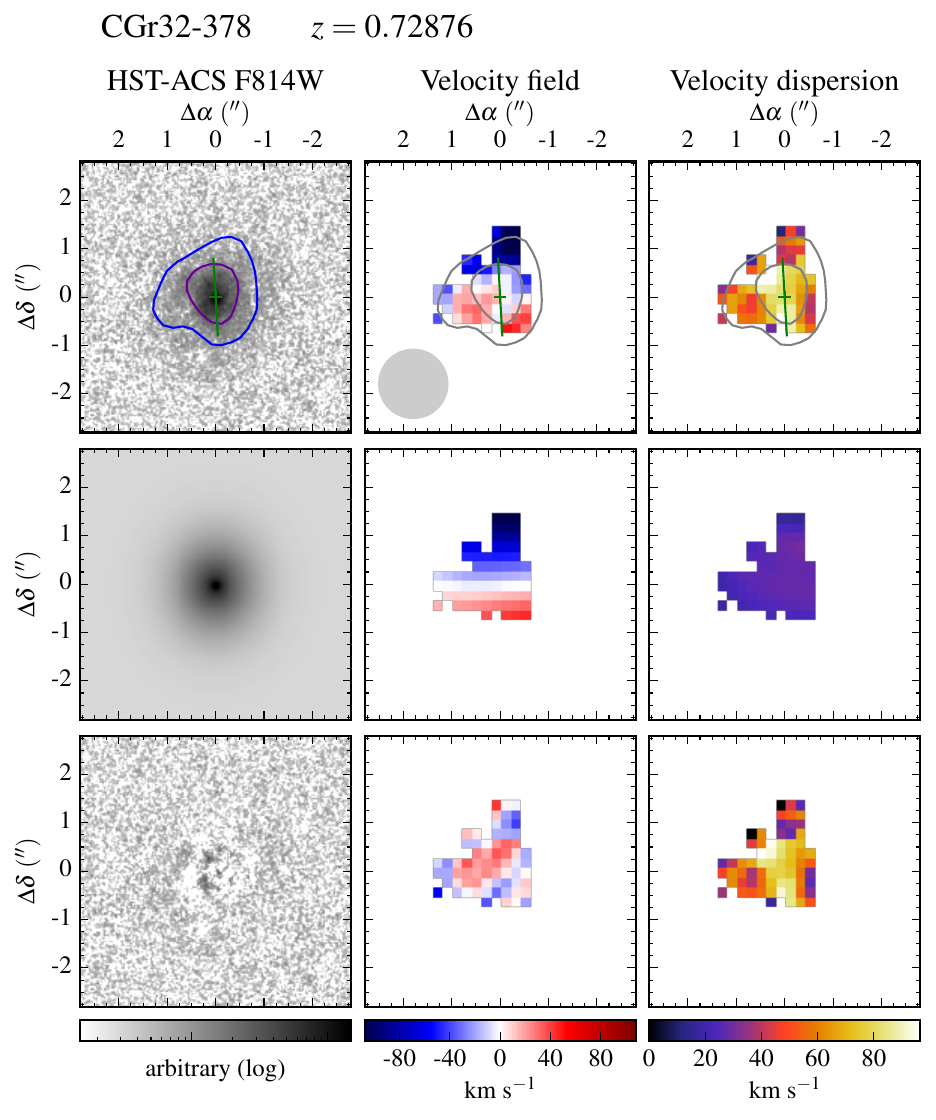}
\caption{Morpho-kinematics maps for galaxy CGr32-378. See caption of Fig. \ref{Mopho_KINMap_CGr28-41} for the description of figure.} 
\label{Mopho_KINMap_CGr32-378} 
\end{figure} 
 
\begin{figure}
\includegraphics[width=0.5\textwidth]{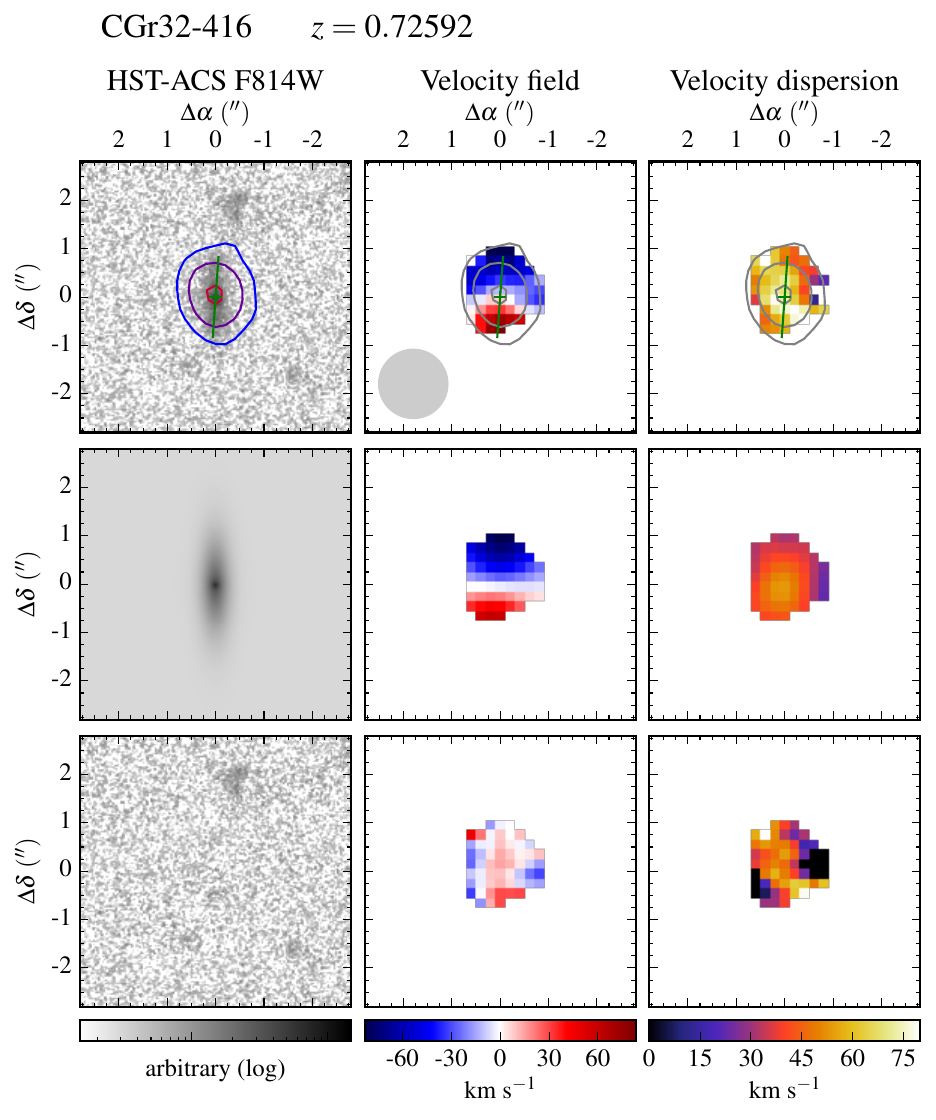}
\caption{Morpho-kinematics maps for galaxy CGr32-416. See caption of Fig. \ref{Mopho_KINMap_CGr28-41} for the description of figure.} 
\label{Mopho_KINMap_CGr32-416} 
\end{figure} 

\clearpage

\begin{figure}
\includegraphics[width=0.5\textwidth]{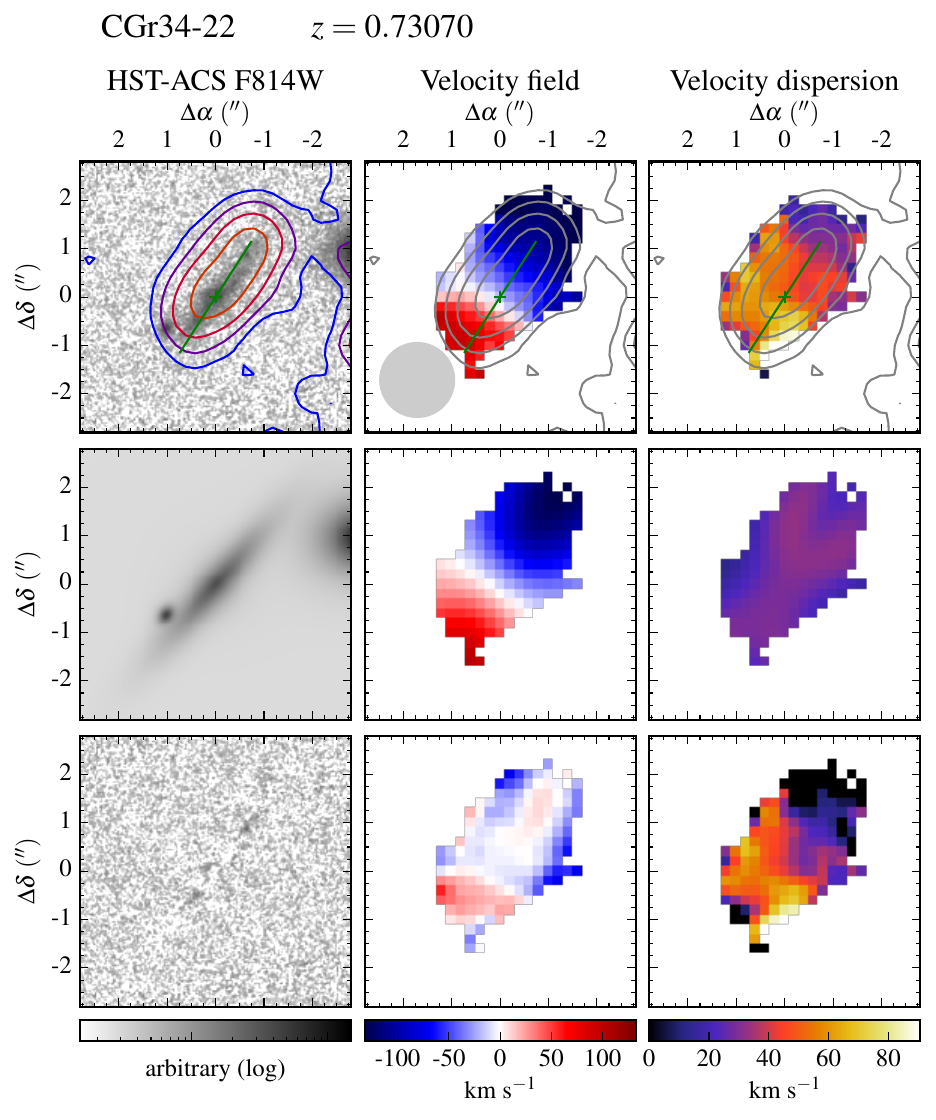}
\caption{Morpho-kinematics maps for galaxy CGr34-22. See caption of Fig. \ref{Mopho_KINMap_CGr28-41} for the description of figure.} 
\label{Mopho_KINMap_CGr34-22} 
\end{figure}

\begin{figure}
\includegraphics[width=0.5\textwidth]{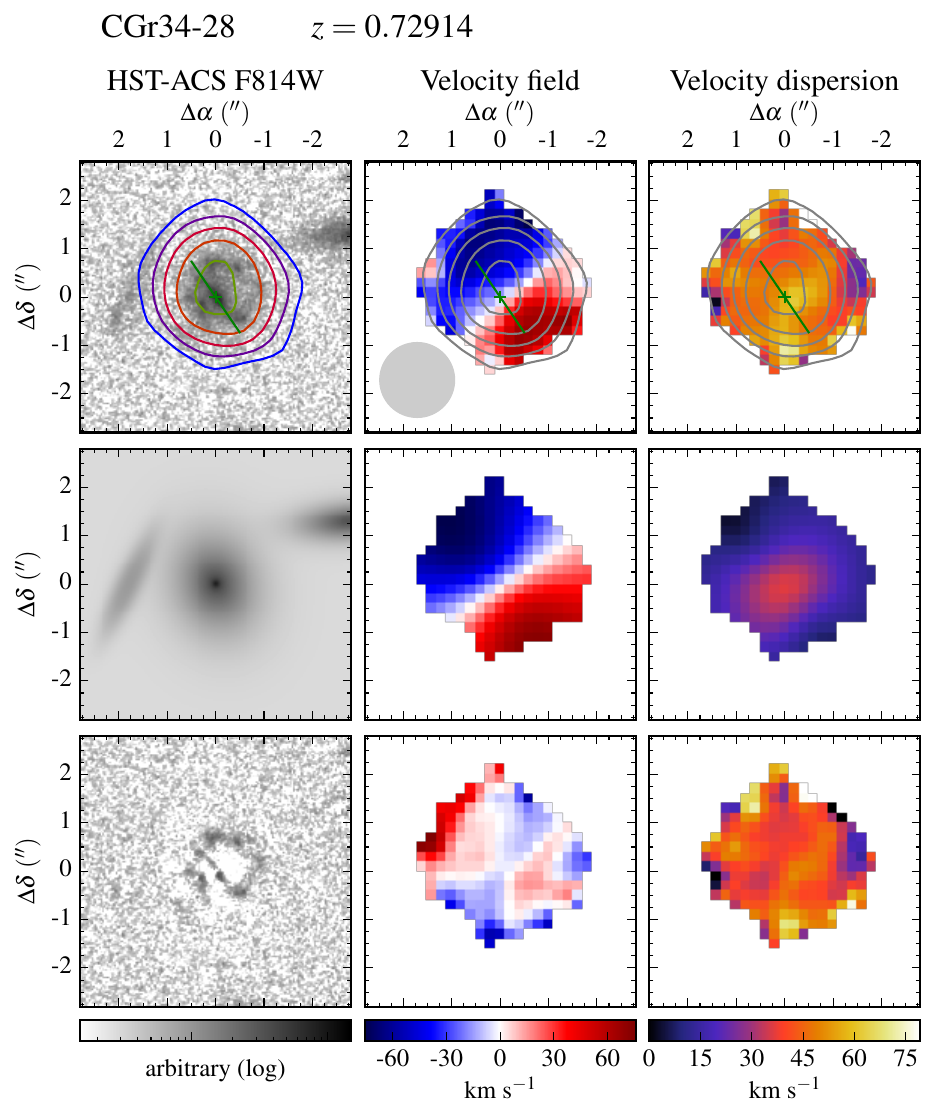}
\caption{Morpho-kinematics maps for galaxy CGr34-28. See caption of Fig. \ref{Mopho_KINMap_CGr28-41} for the description of figure.} 
\label{Mopho_KINMap_CGr34-28} 
\end{figure} 
 
\begin{figure}
\includegraphics[width=0.5\textwidth]{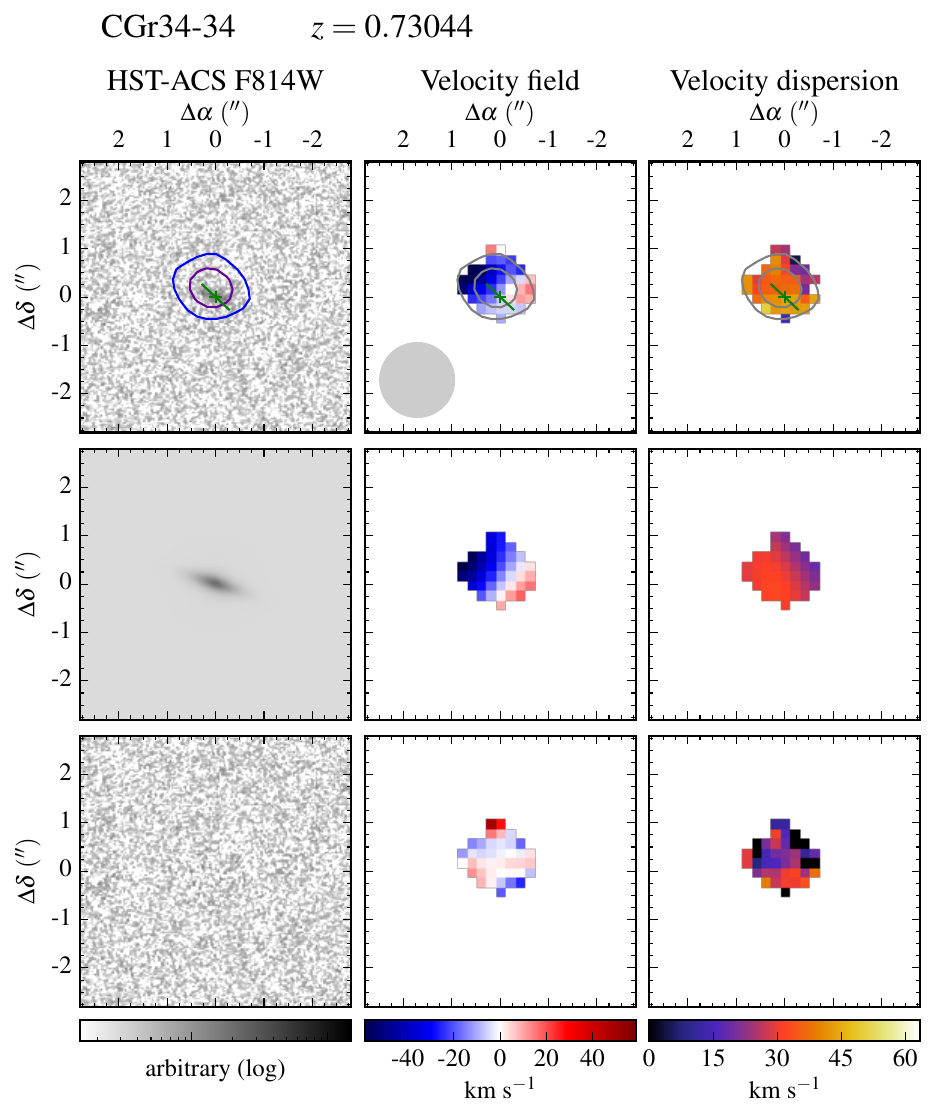}
\caption{Morpho-kinematics maps for galaxy CGr34-34. See caption of Fig. \ref{Mopho_KINMap_CGr28-41} for the description of figure.} 
\label{Mopho_KINMap_CGr34-34} 
\end{figure} 
 
\begin{figure}
\includegraphics[width=0.5\textwidth]{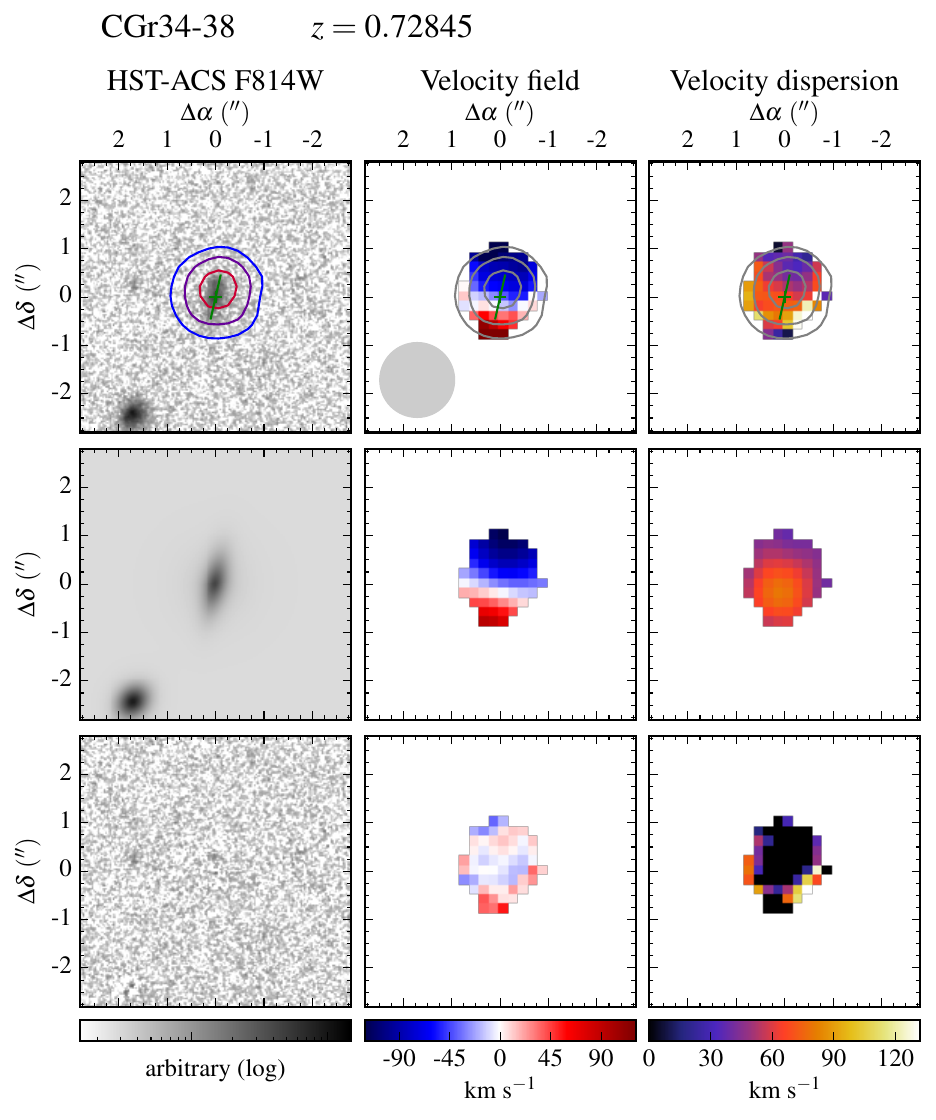}
\caption{Morpho-kinematics maps for galaxy CGr34-38. See caption of Fig. \ref{Mopho_KINMap_CGr28-41} for the description of figure.} 
\label{Mopho_KINMap_CGr34-38} 
\end{figure} 
 
\begin{figure}
\includegraphics[width=0.5\textwidth]{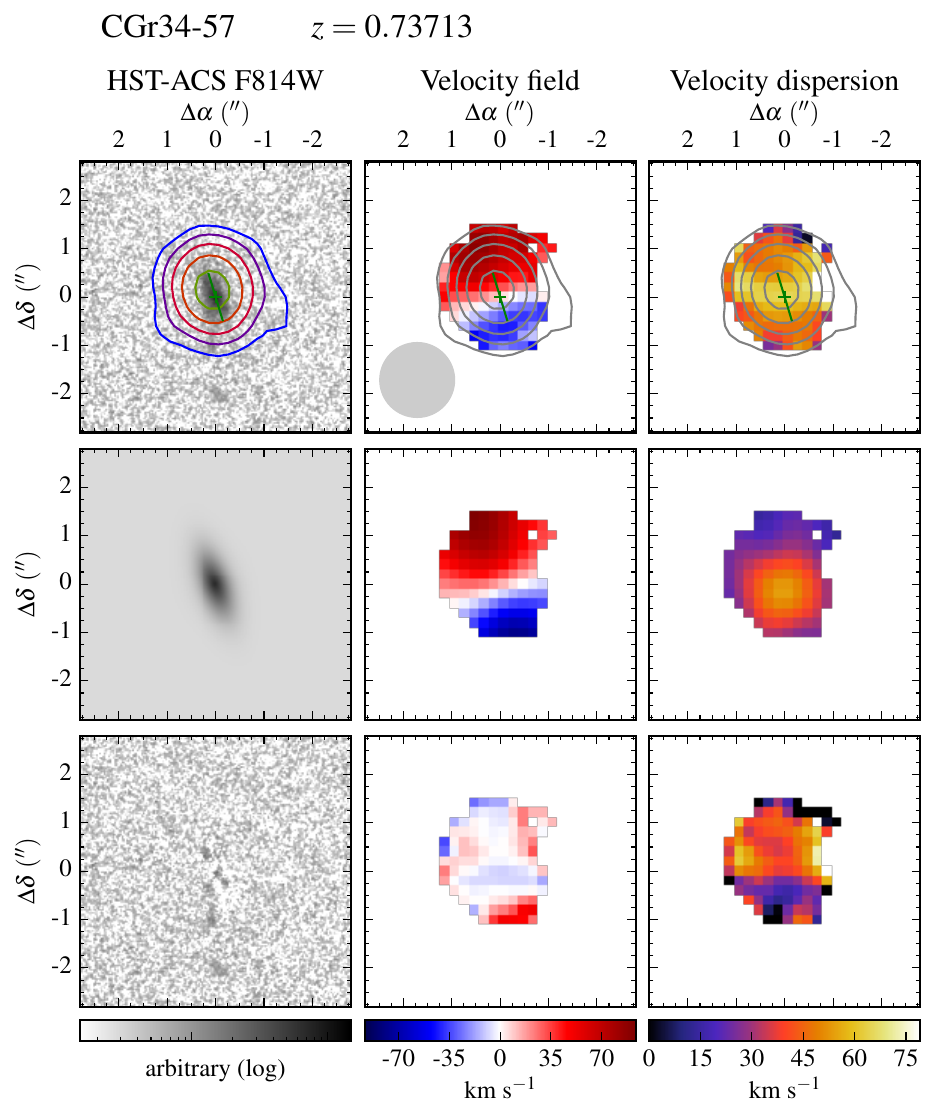}
\caption{Morpho-kinematics maps for galaxy CGr34-57. See caption of Fig. \ref{Mopho_KINMap_CGr28-41} for the description of figure.} 
\label{Mopho_KINMap_CGr34-57} 
\end{figure} 
 
\begin{figure}
\includegraphics[width=0.5\textwidth]{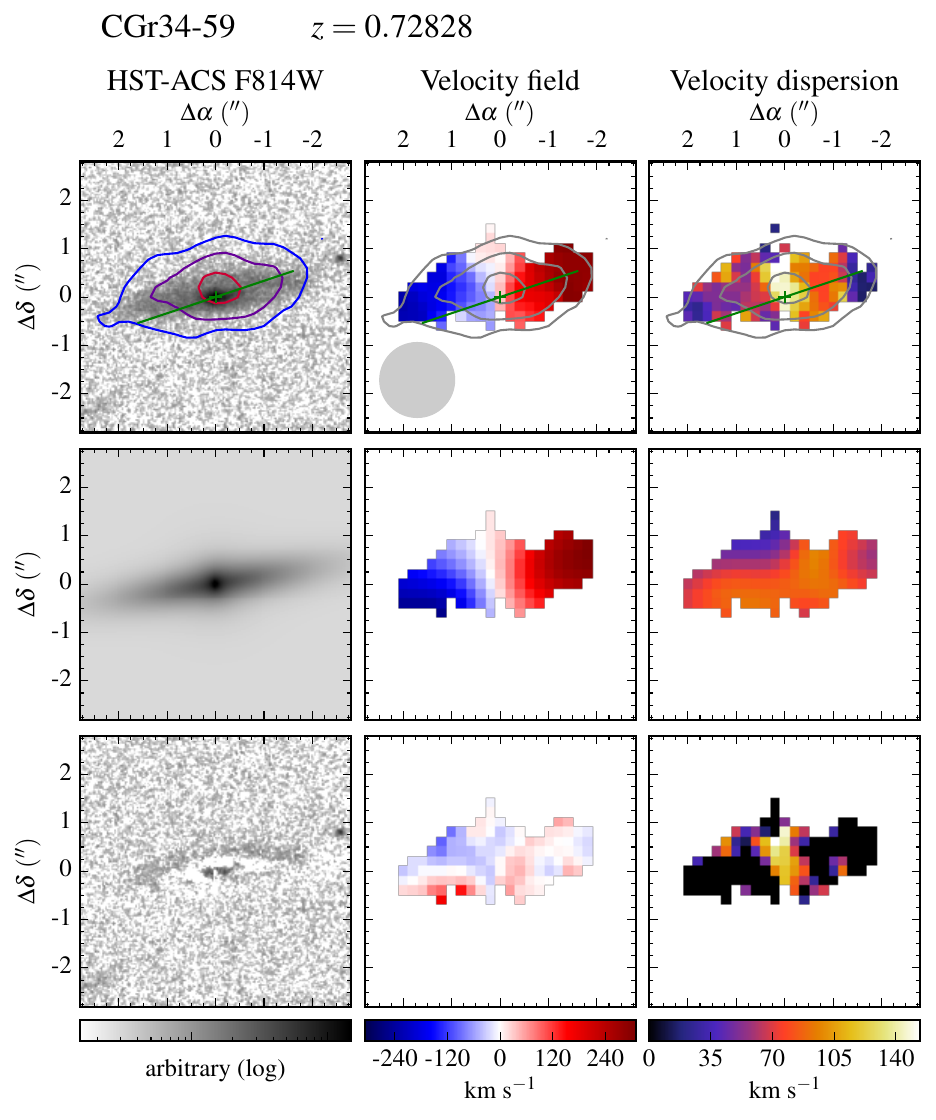}
\caption{Morpho-kinematics maps for galaxy CGr34-59. See caption of Fig. \ref{Mopho_KINMap_CGr28-41} for the description of figure.} 
\label{Mopho_KINMap_CGr34-59} 
\end{figure} 
 
\begin{figure}
\includegraphics[width=0.5\textwidth]{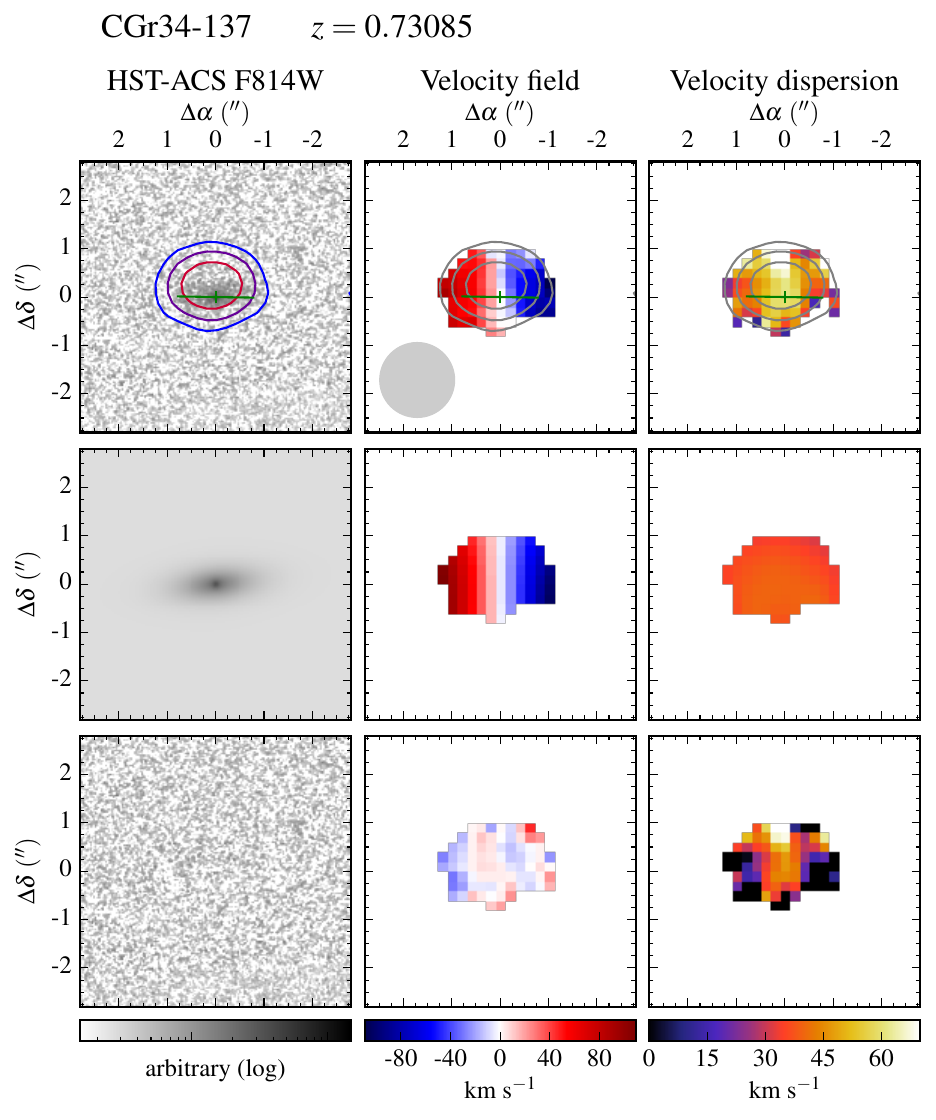}
\caption{Morpho-kinematics maps for galaxy CGr34-137. See caption of Fig. \ref{Mopho_KINMap_CGr28-41} for the description of figure.} 
\label{Mopho_KINMap_CGr34-137} 
\end{figure} 
 
\begin{figure}
\includegraphics[width=0.5\textwidth]{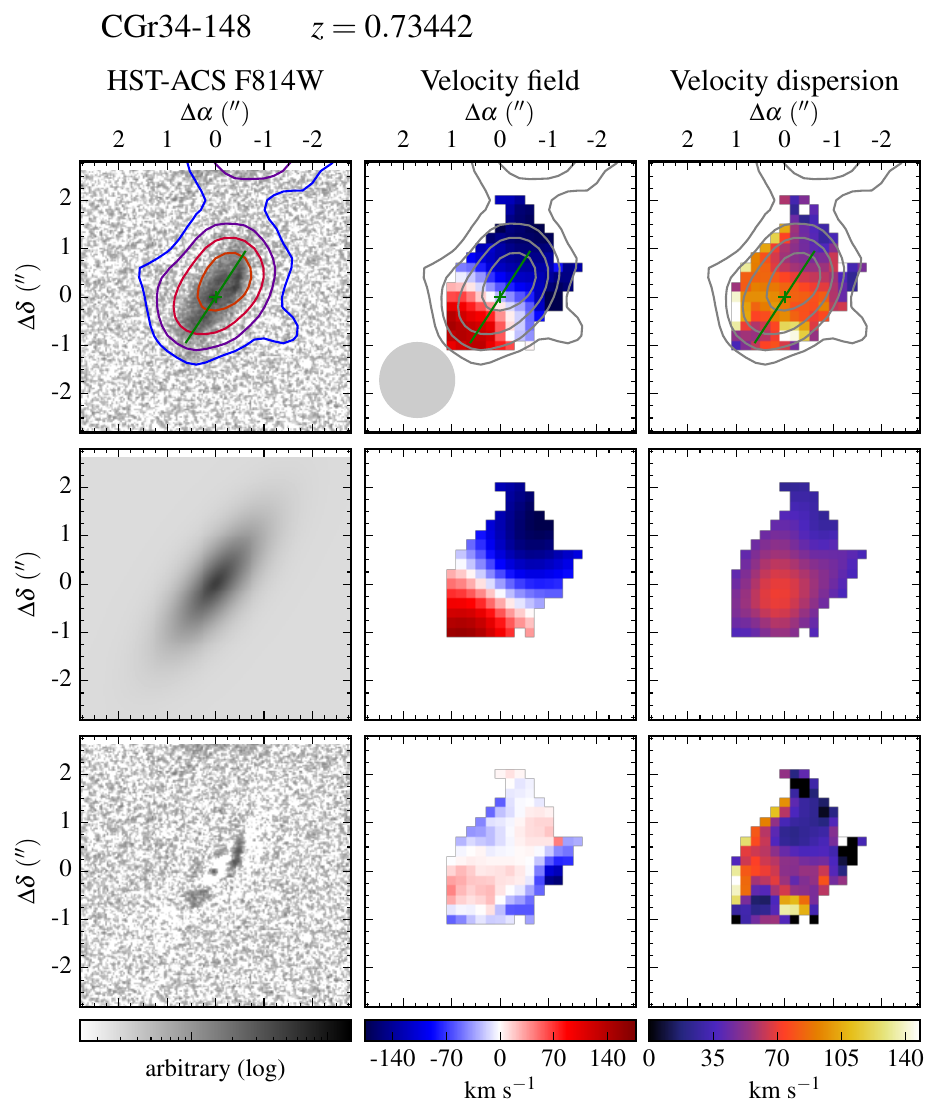}
\caption{Morpho-kinematics maps for galaxy CGr34-148. See caption of Fig. \ref{Mopho_KINMap_CGr28-41} for the description of figure.} 
\label{Mopho_KINMap_CGr34-148} 
\end{figure} 
 
\begin{figure}
\includegraphics[width=0.5\textwidth]{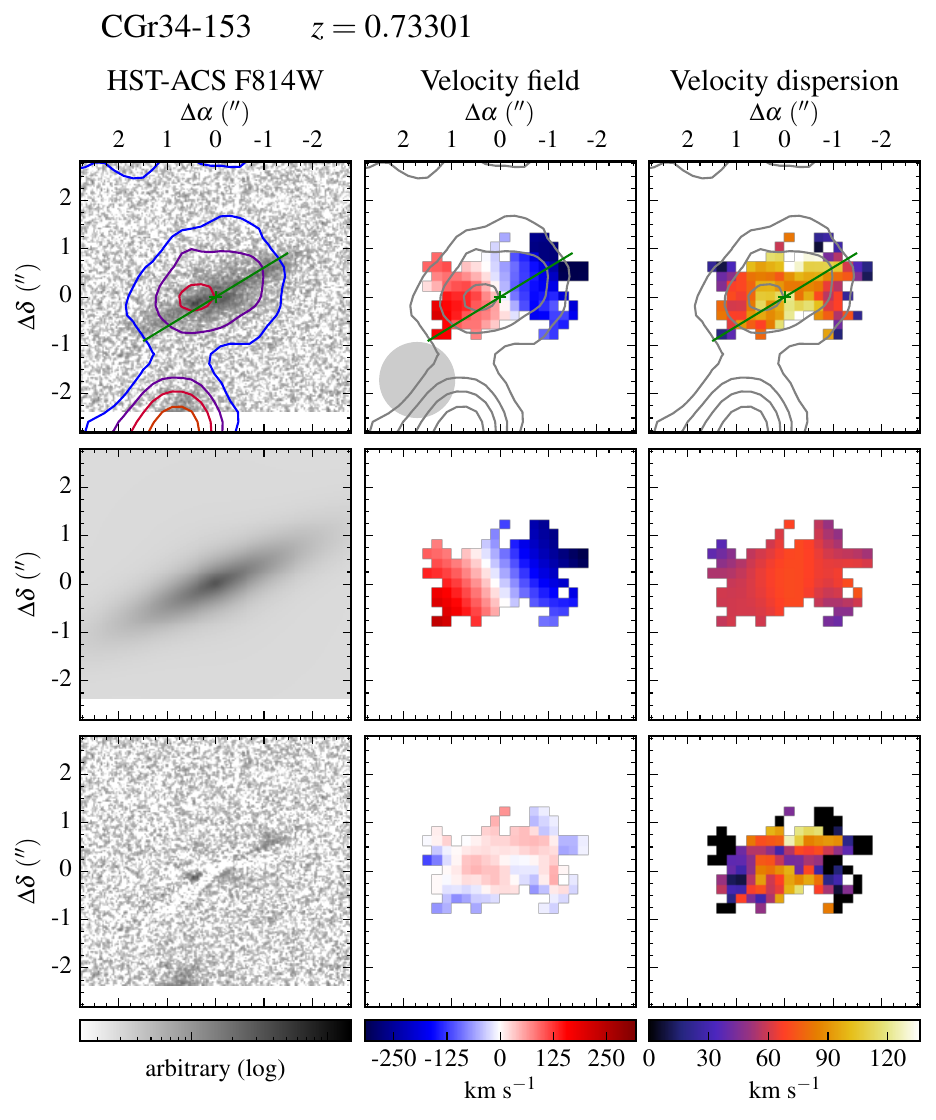}
\caption{Morpho-kinematics maps for galaxy CGr34-153. See caption of Fig. \ref{Mopho_KINMap_CGr28-41} for the description of figure.} 
\label{Mopho_KINMap_CGr34-153} 
\end{figure} 

\clearpage

\begin{figure}
\includegraphics[width=0.5\textwidth]{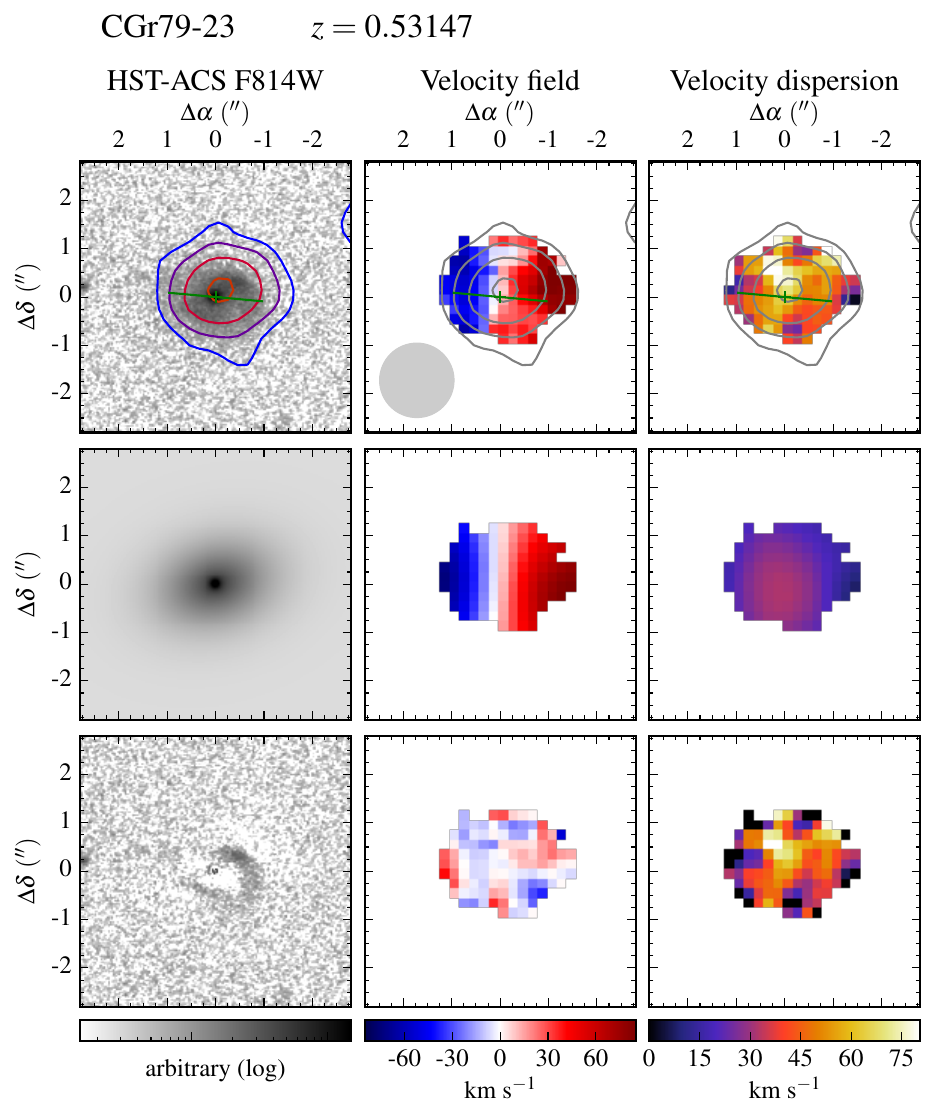}
\caption{Morpho-kinematics maps for galaxy CGr79-23. See caption of Fig. \ref{Mopho_KINMap_CGr28-41} for the description of figure.} 
\label{Mopho_KINMap_CGr79-23} 
\end{figure} 
 
\begin{figure}
\includegraphics[width=0.5\textwidth]{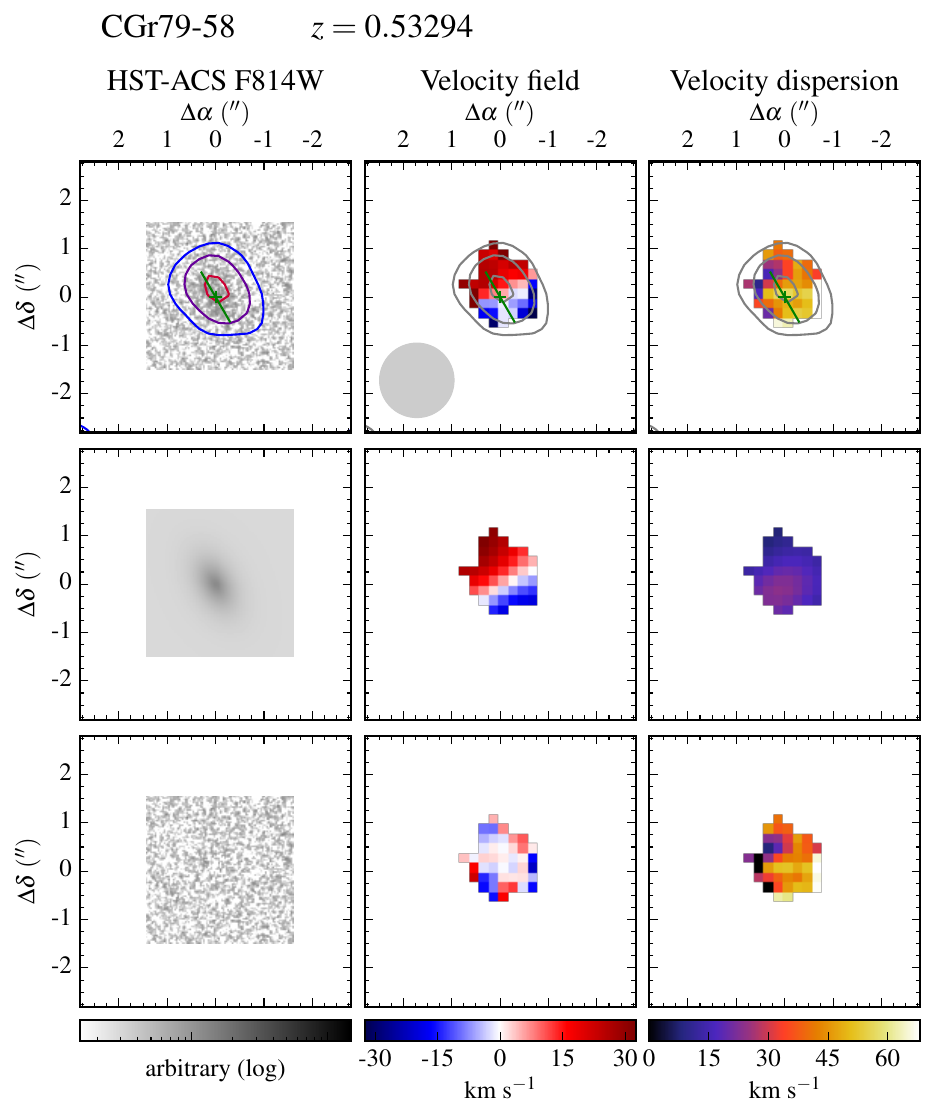}
\caption{Morpho-kinematics maps for galaxy CGr79-58. See caption of Fig. \ref{Mopho_KINMap_CGr28-41} for the description of figure.} 
\label{Mopho_KINMap_CGr79-58} 
\end{figure} 
 
\begin{figure}
\includegraphics[width=0.5\textwidth]{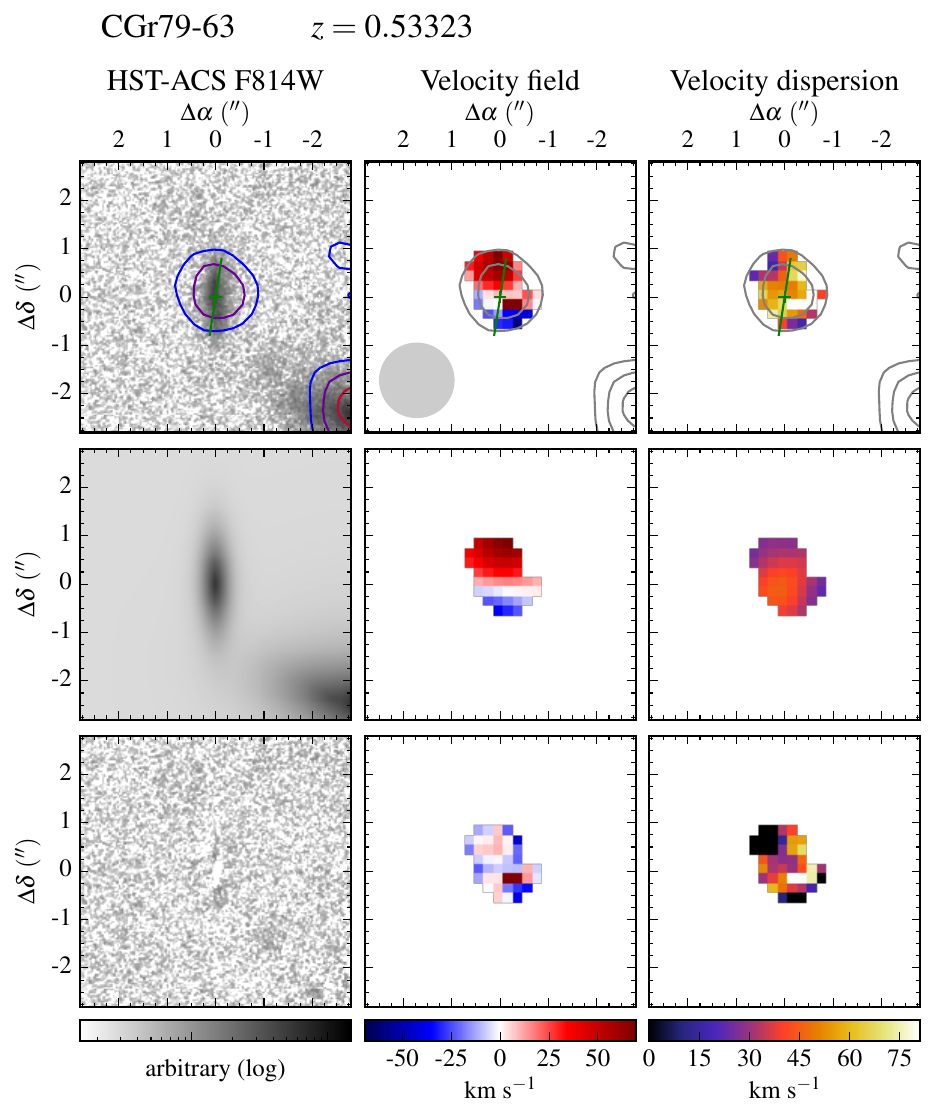}
\caption{Morpho-kinematics maps for galaxy CGr79-63. See caption of Fig. \ref{Mopho_KINMap_CGr28-41} for the description of figure.} 
\label{Mopho_KINMap_CGr79-63} 
\end{figure} 
 
\begin{figure}
\includegraphics[width=0.5\textwidth]{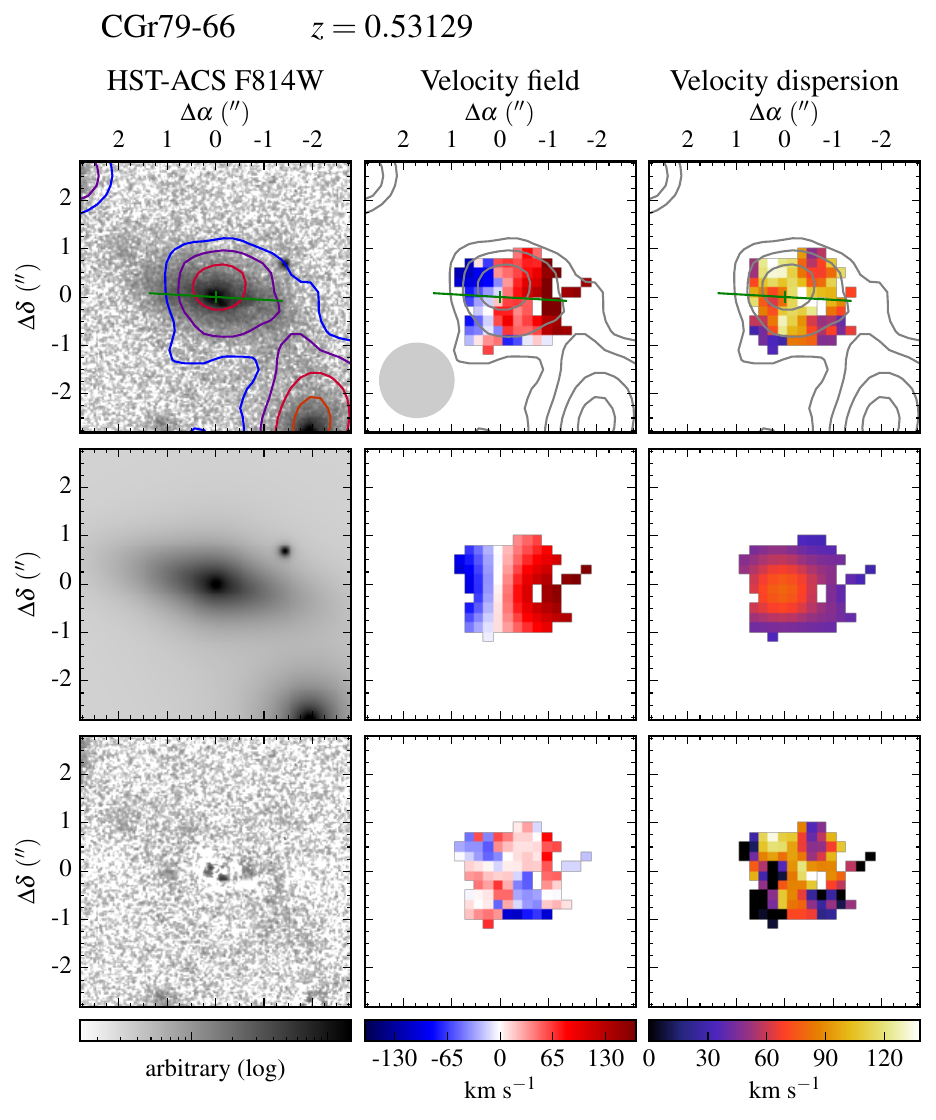}
\caption{Morpho-kinematics maps for galaxy CGr79-66. See caption of Fig. \ref{Mopho_KINMap_CGr28-41} for the description of figure.} 
\label{Mopho_KINMap_CGr79-66} 
\end{figure} 
 
\begin{figure}
\includegraphics[width=0.5\textwidth]{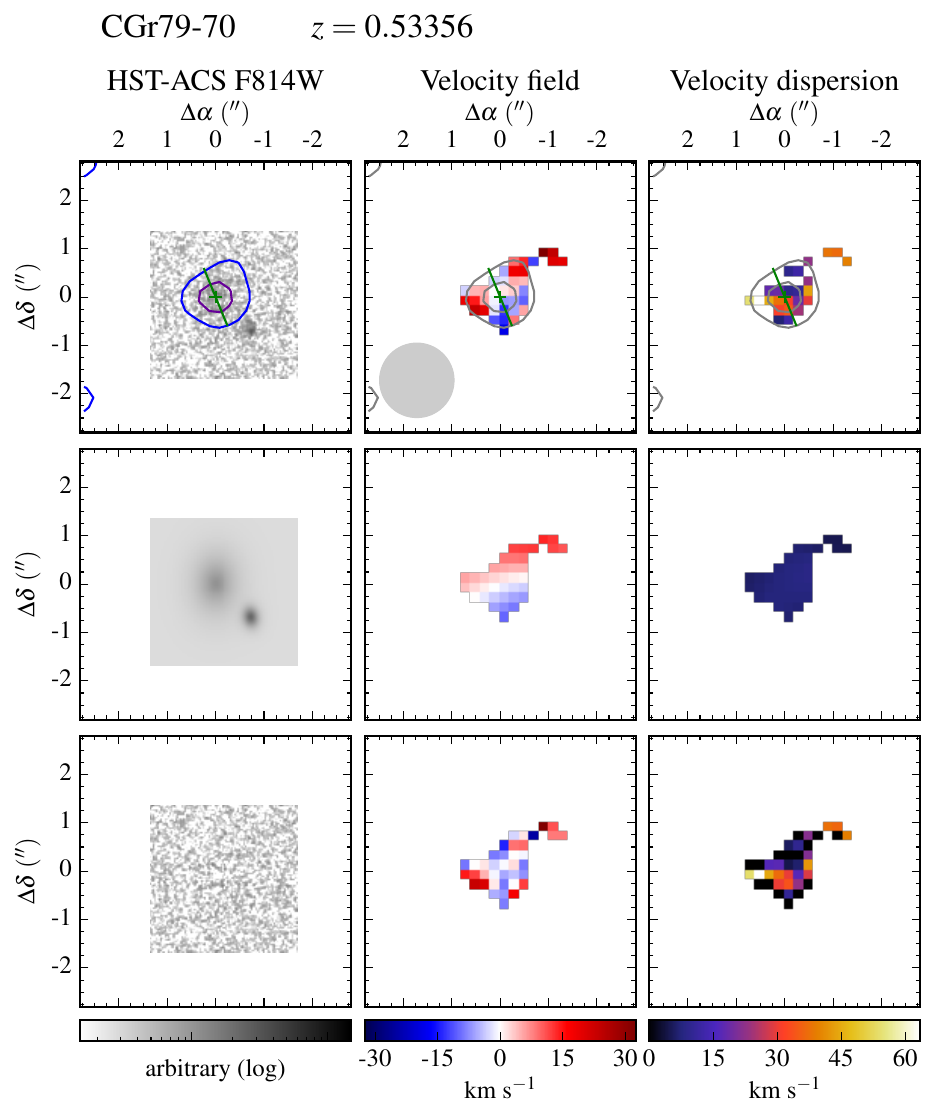}
\caption{Morpho-kinematics maps for galaxy CGr79-70. See caption of Fig. \ref{Mopho_KINMap_CGr28-41} for the description of figure.} 
\label{Mopho_KINMap_CGr79-70} 
\end{figure} 
 
\begin{figure}
\includegraphics[width=0.5\textwidth]{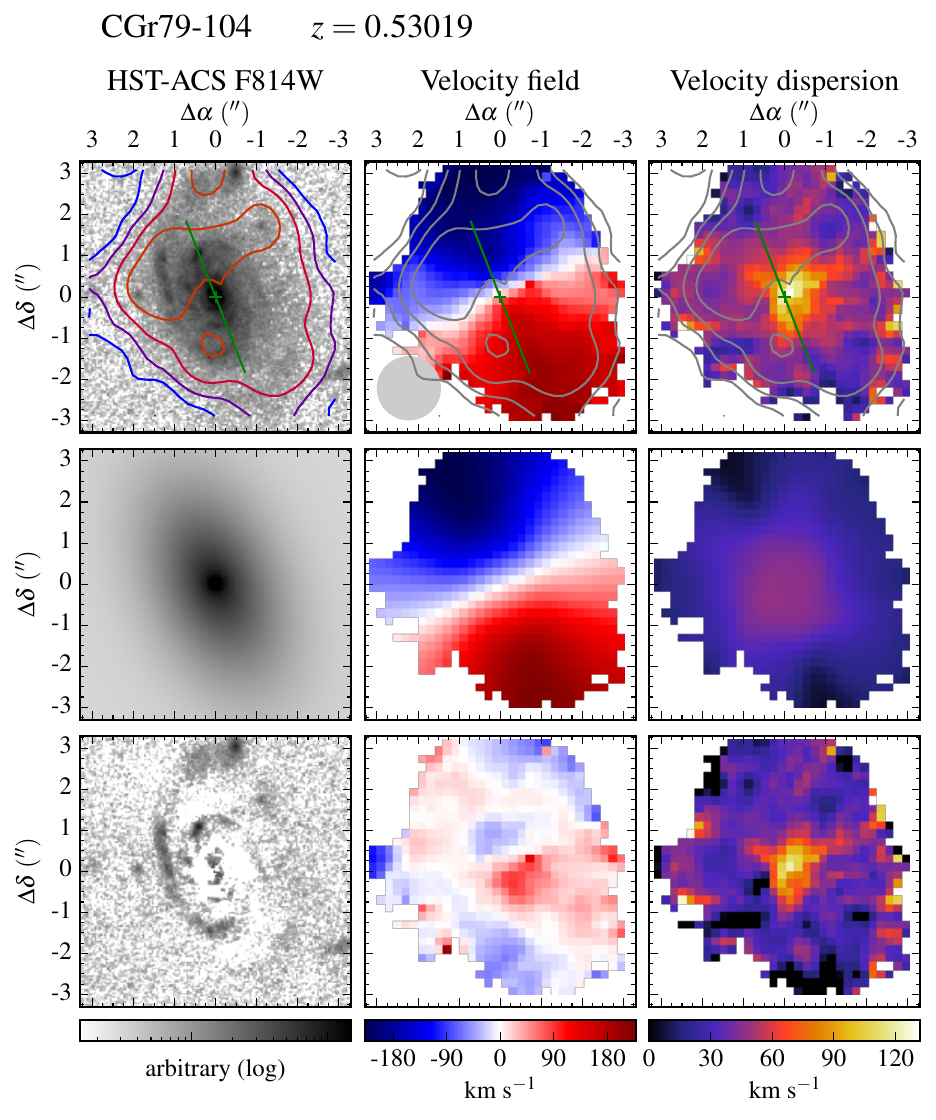}
\caption{Morpho-kinematics maps for galaxy CGr79-104. See caption of Fig. \ref{Mopho_KINMap_CGr28-41} for the description of figure.} 
\label{Mopho_KINMap_CGr79-104} 
\end{figure} 
 
\begin{figure}
\includegraphics[width=0.5\textwidth]{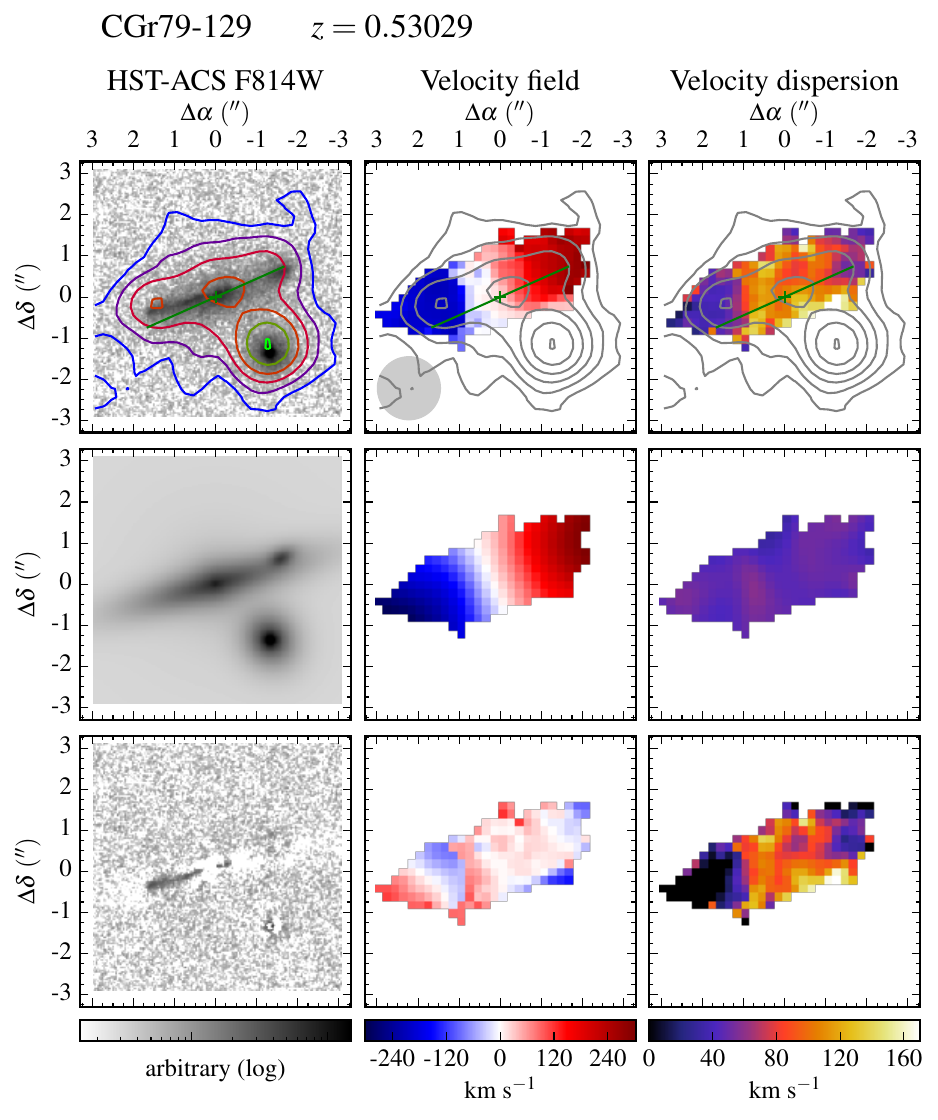}
\caption{Morpho-kinematics maps for galaxy CGr79-129. See caption of Fig. \ref{Mopho_KINMap_CGr28-41} for the description of figure.} 
\label{Mopho_KINMap_CGr79-129} 
\end{figure} 
 
\begin{figure}
\includegraphics[width=0.5\textwidth]{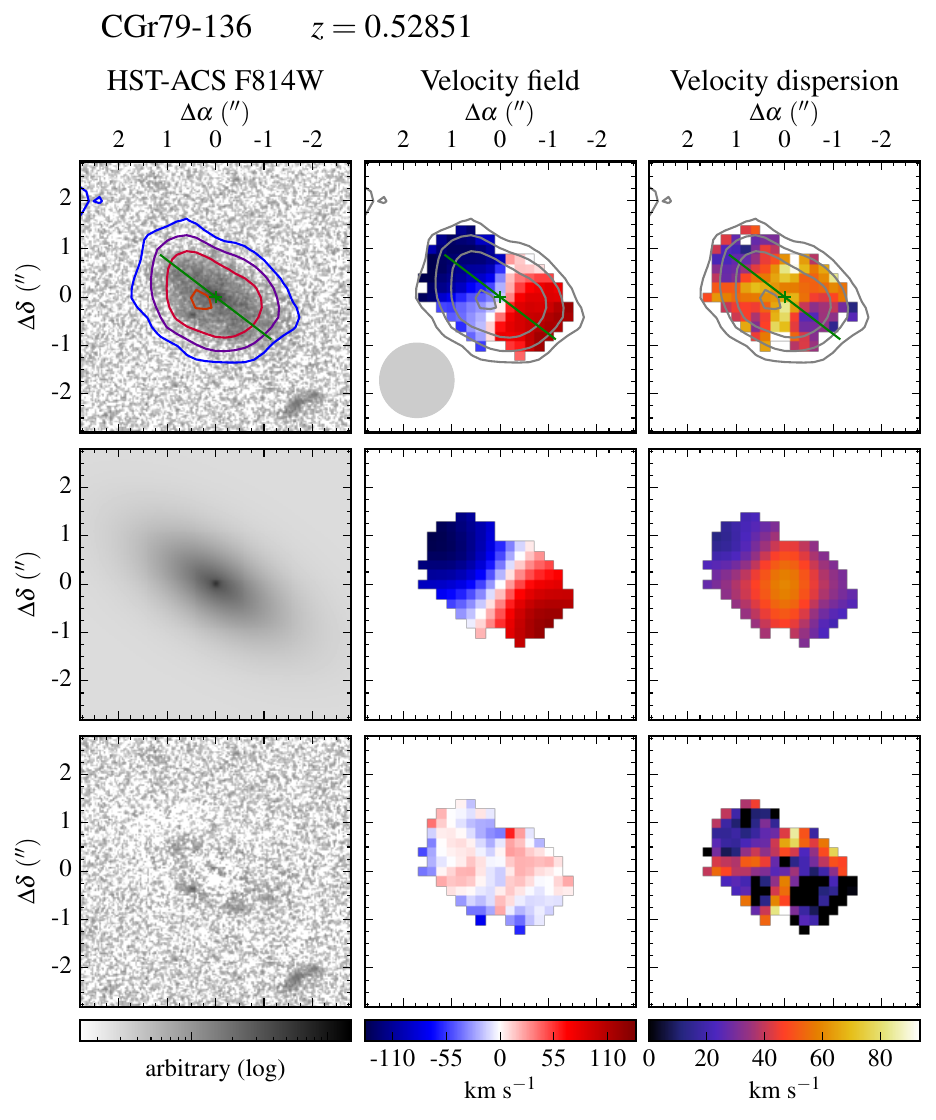}
\caption{Morpho-kinematics maps for galaxy CGr79-136. See caption of Fig. \ref{Mopho_KINMap_CGr28-41} for the description of figure.} 
\label{Mopho_KINMap_CGr79-136} 
\end{figure} 

\clearpage

\begin{figure}
\includegraphics[width=0.5\textwidth]{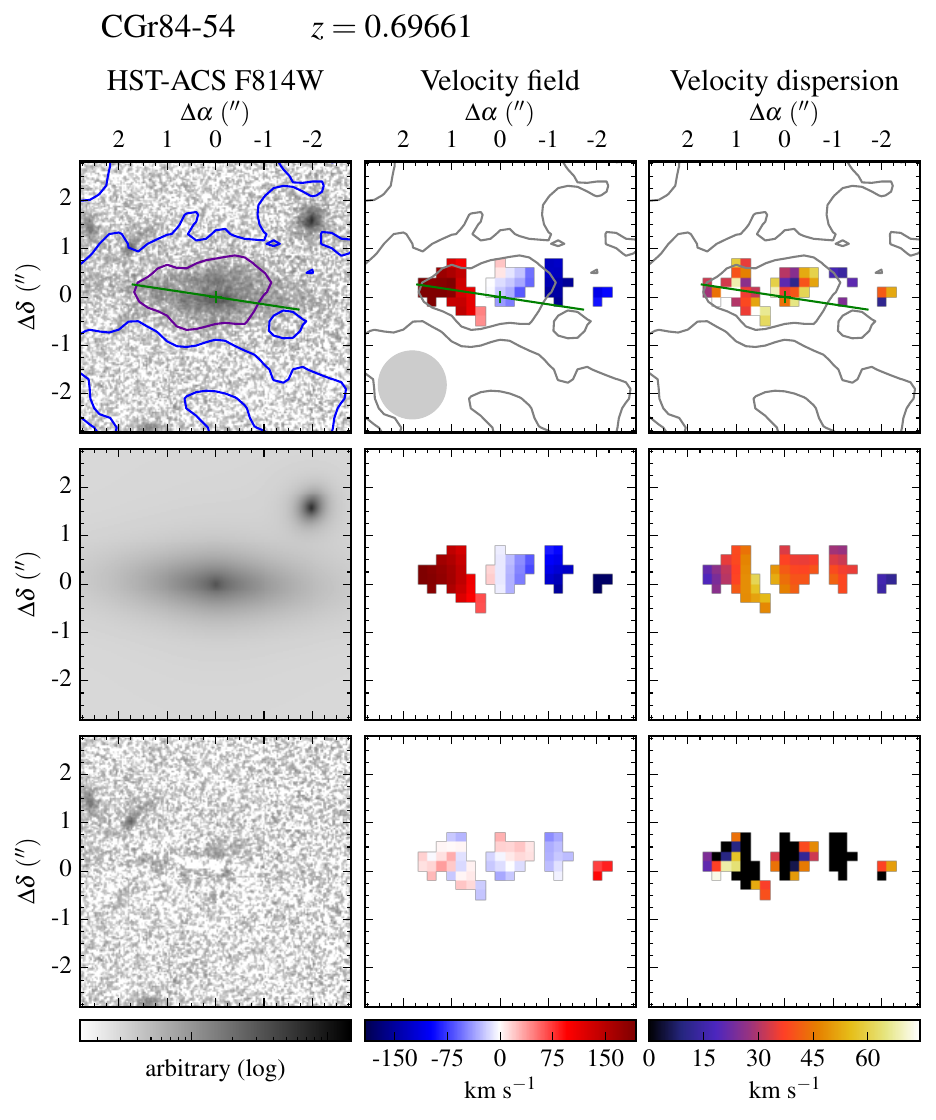}
\caption{Morpho-kinematics maps for galaxy CGr84-54. See caption of Fig. \ref{Mopho_KINMap_CGr28-41} for the description of figure.} 
\label{Mopho_KINMap_CGr84-55} 
\end{figure} 

\begin{figure}
\includegraphics[width=0.5\textwidth]{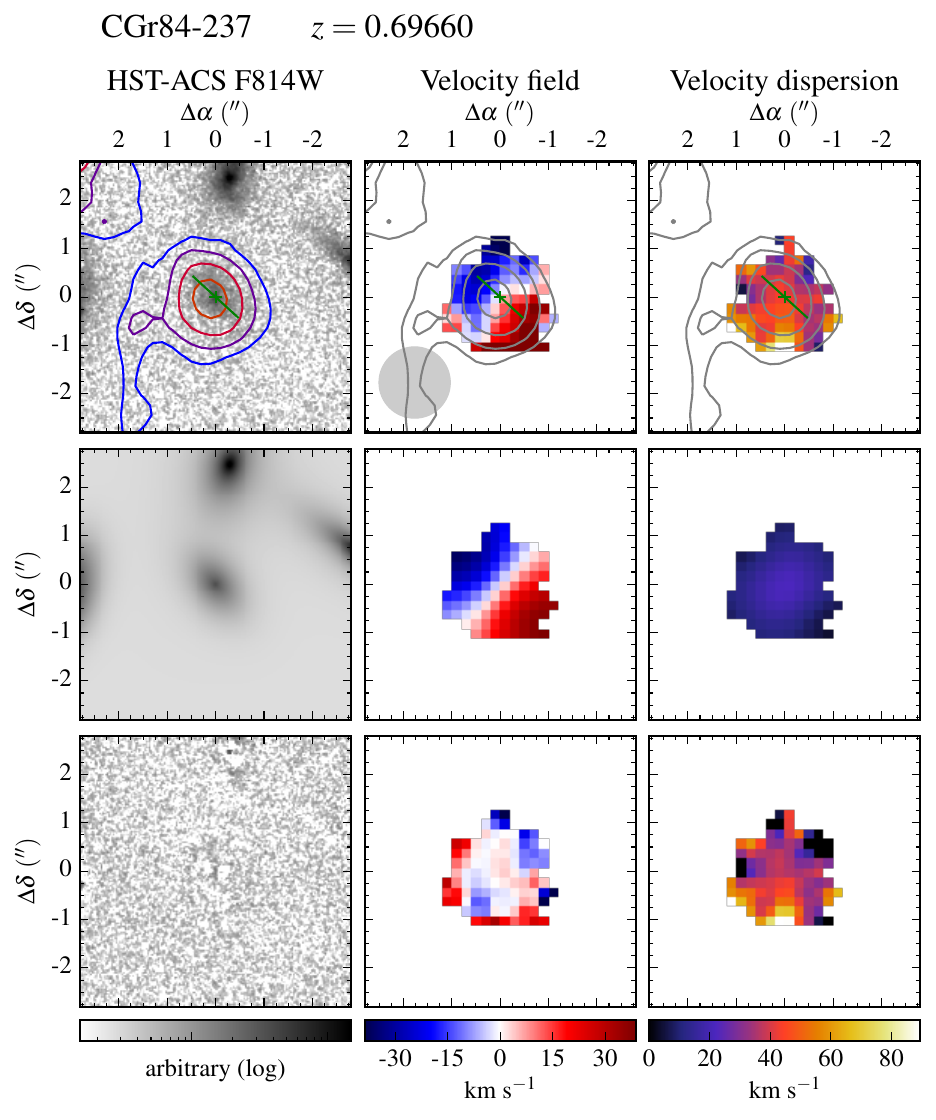}
\caption{Morpho-kinematics maps for galaxy CGr84-237. See caption of Fig. \ref{Mopho_KINMap_CGr28-41} for the description of figure.} 
\label{Mopho_KINMap_CGr84-237} 
\end{figure} 

\begin{figure}
\includegraphics[width=0.5\textwidth]{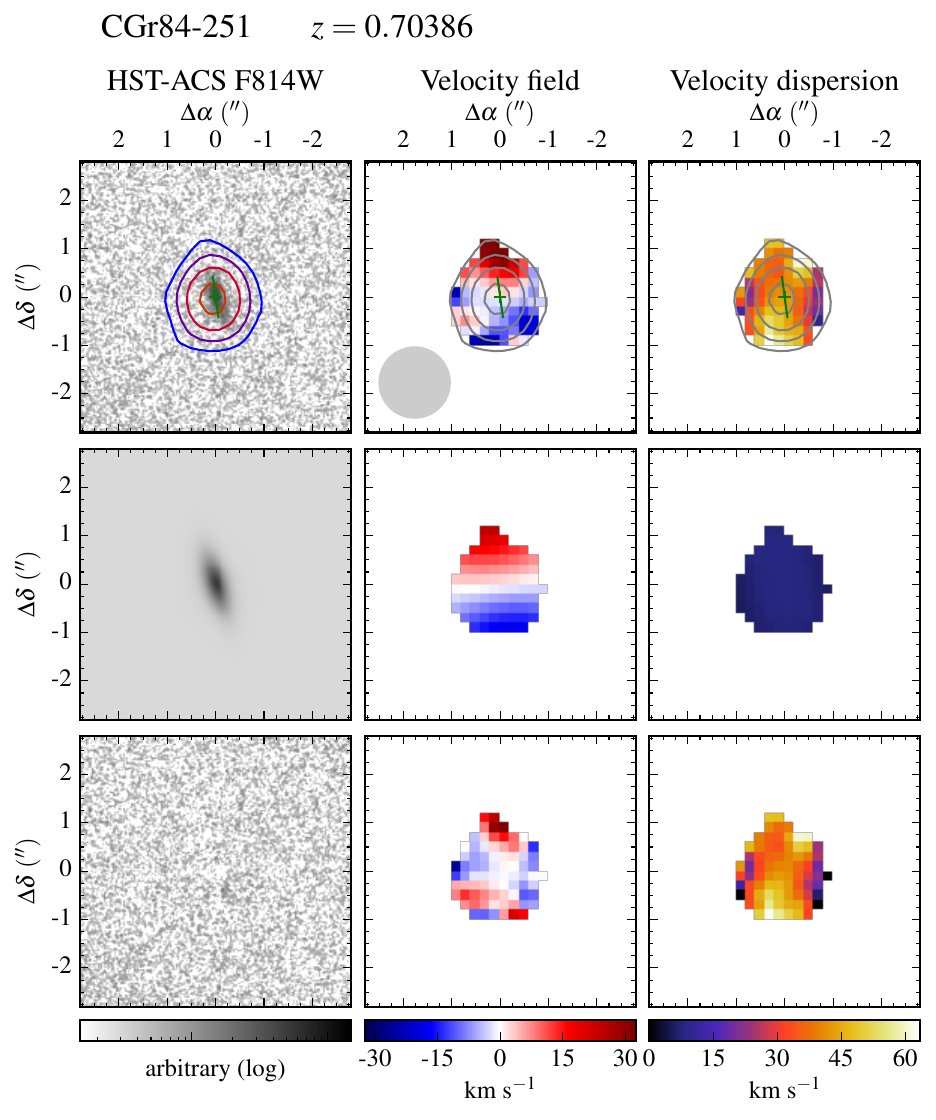}
\caption{Morpho-kinematics maps for galaxy CGr84-251. See caption of Fig. \ref{Mopho_KINMap_CGr28-41} for the description of figure.} 
\label{Mopho_KINMap_CGr84-251} 
\end{figure}

\begin{figure}
\includegraphics[width=0.5\textwidth]{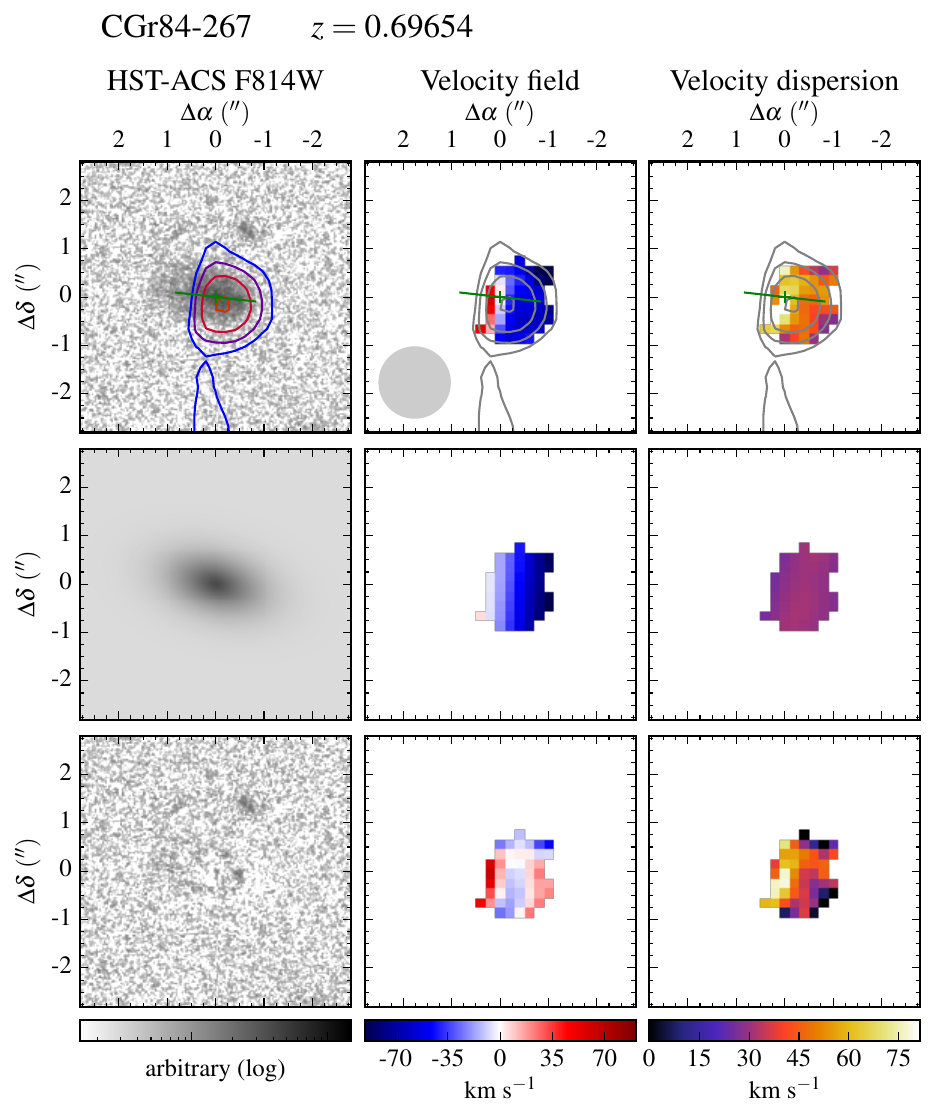}
\caption{Morpho-kinematics maps for galaxy CGr84-267. See caption of Fig. \ref{Mopho_KINMap_CGr28-41} for the description of figure.} 
\label{Mopho_KINMap_CGr84-267} 
\end{figure} 
 
\begin{figure}
\includegraphics[width=0.5\textwidth]{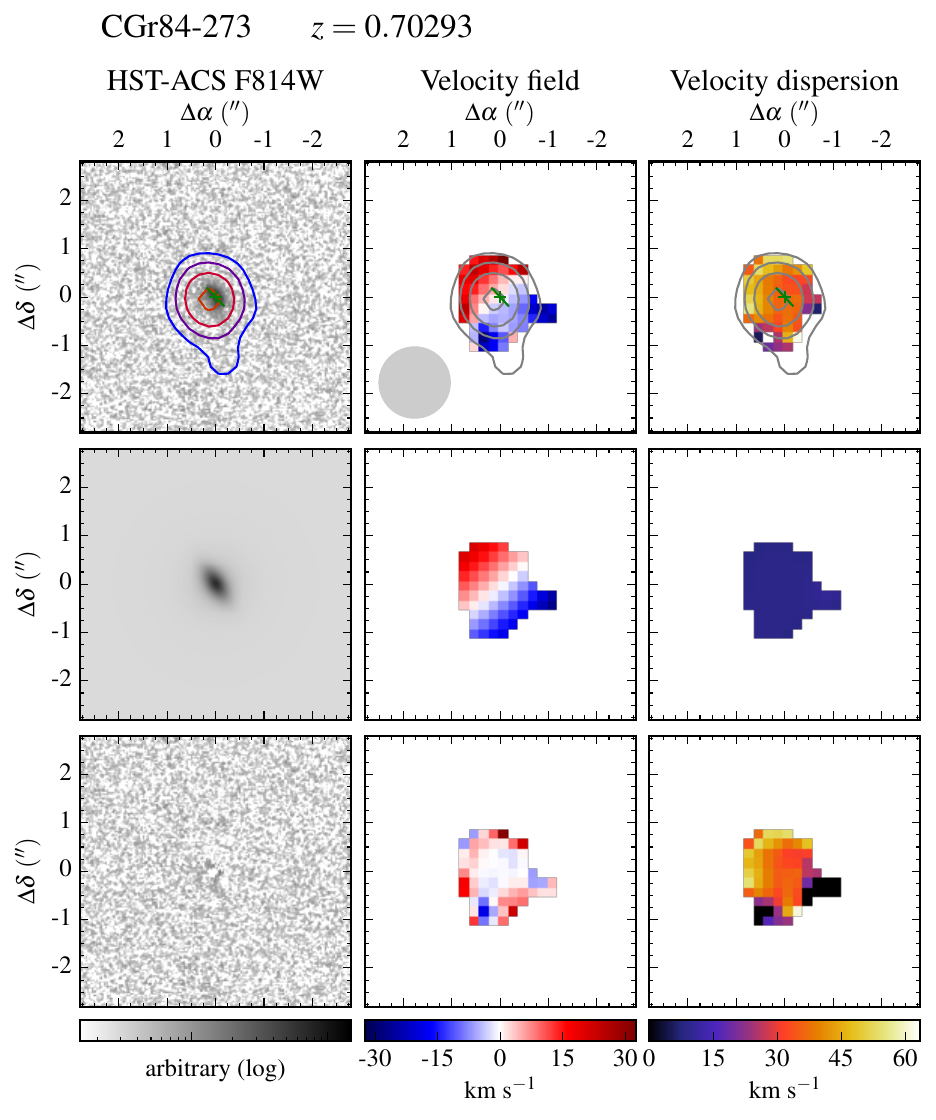}
\caption{Morpho-kinematics maps for galaxy CGr84-273. See caption of Fig. \ref{Mopho_KINMap_CGr28-41} for the description of figure.} 
\label{Mopho_KINMap_CGr84-273} 
\end{figure} 
 
\begin{figure}
\includegraphics[width=0.5\textwidth]{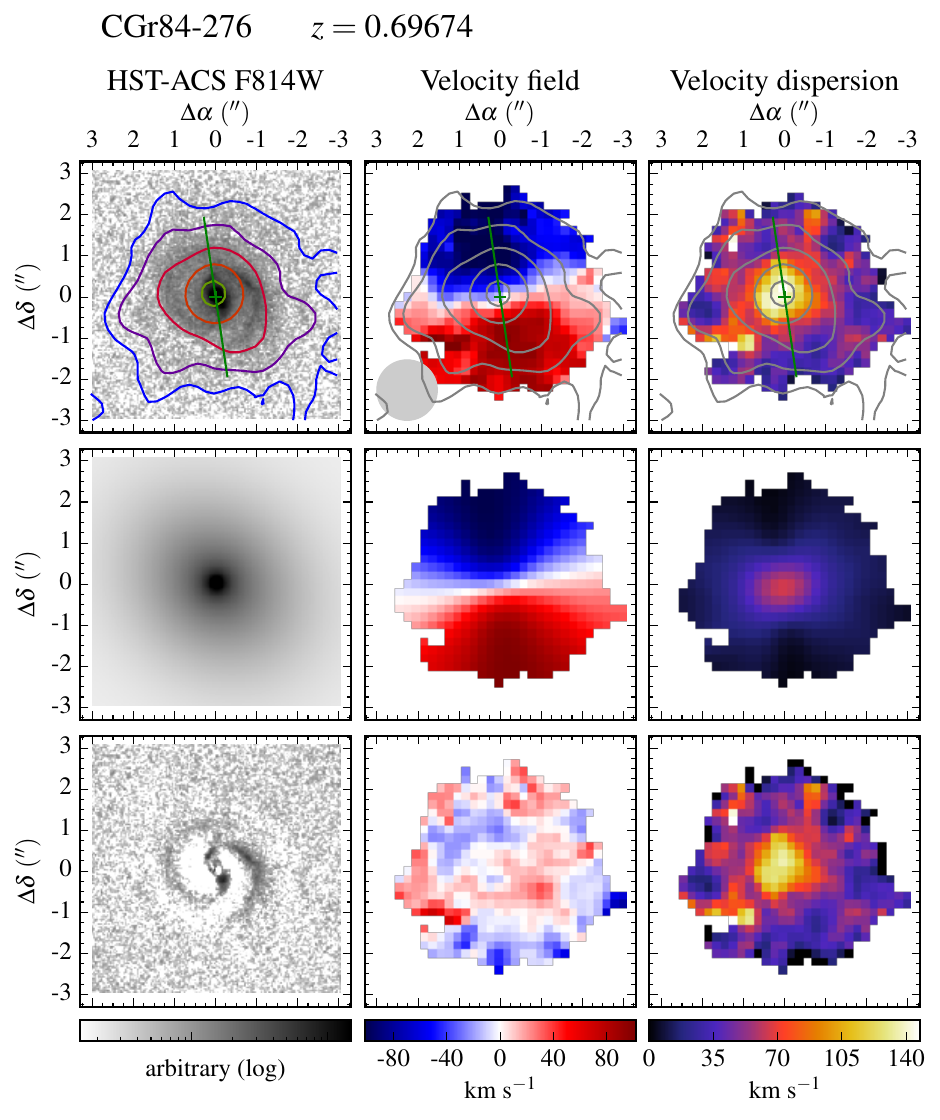}
\caption{Morpho-kinematics maps for galaxy CGr84-276. See caption of Fig. \ref{Mopho_KINMap_CGr28-41} for the description of figure.} 
\label{Mopho_KINMap_CGr84-276} 
\end{figure} 
 
\begin{figure}
\includegraphics[width=0.5\textwidth]{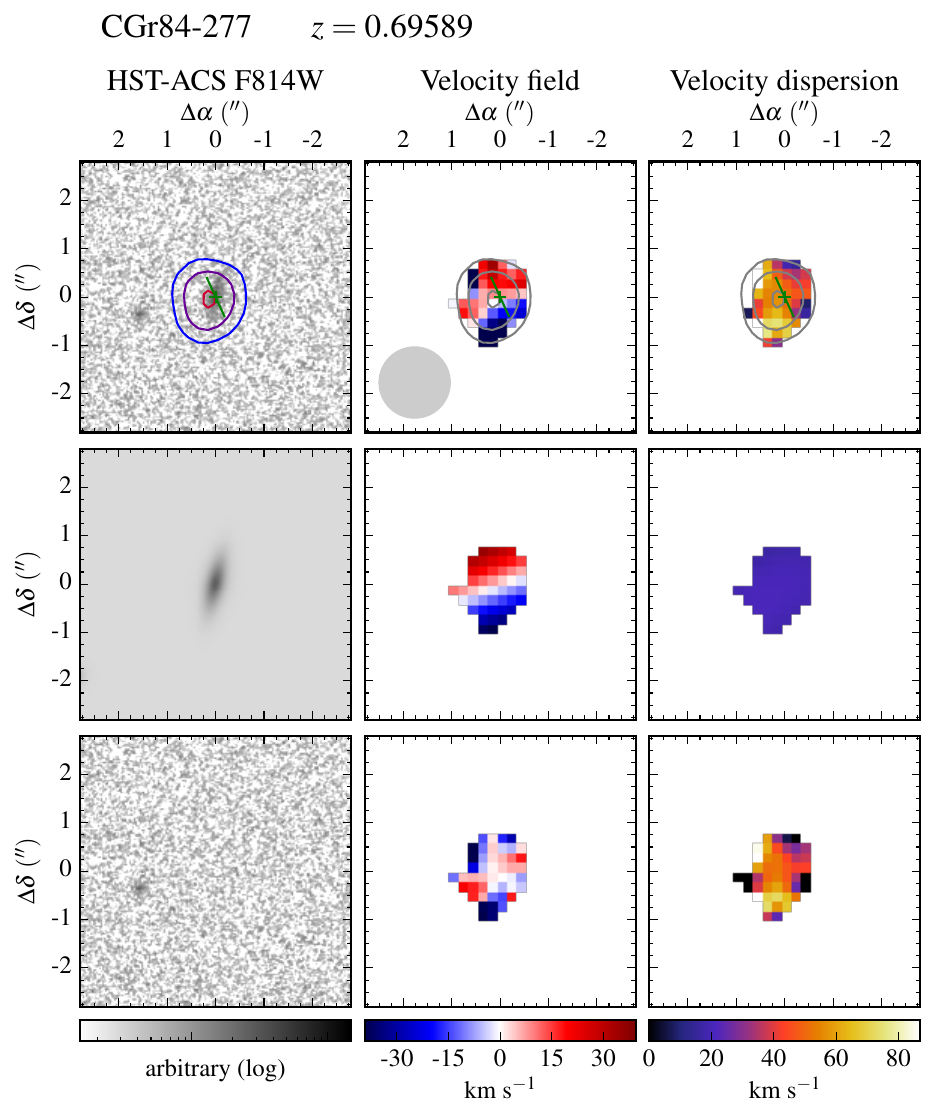}
\caption{Morpho-kinematics maps for galaxy CGr84-277. See caption of Fig. \ref{Mopho_KINMap_CGr28-41} for the description of figure.} 
\label{Mopho_KINMap_CGr84-277} 
\end{figure} 
 
\begin{figure}
\includegraphics[width=0.5\textwidth]{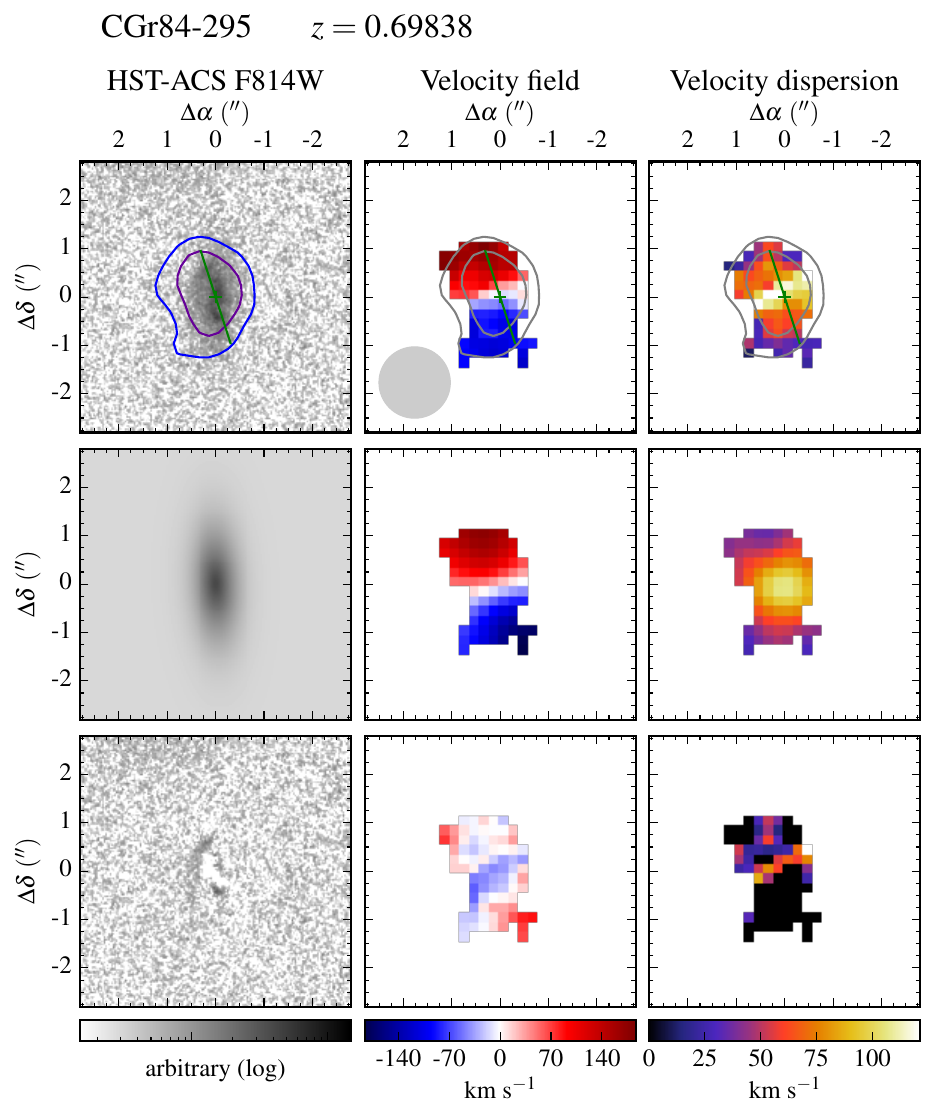}
\caption{Morpho-kinematics maps for galaxy CGr84-295. See caption of Fig. \ref{Mopho_KINMap_CGr28-41} for the description of figure.} 
\label{Mopho_KINMap_CGr84-295} 
\end{figure}

\clearpage

\begin{figure}
\includegraphics[width=0.5\textwidth]{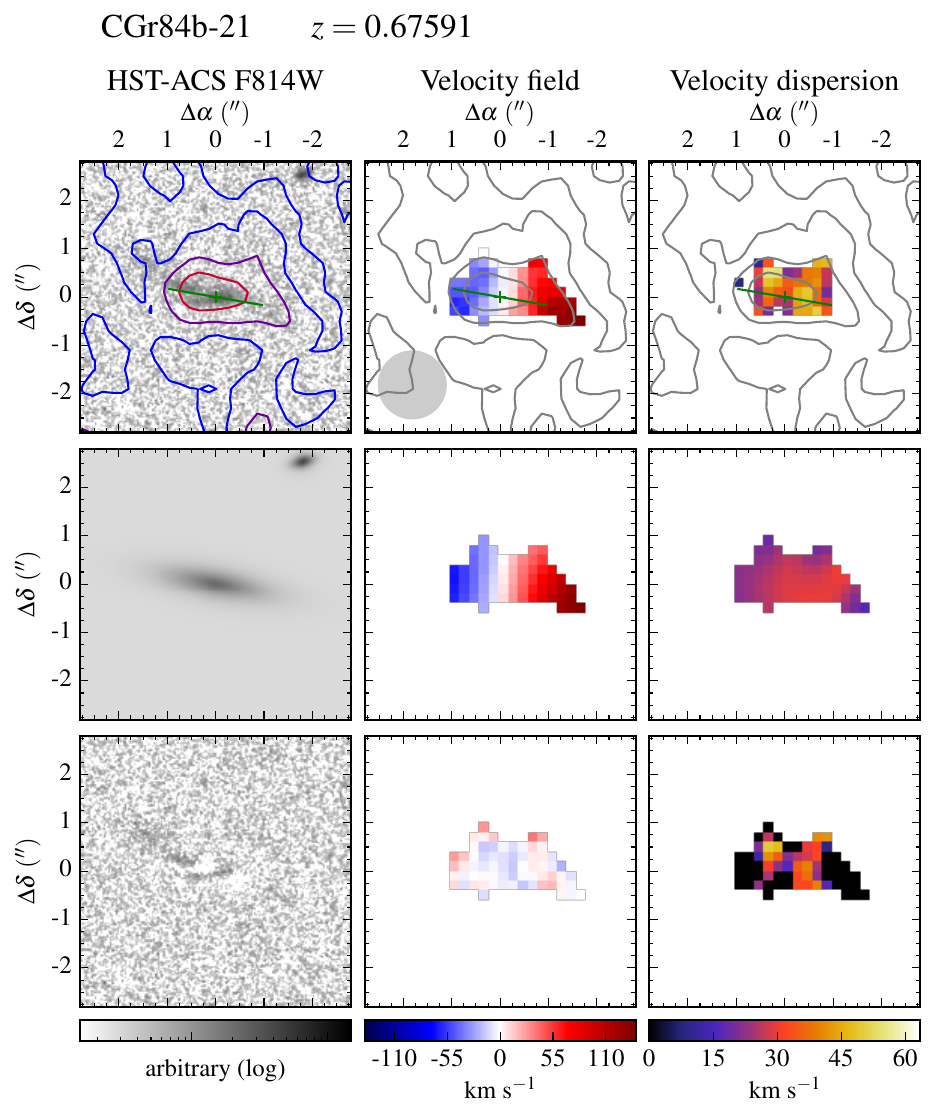}
\caption{Morpho-kinematics maps for galaxy CGr84b-21. See caption of Fig. \ref{Mopho_KINMap_CGr28-41} for the description of figure.} 
\label{Mopho_KINMap_CGr84b-22} 
\end{figure} 
 
\begin{figure}
\includegraphics[width=0.5\textwidth]{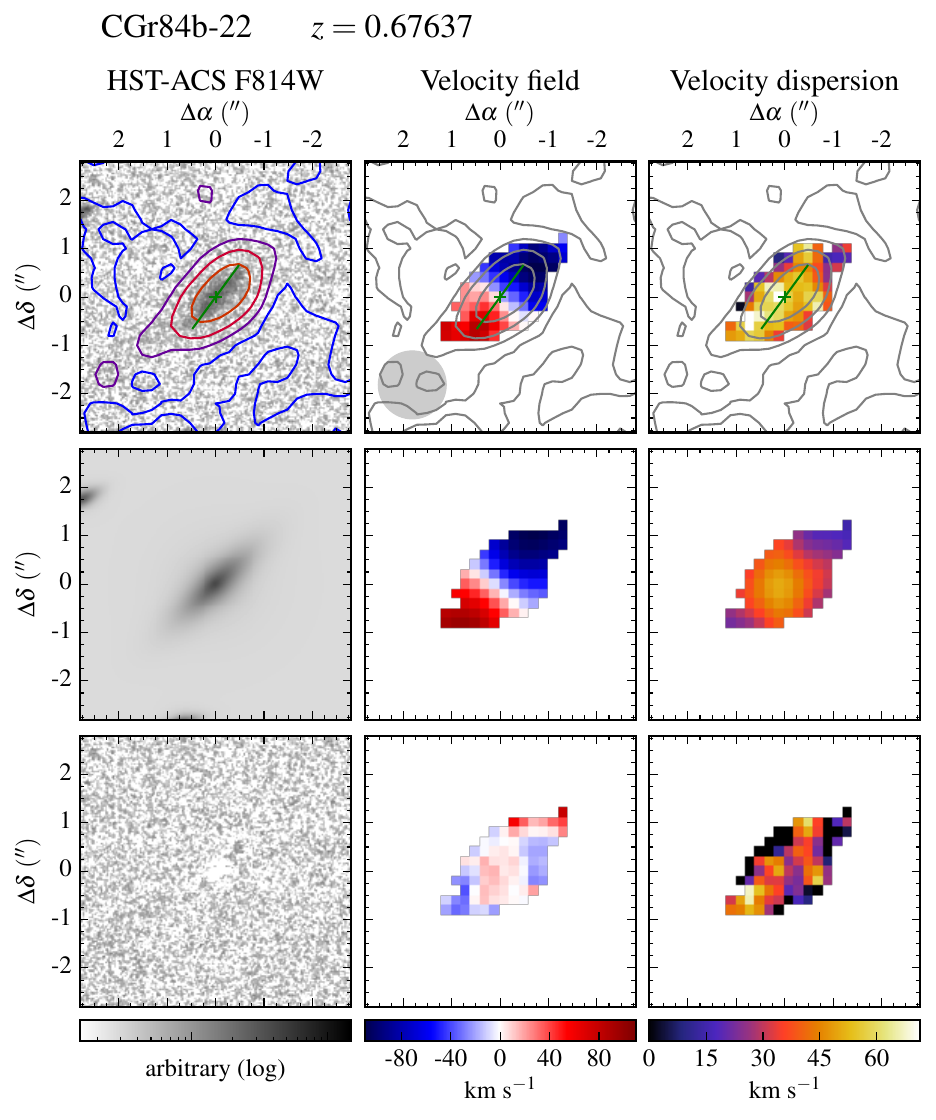}
\caption{Morpho-kinematics maps for galaxy CGr84b-22. See caption of Fig. \ref{Mopho_KINMap_CGr28-41} for the description of figure.} 
\label{Mopho_KINMap_CGr84b-23} 
\end{figure} 
 
\begin{figure}
\includegraphics[width=0.5\textwidth]{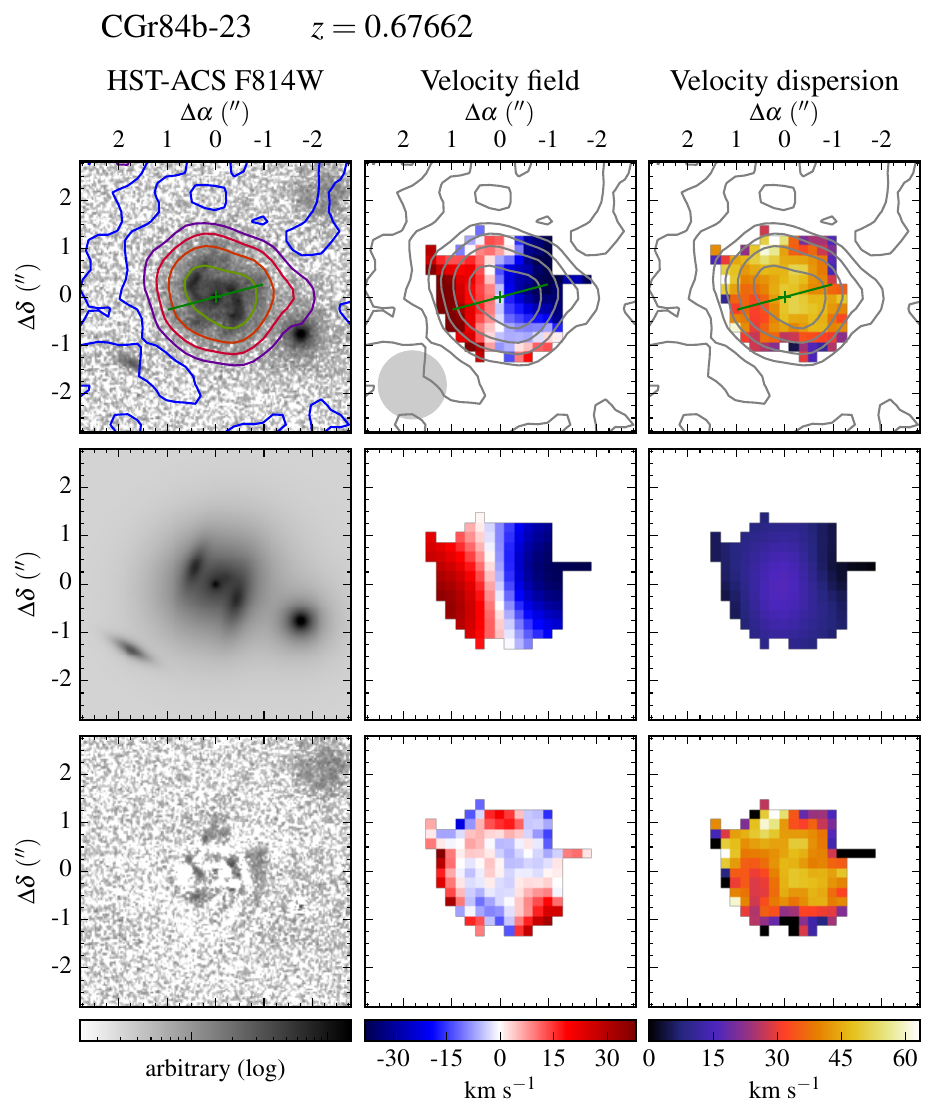}
\caption{Morpho-kinematics maps for galaxy CGr84b-23. See caption of Fig. \ref{Mopho_KINMap_CGr28-41} for the description of figure.} 
\label{Mopho_KINMap_CGr84b-24} 
\end{figure} 
 
\begin{figure}
\includegraphics[width=0.5\textwidth]{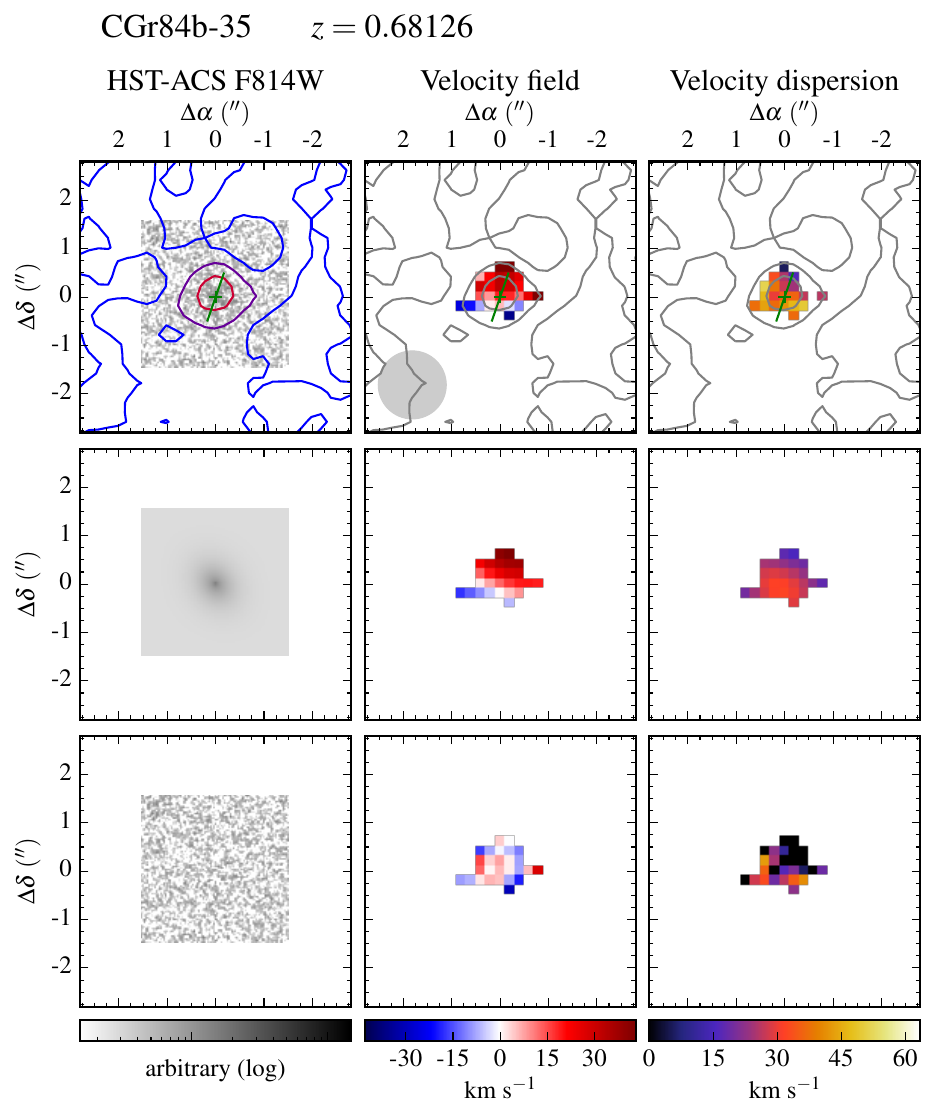}
\caption{Morpho-kinematics maps for galaxy CGr84b-35. See caption of Fig. \ref{Mopho_KINMap_CGr28-41} for the description of figure.} 
\label{Mopho_KINMap_CGr84b-36} 
\end{figure} 
 
\begin{figure}
\includegraphics[width=0.5\textwidth]{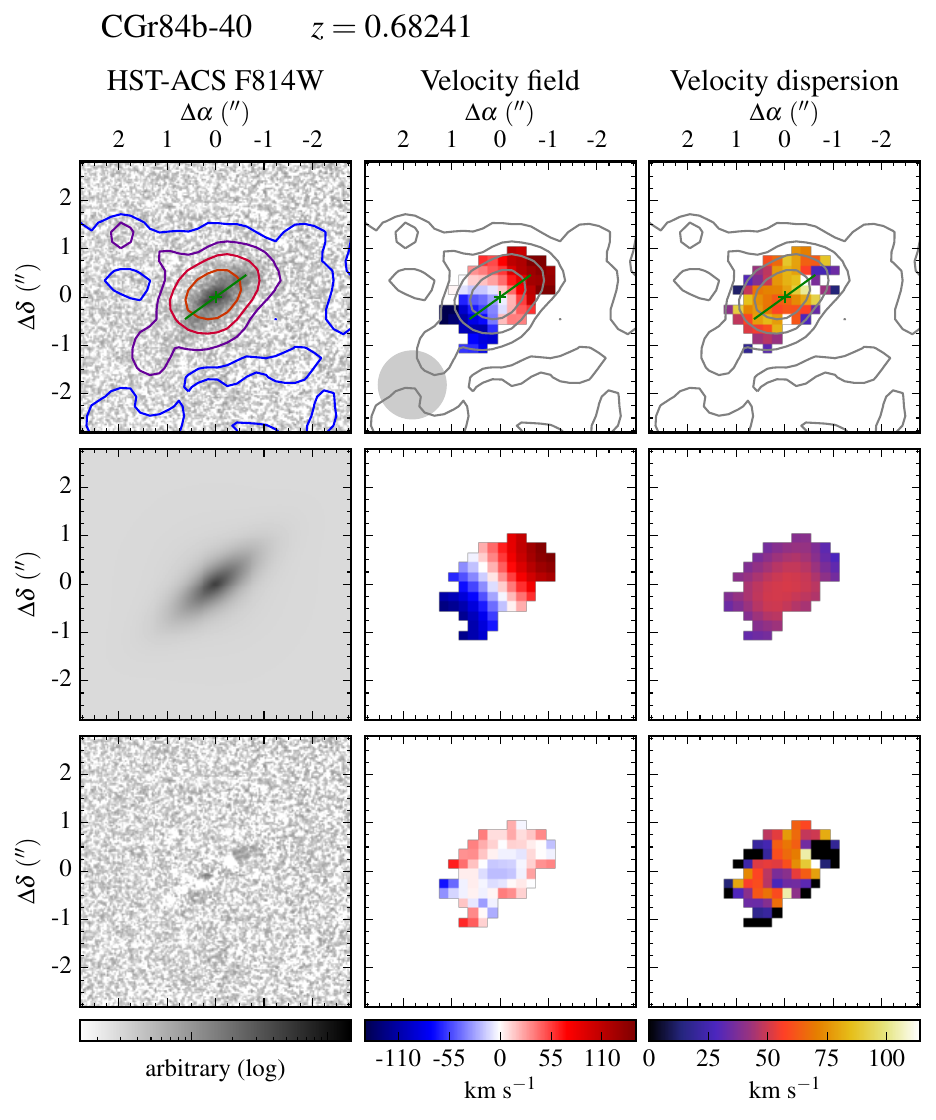}
\caption{Morpho-kinematics maps for galaxy CGr84b-40. See caption of Fig. \ref{Mopho_KINMap_CGr28-41} for the description of figure.} 
\label{Mopho_KINMap_CGr84b-41} 
\end{figure}

\begin{figure}
\includegraphics[width=0.5\textwidth]{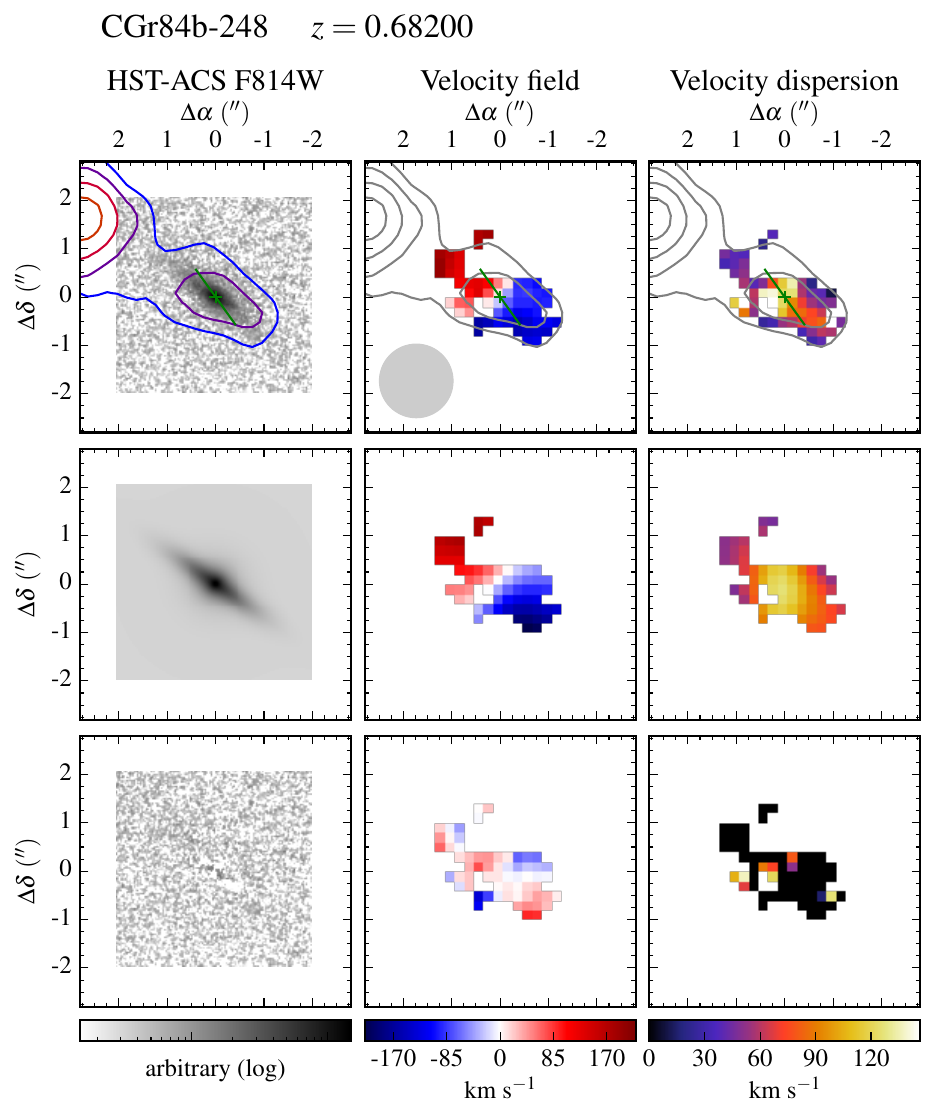}
\caption{Morpho-kinematics maps for galaxy CGr84b-248. See caption of Fig. \ref{Mopho_KINMap_CGr28-41} for the description of figure.} 
\label{Mopho_KINMap_CGr84b-248} 
\end{figure} 
 
\begin{figure}
\includegraphics[width=0.5\textwidth]{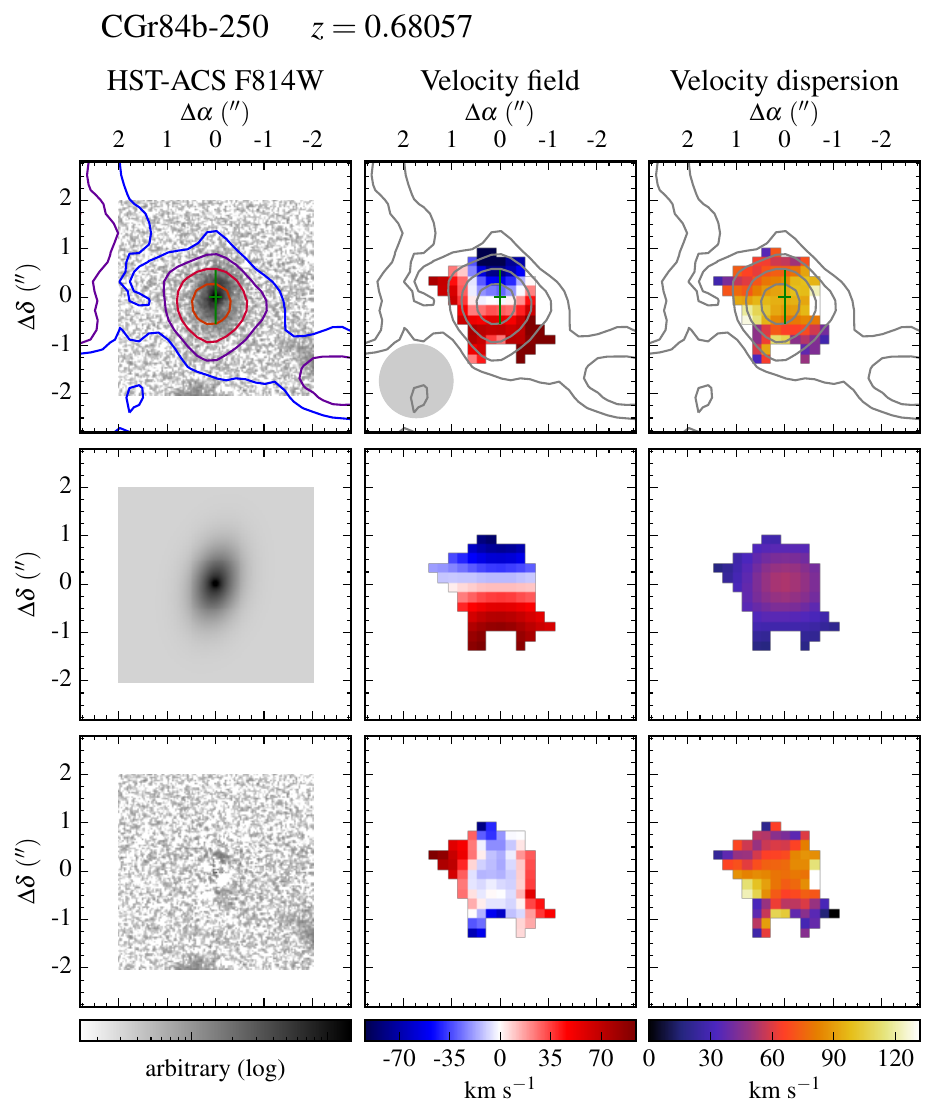}
\caption{Morpho-kinematics maps for galaxy CGr84b-250. See caption of Fig. \ref{Mopho_KINMap_CGr28-41} for the description of figure.} 
\label{Mopho_KINMap_CGr84b-250} 
\end{figure}

\begin{figure}
\includegraphics[width=0.5\textwidth]{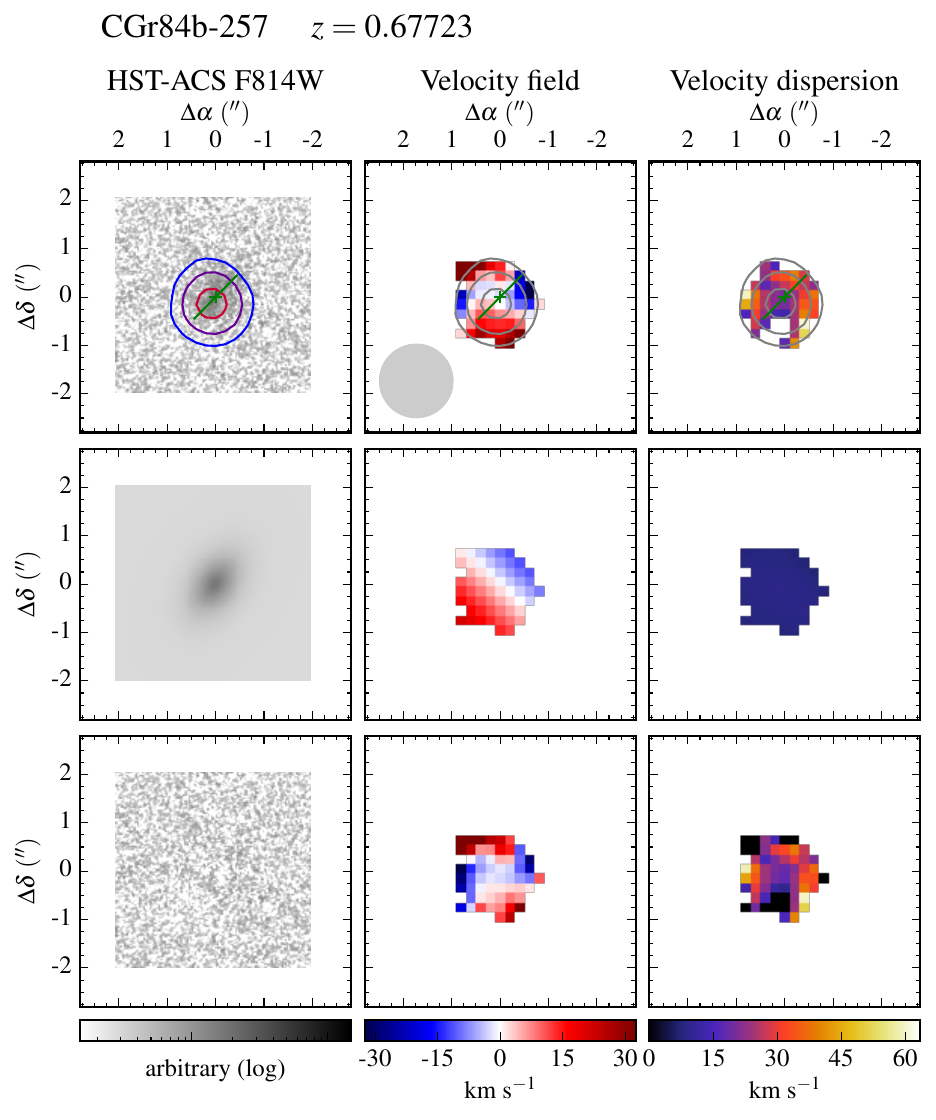}
\caption{Morpho-kinematics maps for galaxy CGr84b-257. See caption of Fig. \ref{Mopho_KINMap_CGr28-41} for the description of figure.} 
\label{Mopho_KINMap_CGr84b-257} 
\end{figure} 

\begin{figure}
\includegraphics[width=0.5\textwidth]{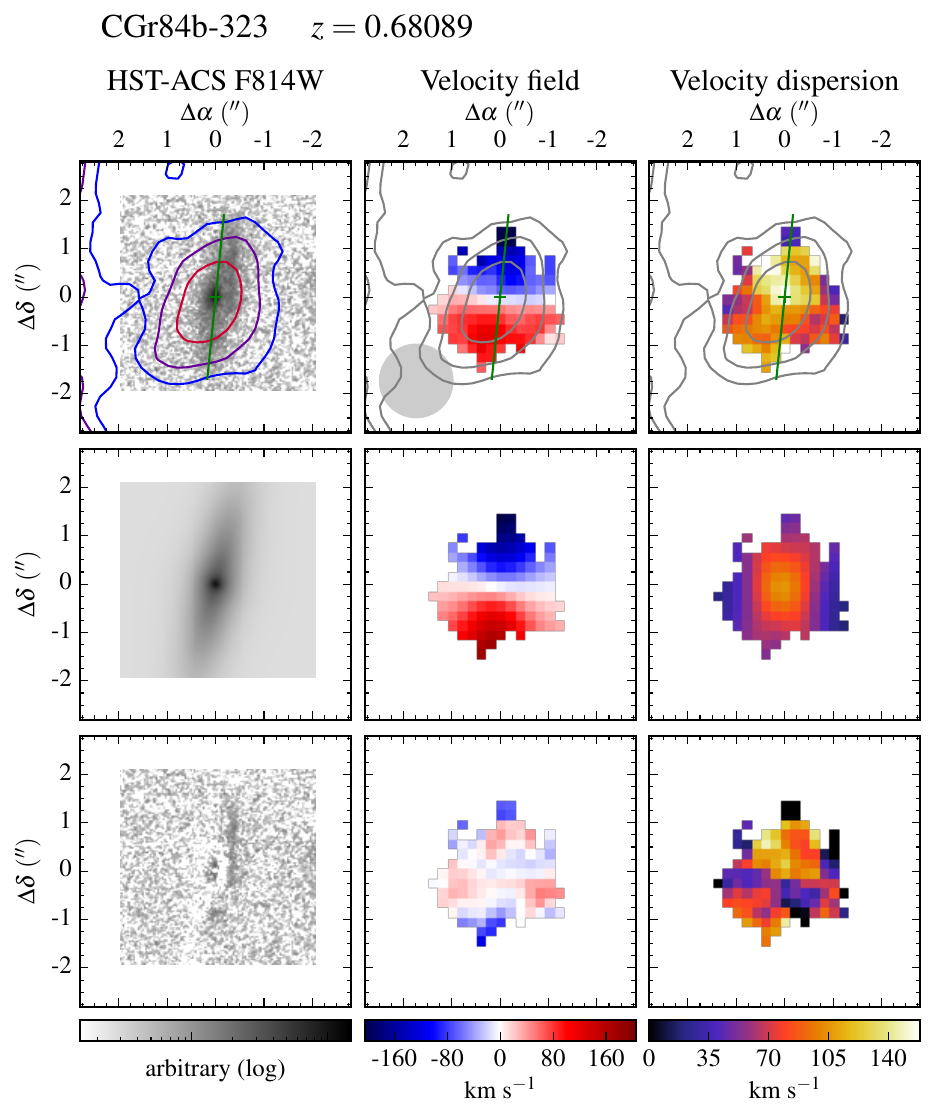}
\caption{Morpho-kinematics maps for galaxy CGr84b-323. See caption of Fig. \ref{Mopho_KINMap_CGr28-41} for the description of figure.} 
\label{Mopho_KINMap_CGr84b-323} 
\end{figure}

\clearpage

\begin{figure}
\includegraphics[width=0.5\textwidth]{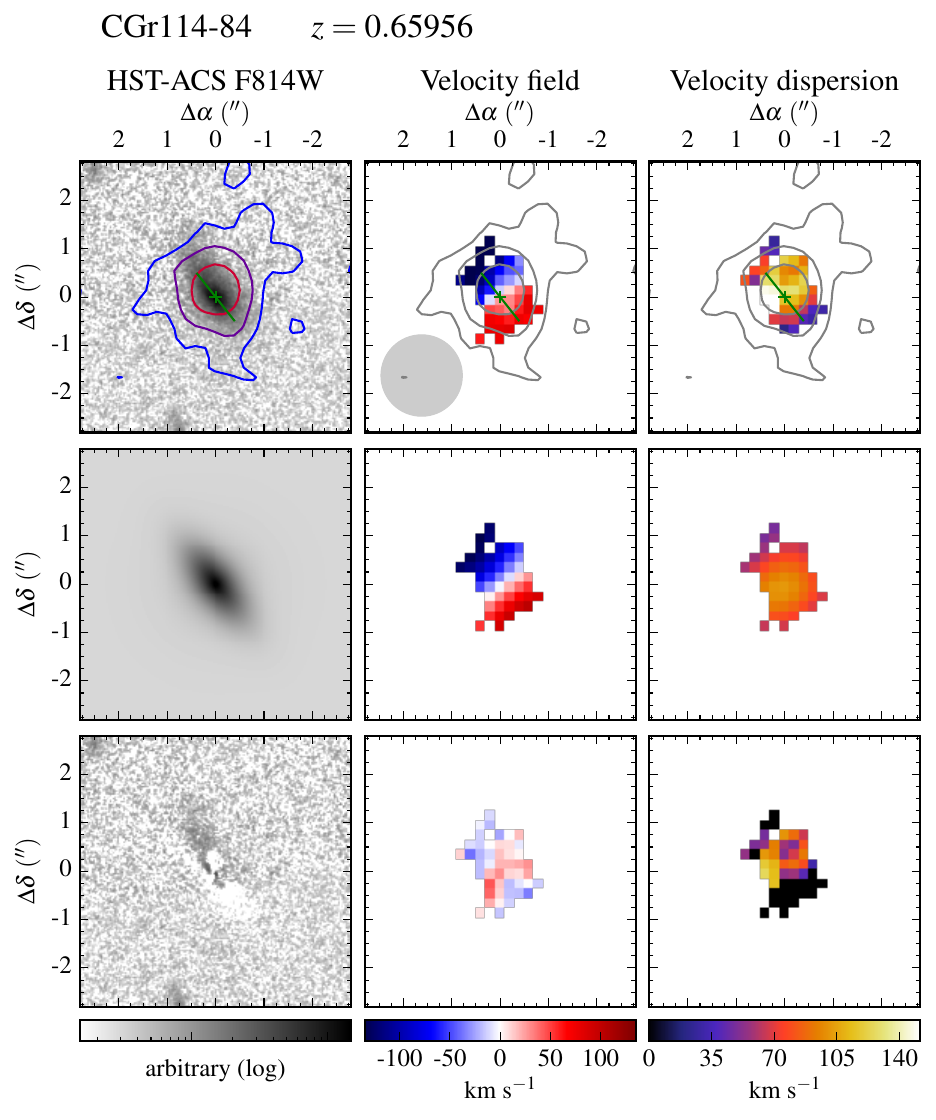}
\caption{Morpho-kinematics maps for galaxy CGr114-84. See caption of Fig. \ref{Mopho_KINMap_CGr28-41} for the description of figure.} 
\label{Mopho_KINMap_CGr114-84} 
\end{figure} 
 
\begin{figure}
\includegraphics[width=0.5\textwidth]{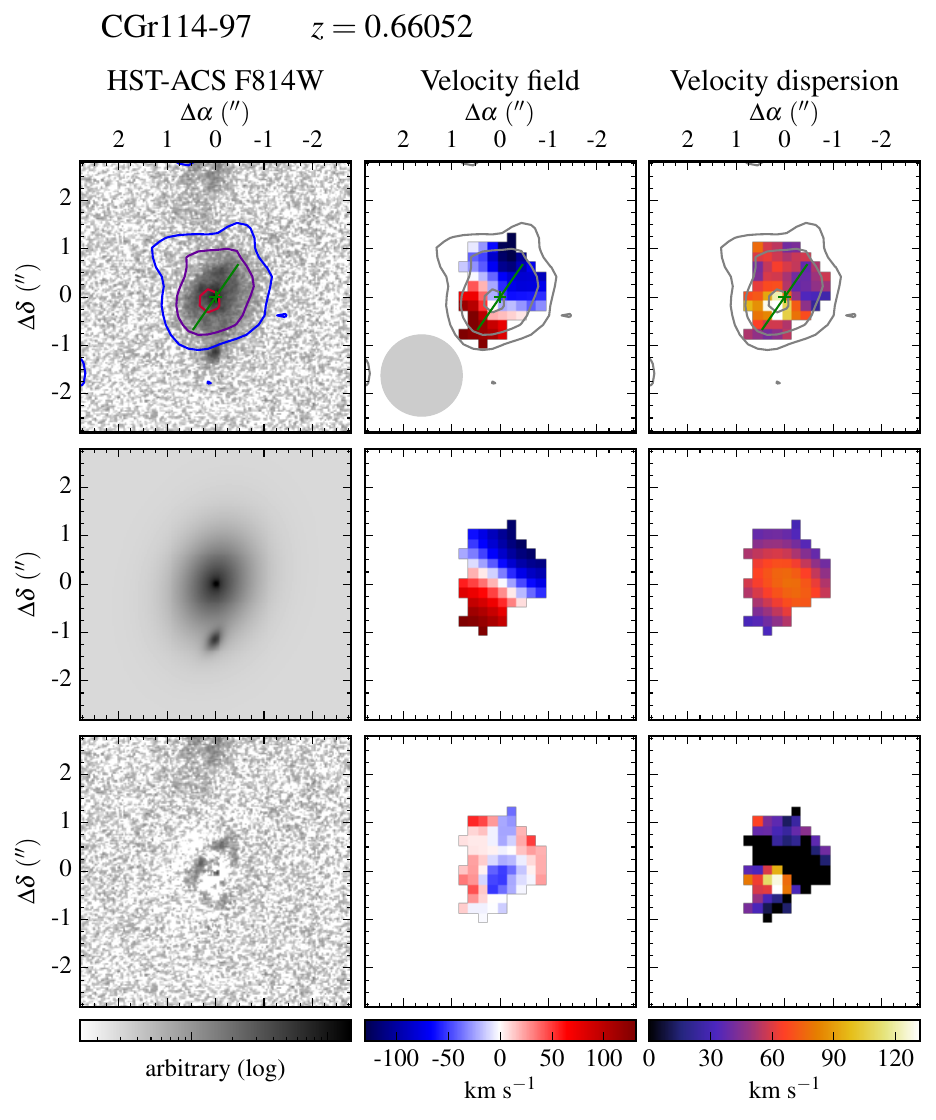}
\caption{Morpho-kinematics maps for galaxy CGr114-97. See caption of Fig. \ref{Mopho_KINMap_CGr28-41} for the description of figure.} 
\label{Mopho_KINMap_CGr114-97} 
\end{figure}  

\clearpage

\subsection{Galaxies with kinematics biased by an AGN}
\label{app:agn}

\begin{figure}[h]
\includegraphics[width=0.5\textwidth]{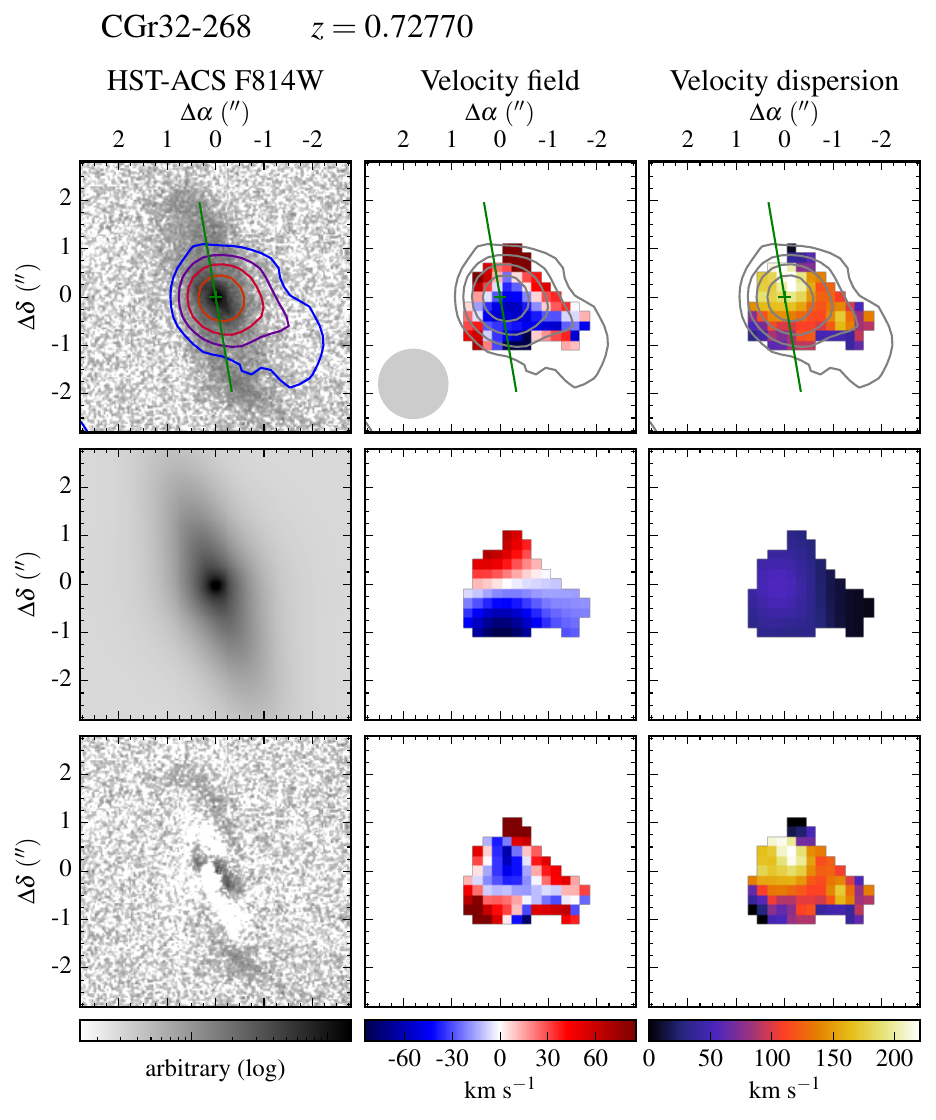}
\caption{Morpho-kinematics maps for galaxy CGr32-268. See caption of Fig. \ref{Mopho_KINMap_CGr28-41} for the description of figure.} 
\label{Mopho_KINMap_CGr32-268} 
\end{figure}
 
\begin{figure}
\includegraphics[width=0.5\textwidth]{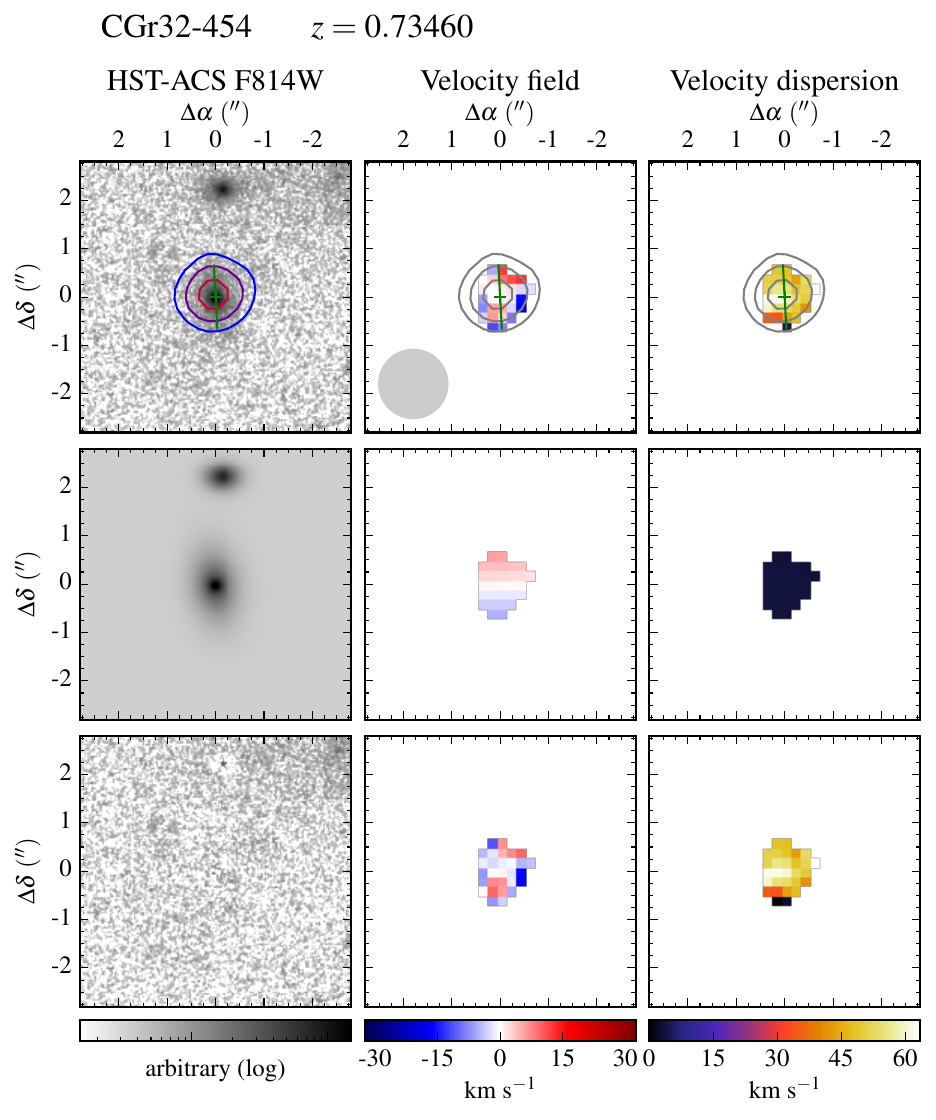}
\caption{Morpho-kinematics maps for galaxy CGr32-454. See caption of Fig. \ref{Mopho_KINMap_CGr28-41} for the description of figure.} 
\label{Mopho_KINMap_CGr32-454} 
\end{figure} 

\clearpage

\subsection{Galaxies with a dominant bulge within the effective radius}
\label{app:bd}

\begin{figure}[h]
\includegraphics[width=0.5\textwidth]{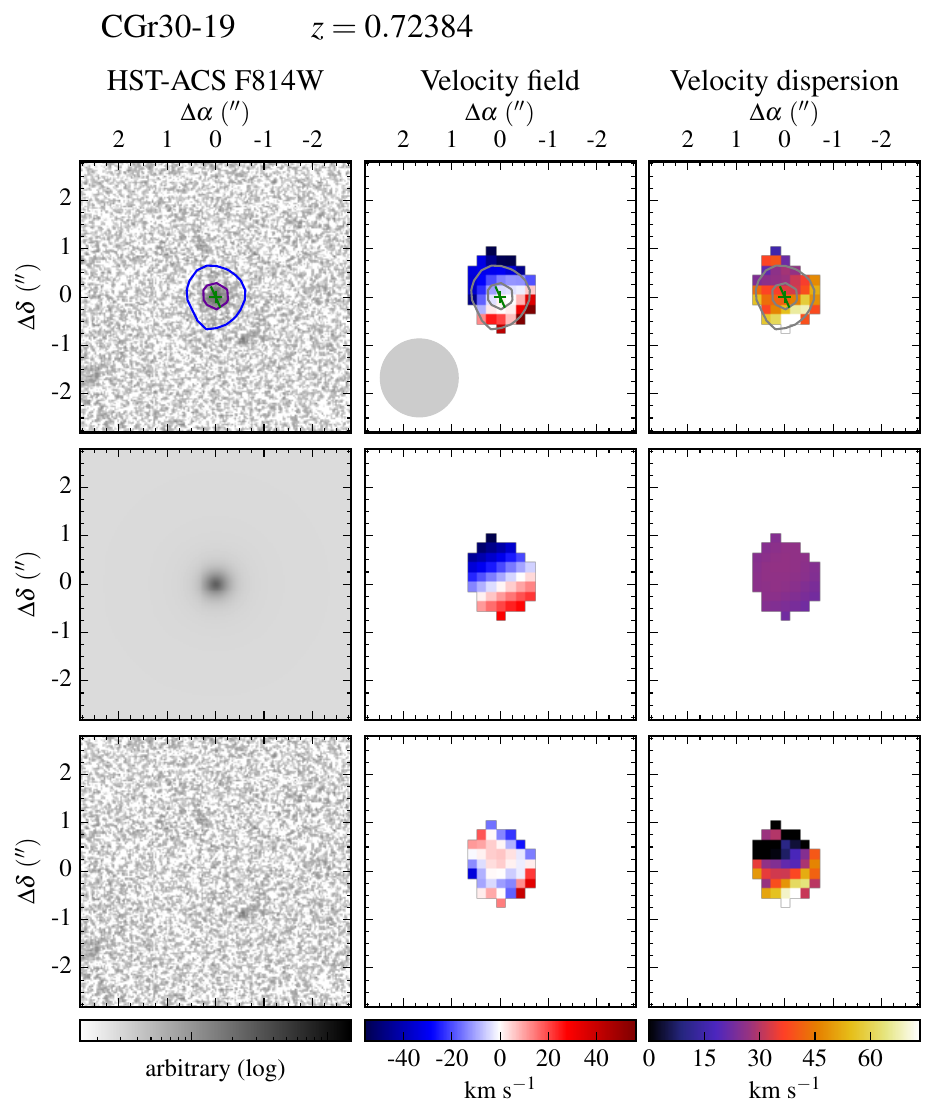}
\caption{Morpho-kinematics maps for galaxy CGr30-19. See caption of Fig. \ref{Mopho_KINMap_CGr28-41} for the description of figure.} 
\label{Mopho_KINMap_CGr30-19} 
\end{figure} 
 
\begin{figure}
\includegraphics[width=0.5\textwidth]{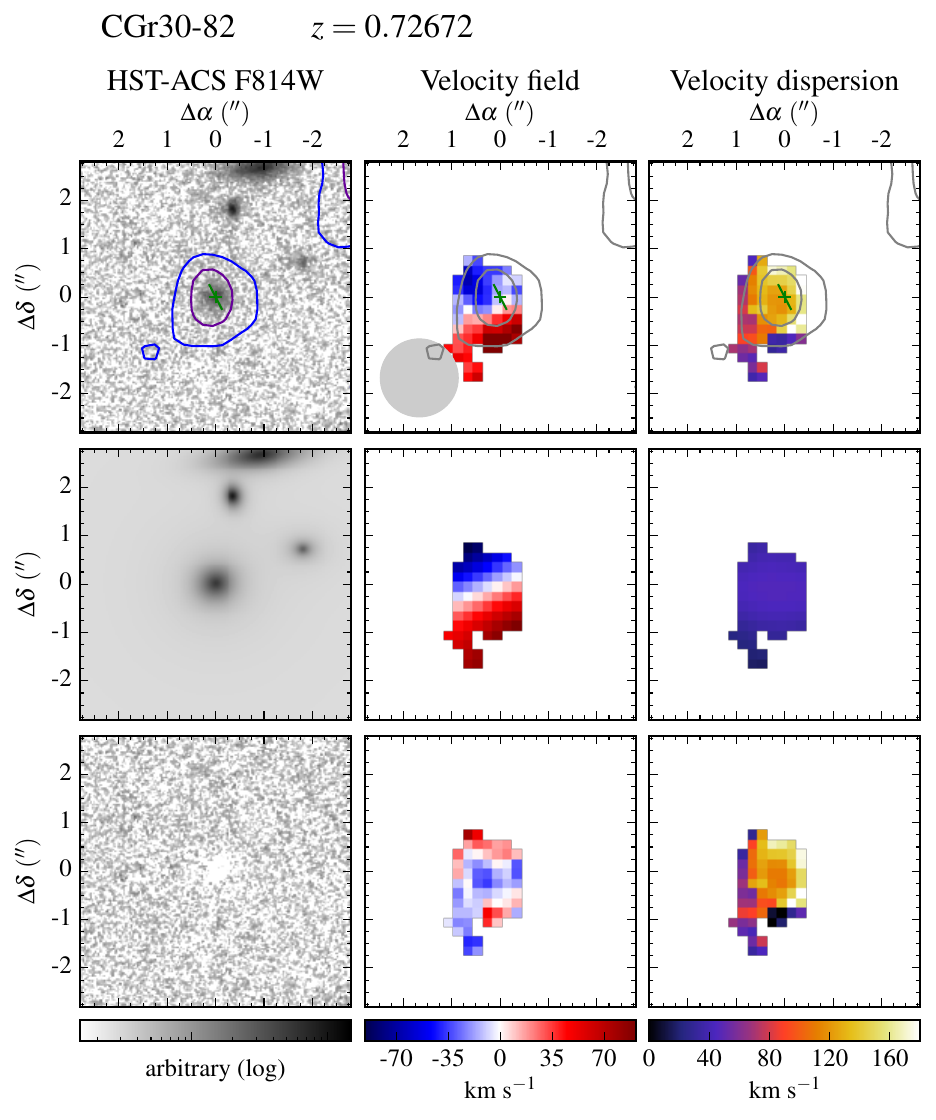}
\caption{Morpho-kinematics maps for galaxy CGr30-82. See caption of Fig. \ref{Mopho_KINMap_CGr28-41} for the description of figure.} 
\label{Mopho_KINMap_CGr30-82} 
\end{figure} 
 
\begin{figure}
\includegraphics[width=0.5\textwidth]{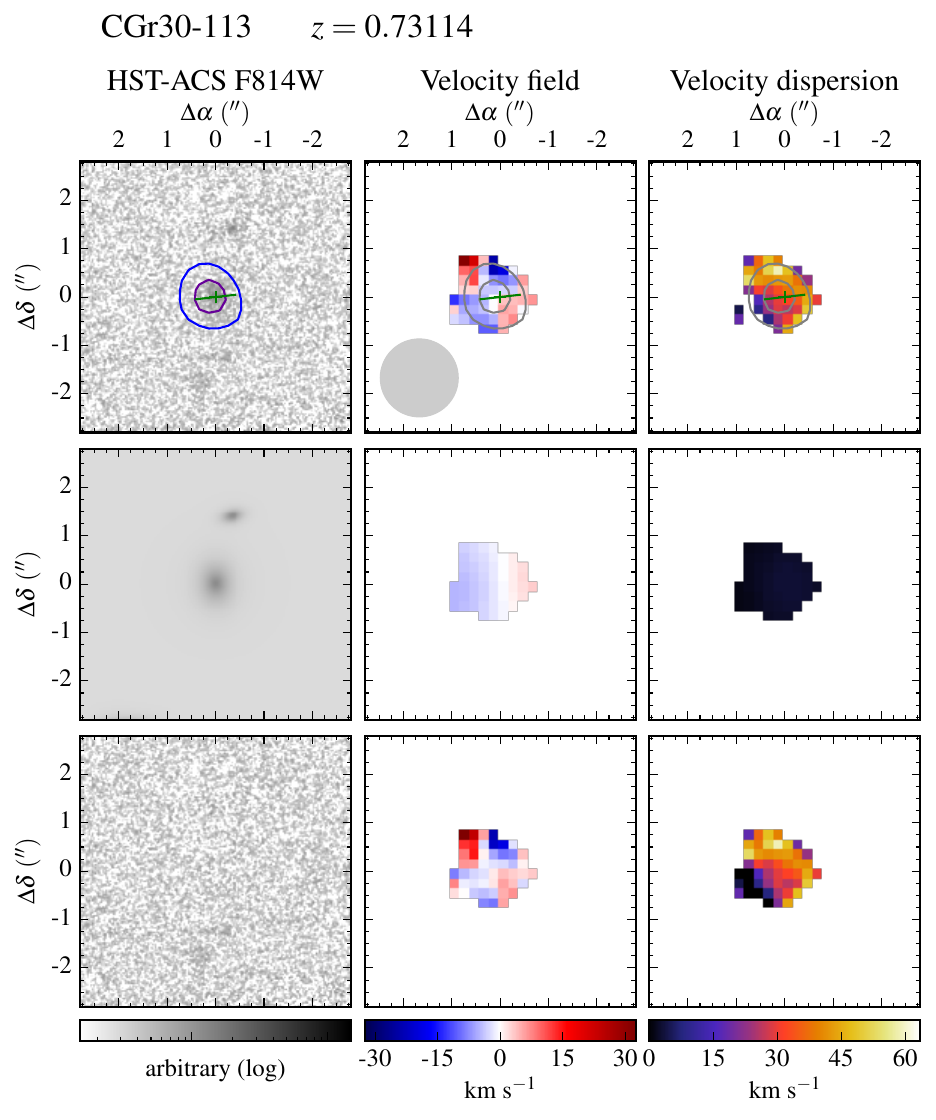}
\caption{Morpho-kinematics maps for galaxy CGr30-113. See caption of Fig. \ref{Mopho_KINMap_CGr28-41} for the description of figure.} 
\label{Mopho_KINMap_CGr30-113} 
\end{figure} 
 
\begin{figure}
\includegraphics[width=0.5\textwidth]{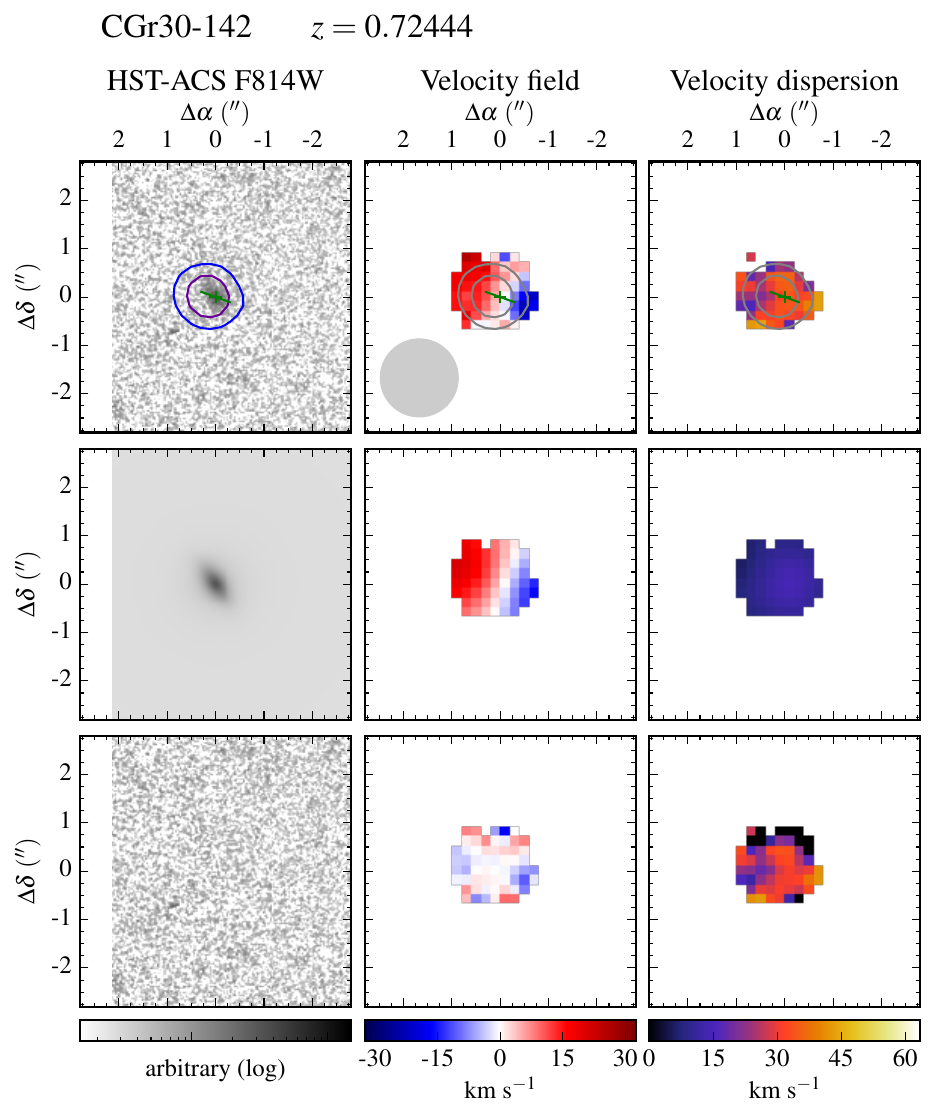}
\caption{Morpho-kinematics maps for galaxy CGr30-142. See caption of Fig. \ref{Mopho_KINMap_CGr28-41} for the description of figure.} 
\label{Mopho_KINMap_CGr30-142} 
\end{figure} 
 
\begin{figure}
\includegraphics[width=0.5\textwidth]{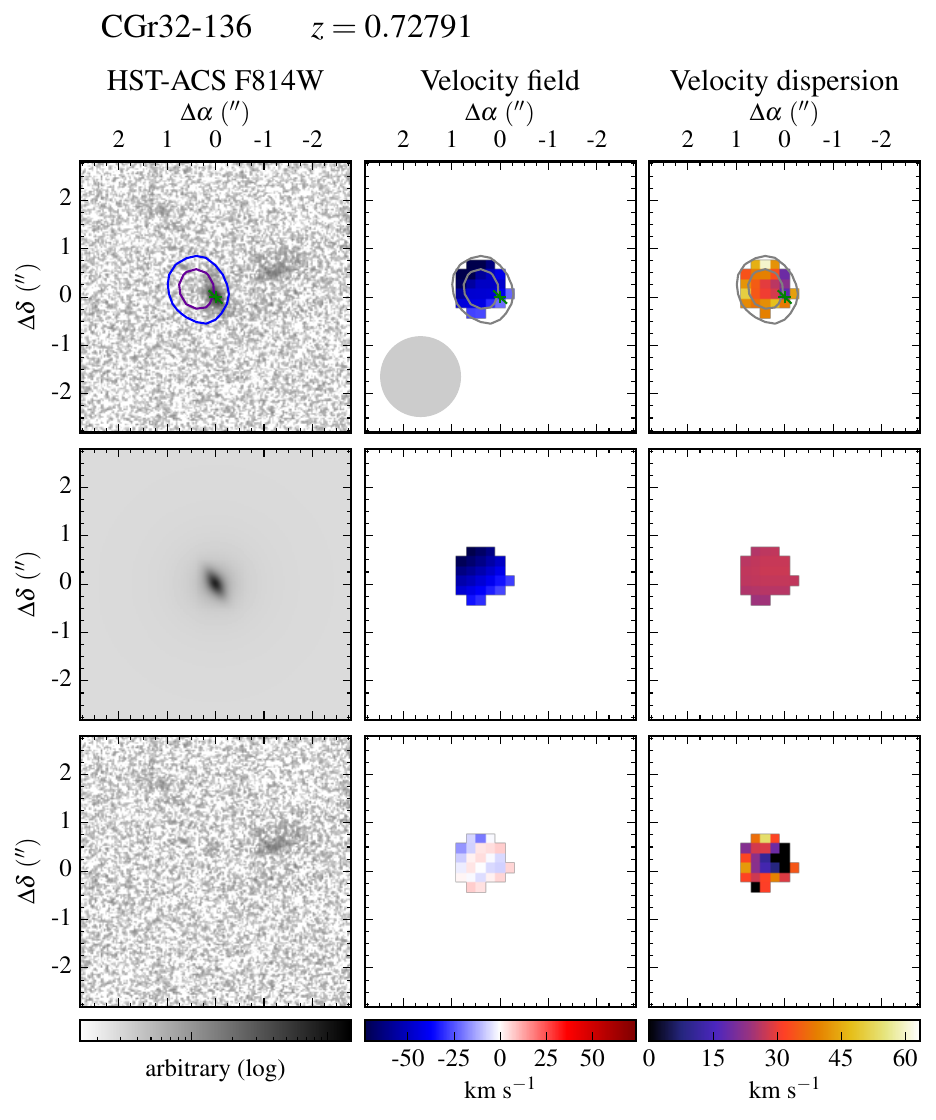}
\caption{Morpho-kinematics maps for galaxy CGr32-136. See caption of Fig. \ref{Mopho_KINMap_CGr28-41} for the description of figure.} 
\label{Mopho_KINMap_CGr32-136} 
\end{figure} 

\begin{figure}
\includegraphics[width=0.5\textwidth]{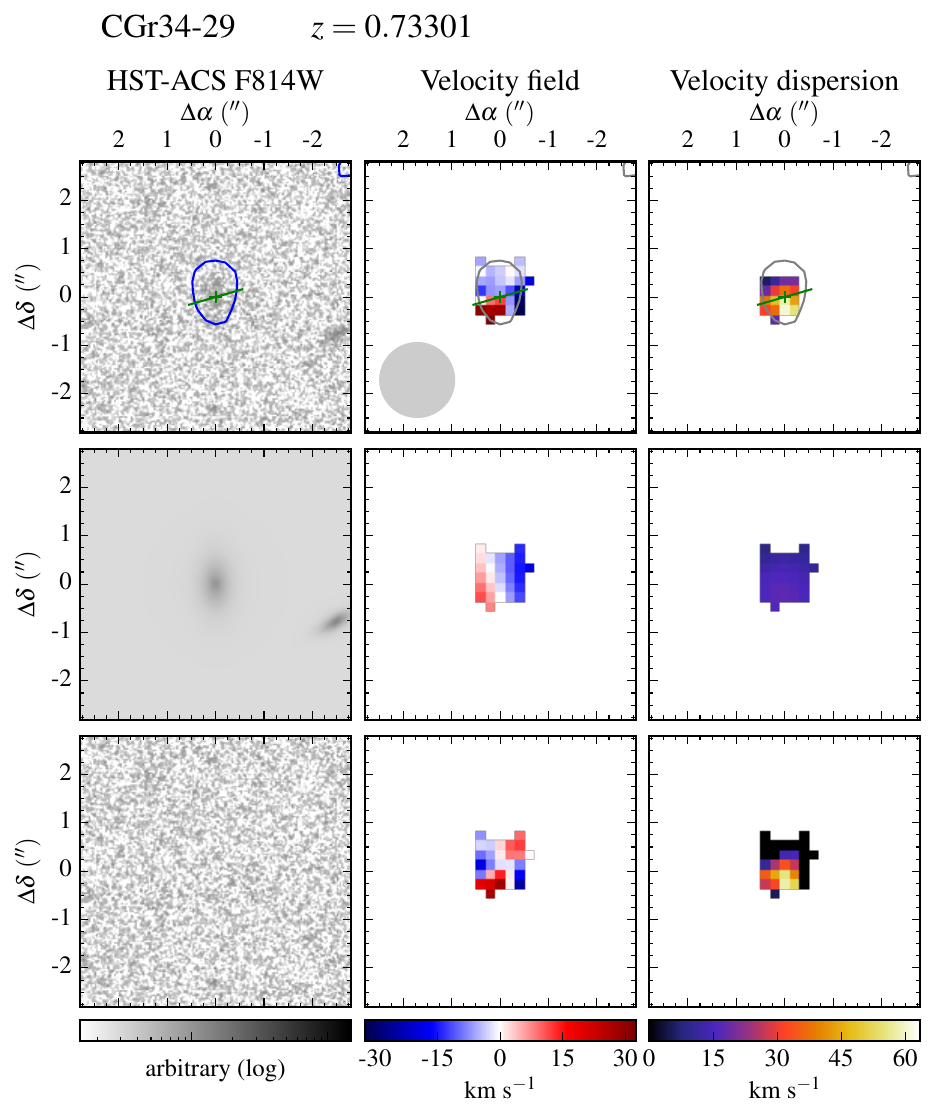}
\caption{Morpho-kinematics maps for galaxy CGr34-29. See caption of Fig. \ref{Mopho_KINMap_CGr28-41} for the description of figure.} 
\label{Mopho_KINMap_CGr34-29} 
\end{figure} 
 
\begin{figure}
\includegraphics[width=0.5\textwidth]{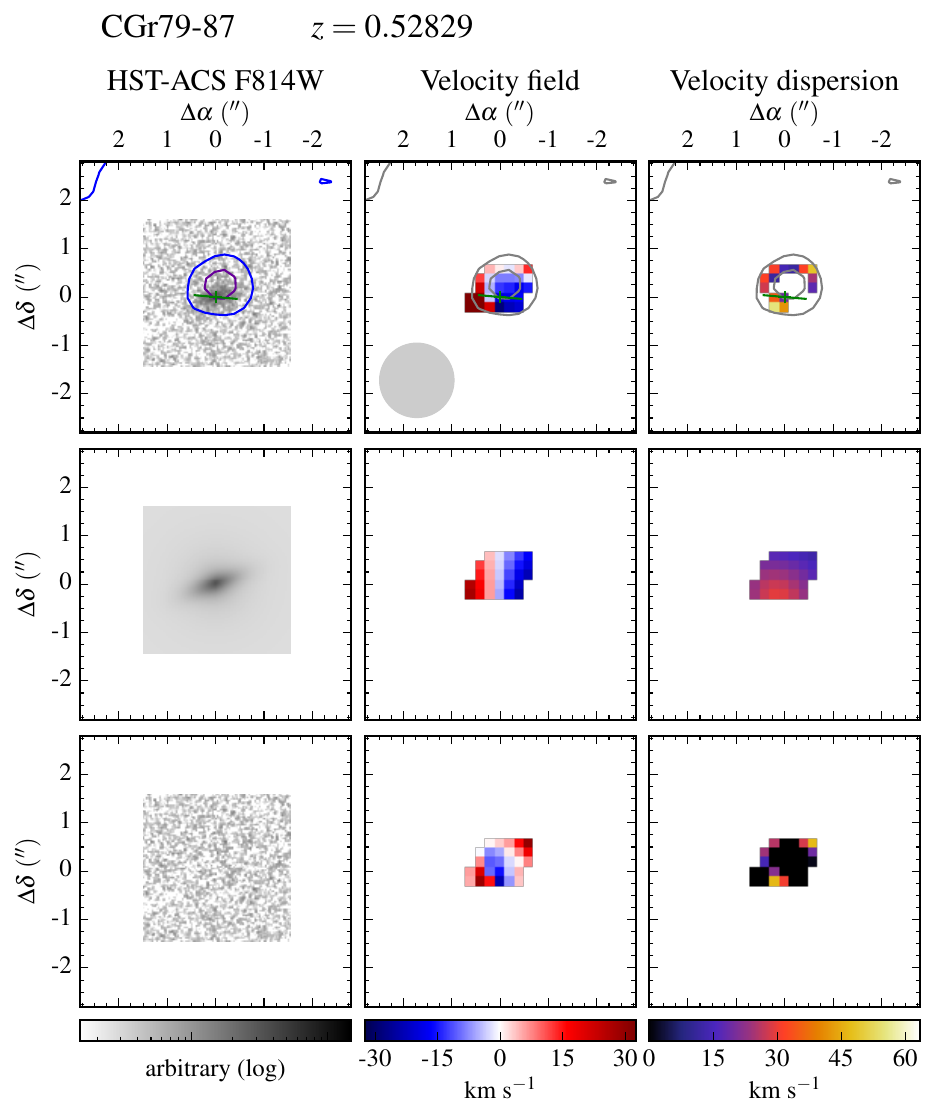}
\caption{Morpho-kinematics maps for galaxy CGr79-87. See caption of Fig. \ref{Mopho_KINMap_CGr28-41} for the description of figure.} 
\label{Mopho_KINMap_CGr79-87} 
\end{figure} 
 
\begin{figure}
\includegraphics[width=0.5\textwidth]{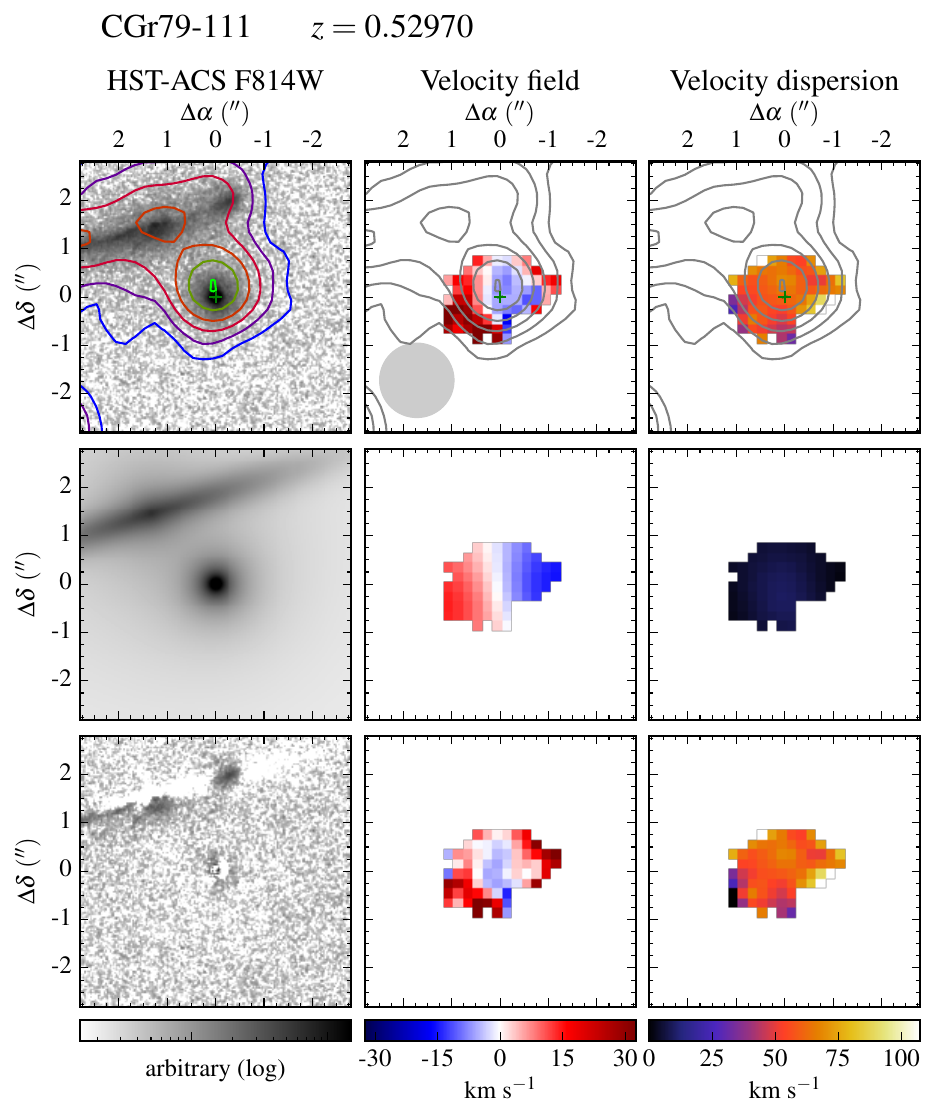}
\caption{Morpho-kinematics maps for galaxy CGr79-111. See caption of Fig. \ref{Mopho_KINMap_CGr28-41} for the description of figure.} 
\label{Mopho_KINMap_CGr79-111} 
\end{figure}

\clearpage
\section{Tables: Physical and morpho-kinematic properties of the kinematics sample}
\label{app:tables}

\onecolumn
\begin{landscape}
\begin{longtable}[ht!]{ccccccccccccc}
\caption{\label{tab:morphokin} Observational information, morphological and kinematic model parameters for the 77 galaxies with S/N $\ge$ 40 and R$_{eff}$/FWHM $\ge$ 0.5.}\\
\hline
\hline
Gr ID&   ID &   z&   R.A. &   Dec.   &   FWHM &   [\ion{O}{ii}] flux &  b/a &   PA$_{m}$ &   R$_{d}$ &   PA$_{k}$ &   $r_{t}$ &   $V_{t}$\\ 
       &    &     &   [J2000] &   [J2000] &   [\arcsec]  & [10$^{-20}$ erg s$^{-1}$ cm$^{-2}$]  &    &   [\degr] &   [kpc] &   [\degr] &   [kpc] &   [km s$^{-1}$]\\
(1)  &   (2) &   (3)  &   (4)  &   (5)  &   (6) &   (7) &   (8)  &   (9)  &   (10)  &   (11)  &   (12)  &   (13) \\ 
\hline\endfirsthead
\caption{continued}\\
\hline
\hline
Gr ID&   ID &   z&   R.A. &   Dec.   &   FWHM &   [\ion{O}{ii}] flux &  b/a &   PA$_{m}$ &   R$_{d}$ &   PA$_{k}$ &   $r_{t}$ &   $V_{t}$\\ 
       &    &     &   [J2000] &   [J2000] &   [\arcsec]  & [10$^{-20}$ erg s$^{-1}$ cm$^{-2}$]  &    &   [\degr] &   [kpc] &   [\degr] &   [kpc] &   [km s$^{-1}$]\\
(1)  &   (2) &   (3)  &   (4)  &   (5)  &   (6) &   (7) &   (8)  &   (9)  &   (10)  &   (11)  &   (12)  &   (13) \\ 
\hline\endhead
\hline
\endfoot
CGr28      &         41 &    0.52975 & 150\degr 13\arcmin 48.7\arcsec  & 1\degr 48\arcmin 19.4\arcsec  &      0.654 &    24421 $\pm$      94 &     0.86 $\pm$    0.01 &      175 $\pm$       3 &     2.10 $\pm$    0.04 &    303 $\pm$     1 &      1.3 $\pm$     0.1 &      170 $\pm$       2 \\ 
CGr28      &         85 &    0.52817 & 150\degr 13\arcmin 06.3\arcsec  & 1\degr 48\arcmin 32.8\arcsec  &      0.654 &    17617 $\pm$     115 &     0.59 $\pm$    0.01 &       96 $\pm$       0 &     3.72 $\pm$    0.01 &     99 $\pm$     1 &      1.3 $\pm$     0.1 &      232 $\pm$       2 \\ 
CGr28      &        145 &    0.52450 & 150\degr 13\arcmin 49.8\arcsec  & 1\degr 49\arcmin 15.7\arcsec  &      0.654 &     5219 $\pm$      74 &     0.42 $\pm$    0.01 &      151 $\pm$       1 &     2.81 $\pm$    0.02 &    144 $\pm$     5 &      5.5 $\pm$     1.5 &      155 $\pm$      22 \\ 
CGr30      &         69 &    0.72554 & 150\degr 08\arcmin 37.3\arcsec  & 2\degr 03\arcmin 49.0\arcsec  &      0.700 &      341 $\pm$       6 &     0.19 $\pm$    0.03 &       16 $\pm$       2 &     1.79 $\pm$    0.19 &    202 $\pm$    11 &      1.4 $\pm$     0.1 &       60 $\pm$       7 \\ 
CGr30      &         71 &    0.72460 & 150\degr 08\arcmin 55.5\arcsec  & 2\degr 03\arcmin 38.1\arcsec  &      0.700 &     5461 $\pm$      34 &     0.50 $\pm$    0.01 &      150 $\pm$       1 &     2.95 $\pm$    0.04 &    353 $\pm$     1 &      3.5 $\pm$     0.6 &      215 $\pm$       5 \\ 
CGr30      &        105 &    0.72693 & 150\degr 08\arcmin 37.0\arcsec  & 2\degr 03\arcmin 53.1\arcsec  &      0.700 &     2552 $\pm$      17 &     0.72 $\pm$    0.01 &       81 $\pm$       1 &     2.48 $\pm$    0.01 &    266 $\pm$     1 &      1.6 $\pm$     1.6 &      203 $\pm$       2 \\ 
CGr30      &        110 &    0.72656 & 150\degr 08\arcmin 43.2\arcsec  & 2\degr 03\arcmin 54.2\arcsec  &      0.700 &     1000 $\pm$      10 &     0.73 $\pm$    0.01 &      102 $\pm$       2 &     2.36 $\pm$    0.02 &    293 $\pm$     5 &      5.4 $\pm$     2.8 &       69 $\pm$      22 \\ 
CGr30      &        137 &    0.73015 & 150\degr 08\arcmin 28.9\arcsec  & 2\degr 04\arcmin 06.1\arcsec  &      0.700 &     2170 $\pm$      15 &     0.16 $\pm$    0.01 &      121 $\pm$       1 &     3.78 $\pm$    0.06 &    298 $\pm$     2 &      9.9 $\pm$     1.2 &      243 $\pm$      14 \\ 
CGr30      &        158 &    0.72257 & 150\degr 07\arcmin 59.7\arcsec  & 2\degr 04\arcmin 14.9\arcsec  &      0.700 &     1542 $\pm$      25 &     0.55 $\pm$    0.01 &       65 $\pm$       1 &     1.65 $\pm$    0.02 &     54 $\pm$     3 &      5.2 $\pm$     0.9 &      240 $\pm$      19 \\ 
CGr30      &        170 &    0.72622 & 150\degr 08\arcmin 19.4\arcsec  & 2\degr 04\arcmin 10.6\arcsec  &      0.700 &     4314 $\pm$      20 &     0.74 $\pm$    0.01 &       41 $\pm$       1 &     3.32 $\pm$    0.03 &    230 $\pm$     1 &      1.4 $\pm$     0.1 &      165 $\pm$       1 \\ 
CGr30      &        174 &    0.72867 & 150\degr 08\arcmin 24.5\arcsec  & 2\degr 04\arcmin 15.3\arcsec  &      0.700 &      842 $\pm$      14 &     0.70 $\pm$    0.01 &       23 $\pm$       1 &     2.39 $\pm$    0.02 &    185 $\pm$     4 &      1.5 $\pm$     0.1 &      169 $\pm$       5 \\ 
CGr30      &        185 &    0.72559 & 150\degr 08\arcmin 05.2\arcsec  & 2\degr 04\arcmin 20.7\arcsec  &      0.700 &      413 $\pm$       9 &     0.89 $\pm$    0.01 &       25 $\pm$       5 &     2.46 $\pm$    0.05 &     42 $\pm$    10 &     14.3 $\pm$ 1$\times$10$^8$ &      381 $\pm$ 3$\times$10$^9$ \\ 
CGr30      &        186 &    0.72509 & 150\degr 08\arcmin 18.8\arcsec  & 2\degr 04\arcmin 17.1\arcsec  &      0.700 &    17436 $\pm$      30 &     0.84 $\pm$    0.01 &       92 $\pm$       2 &     4.79 $\pm$    0.05 &    259 $\pm$     1 &      1.4 $\pm$     0.1 &      218 $\pm$       0 \\ 
CGr30      &        188 &    0.72388 & 150\degr 08\arcmin 50.5\arcsec  & 2\degr 04\arcmin 26.7\arcsec  &      0.700 &     1781 $\pm$      11 &     0.55 $\pm$    0.01 &       43 $\pm$       2 &     1.78 $\pm$    0.06 &    226 $\pm$     3 &      1.4 $\pm$     0.1 &       63 $\pm$       3 \\ 
CGr30      &        189 &    0.72059 & 150\degr 08\arcmin 27.6\arcsec  & 2\degr 04\arcmin 28.1\arcsec  &      0.700 &     4605 $\pm$      13 &     0.82 $\pm$    0.02 &      136 $\pm$       6 &     2.40 $\pm$    0.09 &    227 $\pm$     2 &      7.1 $\pm$     0.7 &       58 $\pm$       4 \\ 
CGr30      &        193 &    0.72686 & 150\degr 08\arcmin 07.0\arcsec  & 2\degr 04\arcmin 29.1\arcsec  &      0.700 &     1219 $\pm$      13 &     0.39 $\pm$    0.01 &      154 $\pm$       1 &     1.83 $\pm$    0.03 &    154 $\pm$     3 &      1.5 $\pm$     0.1 &      134 $\pm$       4 \\ 
CGr30      &        195 &    0.72677 & 150\degr 08\arcmin 12.4\arcsec  & 2\degr 04\arcmin 27.0\arcsec  &      0.700 &     1783 $\pm$      15 &     0.25 $\pm$    0.01 &       44 $\pm$       1 &     3.24 $\pm$    0.06 &     48 $\pm$     2 &      1.5 $\pm$     0.1 &      159 $\pm$       4 \\ 
CGr30      &        196 &    0.72279 & 150\degr 08\arcmin 07.4\arcsec  & 2\degr 04\arcmin 32.4\arcsec  &      0.700 &     5532 $\pm$      36 &     0.88 $\pm$    0.01 &       74 $\pm$       3 &     1.75 $\pm$    0.01 &    266 $\pm$     2 &      6.5 $\pm$     0.3 &      222 $\pm$       6 \\ 
CGr32      &         10 &    0.71685 & 149\degr 55\arcmin 15.5\arcsec  & 2\degr 30\arcmin 10.1\arcsec  &      0.722 &    12461 $\pm$      62 &     0.46 $\pm$    0.01 &      104 $\pm$       1 &     4.06 $\pm$    0.03 &    108 $\pm$     1 &      1.4 $\pm$     0.1 &      179 $\pm$       2 \\ 
CGr32      &         36 &    0.72580 & 149\degr 54\arcmin 54.2\arcsec  & 2\degr 30\arcmin 22.9\arcsec  &      0.722 &     5981 $\pm$      60 &     0.39 $\pm$    0.01 &       87 $\pm$       1 &     4.89 $\pm$    0.03 &     96 $\pm$     2 &      9.9 $\pm$     0.7 &      364 $\pm$      16 \\ 
CGr32      &        109 &    0.72863 & 149\degr 55\arcmin 35.1\arcsec  & 2\degr 30\arcmin 45.4\arcsec  &      0.722 &     8152 $\pm$      42 &     0.30 $\pm$    0.01 &        7 $\pm$       1 &     2.44 $\pm$    0.02 &    210 $\pm$     1 &      1.5 $\pm$     0.1 &      287 $\pm$       3 \\ 
CGr32      &        112 &    0.72891 & 149\degr 55\arcmin 46.1\arcsec  & 2\degr 30\arcmin 52.5\arcsec  &      0.722 &     5353 $\pm$      27 &     0.56 $\pm$    0.02 &       57 $\pm$       2 &     1.94 $\pm$    0.06 &    215 $\pm$     2 &     22.5 $\pm$     0.1 &      169 $\pm$       6 \\ 
CGr32      &        132 &    0.73408 & 149\degr 56\arcmin 20.1\arcsec  & 2\degr 30\arcmin 56.1\arcsec  &      0.624 &     7683 $\pm$      32 &     0.53 $\pm$    0.01 &       99 $\pm$       1 &     1.99 $\pm$    0.04 &    284 $\pm$     1 &      4.2 $\pm$     0.4 &      112 $\pm$       4 \\ 
CGr32      &        183 &    0.73395 & 149\degr 55\arcmin 57.8\arcsec  & 2\degr 31\arcmin 07.0\arcsec  &      0.624 &     1206 $\pm$      19 &     0.62 $\pm$    0.01 &       74 $\pm$       1 &     1.65 $\pm$    0.03 &    220 $\pm$     7 &     22.6 $\pm$     0.1 &      392 $\pm$      48 \\ 
CGr32      &        198 &    0.72208 & 149\degr 55\arcmin 45.7\arcsec  & 2\degr 31\arcmin 06.5\arcsec  &      0.624 &     4461 $\pm$      37 &     0.23 $\pm$    0.01 &       53 $\pm$       1 &     3.08 $\pm$    0.03 &    226 $\pm$     1 &      8.8 $\pm$     0.4 &      344 $\pm$       8 \\ 
CGr32      &        295 &    0.73347 & 149\degr 56\arcmin 25.0\arcsec  & 2\degr 31\arcmin 33.1\arcsec  &      0.624 &     4412 $\pm$      25 &     0.44 $\pm$    0.01 &       22 $\pm$       1 &     2.93 $\pm$    0.06 &     21 $\pm$     2 &      7.6 $\pm$     0.6 &      113 $\pm$       6 \\ 
CGr32      &        325 &    0.73652 & 149\degr 54\arcmin 51.1\arcsec  & 2\degr 31\arcmin 39.8\arcsec  &      0.596 &    18088 $\pm$      46 &     0.88 $\pm$    0.01 &       91 $\pm$       3 &     2.86 $\pm$    0.02 &     67 $\pm$     1 &      1.3 $\pm$     0.1 &      192 $\pm$       1 \\ 
CGr32      &        340 &    0.73653 & 149\degr 54\arcmin 52.5\arcsec  & 2\degr 31\arcmin 44.6\arcsec  &      0.596 &     3095 $\pm$      28 &     0.22 $\pm$    0.01 &       91 $\pm$       1 &     4.09 $\pm$    0.07 &     92 $\pm$     1 &      4.4 $\pm$     0.7 &      190 $\pm$       5 \\ 
CGr32      &        345 &    0.72556 & 149\degr 55\arcmin 08.6\arcsec  & 2\degr 31\arcmin 41.2\arcsec  &      0.596 &    12586 $\pm$      78 &     0.32 $\pm$    0.02 &      104 $\pm$       1 &     1.65 $\pm$    0.16 &     95 $\pm$     1 &      1.3 $\pm$     0.1 &      176 $\pm$       3 \\ 
CGr32      &        378 &    0.72876 & 149\degr 54\arcmin 48.8\arcsec  & 2\degr 31\arcmin 48.6\arcsec  &      0.596 &     1293 $\pm$      22 &     0.87 $\pm$    0.01 &      179 $\pm$       2 &     2.62 $\pm$    0.02 &    183 $\pm$     4 &      8.6 $\pm$     1.9 &      249 $\pm$      45 \\ 
CGr32      &        416 &    0.72592 & 149\degr 55\arcmin 01.9\arcsec  & 2\degr 32\arcmin 00.8\arcsec  &      0.596 &     1353 $\pm$      17 &     0.25 $\pm$    0.01 &        0 $\pm$       1 &     2.73 $\pm$    0.06 &    176 $\pm$     3 &      1.3 $\pm$     0.1 &      128 $\pm$       7 \\ 
CGr34      &         22 &    0.73070 & 149\degr 50\arcmin 57.4\arcsec  & 2\degr 28\arcmin 58.2\arcsec  &      0.664 &     9297 $\pm$      30 &     0.19 $\pm$    0.01 &      139 $\pm$       1 &     4.47 $\pm$    0.06 &    147 $\pm$     1 &     10.8 $\pm$     0.3 &      157 $\pm$       1 \\ 
CGr34      &         28 &    0.72914 & 149\degr 51\arcmin 51.0\arcsec  & 2\degr 29\arcmin 02.4\arcsec  &      0.664 &    13863 $\pm$      29 &     0.80 $\pm$    0.01 &       17 $\pm$       2 &     2.91 $\pm$    0.03 &    214 $\pm$     1 &      4.2 $\pm$     0.1 &      131 $\pm$       1 \\ 
CGr34      &         34 &    0.73044 & 149\degr 51\arcmin 13.6\arcsec  & 2\degr 29\arcmin 11.8\arcsec  &      0.664 &      784 $\pm$      11 &     0.21 $\pm$    0.02 &       73 $\pm$       2 &     1.24 $\pm$    0.07 &    227 $\pm$     6 &      5.2 $\pm$     4.9 &      119 $\pm$      42 \\ 
CGr34      &         38 &    0.72845 & 149\degr 51\arcmin 23.5\arcsec  & 2\degr 29\arcmin 10.4\arcsec  &      0.664 &     1758 $\pm$      19 &     0.33 $\pm$    0.01 &      168 $\pm$       1 &     1.50 $\pm$    0.04 &    167 $\pm$     2 &      1.5 $\pm$     0.1 &      175 $\pm$       5 \\ 
CGr34      &         57 &    0.73713 & 149\degr 51\arcmin 36.3\arcsec  & 2\degr 29\arcmin 18.4\arcsec  &      0.664 &     7791 $\pm$      28 &     0.39 $\pm$    0.01 &       19 $\pm$       1 &     1.66 $\pm$    0.02 &     17 $\pm$     1 &      1.5 $\pm$     1.5 &      116 $\pm$       2 \\ 
CGr34      &         59 &    0.72828 & 149\degr 51\arcmin 26.6\arcsec  & 2\degr 29\arcmin 18.3\arcsec  &      0.664 &     2933 $\pm$      32 &     0.17 $\pm$    0.01 &       99 $\pm$       1 &     5.55 $\pm$    0.06 &    288 $\pm$     1 &      7.7 $\pm$     0.4 &      377 $\pm$       4 \\ 
CGr34      &        137 &    0.73085 & 149\degr 51\arcmin 09.8\arcsec  & 2\degr 29\arcmin 51.8\arcsec  &      0.664 &     2673 $\pm$      18 &     0.35 $\pm$    0.01 &       96 $\pm$       1 &     2.57 $\pm$    0.12 &     89 $\pm$     2 &     11.8 $\pm$     2.0 &      248 $\pm$      36 \\ 
CGr34      &        148 &    0.73442 & 149\degr 51\arcmin 35.2\arcsec  & 2\degr 29\arcmin 51.9\arcsec  &      0.664 &     6836 $\pm$      36 &     0.36 $\pm$    0.01 &      144 $\pm$       1 &     3.70 $\pm$    0.03 &    147 $\pm$     1 &      5.7 $\pm$     0.3 &      217 $\pm$       3 \\ 
CGr34      &        153 &    0.73301 & 149\degr 51\arcmin 34.0\arcsec  & 2\degr 29\arcmin 55.1\arcsec  &      0.664 &     2763 $\pm$      32 &     0.22 $\pm$    0.01 &      114 $\pm$       1 &     5.72 $\pm$    0.11 &    121 $\pm$     2 &     18.8 $\pm$     1.5 &      581 $\pm$      40 \\ 
CGr79      &         23 &    0.53147 & 149\degr 49\arcmin 03.2\arcsec  & 1\degr 48\arcmin 51.0\arcsec  &      0.658 &     4392 $\pm$      32 &     0.70 $\pm$    0.01 &      105 $\pm$       1 &     2.76 $\pm$    0.02 &    265 $\pm$     3 &      5.0 $\pm$     0.6 &      127 $\pm$       7 \\ 
CGr79      &         58 &    0.53294 & 149\degr 49\arcmin 28.9\arcsec  & 1\degr 49\arcmin 11.5\arcsec  &      0.658 &     1349 $\pm$      17 &     0.59 $\pm$    0.04 &       28 $\pm$       4 &     1.68 $\pm$    0.10 &     30 $\pm$    10 &      1.3 $\pm$     0.1 &       50 $\pm$       6 \\ 
CGr79      &         63 &    0.53323 & 149\degr 49\arcmin 08.1\arcsec  & 1\degr 49\arcmin 09.5\arcsec  &      0.658 &      998 $\pm$      19 &     0.27 $\pm$    0.01 &        1 $\pm$       1 &     2.28 $\pm$    0.02 &    351 $\pm$     7 &      2.1 $\pm$     7.3 &      111 $\pm$      28 \\ 
CGr79      &         66 &    0.53129 & 149\degr 49\arcmin 05.0\arcsec  & 1\degr 49\arcmin 07.1\arcsec  &      0.658 &     2878 $\pm$      37 &     0.35 $\pm$    0.01 &       78 $\pm$       1 &     3.91 $\pm$    0.03 &    267 $\pm$     2 &      3.5 $\pm$     1.4 &      212 $\pm$      10 \\ 
CGr79      &         70 &    0.53356 & 149\degr 49\arcmin 13.5\arcsec  & 1\degr 49\arcmin 18.8\arcsec  &      0.658 &      662 $\pm$      13 &     0.72 $\pm$    0.05 &      178 $\pm$       6 &     1.80 $\pm$    0.13 &     22 $\pm$    14 &     14.8 $\pm$ 2$\times$10$^8$ &      119 $\pm$ 2$\times$10$^9$ \\ 
CGr79      &        104 &    0.53019 & 149\degr 49\arcmin 25.7\arcsec  & 1\degr 49\arcmin 31.0\arcsec  &      0.658 &    41536 $\pm$      84 &     0.58 $\pm$    0.01 &       22 $\pm$       1 &     5.63 $\pm$    0.03 &    201 $\pm$     1 &     10.4 $\pm$     0.1 &      284 $\pm$       1 \\ 
CGr79      &        129 &    0.53029 & 149\degr 49\arcmin 12.6\arcsec  & 1\degr 49\arcmin 37.0\arcsec  &      0.658 &    10166 $\pm$      58 &     0.20 $\pm$    0.01 &      105 $\pm$       1 &     5.18 $\pm$    0.06 &    294 $\pm$     1 &     18.0 $\pm$     0.5 &      459 $\pm$       9 \\ 
CGr79      &        136 &    0.52851 & 149\degr 49\arcmin 24.3\arcsec  & 1\degr 49\arcmin 50.7\arcsec  &      0.658 &     5792 $\pm$      34 &     0.48 $\pm$    0.01 &       58 $\pm$       1 &     4.07 $\pm$    0.05 &    233 $\pm$     1 &      4.4 $\pm$     0.4 &      170 $\pm$       2 \\ 
CGr84      &         54 &    0.69648 & 150\degr 03\arcmin 13.3\arcsec  & 2\degr 36\arcmin 20.5\arcsec  &      0.578 &     1467 $\pm$      34 &     0.32 $\pm$    0.01 &       89 $\pm$       1 &     5.62 $\pm$    0.13 &     81 $\pm$     3 &      7.4 $\pm$     1.1 &      220 $\pm$       8 \\ 
CGr84      &        237 &    0.69660 & 150\degr 03\arcmin 31.2\arcsec  & 2\degr 35\arcmin 38.5\arcsec  &      0.620 &     4118 $\pm$      22 &     0.65 $\pm$    0.01 &       35 $\pm$       2 &     2.02 $\pm$    0.04 &    228 $\pm$     2 &      2.6 $\pm$     1.3 &       56 $\pm$       4 \\ 
CGr84      &        251 &    0.70386 & 150\degr 03\arcmin 27.5\arcsec  & 2\degr 35\arcmin 42.8\arcsec  &      0.620 &     3074 $\pm$      16 &     0.32 $\pm$    0.01 &       17 $\pm$       1 &     1.37 $\pm$    0.03 &      8 $\pm$     6 &      7.6 $\pm$     6.3 &       38 $\pm$      23 \\ 
CGr84      &        267 &    0.69654 & 150\degr 03\arcmin 33.0\arcsec  & 2\degr 35\arcmin 44.6\arcsec  &      0.620 &     2395 $\pm$      29 &     0.56 $\pm$    0.01 &       75 $\pm$       1 &     2.68 $\pm$    0.03 &     84 $\pm$     3 &     13.6 $\pm$ 2$\times$10$^7$ &      294 $\pm$ 4$\times$10$^8$ \\ 
CGr84      &        273 &    0.70293 & 150\degr 03\arcmin 10.6\arcsec  & 2\degr 35\arcmin 53.3\arcsec  &      0.620 &     2477 $\pm$      15 &     0.44 $\pm$    0.01 &       32 $\pm$       1 &     0.80 $\pm$    0.01 &     43 $\pm$     6 &     22.2 $\pm$     0.1 &      129 $\pm$      11 \\ 
CGr84      &        276 &    0.69674 & 150\degr 02\arcmin 47.6\arcsec  & 2\degr 35\arcmin 41.5\arcsec  &      0.620 &    16947 $\pm$      62 &     0.84 $\pm$    0.01 &       48 $\pm$       2 &     6.32 $\pm$    0.10 &    188 $\pm$     1 &      1.3 $\pm$     0.1 &      191 $\pm$       1 \\ 
CGr84      &        277 &    0.69589 & 150\degr 03\arcmin 24.4\arcsec  & 2\degr 35\arcmin 56.0\arcsec  &      0.620 &     1158 $\pm$      15 &     0.26 $\pm$    0.01 &      168 $\pm$       1 &     1.41 $\pm$    0.07 &     24 $\pm$     6 &     22.1 $\pm$     0.1 &      276 $\pm$      28 \\ 
CGr84      &        295 &    0.69838 & 150\degr 02\arcmin 54.8\arcsec  & 2\degr 35\arcmin 51.3\arcsec  &      0.620 &     1874 $\pm$      25 &     0.37 $\pm$    0.01 &        2 $\pm$       1 &     3.25 $\pm$    0.03 &     18 $\pm$     1 &      1.3 $\pm$     0.1 &      227 $\pm$       3 \\ 
CGr84b     &         21 &    0.66875 & 150\degr 03\arcmin 15.7\arcsec  & 2\degr 37\arcmin 05.4\arcsec  &      0.578 &     2531 $\pm$      36 &     0.23 $\pm$    0.01 &       80 $\pm$       1 &     3.10 $\pm$    0.07 &    260 $\pm$     3 &     12.2 $\pm$     2.7 &      181 $\pm$      32 \\ 
CGr84b     &         22 &    0.67591 & 150\degr 03\arcmin 58.4\arcsec  & 2\degr 36\arcmin 46.7\arcsec  &      0.578 &     5483 $\pm$      49 &     0.31 $\pm$    0.01 &      133 $\pm$       1 &     2.55 $\pm$    0.06 &    144 $\pm$     2 &      4.1 $\pm$     0.7 &      136 $\pm$       5 \\ 
CGr84b     &         23 &    0.67637 & 150\degr 03\arcmin 49.2\arcsec  & 2\degr 36\arcmin 56.3\arcsec  &      0.578 &    16211 $\pm$      65 &     0.90 $\pm$    0.01 &      128 $\pm$       3 &     3.19 $\pm$    0.03 &    105 $\pm$     2 &      4.5 $\pm$     0.6 &       88 $\pm$       3 \\ 
CGr84b     &         35 &    0.68086 & 150\degr 03\arcmin 39.9\arcsec  & 2\degr 36\arcmin 54.9\arcsec  &      0.578 &     1143 $\pm$      27 &     0.79 $\pm$    0.06 &       36 $\pm$      14 &     1.65 $\pm$    0.19 &    341 $\pm$     8 &      1.3 $\pm$     0.1 &       89 $\pm$      14 \\ 
CGr84b     &         40 &    0.68223 & 150\degr 03\arcmin 23.3\arcsec  & 2\degr 37\arcmin 12.5\arcsec  &      0.578 &     5111 $\pm$      55 &     0.35 $\pm$    0.01 &      128 $\pm$       1 &     2.44 $\pm$    0.03 &    306 $\pm$     2 &      6.1 $\pm$     0.7 &      189 $\pm$       9 \\ 
CGr84b     &        248 &    0.68200 & 150\degr 03\arcmin 28.3\arcsec  & 2\degr 35\arcmin 39.2\arcsec  &      0.648 &     1247 $\pm$      25 &     0.20 $\pm$    0.01 &       55 $\pm$       1 &     2.23 $\pm$    0.03 &     36 $\pm$     1 &      1.4 $\pm$     0.1 &      328 $\pm$       8 \\ 
CGr84b     &        250 &    0.68057 & 150\degr 03\arcmin 30.9\arcsec  & 2\degr 35\arcmin 40.9\arcsec  &      0.648 &     4168 $\pm$      36 &     0.59 $\pm$    0.01 &      166 $\pm$       1 &     1.73 $\pm$    0.02 &    180 $\pm$     2 &      3.2 $\pm$     1.0 &      141 $\pm$       9 \\ 
CGr84b     &        257 &    0.67723 & 150\degr 03\arcmin 13.5\arcsec  & 2\degr 35\arcmin 47.8\arcsec  &      0.648 &     1503 $\pm$      13 &     0.62 $\pm$    0.01 &      147 $\pm$       3 &     2.00 $\pm$    0.06 &    135 $\pm$     7 &     21.4 $\pm$ 1$\times$10$^8$ &      137 $\pm$ 7$\times$10$^8$ \\ 
CGr84b     &        323 &    0.68089 & 150\degr 03\arcmin 30.9\arcsec  & 2\degr 36\arcmin 07.4\arcsec  &      0.648 &     4243 $\pm$      43 &     0.24 $\pm$    0.01 &      168 $\pm$       1 &     5.46 $\pm$    0.13 &    174 $\pm$     1 &      1.4 $\pm$     0.1 &      252 $\pm$       5 \\ 
CGr114     &         84 &    0.65956 & 149\degr 59\arcmin 33.1\arcsec  & 2\degr 15\arcmin 26.8\arcsec  &      0.740 &     1934 $\pm$      38 &     0.42 $\pm$    0.01 &       37 $\pm$       1 &     1.95 $\pm$    0.01 &    218 $\pm$     3 &      1.4 $\pm$     0.1 &      210 $\pm$       8 \\ 
CGr114     &         97 &    0.66052 & 149\degr 59\arcmin 54.4\arcsec  & 2\degr 15\arcmin 39.0\arcsec  &      0.740 &     1977 $\pm$      35 &     0.73 $\pm$    0.01 &      163 $\pm$       1 &     2.56 $\pm$    0.02 &    145 $\pm$     3 &      3.8 $\pm$     1.4 &      225 $\pm$      17 \\ 
\hline
CGr32      &        268 &    0.72770 & 149\degr 55\arcmin 33.8\arcsec  & 2\degr 31\arcmin 34.6\arcsec  &      0.596 &     4812 $\pm$      51 &     0.30 $\pm$    0.01 &       19 $\pm$       1 &     6.48 $\pm$    0.07 &     10 $\pm$     3 &      1.6 $\pm$     7.0 &      132 $\pm$      11 \\ 
CGr32      &        454 &    0.73460 & 149\degr 55\arcmin 11.7\arcsec  & 2\degr 32\arcmin 26.2\arcsec  &      0.596 &     1036 $\pm$      16 &     0.55 $\pm$    0.01 &        8 $\pm$       2 &     2.17 $\pm$    0.08 &      3 $\pm$    89 &      7.9 $\pm$ 7$\times$10$^2$&       24 $\pm$ 2$\times$10$^3$ \\ 
\hline
CGr30      &         19 &    0.72384 & 150\degr 08\arcmin 12.0\arcsec  & 2\degr 03\arcmin 33.1\arcsec  &      0.700 &      639 $\pm$       9 &     0.96 $\pm$    0.04 &      118 $\pm$      49 &     0.72 $\pm$    0.03 &    204 $\pm$     6 &     11.0 $\pm$ 4$\times$10$^7$ &      549 $\pm$ 2$\times$10$^9$ \\ 
CGr30      &         82 &    0.72672 & 150\degr 08\arcmin 53.3\arcsec  & 2\degr 03\arcmin 45.4\arcsec  &      0.700 &     1131 $\pm$      18 &     0.89 $\pm$    0.02 &      106 $\pm$       9 &     0.91 $\pm$    0.03 &    207 $\pm$     5 &      6.1 $\pm$     1.2 &      304 $\pm$      45 \\ 
CGr30      &        113 &    0.73114 & 150\degr 08\arcmin 19.6\arcsec  & 2\degr 03\arcmin 59.5\arcsec  &      0.700 &      745 $\pm$       8 &     0.69 $\pm$    0.04 &        2 $\pm$       7 &     1.34 $\pm$    0.15 &    276 $\pm$    32 &      1.5 $\pm$     0.1 &        8 $\pm$       5 \\ 
CGr30      &        142 &    0.72444 & 150\degr 08\arcmin 47.9\arcsec  & 2\degr 04\arcmin 12.1\arcsec  &      0.700 &      930 $\pm$       8 &     0.43 $\pm$    0.03 &       35 $\pm$       2 &     1.03 $\pm$    0.06 &     70 $\pm$     6 &      1.4 $\pm$     0.1 &       30 $\pm$       3 \\ 
CGr32      &        136 &    0.72791 & 149\degr 54\arcmin 48.9\arcsec  & 2\degr 30\arcmin 58.5\arcsec  &      0.722 &      597 $\pm$      10 &     0.33 $\pm$    0.02 &       30 $\pm$       1 &     0.58 $\pm$    0.01 &    226 $\pm$    24 &      7.7 $\pm$    10.9 &      144 $\pm$     148 \\ 
CGr34      &         29 &    0.73301 & 149\degr 51\arcmin 50.2\arcsec  & 2\degr 29\arcmin 09.9\arcsec  &      0.664 &      417 $\pm$       8 &     0.46 $\pm$    0.06 &        1 $\pm$       5 &     1.90 $\pm$    0.25 &    106 $\pm$    15 &      1.5 $\pm$     0.1 &       39 $\pm$      10 \\ 
CGr79      &         87 &    0.52829 & 149\degr 49\arcmin 20.5\arcsec  & 1\degr 49\arcmin 25.6\arcsec  &      0.658 &      553 $\pm$      11 &     0.32 $\pm$    0.02 &      110 $\pm$       2 &     1.25 $\pm$    0.07 &     85 $\pm$    17 &      1.3 $\pm$     0.1 &       65 $\pm$      12 \\ 
CGr79      &        111 &    0.52970 & 149\degr 49\arcmin 11.3\arcsec  & 1\degr 49\arcmin 35.7\arcsec  &      0.658 &     9115 $\pm$      35 &     0.83 $\pm$    0.05 &      120 $\pm$      10 &     0.11 $\pm$    0.00 &    100 $\pm$     8 &      1.3 $\pm$     0.1 &       29 $\pm$       3 \\ 
\end{longtable}
\tablefoot{
 (1) COSMOS group ID; (2) galaxy ID; (3) spectroscopic redshift; (4) and (5) J2000 coordinates; (6) median PSF FWHM, corresponding to narrow band [\ion{O}{ii}] MUSE observations; (7) [\ion{O}{ii}] flux derived from MUSE flux maps; (8), (9) and (10) axis ratio, position angle of the major axis and disk scale length from morphological modeling; (11) kinematic position angle of the major axis; (12) radius at which the plateau is reached; (13) velocity of the plateau. Galaxies with large uncertainties on both $r_t$ and $V_t$ are those for which the plateau is not reached within the data, nevertheless their slope might be well constrained (see Sect. \ref{kinematics}).
The table is split in three parts to identify (i) the final kinematic sample, (ii) galaxies having their kinematics biased by an AGN, and (iii) those with a dominant bulge within the effective radius.
}
\end{landscape}


\begin{landscape}
\begin{longtable}[ht!]{cccccccccccccc}
\caption{\label{tab:dynamics} Physical properties of the 77 galaxies with S/N $\ge$ 40 and R$_{eff}$/FWHM $\ge$ 0.5.}\\
\hline
\hline
Gr ID  &  ID  &   R$_{eff}$  & log(B/D) &  $i$  & $V_{r22}$  & $\sigma$  & $V_{c22}$ &  log($M_*$) & log($M_*(R_{22})$) & log(SFR)  &log($M_{g}$)  &  log($M_{dyn}$($V_{r22}$)) &  log($M_{dyn}$($V_{c22}$))   \\ 
    &    &  [kpc] &    & [\degr]  &  [km s$^{-1}$]  &  [km s$^{-1}$] &  [km s$^{-1}$] &  [M$_\sun$] &  [M$_\sun$] &  [M$_\sun$ yr$^{-1}$]  &  [M$_\sun$] &  [M$_\sun$] &  [M$_\sun$]  \\
(1)  &   (2) &   (3)  &   (4)  &   (5)  &  (6)  &  (7) &   (8)  &  (9)  &  (10)  &  (11)  &  (12)  &  (13) &  (14)  \\ 
\hline\endfirsthead
\caption{continued}\\
\hline
\hline
Gr ID  &  ID  &   R$_{eff}$  & log(B/D) &  $i$  & $V_{r22}$  & $\sigma$  & $V_{c22}$ &  log($M_*$) & log($M_*(R_{22})$) & log(SFR)  &log($M_{g}$)  &  log($M_{dyn}$($V_{r22}$)) &  log($M_{dyn}$($V_{c22}$))   \\ 
    &    &  [kpc] &    & [\degr]  &  [km s$^{-1}$]  &  [km s$^{-1}$] &  [km s$^{-1}$] &  [M$_\sun$] &  [M$_\sun$] &  [M$_\sun$ yr$^{-1}$]  &  [M$_\sun$] &  [M$_\sun$] &  [M$_\sun$]  \\
(1)  &   (2) &   (3)  &   (4)  &   (5)  &  (6)  &  (7) &   (8)  &  (9)  &   (10)  &   (11)  &  (12)  &  (13) &  (14)  \\ 
\hline\endhead
\hline \endfoot
CGr28      &         41 &     3.46 &    -1.68 &   31 $\pm$   2 &      170 $\pm$       4 &       45 $\pm$      21 &      194 $\pm$      22 &     9.86$^{+0.06}_{-0.01}$ &     9.70 $\pm$    0.20  &     0.15$^{+0.07}_{-0.01}$  &     9.20 $\pm$    0.04 &    10.49 $\pm$    0.02 &    10.61 $\pm$    0.10 \\
CGr28      &         85 &     5.93 &    -1.27 &   54 $\pm$   1 &      232 $\pm$       4 &       39 $\pm$      32 &      247 $\pm$      23 &    10.67$^{+0.13}_{-0.01}$ &    10.58 $\pm$    0.22  &     0.43$^{+0.01}_{-0.07}$  &     9.54 $\pm$    0.04 &    11.01 $\pm$    0.02 &    11.06 $\pm$    0.08 \\
CGr28      &        145 &     4.01 &    -0.78 &   65 $\pm$   1 &      155 $\pm$      22 &       72 $\pm$      35 &      217 $\pm$      54 &    10.24$^{+0.14}_{-0.07}$ &    10.12 $\pm$    0.23  &     0.18$^{+0.40}_{-0.06}$  &     9.30 $\pm$    0.20 &    10.54 $\pm$    0.12 &    10.83 $\pm$    0.21 \\
CGr30      &         69 &     3.01 &    -4.14 &   79 $\pm$   2 &       60 $\pm$       7 &       20 $\pm$      16 &       74 $\pm$      20 &     8.41$^{+0.32}_{-0.38}$ &     8.23 $\pm$    0.40  &    -1.27$^{+1.54}_{-0.88}$  &     8.15 $\pm$    0.90 &     9.52 $\pm$    0.12 &     9.70 $\pm$    0.24 \\
CGr30      &         71 &     3.04 &    -0.18 &   60 $\pm$   1 &      215 $\pm$       5 &      100 $\pm$      40 &      301 $\pm$      58 &    10.90$^{+0.08}_{-0.01}$ &    10.81 $\pm$    0.21  &     2.05$^{+0.10}_{-0.11}$  &    10.64 $\pm$    0.08 &    10.84 $\pm$    0.02 &    11.14 $\pm$    0.17 \\
CGr30      &        105 &     4.16 &    -5.67 &   44 $\pm$   1 &      203 $\pm$       4 &       36 $\pm$      28 &      217 $\pm$      21 &    10.49$^{+0.01}_{-0.10}$ &    10.33 $\pm$    0.21  &     0.82$^{+0.01}_{-0.13}$  &     9.72 $\pm$    0.07 &    10.72 $\pm$    0.02 &    10.78 $\pm$    0.08 \\
CGr30      &        110 &     3.91 &    -1.89 &   43 $\pm$   1 &       67 $\pm$      40 &       35 $\pm$      16 &       99 $\pm$      37 &    10.08$^{+0.23}_{-0.08}$ &     9.91 $\pm$    0.26  &     0.40$^{+0.26}_{-0.20}$  &     9.41 $\pm$    0.17 &     9.73 $\pm$    0.52 &    10.08 $\pm$    0.33 \\
CGr30      &        137 &     6.35 &    -5.09 &   81 $\pm$   1 &      204 $\pm$      27 &       39 $\pm$      38 &      220 $\pm$      39 &     9.68$^{+0.08}_{-0.01}$ &     9.54 $\pm$    0.21  &     0.52$^{+0.23}_{-0.45}$  &     9.61 $\pm$    0.26 &    10.91 $\pm$    0.11 &    10.97 $\pm$    0.15 \\
CGr30      &        158 &     2.77 &    -5.18 &   57 $\pm$   1 &      168 $\pm$      31 &        0 $\pm$      30 &      168 $\pm$      31 &     9.88$^{+0.28}_{-0.09}$ &     9.69 $\pm$    0.29  &     0.73$^{+0.29}_{-0.52}$  &     9.56 $\pm$    0.30 &    10.38 $\pm$    0.16 &    10.38 $\pm$    0.16 \\
CGr30      &        170 &     5.32 &    -1.32 &   42 $\pm$   1 &      165 $\pm$       4 &       42 $\pm$      21 &      187 $\pm$      21 &    10.35$^{+0.01}_{-0.07}$ &    10.23 $\pm$    0.21  &     1.02$^{+0.01}_{-0.43}$  &     9.94 $\pm$    0.22 &    10.66 $\pm$    0.02 &    10.77 $\pm$    0.10 \\
CGr30      &        174 &     3.88 &    -1.36 &   46 $\pm$   1 &      169 $\pm$       5 &       52 $\pm$      37 &      202 $\pm$      43 &    10.38$^{+0.30}_{-0.01}$ &    10.22 $\pm$    0.29  &    -0.29$^{+0.55}_{-0.01}$  &     8.92 $\pm$    0.28 &    10.54 $\pm$    0.03 &    10.70 $\pm$    0.18 \\
CGr30      &        185 &     3.38 &    -0.64 &   27 $\pm$   2 &      144 $\pm$      31 &       56 $\pm$      25 &      186 $\pm$      41 &    10.24$^{+0.01}_{-0.01}$ &    10.12 $\pm$    0.20  &    -0.43$^{+0.01}_{-0.01}$  &     8.83 $\pm$    0.01 &    10.42 $\pm$    0.19 &    10.64 $\pm$    0.19 \\
CGr30      &        186 &     7.60 &    -1.19 &   33 $\pm$   2 &      218 $\pm$       4 &       40 $\pm$      16 &      234 $\pm$      12 &    10.49$^{+0.01}_{-0.03}$ &    10.47 $\pm$    0.20  &     1.54$^{+0.15}_{-0.22}$  &    10.40 $\pm$    0.13 &    11.07 $\pm$    0.02 &    11.13 $\pm$    0.05 \\
CGr30      &        188 &     2.98 &    -4.95 &   57 $\pm$   1 &       63 $\pm$       4 &       44 $\pm$      19 &      111 $\pm$      33 &     9.29$^{+0.17}_{-0.11}$ &     9.10 $\pm$    0.25  &    -0.42$^{+0.44}_{-0.09}$  &     8.75 $\pm$    0.23 &     9.55 $\pm$    0.06 &    10.05 $\pm$    0.26 \\
CGr30      &        189 &     3.99 &    -1.92 &   35 $\pm$   3 &       43 $\pm$       6 &       27 $\pm$      13 &       71 $\pm$      23 &     8.89$^{+0.18}_{-0.19}$ &     8.73 $\pm$    0.27  &     0.24$^{+0.37}_{-0.47}$  &     9.30 $\pm$    0.30 &     9.36 $\pm$    0.11 &     9.79 $\pm$    0.28 \\
CGr30      &        193 &     3.03 &    -1.88 &   67 $\pm$   1 &      134 $\pm$       4 &       40 $\pm$      30 &      158 $\pm$      34 &     9.59$^{+0.11}_{-0.02}$ &     9.41 $\pm$    0.22  &    -0.10$^{+0.38}_{-0.29}$  &     8.99 $\pm$    0.24 &    10.22 $\pm$    0.03 &    10.37 $\pm$    0.19 \\
CGr30      &        195 &     4.95 &    -0.79 &   76 $\pm$   1 &      159 $\pm$       4 &       55 $\pm$      33 &      196 $\pm$      41 &     9.84$^{+0.07}_{-0.05}$ &     9.71 $\pm$    0.21  &     0.16$^{+0.07}_{-0.31}$  &     9.32 $\pm$    0.16 &    10.62 $\pm$    0.02 &    10.81 $\pm$    0.18 \\
CGr30      &        196 &     2.93 &    -4.87 &   28 $\pm$   2 &      131 $\pm$       7 &       36 $\pm$      19 &      151 $\pm$      21 &     9.97$^{+0.08}_{-0.01}$ &     9.79 $\pm$    0.21  &     0.81$^{+0.22}_{-0.01}$  &     9.63 $\pm$    0.11 &    10.18 $\pm$    0.05 &    10.31 $\pm$    0.12 \\
CGr32      &         10 &     6.81 &    -3.11 &   63 $\pm$   1 &      179 $\pm$       4 &       56 $\pm$      32 &      214 $\pm$      37 &    10.18$^{+0.09}_{-0.01}$ &    10.07 $\pm$    0.21  &     1.41$^{+0.26}_{-0.13}$  &    10.27 $\pm$    0.15 &    10.82 $\pm$    0.02 &    10.98 $\pm$    0.15 \\
CGr32      &         36 &     8.01 &    -1.57 &   67 $\pm$   1 &      364 $\pm$      16 &       78 $\pm$      52 &      399 $\pm$      47 &    10.69$^{+0.07}_{-0.04}$ &    10.61 $\pm$    0.21  &     1.92$^{+0.07}_{-0.51}$  &    10.68 $\pm$    0.26 &    11.52 $\pm$    0.04 &    11.60 $\pm$    0.10 \\
CGr32      &        109 &     4.10 &    -5.61 &   73 $\pm$   1 &      287 $\pm$       4 &        0 $\pm$      31 &      287 $\pm$       4 &     9.92$^{+0.10}_{-0.02}$ &     9.74 $\pm$    0.21  &     1.15$^{+0.23}_{-0.28}$  &     9.95 $\pm$    0.18 &    11.01 $\pm$    0.01 &    11.01 $\pm$    0.01 \\
CGr32      &        112 &     3.25 &    -4.86 &   56 $\pm$   2 &       32 $\pm$       9 &       33 $\pm$      15 &       76 $\pm$      29 &     8.63$^{+0.01}_{-0.01}$ &     8.45 $\pm$    0.20  &    -0.05$^{+0.01}_{-0.01}$  &     9.04 $\pm$    0.01 &     9.00 $\pm$    0.24 &     9.76 $\pm$    0.33 \\
CGr32      &        132 &     3.23 &    -1.45 &   58 $\pm$   1 &      112 $\pm$       4 &       39 $\pm$      14 &      139 $\pm$      18 &     9.77$^{+0.15}_{-0.44}$ &     9.60 $\pm$    0.38  &     0.25$^{+1.16}_{-0.30}$  &     9.26 $\pm$    0.61 &    10.11 $\pm$    0.03 &    10.30 $\pm$    0.11 \\
CGr32      &        183 &     2.71 &    -1.62 &   52 $\pm$   1 &       63 $\pm$      17 &       45 $\pm$      13 &      113 $\pm$      24 &     9.58$^{+0.08}_{-0.01}$ &     9.40 $\pm$    0.21  &     0.43$^{+0.09}_{-0.11}$  &     9.34 $\pm$    0.07 &     9.53 $\pm$    0.24 &    10.03 $\pm$    0.18 \\
CGr32      &        198 &     5.48 &    -0.76 &   77 $\pm$   1 &      264 $\pm$      15 &       43 $\pm$      41 &      279 $\pm$      31 &     9.79$^{+0.02}_{-0.10}$ &     9.64 $\pm$    0.21  &     1.49$^{+0.11}_{-0.02}$  &    10.25 $\pm$    0.06 &    11.04 $\pm$    0.05 &    11.09 $\pm$    0.10 \\
CGr32      &        295 &     4.74 &    -1.42 &   64 $\pm$   1 &       96 $\pm$       9 &       30 $\pm$      16 &      115 $\pm$      20 &     9.62$^{+0.01}_{-0.32}$ &     9.47 $\pm$    0.30  &    -0.26$^{+0.17}_{-0.03}$  &     8.99 $\pm$    0.09 &    10.14 $\pm$    0.08 &    10.30 $\pm$    0.15 \\
CGr32      &        325 &     4.72 &    -1.74 &   28 $\pm$   2 &      192 $\pm$       4 &       39 $\pm$      15 &      209 $\pm$      13 &     9.89$^{+0.17}_{-0.01}$ &     9.75 $\pm$    0.23  &     0.36$^{+0.01}_{-0.01}$  &     9.43 $\pm$    0.00 &    10.73 $\pm$    0.02 &    10.81 $\pm$    0.05 \\
CGr32      &        340 &     6.86 &    -5.24 &   77 $\pm$   1 &      190 $\pm$       5 &       42 $\pm$      30 &      209 $\pm$      27 &    10.51$^{+0.01}_{-0.36}$ &    10.38 $\pm$    0.32  &     0.44$^{+0.13}_{-0.34}$  &     9.57 $\pm$    0.18 &    10.88 $\pm$    0.02 &    10.96 $\pm$    0.11 \\
CGr32      &        345 &     2.75 &    -2.06 &   71 $\pm$   2 &      176 $\pm$       4 &      131 $\pm$      48 &      326 $\pm$      84 &    10.32$^{+0.01}_{-0.05}$ &    10.14 $\pm$    0.20  &     1.58$^{+0.03}_{-0.10}$  &    10.16 $\pm$    0.06 &    10.42 $\pm$    0.05 &    10.95 $\pm$    0.23 \\
CGr32      &        378 &     3.99 &    -0.94 &   30 $\pm$   2 &      168 $\pm$      48 &       61 $\pm$      22 &      211 $\pm$      48 &    10.74$^{+0.17}_{-0.11}$ &    10.61 $\pm$    0.25  &     0.51$^{+0.42}_{-0.07}$  &     9.51 $\pm$    0.22 &    10.58 $\pm$    0.25 &    10.78 $\pm$    0.20 \\
CGr32      &        416 &     4.30 &    -1.18 &   76 $\pm$   1 &      128 $\pm$       7 &       39 $\pm$      22 &      152 $\pm$      26 &     9.85$^{+0.01}_{-0.09}$ &     9.69 $\pm$    0.21  &     0.52$^{+0.19}_{-0.39}$  &     9.53 $\pm$    0.22 &    10.36 $\pm$    0.05 &    10.51 $\pm$    0.15 \\
CGr34      &         22 &     7.81 &    -0.62 &   79 $\pm$   1 &      144 $\pm$       5 &       33 $\pm$      24 &      159 $\pm$      22 &     9.64$^{+0.16}_{-0.07}$ &     9.54 $\pm$    0.24  &     0.29$^{+0.62}_{-0.01}$  &     9.49 $\pm$    0.31 &    10.67 $\pm$    0.03 &    10.76 $\pm$    0.12 \\
CGr34      &         28 &     4.74 &    -1.51 &   37 $\pm$   1 &      131 $\pm$       4 &       41 $\pm$      14 &      157 $\pm$      16 &     9.86$^{+0.05}_{-0.04}$ &     9.72 $\pm$    0.21  &     0.90$^{+0.14}_{-0.14}$  &     9.82 $\pm$    0.10 &    10.41 $\pm$    0.03 &    10.56 $\pm$    0.09 \\
CGr34      &         34 &     2.58 &    -0.48 &   78 $\pm$   2 &       62 $\pm$      63 &       21 $\pm$      12 &       76 $\pm$      53 &     8.39$^{+0.17}_{-0.19}$ &     8.17 $\pm$    0.27  &    -1.28$^{+0.49}_{-0.01}$  &     8.05 $\pm$    0.25 &     9.39 $\pm$    0.88 &     9.57 $\pm$    0.61 \\
CGr34      &         38 &     3.41 &    -0.25 &   71 $\pm$   1 &      175 $\pm$       5 &        0 $\pm$      41 &      175 $\pm$       5 &     9.48$^{+0.03}_{-0.14}$ &     9.26 $\pm$    0.22  &     0.13$^{+0.58}_{-0.03}$  &     9.10 $\pm$    0.29 &    10.37 $\pm$    0.03 &    10.37 $\pm$    0.03 \\
CGr34      &         57 &     2.74 &    -1.84 &   67 $\pm$   1 &      116 $\pm$       4 &       37 $\pm$      19 &      140 $\pm$      23 &     9.45$^{+0.13}_{-0.13}$ &     9.27 $\pm$    0.24  &    -0.02$^{+0.19}_{-0.09}$  &     9.02 $\pm$    0.11 &    10.05 $\pm$    0.03 &    10.22 $\pm$    0.14 \\
CGr34      &         59 &     7.29 &    -0.52 &   80 $\pm$   1 &      377 $\pm$       4 &        0 $\pm$      44 &      377 $\pm$       4 &    10.80$^{+0.07}_{-0.10}$ &    10.75 $\pm$    0.22  &     0.74$^{+0.01}_{-0.72}$  &     9.86 $\pm$    0.36 &    11.61 $\pm$    0.01 &    11.61 $\pm$    0.01 \\
CGr34      &        137 &     4.04 &    -1.16 &   70 $\pm$   1 &      119 $\pm$      24 &       29 $\pm$      22 &      134 $\pm$      30 &     9.20$^{+0.04}_{-0.13}$ &     9.04 $\pm$    0.22  &    -0.49$^{+0.01}_{-0.14}$  &     8.79 $\pm$    0.07 &    10.27 $\pm$    0.18 &    10.37 $\pm$    0.20 \\
CGr34      &        148 &     6.20 &    -5.61 &   69 $\pm$   1 &      217 $\pm$       4 &       46 $\pm$      33 &      237 $\pm$      28 &    10.21$^{+0.11}_{-0.09}$ &    10.08 $\pm$    0.22  &     0.15$^{+0.88}_{-0.07}$  &     9.34 $\pm$    0.45 &    10.95 $\pm$    0.02 &    11.03 $\pm$    0.10 \\
CGr34      &        153 &     9.71 &    -0.76 &   77 $\pm$   1 &      390 $\pm$      35 &       48 $\pm$      37 &      403 $\pm$      39 &    10.49$^{+0.05}_{-0.01}$ &    10.44 $\pm$    0.20  &     1.33$^{+0.01}_{-0.13}$  &    10.29 $\pm$    0.07 &    11.65 $\pm$    0.08 &    11.68 $\pm$    0.09 \\
CGr79      &         23 &     4.07 &    -0.88 &   46 $\pm$   1 &      127 $\pm$       7 &       40 $\pm$      22 &      152 $\pm$      26 &    10.16$^{+0.12}_{-0.01}$ &    10.04 $\pm$    0.22  &    -0.09$^{+0.02}_{-0.06}$  &     9.10 $\pm$    0.03 &    10.36 $\pm$    0.05 &    10.52 $\pm$    0.15 \\
CGr79      &         58 &     2.82 &    -4.64 &   54 $\pm$   3 &       50 $\pm$       6 &       41 $\pm$      18 &       99 $\pm$      32 &     8.68$^{+0.31}_{-0.23}$ &     8.50 $\pm$    0.34  &    -0.28$^{+0.52}_{-0.74}$  &     8.84 $\pm$    0.46 &     9.33 $\pm$    0.11 &     9.92 $\pm$    0.28 \\
CGr79      &         63 &     3.82 &    -5.32 &   74 $\pm$   1 &      111 $\pm$      28 &       32 $\pm$      28 &      130 $\pm$      39 &     9.88$^{+0.15}_{-0.01}$ &     9.71 $\pm$    0.23  &    -0.79$^{+0.67}_{-0.01}$  &     8.55 $\pm$    0.34 &    10.16 $\pm$    0.22 &    10.29 $\pm$    0.26 \\
CGr79      &         66 &     5.87 &    -0.26 &   70 $\pm$   1 &      212 $\pm$      10 &       63 $\pm$      40 &      250 $\pm$      46 &    10.77$^{+0.01}_{-0.05}$ &    10.70 $\pm$    0.20  &     0.10$^{+0.01}_{-0.05}$  &     9.32 $\pm$    0.03 &    10.95 $\pm$    0.04 &    11.10 $\pm$    0.16 \\
CGr79      &         70 &     3.02 &    -4.68 &   44 $\pm$   5 &       32 $\pm$       8 &       11 $\pm$      18 &       39 $\pm$      23 &     8.04$^{+0.17}_{-0.06}$ &     7.86 $\pm$    0.24  &     0.05$^{+0.04}_{-0.16}$  &     9.09 $\pm$    0.09 &     8.97 $\pm$    0.23 &     9.15 $\pm$    0.51 \\
CGr79      &        104 &     8.49 &    -0.90 &   55 $\pm$   1 &      284 $\pm$       4 &       36 $\pm$      21 &      295 $\pm$      12 &    10.99$^{+0.01}_{-0.01}$ &    11.01 $\pm$    0.20  &     0.55$^{+0.01}_{-0.01}$  &     9.73 $\pm$    0.00 &    11.37 $\pm$    0.01 &    11.40 $\pm$    0.04 \\
CGr79      &        129 &     8.84 &    -0.32 &   78 $\pm$   1 &      290 $\pm$      24 &       70 $\pm$      45 &      325 $\pm$      48 &    10.38$^{+0.01}_{-0.05}$ &    10.35 $\pm$    0.20  &     1.04$^{+0.13}_{-0.01}$  &    10.06 $\pm$    0.07 &    11.35 $\pm$    0.07 &    11.45 $\pm$    0.13 \\
CGr79      &        136 &     6.74 &    -1.84 &   61 $\pm$   1 &      170 $\pm$       4 &       26 $\pm$      22 &      178 $\pm$      14 &     9.81$^{+0.20}_{-0.01}$ &     9.73 $\pm$    0.25  &    -0.25$^{+0.39}_{-0.06}$  &     9.08 $\pm$    0.20 &    10.78 $\pm$    0.02 &    10.82 $\pm$    0.07 \\
CGr84      &         54 &     9.48 &    -0.56 &   71 $\pm$   1 &      220 $\pm$       8 &        0 $\pm$      23 &      220 $\pm$       8 &    10.48$^{+0.01}_{-0.05}$ &    10.44 $\pm$    0.20  &     0.78$^{+0.49}_{-0.01}$  &     9.90 $\pm$    0.25 &    11.14 $\pm$    0.03 &    11.14 $\pm$    0.03 \\
CGr84      &        237 &     3.34 &    -1.79 &   49 $\pm$   1 &       56 $\pm$       4 &       37 $\pm$      20 &       95 $\pm$      34 &     9.39$^{+0.01}_{-0.01}$ &     9.22 $\pm$    0.20  &     0.24$^{+0.01}_{-0.01}$  &     9.26 $\pm$    0.01 &     9.51 $\pm$    0.06 &     9.97 $\pm$    0.31 \\
CGr84      &        251 &     2.29 &    -4.96 &   71 $\pm$   1 &       15 $\pm$      15 &       39 $\pm$      13 &       83 $\pm$      27 &     8.98$^{+0.20}_{-0.04}$ &     8.79 $\pm$    0.25  &    -0.36$^{+0.36}_{-0.14}$  &     8.73 $\pm$    0.20 &     8.19 $\pm$    0.90 &     9.68 $\pm$    0.28 \\
CGr84      &        267 &     4.50 &    -5.43 &   56 $\pm$   1 &      128 $\pm$      21 &       42 $\pm$      21 &      155 $\pm$      31 &     9.85$^{+0.13}_{-0.02}$ &     9.69 $\pm$    0.22  &     0.70$^{+0.18}_{-0.83}$  &     9.66 $\pm$    0.43 &    10.35 $\pm$    0.14 &    10.52 $\pm$    0.17 \\
CGr84      &        273 &     2.36 &    -0.09 &   64 $\pm$   1 &       10 $\pm$       6 &       35 $\pm$      16 &       75 $\pm$      34 &     9.03$^{+0.01}_{-0.10}$ &     8.75 $\pm$    0.21  &    -0.83$^{+0.08}_{-0.01}$  &     8.26 $\pm$    0.04 &     7.63 $\pm$    0.47 &     9.36 $\pm$    0.40 \\
CGr84      &        276 &     7.63 &    -0.40 &   33 $\pm$   2 &      191 $\pm$       4 &       47 $\pm$      29 &      215 $\pm$      28 &    11.02$^{+0.01}_{-0.01}$ &    11.07 $\pm$    0.20  &     0.78$^{+0.01}_{-0.01}$  &     9.93 $\pm$    0.00 &    11.07 $\pm$    0.02 &    11.17 $\pm$    0.11 \\
CGr84      &        277 &     2.36 &    -4.63 &   75 $\pm$   1 &       39 $\pm$      13 &       47 $\pm$      25 &      106 $\pm$      49 &     8.97$^{+0.26}_{-0.32}$ &     8.78 $\pm$    0.35  &     0.28$^{+0.45}_{-0.67}$  &     9.20 $\pm$    0.41 &     9.03 $\pm$    0.28 &     9.91 $\pm$    0.40 \\
CGr84      &        295 &     5.45 &    -5.45 &   68 $\pm$   1 &      227 $\pm$       4 &        0 $\pm$      27 &      227 $\pm$       4 &    10.20$^{+0.03}_{-0.04}$ &    10.05 $\pm$    0.20  &     0.50$^{+0.05}_{-0.03}$  &     9.56 $\pm$    0.03 &    10.93 $\pm$    0.02 &    10.93 $\pm$    0.02 \\
CGr84b     &         21 &     5.20 &    -4.98 &   77 $\pm$   1 &      101 $\pm$      14 &        0 $\pm$      16 &      101 $\pm$      14 &     8.76$^{+0.37}_{-0.10}$ &     8.60 $\pm$    0.34  &     0.36$^{+0.17}_{-0.70}$  &     9.45 $\pm$    0.36 &    10.21 $\pm$    0.12 &    10.21 $\pm$    0.12 \\
CGr84b     &         22 &     5.20 &    -0.25 &   72 $\pm$   1 &      136 $\pm$       5 &       30 $\pm$      19 &      150 $\pm$      17 &     9.38$^{+0.12}_{-0.06}$ &     9.21 $\pm$    0.22  &     0.23$^{+0.33}_{-0.54}$  &     9.31 $\pm$    0.32 &    10.38 $\pm$    0.03 &    10.47 $\pm$    0.10 \\
CGr84b     &         23 &     5.23 &    -1.61 &   26 $\pm$   2 &       88 $\pm$       4 &       39 $\pm$      14 &      120 $\pm$      21 &    10.56$^{+0.01}_{-0.01}$ &    10.45 $\pm$    0.20  &     0.48$^{+0.01}_{-0.01}$  &     9.54 $\pm$    0.00 &    10.10 $\pm$    0.04 &    10.37 $\pm$    0.15 \\
CGr84b     &         35 &     2.66 &    -1.37 &   38 $\pm$   6 &       89 $\pm$      14 &       17 $\pm$      15 &       95 $\pm$      17 &     7.97$^{+0.04}_{-0.08}$ &     7.80 $\pm$    0.21  &    -0.70$^{+0.01}_{-0.05}$  &     8.54 $\pm$    0.04 &     9.82 $\pm$    0.15 &     9.88 $\pm$    0.17 \\
CGr84b     &         40 &     4.53 &    -0.47 &   70 $\pm$   1 &      167 $\pm$      21 &       47 $\pm$      30 &      194 $\pm$      37 &     9.70$^{+0.14}_{-0.01}$ &     9.53 $\pm$    0.22  &     0.93$^{+0.01}_{-0.47}$  &     9.80 $\pm$    0.24 &    10.54 $\pm$    0.11 &    10.67 $\pm$    0.17 \\
CGr84b     &        248 &     2.87 &    -0.26 &   78 $\pm$   1 &      328 $\pm$       8 &        0 $\pm$      35 &      328 $\pm$       8 &    10.27$^{+0.04}_{-0.01}$ &    10.14 $\pm$    0.20  &    -0.41$^{+0.05}_{-0.01}$  &     8.82 $\pm$    0.03 &    11.09 $\pm$    0.02 &    11.09 $\pm$    0.02 \\
CGr84b     &        250 &     2.60 &    -0.92 &   54 $\pm$   1 &      141 $\pm$       9 &       67 $\pm$      27 &      199 $\pm$      40 &    10.13$^{+0.09}_{-0.04}$ &     9.97 $\pm$    0.21  &    -0.54$^{+0.01}_{-0.10}$  &     8.66 $\pm$    0.05 &    10.25 $\pm$    0.05 &    10.54 $\pm$    0.17 \\
CGr84b     &        257 &     3.36 &    -4.88 &   52 $\pm$   1 &       28 $\pm$       7 &       23 $\pm$      15 &       55 $\pm$      27 &     9.12$^{+0.18}_{-0.10}$ &     8.94 $\pm$    0.25  &    -0.93$^{+0.25}_{-0.01}$  &     8.42 $\pm$    0.13 &     8.91 $\pm$    0.20 &     9.50 $\pm$    0.43 \\
CGr84b     &        323 &     7.79 &    -0.72 &   76 $\pm$   1 &      252 $\pm$       5 &       74 $\pm$      38 &      295 $\pm$      42 &    10.59$^{+0.01}_{-0.01}$ &    10.54 $\pm$    0.20  &     0.90$^{+0.01}_{-0.01}$  &     9.97 $\pm$    0.01 &    11.25 $\pm$    0.02 &    11.39 $\pm$    0.12 \\
CGr114     &         84 &     3.26 &    -0.42 &   65 $\pm$   1 &      210 $\pm$       8 &       51 $\pm$      44 &      236 $\pm$      42 &    10.50$^{+0.02}_{-0.01}$ &    10.33 $\pm$    0.20  &    -0.17$^{+0.01}_{-0.01}$  &     8.96 $\pm$    0.01 &    10.64 $\pm$    0.03 &    10.74 $\pm$    0.15 \\
CGr114     &         97 &     4.10 &    -1.31 &   43 $\pm$   1 &      225 $\pm$      17 &       15 $\pm$      34 &      227 $\pm$      20 &    10.33$^{+0.19}_{-0.05}$ &    10.18 $\pm$    0.24  &     1.20$^{+0.11}_{-0.98}$  &    10.00 $\pm$    0.50 &    10.82 $\pm$    0.07 &    10.83 $\pm$    0.08 \\
\hline
CGr32      &        268 &     8.31 &    -0.26 &   73 $\pm$   1 &      132 $\pm$      11 &      118 $\pm$      56 &      280 $\pm$     104 &    10.74$^{+0.02}_{-0.01}$ &    10.74 $\pm$    0.20  &     0.07$^{+0.98}_{-0.01}$  &     9.42 $\pm$    0.49 &    10.76 $\pm$    0.07 &    11.42 $\pm$    0.32 \\
CGr32      &        454 &     2.23 &    -0.20 &   57 $\pm$   1 &       15 $\pm$       4 &       51 $\pm$      15 &      108 $\pm$      31 &     9.91$^{+0.08}_{-0.11}$ &     9.80 $\pm$    0.22  &    -0.57$^{+0.01}_{-0.11}$  &     8.70 $\pm$    0.06 &     8.39 $\pm$    0.24 &    10.11 $\pm$    0.25 \\
\hline
CGr30      &         19 &     3.22 &     0.16 &   16 $\pm$   9 &       79 $\pm$      56 &       30 $\pm$      27 &      101 $\pm$      56 &     8.60$^{+0.35}_{-0.12}$ &     8.27 $\pm$    0.33  &     0.49$^{+0.20}_{-0.95}$  &     9.18 $\pm$    0.49 &     9.36 $\pm$    0.61 &     9.57 $\pm$    0.48 \\
CGr30      &         82 &     4.60 &     0.32 &   27 $\pm$   3 &      100 $\pm$      24 &      109 $\pm$      47 &      250 $\pm$      91 &     9.56$^{+0.35}_{-0.10}$ &     9.22 $\pm$    0.33  &     1.26$^{+0.29}_{-0.65}$  &     9.79 $\pm$    0.36 &     9.67 $\pm$    0.21 &    10.47 $\pm$    0.32 \\
CGr30      &        113 &     4.17 &     0.11 &   46 $\pm$   4 &        8 $\pm$       5 &       30 $\pm$      16 &       64 $\pm$      34 &     8.27$^{+0.23}_{-0.04}$ &     8.01 $\pm$    0.26  &    -0.89$^{+0.16}_{-0.07}$  &     8.35 $\pm$    0.09 &     7.64 $\pm$    0.52 &     9.45 $\pm$    0.46 \\
CGr30      &        142 &     3.81 &     0.16 &   65 $\pm$   2 &       30 $\pm$       4 &       27 $\pm$      13 &       65 $\pm$      23 &     8.91$^{+0.35}_{-0.09}$ &     8.61 $\pm$    0.32  &    -0.24$^{+0.39}_{-0.53}$  &     8.75 $\pm$    0.33 &     8.69 $\pm$    0.12 &     9.35 $\pm$    0.31 \\
CGr32      &        136 &     3.67 &     0.25 &   71 $\pm$   2 &       24 $\pm$      21 &       28 $\pm$      15 &       64 $\pm$      30 &     9.45$^{+0.04}_{-0.36}$ &     9.07 $\pm$    0.32  &    -0.62$^{+0.31}_{-0.16}$  &     8.33 $\pm$    0.18 &     8.22 $\pm$    0.77 &     9.08 $\pm$    0.41 \\
CGr34      &         29 &     4.84 &     0.02 &   63 $\pm$   4 &       39 $\pm$      10 &        6 $\pm$      20 &       41 $\pm$      16 &     8.78$^{+0.14}_{-0.39}$ &     8.57 $\pm$    0.35  &    -1.46$^{+1.00}_{-1.61}$  &     8.03 $\pm$    0.96 &     9.18 $\pm$    0.23 &     9.22 $\pm$    0.34 \\
CGr79      &         87 &     4.73 &     0.33 &   71 $\pm$   2 &       65 $\pm$      12 &        0 $\pm$      15 &       65 $\pm$      12 &     8.70$^{+0.09}_{-0.05}$ &     8.42 $\pm$    0.21  &    -0.99$^{+0.04}_{-0.27}$  &     8.26 $\pm$    0.14 &     9.43 $\pm$    0.16 &     9.43 $\pm$    0.16 \\
CGr79      &        111 &     2.10 &     0.84 &   34 $\pm$   6 &        6 $\pm$       4 &       59 $\pm$      18 &      124 $\pm$      38 &    10.03$^{+0.01}_{-0.08}$ &     9.24 $\pm$    0.21  &    -0.21$^{+0.10}_{-0.01}$  &     8.21 $\pm$    0.05 &     6.25 $\pm$    0.62 &     8.94 $\pm$    0.27 \\
\end{longtable}
\tablefoot{
 (1) COSMOS group ID; (2) galaxy ID; (3) global effective radius; (4) logarithm of the bulge-to-disk ratio at $R_{22}$; (5) disk inclination; (6) rotation velocity at $R_{22}$; (7) median velocity dispersion; (8) corrected velocity at $R_{22}$ as described in Eq. \ref{eq:vcirc22}; at logarithmic scale: (9) stellar mass within an aperture of 3''; (10) corrected stellar mass inside $R_{22}$; (11) SFR from the SED fitting; (12) gas mass computed using the Kennicutt-Schmidt law, from SED SFRs; (13) and (14) dynamical masses computed from the rotation velocity $V_{r22}$ and the corrected velocity $V_{c22}$ respectively. The table is split in three parts to identify (i) the final kinematic sample, (ii) galaxies having their kinematics biased by an AGN, and (iii) those with a dominant bulge within the effective radius.
 }
\end{landscape}

\end{document}